\documentclass[11pt,a4paper]{article}
\usepackage[utf8]{inputenc}
\usepackage{a4wide}
\usepackage{amsmath,amssymb}
\usepackage{amsfonts}
\usepackage{cite}
\usepackage{color}
\usepackage{adjustbox}
\usepackage{multirow}
\usepackage{pdflscape}
\usepackage{textcomp}
\usepackage{gensymb}
\usepackage{graphicx}
\graphicspath{{figures/}}
\usepackage{diagbox}
\usepackage{pifont}
\usepackage{bigstrut}
\usepackage[small,bf]{caption}
\usepackage{tabulary}
\usepackage{bm}
\usepackage{booktabs}
\usepackage{stackengine}
\usepackage{mathtools}
\usepackage{enumerate}
\usepackage{makecell}
\usepackage{listings}
\usepackage{bbm}
\usepackage{mathrsfs}
\usepackage[unicode]{hyperref} 
\graphicspath{{figures/}}
\usepackage{graphicx}
\usepackage{subcaption}
\usepackage{placeins}
\usepackage{multirow}

\definecolor{greenLinks}{rgb}{0, 0.6, 0} 
\definecolor{blueLinks}{rgb}{0, 0, 0.6}
\definecolor{redLinks}{rgb}{0.6, 0, 0}
\definecolor{eprintLinks}{rgb}{0.4, 0.4, 0.4}
\definecolor{journalLinks}{rgb}{0.6, 0, 0}
\begin{document}

\pagenumbering{Alph}
\begin{titlepage}

\vspace*{15mm}

\begin{center}
{ \bf\LARGE {Inverse Seesaw Model in  Non-holomorphic Modular \\[2mm] $A_4$ Flavor Symmetry }}\\[8mm]
Xianshuo Zhang$^{\,a,}$\footnote{Email: \href{mailto:zxianshuo@yeah.net}{\texttt{zxianshuo@yeah.net}}},
Yakefu Reyimuaji$^{\,a,}$\footnote{Email: \href{mailto:yreyi@hotmail.com}{\texttt{yreyi@hotmail.com}}} \\
\vspace{8mm}
$^{a}$\,{\it School of Physical Science and Technology, Xinjiang University, Urumqi 830017, China} \\
\vspace{2mm}
\end{center}
\setcounter{footnote}{0} 

\vspace{8mm}

\begin{abstract}
This paper investigates an inverse seesaw model of neutrino masses based on non-holomorphic modular $A_4$ symmetry, extending the framework of modular-invariant flavor models beyond the conventional holomorphic paradigm. After the general theoretical framework is established, three concrete model realizations distinguished by their $A_4$ representation assignments and modular weight configurations for the matter fields are analyzed. Focusing on these three specific realizations, a comprehensive analysis of neutrino phenomenology is performed. By constraining the modulus parameter $\tau$ to the fundamental domain and systematically scanning the parameter space, regions compatible with current neutrino oscillation data are identified. The numerical results provide predictions for currently unmeasured quantities, including the absolute neutrino mass scale, Dirac CP-violating phase, and Majorana phases. These predictions establish specific, testable signatures for upcoming neutrino experiments, particularly in neutrinoless double beta decay and precision oscillation measurements. The framework offers a well-defined target for future experimental verification or exclusion, while demonstrating the phenomenological viability of non-holomorphic modular symmetry approaches to flavor structure.

\end{abstract}


\end{titlepage}
\pagenumbering{arabic}



\section{Introduction}
\label{sec:intro}

The Standard Model (SM) successfully describes electromagnetic, weak, and strong interactions, yet it fails to explain neutrino masses, a clear sign of beyond-SM (BSM) physics. Neutrino oscillation data confirms that neutrinos are massive, motivating extensions of the SM through seesaw mechanisms. Among these, the inverse seesaw~\cite{Mohapatra:1986aw,Mohapatra:1986bd,Deppisch:2004fa,Dev:2009aw,CentellesChulia:2020dfh} offers a particularly attractive framework, as it accommodates small neutrino masses at experimentally accessible scales while preserving lepton number violation as a perturbative effect.

A major challenge in BSM model-building is explaining the origin of flavor structure. Modular symmetry has emerged as a powerful approach, where Yukawa couplings are constrained by modular forms, holomorphic functions obeying specific transformation rules under the modular group~\cite{Feruglio:2017spp} (see~\cite{Feruglio:2019ybq,Ding:2023htn,Kobayashi:2023zzc} for reviews). This approach naturally restricts the form of Lagrangian, without the need for ad hoc flavon fields or extra dimensions. In particular, the $A_4$ modular group provides a minimal and predictive framework for lepton masses and mixing angles~\cite{Feruglio:2017spp,Kobayashi:2018vbk}. Conventionally, these models rely on holomorphic modular forms, often embedded in supersymmetry (SUSY)~\cite{Kobayashi:2018scp,Novichkov:2018yse,Nomura:2019yft,Ding:2019zxk,Nomura:2019lnr,Ding:2019gof,Zhang:2019ngf,Nomura:2019xsb,Kobayashi:2019gtp,Wang:2019xbo,Ding:2020yen,Nomura:2020cog,Aoki:2020eqf,Asaka:2020tmo,Okada:2020brs,Okada:2021qdf,deMedeirosVarzielas:2021pug,Nomura:2021pld,Kobayashi:2021pav,Kobayashi:2022jvy,Kang:2022psa,Gogoi:2022jwf,Abbas:2022slb,Devi:2023vpe,Nomura:2023usj,Kobayashi:2023qzt,Kumar:2023moh,Ding:2024fsf,Nomura:2024ghc,Singh:2024imk,Kalita:2024vlt,Nomura:2024ctl,Nomura:2025bph,Petcov:2024vph,Pathak:2024sei,Moreno-Sanchez:2025bzz,Pathak:2025zdp}. However, the absence of SUSY signatures at colliders motivates alternative realizations of $A_4$ modular symmetry~\cite{Qu:2024rns,Nomura:2024vzw,Nomura:2024atp,Nomura:2024nwh,Kobayashi:2025hnc,Nomura:2025ovm,Loualidi:2025tgw,Nomura:2025raf}.\footnote{Similar non-holomorphic constructions have been explored for other modular symmetries, including $S_3$~\cite{Okada:2025jjo}, $S_4$~\cite{Ding:2024inn} and $A_5$~\cite{Li:2024svh}.} The main features of this framework include eliminating the need for superpartners while preserving constrained flavor structures, retaining predictive power through automorphic forms, and allowing for direct connections to fundamental physics through the modular group. On the other side, theoretical challenges include understanding the stabilization of modulus vacuum expectation values (VEVs) without SUSY and understanding the physical significance of non-holomorphic forms. There is still room for exploring the phenomenological consequences of these models, particularly their predictions for neutrino masses and mixing parameters.

Recent work has shown that non-holomorphic modular symmetry can replace the holomorphy condition by a Laplacian constraint, enabling viable flavor models without SUSY \cite{Ding:2020zxw}. This approach has been successfully applied to $A_4$ symmetry, yielding testable predictions for neutrino observables \cite{Qu:2024rns,Nomura:2024nwh}. However, most of the existing studies focus on conventional seesaw mechanisms, while the inverse seesaw is less explored in this context~\cite{Kumar:2025bfe}.

In this work, we present a non-holomorphic modular $A_4$ model with an inverse seesaw mechanism. Our framework retains the predictive power of modular symmetry while generating neutrino masses through a low-scale seesaw, offering distinct collider and flavor signatures. We systematically analyze three benchmark scenarios with different modular weight assignments, demonstrating compatibility with current neutrino data and highlighting distinct phenomenological consequences. The paper is organized as follows. Section~\ref{sec:model} introduces the theoretical framework for $A_4$ modular symmetry in inverse seesaw models. Three model realizations with varying weights and representations are presented in section~\ref{sec:rezmod}. Numerical analyses and experimental constraints are discussed in section~\ref{sec:phenom}. Section~\ref{sec:concluds} summarizes our findings and discusses future directions. The appendices contain some basic and technical details.

\section{Model construction }
\label{sec:model}

We extend the SM by incorporating non-holomorphic modular $A_4$ symmetry to generate neutrino masses via the inverse seesaw mechanism~\footnote{The mathematical foundations of modular symmetry and the relevant group theory for our analysis are presented in Appendices~\ref{app:modelularsymandform} and~\ref{app:a4modsym}.}. The model introduces three right-handed neutrino singlets $N_i$, three left-handed fermion singlets $S_i$, and a gauge-singlet scalar $\Phi$ that couples $S$ and $N$ to facilitate the inverse seesaw structure. The field content and their transformations under the SM electroweak gauge group $SU(2)_L \times U(1)_Y$, modular $A_4$ symmetry, and modular weights $k$ are summarized in table~\ref{tab:fieldcont}. The $A_4$ representations ($\rho_L, \rho_E, \rho_N, \rho_S$) and the modular weights ($k_L, k_E, \dots, k_\Phi$) are initially kept general and will be fixed in section~\ref{sec:rezmod} to ensure consistency with the observed neutrino phenomenology\footnote{In this context, these so-called modular weights are used in the formal sense, with no constraints imposed on their possible values~\cite{Feruglio:2017spp}. }.  Modular symmetry is spontaneously broken when the complex modulus $\tau$ acquires a VEV $\langle\tau\rangle$ in the fundamental domain, fixing numerical values of modular forms $Y^{(k)}_{\mathbf{r}}(\langle\tau\rangle)$ that determine flavor structures.  
\begin{table}[h!]
    \centering
    \renewcommand{\arraystretch}{1.3}
    \setlength{\tabcolsep}{20pt}
       \begin{tabular}{c||c|c|c|c||c|c}
       \toprule
        \text{Fields}&$L$&$E_{R}$&$N$&$S$&$H$ & $\Phi$\\
        \hline  
        $SU(2)_{L}$&$\mathbf{2}$&$\mathbf{1}$&$\mathbf{1}$&$\mathbf{1}$&$\mathbf{2} $ & $ \mathbf{1}$\\
        $U(1)_{Y}$&$-\frac{1}{2}$&-1&0&0&$\frac{1}{2}$ & $0$\\
        $A_4$&$\rho_L$ &$\rho_E$&$\rho_N$ &$\rho_S$&$\mathbf{1}$ & $\mathbf{1}$\\
        $k$&$k_L$&$k_E$&$k_N$ &$k_{S}$&$0 $ & $ k_{\Phi}$\\ 
        \bottomrule
        \end{tabular}   
     \caption{Field content and their transformations under the electroweak gauge group $ SU(2)_L \times U(1)_Y $, modular $ A_4 $ , and modular weights $ k $. 
       The $A_4$ representations ($\rho_L, \dots, \rho_S$) and weights ($k_I$) will be specified in section~\ref{sec:rezmod}.  }
     \label{tab:fieldcont}
\end{table} 

The modular invariant Lagrangians for charged leptons ($\mathscr{L}_{\rm CL}$) and neutrinos ($\mathscr{L}_{\nu}$) are
\begin{align} 	   
-\mathscr{L}_{\rm CL} =& \sum_{\mathbf{r}}a_\mathbf{r}\left(\bar{L}HE_R\right)Y_{\mathbf{r}}^{(-k_L-k_E)}+\mathrm{h.c.}, \\ \label{eq:yukawaterms}
-\mathscr{L}_{\nu} = & \sum_{\mathbf{r}}b_\mathbf{r}\left(\bar{L}\tilde{H}N\right)Y_{\mathbf{r}}^{(-k_L-k_N)}+\sum_{\mathbf{r}}c_\mathbf{r}\left(\bar{L}\tilde{H}S\right)Y_{\mathbf{r}}^{(-k_L-k_S)}  +\sum_{\mathbf{r}}d_\mathbf{r}\left(\bar{S}N\right)\Phi Y_{\mathbf{r}}^{(-k_S-k_N-k_{\Phi})}\\ \nonumber
&+\sum_{\mathbf{r}}e_\mathbf{r}\left(\bar{N}N\right)Y_{\mathbf{r}}^{(-2k_N)}+\sum_{\mathbf{r}}f_\mathbf{r}\left(\bar{S}S\right)Y_{\mathbf{r}}^{(-2k_S)}+\sum_{\mathbf{r}}g_\mathbf{r}\left(\bar{N}N\right)\Phi Y_{\mathbf{r}}^{(-2k_N-k_{\Phi})}\\ \nonumber
& +\sum_{\mathbf{r}}h_\mathbf{r}\left(\bar{S}S\right)\Phi Y_{\mathbf{r}}^{(-2k_S-k_{\Phi})} +\mathrm{h.c.},
\end{align}
where $\mathbf{r}$ denotes $A_4$ representations, and $Y^{(k)}_{\mathbf{r}}$ are polyharmonic Maa{\ss} forms (see Appendix~\ref{app:polmaasa4} for explicit expressions with weights $k = -4, -2, 0$). Non-holomorphic modular forms $Y^{(k)}_{\mathbf{r}}$ satisfy the weight-$k$ hyperbolic Laplacian equation $\Delta_k Y^{(k)}_{\mathbf{r}} = 0$, different from holomorphic realizations. From a bottom-up approach, in the context of non-holomorphic modular flavor symmetries, as introduced in Ref.~\cite{Qu:2024rns}, we focus on the minimal kinetic terms of the fields. After the modulus $\tau$ acquires a VEV, the kinetic terms can be written as
\begin{equation}
	\begin{aligned}
			\mathscr{L}_K&=\langle-i\tau+i\bar{\tau}\rangle^{-k_{L}}\,i\,L^{\dagger} \, \overline{\sigma}^{\mu}\partial_{\mu}L+\langle-i\tau+i\bar{\tau}\rangle^{-k_{E}}\,i\,E^{\dagger}_R \, \overline{\sigma}^{\mu}\partial_{\mu}E_R\\
			&+\langle-i\tau+i\bar{\tau}\rangle^{-k_{N}}\,i\,N^{\dagger} \, \overline{\sigma}^{\mu}\partial_{\mu}N
			+\langle-i\tau+i\bar{\tau}\rangle^{-k_{S}}\,i\,S^{\dagger} \, \overline{\sigma}^{\mu}\partial_{\mu}S\\
			&+\langle-i\tau+i\bar{\tau}\rangle^{-k_{H}}\,(\partial_{\mu}H)^{\dagger} (\partial^{\mu}H)+\langle-i\tau+i\bar{\tau}\rangle^{-k_{\Phi}}\,(\partial_{\mu}\Phi)^{\dagger} (\partial^{\mu}\Phi)+\text{h.c.}\,.
			\label{eq:L-kinetic}
	\end{aligned}
\end{equation}
The canonical kinetic terms for the physical fields can be obtained through field rescaling:
$\psi\rightarrow \langle-i\tau+i\bar{\tau}\rangle^{k_{\psi}/2}\psi.$
The effect of this rescaling can be absorbed into the unknown couplings of the Yukawa interactions and does not affect the numerical study.

However, from a top-down perspective, the assignment of negative modular weights to physical fields may lead to uncontrolled kinetic terms. This issue has not been addressed in the current literature on non-holomorphic flavor symmetries. For theoretical consistency, we can adopt the approach proposed in Ref.~\cite{Kobayashi:2023qzt} by introducing an additional modulus $T$ that determines the overall volume, while the modular flavor symmetry remains governed by $\tau$. In this framework, the kinetic terms are given by
\begin{equation}
	\begin{aligned}
		\mathscr{L}_K=&\sum_\psi \langle-iT+i\bar{T}\rangle^{-n}\langle-i\tau+i\bar{\tau}\rangle^{-k}\,i\,\psi^{\dagger} \, \overline{\sigma}^{\mu}\partial_{\mu}\psi\\
		&+\sum_\phi \langle-iT+i\bar{T}\rangle^{-n}\langle-i\tau+i\bar{\tau}\rangle^{-k}\,(\partial_{\mu}\phi)^{\dagger} (\partial^{\mu}\phi)+\text{h.c.}\,,
	\end{aligned}
\end{equation}
where $\psi$ and $\phi$ collectively denote the spinor and scalar fields, respectively. For negative values of the modular weight $k$, the potentially divergent behavior of the kinetic terms as  $\text{Im}(\tau)\to \infty$ can be controlled by selecting a positive value for $n$. To obtain canonical kinetic terms, as before, the fields are rescaled by absorbing the factor $\left[\langle-iT+i\bar{T}\rangle^{-n}\langle-i\tau+i\bar{\tau}\rangle^{-k}\right]^{1/2}$ into the free parameters, yielding the canonical kinetic terms,
\begin{eqnarray}
	\mathscr{L}_K=\sum i\,\psi^{\dagger} \, \overline{\sigma}^{\mu}\partial_{\mu}\psi+\sum \,(\partial_{\mu}\phi)^{\dagger} (\partial^{\mu}\phi)+\text{h.c.}\,.
\end{eqnarray}
We present these preliminary observations to highlight the mechanism, though the detailed theoretical aspects require further investigation. The control of kinetic terms in the non-holomorphic framework demands more thorough study from top-down approaches.

The neutrino sector includes Dirac mass terms for $N$ and $S$, $N$-$S$ mixing mediated by $\Phi$, and Majorana mass terms for $N$ and $S$. Importantly, the $\bar{N}N$, $\bar{L}\tilde{H}S$, $\bar{N}N\Phi$ and $\bar{S}S\Phi$ terms in eq.~\eqref{eq:yukawaterms} can be forbidden by appropriate modular weight assignments without introducing additional symmetries, as demonstrated later. The lepton part of the Lagrangian is
\begin{equation}
\mathscr{L}_{\rm lepton}=\mathscr{L}_{\rm CL}+\mathscr{L}_{M_{D}}+\mathscr{L}_{M_{NS}}+\mathscr{L}_{M_{S}},
\label{eq:lagleptons}
\end{equation}
in which $\mathscr{L}_{M_{D}}$ is the Dirac  mass term of  $N$, $\mathscr{L}_{M_{NS}}$ includes the mixing term of  $N$ and $S$, $\mathscr{L}_{M_{S}}$ is the Majorana mass term of $S$. In the flavor basis of ($\nu_L$, $\bar{N_i}$, $S_i$), the $9\times9$ neutrino mass matrix $M$ exhibits a block structure 
\begin{equation}
M=\left(\begin{array}{ccc}
0 & M_{D} & 0 \\
M_{D}^{T} & 0 & M_{NS}^{T}\\
0 & M_{NS} &M_{S}\\
\end{array}\right),
\end{equation}
which realizes the inverse seesaw mechanism. The hierarchy $M_S \ll M_D \ll M_{NS}$ is essential for generating small neutrino masses while keeping new physics at experimentally accessible scales. Block-diagonalizing $M$ (following Ref.~\cite{Wang:2024qhe}) yields the light neutrino mass matrix
\begin{equation}  
	M_\nu = M_D M_{NS}^{-1} M_S \left(M_{NS}^T\right)^{-1} M_D^T.  
		\label{eq:numasmatinvss}
\end{equation}  

For Majorana neutrinos, $M_\nu$ is a symmetric matrix and diagonalized via 
\begin{equation}  
	U_\nu^T M_\nu U_\nu = \mathrm{diag}(m_1, m_2, m_3),  
\end{equation}  
where $m_i$ are the physical neutrino masses, and $U_\nu$ is a unitary matrix. The Pontecorvo-Maki-Nakagawa-Sakata (PMNS) matrix $U_{\text{PMNS}} \simeq U_{l}^\dagger U_\nu$ includes neutrino mixing observables, with $U_{l}$ diagonalizing the charged lepton mass matrix, $U^{\dagger}_{l}M_{\rm CL}M^{\dagger}_{\rm CL}U_{l}=\mathrm{diag}(m^2_{e},m^2_{\mu},m^2_{\tau})$. Its standard parameterization 
\begin{equation}  
	U_{\text{PMNS}} = \begin{pmatrix}  
		c_{12}c_{13} & s_{12}c_{13} & s_{13}e^{-i\delta_{\rm CP}} \\  
		-s_{12}c_{23} - c_{12}s_{13}s_{23}e^{i\delta_{\rm CP}} & c_{12}c_{23} - s_{12}s_{13}s_{23}e^{i\delta_{\rm CP}} & c_{13}s_{23} \\  
		s_{12}s_{23} - c_{12}s_{13}c_{23}e^{i\delta_{\rm CP}} & -c_{12}s_{23} - s_{12}s_{13}c_{23}e^{i\delta_{\rm CP}} & c_{13}c_{23}  
	\end{pmatrix}  
	\begin{pmatrix}  
		e^{i\eta_1} & 0 & 0 \\  
		0 & e^{i\eta_2} & 0 \\  
		0 & 0 & 1  
	\end{pmatrix},  
\end{equation}  
involves three mixing angles $\theta_{12}, \theta_{13}, \theta_{23}$, a Dirac CP-violating phase $\delta_{\rm CP}$, and two Majorana phases $\eta_1, \eta_2$.  The modular $A_4$ symmetry constraints on $Y^{(k)}_{\mathbf{r}}$ and the chosen weight assignments will correlate these parameters, offering testable predictions for oscillation experiments.

\section{Realization of the model}
\label{sec:rezmod}

Building upon the modular symmetry framework established in previous sections, we construct three concrete models for lepton mass generation through specific assignments of group representations and modular weights. All models implement the fermion singlets $ S_i $ as an $ A_4 $ triplet while suppressing extraneous couplings through optimized modular weight selection. Model 1 presents a minimal scenario where the remaining fields are assigned to $ A_4 $ singlets, yielding diagonal charged lepton mass matrices. Model 2 adopts identical singlet representations for both $ E_R $ and $ N $ fields, while model 3 introduces distinct singlet representations for the right-handed charged leptons $ E_R $. Given $k > 2$, the non-holomorphic modular forms reduce to the holomorphic case; for $k = 2$, only $Y^{(2)}_{\mathbf{1}}$ differs from the holomorphic scenario. In this work, we focus on the extreme case where all modular weights are negative, corresponding to Yukawa couplings described entirely by non-holomorphic modular forms, with no holomorphic contributions.

\subsection{Model 1: minimal singlet assignments}

In this minimal construction, three generations of the matter fields $ L_i $, $ E_{R_i} $, and $ N_i $ are assigned to different $ A_4 $ singlets ($ \mathbf{1} $, $ \mathbf{1^\prime} $, $ \mathbf{1^{\prime\prime}} $), ensuring non-degenerate lepton masses even with identical modular weights. The left-handed fermions $ S_i $ transform as an $ A_4 $ triplet, while the both scalar fields $ H $ and $ \Phi $ are $ A_4 $ trivial singlet $\mathbf{1}$. Modular invariance of the Lagrangian is ensured by selecting weights such that the overall modular weight of each term in eq.~\eqref{eq:lagleptons} vanishes, irrespective of whether individual weights are integers. The field assignments and weights are summarized in table~\ref{tab:fieldcont1}. In turn, these weights are chosen such that the terms $\bar{L}\tilde{H}S$, $\bar{N}N$, $\bar{N}N\Phi$ and $\bar{S}S\Phi$ are forbidden since the modular forms only have even weights.
\begin{table}[h!]
    \begin{center}
      \renewcommand{\arraystretch}{1.3}
  \setlength{\tabcolsep}{12pt}
     \begin{tabular}{c||c|c|c|c||c|c}
       \toprule
        \text{Fields}  &$L_{1}\quad L_{2}\quad L_{3}$&$e_{R}\quad\mu_{R}\quad\tau_{R}$&$N_1$\quad $N_2$\quad $N_3$&$S$&$H$&$\Phi$\\
        \hline  
        $SU(2)_{L}$&$\mathbf{2}$&$\mathbf{1}$&$\mathbf{1}$&$\mathbf{1}$&$\mathbf{2}$&$\mathbf{1}$\\
        $U(1)_{Y}$&$-\frac{1}{2}$&-1&0&0&$\frac{1}{2}$& $0$\\
        $A_4$&$\mathbf{1\quad1^\prime\quad1^{\prime\prime}}$ &$\mathbf{1\quad1^{\prime\prime}\quad1^{\prime}}$&$\mathbf{1\quad1^{\prime\prime}\quad1^{\prime}}$ &$\mathbf{3}$&$\mathbf{1}$&$\mathbf{1}$\\
        $k$&$\frac{1}{2}$&$-\frac{1}{2}$&$-\frac{1}{2}$&$0$&$0$&$ \frac{1}{2} $\\ \bottomrule
        \end{tabular}
     \caption{Field assignments to  $A_4$ irreducible representation and their weights in model 1. The $A_4$ representations are specified as singlets ($\mathbf{1}$, $\mathbf{1}^\prime$, $\mathbf{1}^{\prime\prime}$) or triplet ($\mathbf{3}$), with fractional modular weights chosen to ensure invariant Lagrangian terms.}
   \label{tab:fieldcont1}  
     \end{center}   
\end{table}

The Lagrangians for charged leptons and neutrinos are given by
\begin{align}
-\mathscr{L}_{\rm CL}=& \alpha_{\rm CL}\bar{L}_{1}He_RY_{\mathbf{1}}^{(0)}+\beta_{\rm CL}\bar{L}_{2}H\mu_{R}Y_{\mathbf{1}}^{(0)}+\gamma_{\rm CL}\bar{L}_{3}H\tau_{R}Y_{\mathbf{1}}^{(0)}+\mathrm{h.c.},\\
\mathscr{L}_{\nu}=& \mathscr{L}_{M_{D}}+\mathscr{L}_{M_{NS}}+\mathscr{L}_{M_{S}}.
\end{align}
where the modular form $Y_{\mathbf{1}}^{(0)}$ reduces to unity ($Y_{\mathbf{1}}^{(0)} = 1$) as required by its vanishing weight. The neutrino sector contains three distinct contributions, $\mathscr{L}_{M_D}$ generating the Dirac mass term,  $\mathscr{L}_{M_{NS}}$ mediating $N$-$S$ mixing, and $\mathscr{L}_{M_S}$ producing the Majorana mass term,
\begin{align}
-\mathscr{L}_{M_{D}}=& \alpha_{{D}}\bar{L}_{1}\tilde{H}N_{{1}}Y_{\mathbf{1}}^{(0)}+\beta_{{D}}\bar{L}_{2}\tilde{H}N_{{2}}Y_{\mathbf{1}}^{(0)}+\gamma_{{D}}\bar{L}_{3}\tilde{H}N_{{3}}Y_{\mathbf{1}}^{(0)}+\mathrm{h.c.},\\
-\mathscr{L}_{M_{NS}}=& \alpha_{{NS}}\left(\bar{S}Y_{\mathbf{3}}^{(0)}\right)_{\mathbf{1}}N_1\Phi+\beta_{{NS}}\left(\bar{S}Y_{\mathbf{3}}^{(0)}\right)_{\mathbf{1^\prime}}N_2\Phi+\gamma_{{NS}}\left(\bar{S}Y_{\mathbf{3}}^{(0)}\right)_{\mathbf{1^{\prime\prime}}}N_3\Phi+\mathrm{h.c.}, \\
-\mathscr{L}_{M_{S}}=& \alpha_{{S}}\left(\bar{S}S\right)_{\mathbf{3}_S}Y_{\mathbf{3}}^{(0)}+\beta_{{S}}\left(\bar{S}S\right)_{\mathbf{1}}Y_{\mathbf{1}}^{(0)}+\mathrm{h.c.}.
\end{align}
The boldface subscripts in the Lagrangian terms (enclosed in brackets) indicate the decomposition of $A_4$ group products into specified irreducible representations. Flavor structure of the neutrino sector originates entirely from the weight-0 modular forms $Y_{\mathbf{1}}^{(0)}$ and $Y_{\mathbf{3}}^{(0)}=\left(Y_1, Y_2, Y_3 \right)$. After electroweak symmetry breaking, the scalar fields acquire VEVs $ \langle H \rangle = v_H $ and $ \langle \Phi \rangle = v_\Phi $, leading to the following mass matrices: 
\begin{align}
	M_{\rm CL}= & v_H\left(\begin{array}{ccc}
		\alpha_{{\rm CL}}& 0& 0 \\
		0  & \beta_{{\rm CL}}& 0 \\
		0 & 0 & \gamma_{{\rm CL}} \\
	\end{array}\right),\\
	M_{D}=& v_H\left(\begin{array}{ccc}
		\alpha_{{D}}& 0& 0 \\
		0  & \beta_{{D}}& 0 \\
		0 & 0 & \gamma_{{D}} \\
	\end{array}\right),
\end{align}
\begin{align}
M_{NS}=& v_\Phi\left(\begin{array}{ccc}
\alpha_{{NS}}Y_{\mathbf{3}, 1}^{(0)} & \beta_{{NS}}Y_{\mathbf{3}, 2}^{(0)} & \gamma_{{NS}}Y_{\mathbf{3}, 3}^{(0)} \\
\alpha_{{NS}}Y_{\mathbf{3}, 3}^{(0)}  & \beta_{{NS}}Y_{\mathbf{3}, 1}^{(0)} & \gamma_{{NS}}Y_{\mathbf{3}, 2}^{(0)}  \\
\alpha_{{NS}}Y_{\mathbf{3}, 2}^{(0)}  & \beta_{{NS}} Y_{\mathbf{3}, 3}^{(0)} & \gamma_{{NS}}Y_{\mathbf{3}, 1}^{(0)} \\
\end{array}\right),\\
M_{S}=& \left(\begin{array}{ccc}
2\alpha_{{S}}Y_{\mathbf{3}, 1}^{(0)}+\beta_{{S}}& -\alpha_{{S}}Y_{\mathbf{3}, 3}^{(0)} & -\alpha_{{S}}Y_{\mathbf{3}, 2}^{(0)}\\
-\alpha_{{S}}Y_{\mathbf{3}, 3}^{(0)}  & 2\alpha_{{S}}Y_{\mathbf{3}, 2}^{(0)}& -\alpha_{{S}}Y_{\mathbf{3}, 1}^{(0)}+\beta_{{S}}  \\
-\alpha_{{S}}Y_{\mathbf{3}, 2}^{(0)} & -\alpha_{{S}} Y_{\mathbf{3}, 1}^{(0)}+\beta_{{S}} & 2\alpha_{{S}}Y_{\mathbf{3}, 3}^{(0)} \\
\end{array}\right).
\end{align}
The charged lepton mass matrix $M_{\rm CL}$ naturally emerges as diagonal in this construction due to the $A_4$ singlet assignments of the relevant fields. This diagonal structure allows the charged lepton masses to be directly identified with the corresponding Yukawa couplings. While this feature imposes no constraints on the charged lepton sector, it enables unambiguous predictions for the neutrino sector through the modular symmetry framework. Under the inverse seesaw condition $ M_S \ll M_D \ll M_{NS} $, the light neutrino mass matrix is given by eq.~\eqref{eq:numasmatinvss}. For numerical analysis, we express $M_{\nu}=k m_{\nu}$, where $m_{\nu}=m_{D}m_{NS}^{-1}m_{S}\left(m_{NS}^{T}\right)^{-1}m_{D}^{T}$, in terms of dimensionless parameters $ m_D $, $ m_{NS} $, and $ m_S $, with rescaling factor $ k = (v_H \alpha_{D} / v_\Phi \alpha_{{NS}})^2 \alpha_{S} $,
\begin{align}
m_{D}=& \left(\begin{array}{ccc}
1& 0& 0 \\
0  & \frac{\beta_{{D}}}{\alpha_{{D}}}& 0 \\
0 & 0 & \frac{\gamma_{{D}}}{\alpha_{{D}}} \\
\end{array}\right),\label{eq:mdmmatrx}\\
m_{NS}=& \left(\begin{array}{ccc}
Y_{\mathbf{3}, 1}^{(0)} & \frac{\beta_{{NS}}}{\alpha_{{NS}}}Y_{\mathbf{3}, 2}^{(0)} & \frac{\gamma_{{NS}}}{\alpha_{{NS}}}Y_{\mathbf{3}, 3}^{(0)} \\
Y_{\mathbf{3}, 3}^{(0)}  & \frac{\beta_{{NS}}}{\alpha_{{NS}}}Y_{\mathbf{3}, 1}^{(0)} & \frac{\gamma_{{NS}}}{\alpha_{{NS}}}Y_{\mathbf{3}, 2}^{(0)}  \\
Y_{\mathbf{3}, 2}^{(0)}  & \frac{\beta_{{NS}}}{\alpha_{{NS}}} Y_{\mathbf{3}, 3}^{(0)} & \frac{\gamma_{{NS}}}{\alpha_{{NS}}}Y_{\mathbf{3}, 1}^{(0)} \\
\end{array}\right),\label{eq:mnsmixmatrx}\\
m_{S}=& \left(\begin{array}{ccc}
2Y_{\mathbf{3}, 1}^{(0)}+\frac{\beta_{{S}}}{\alpha_{{S}}}& -Y_{\mathbf{3}, 3}^{(0)} & -Y_{\mathbf{3}, 2}^{(0)}\\
-Y_{\mathbf{3}, 3}^{(0)}  & 2Y_{\mathbf{3}, 2}^{(0)}& -Y_{\mathbf{3}, 1}^{(0)}+\frac{\beta_{{S}}}{\alpha_{{S}}}  \\
-Y_{\mathbf{3}, 2}^{(0)} & -Y_{\mathbf{3}, 1}^{(0)}+\frac{\beta_{{S}}}{\alpha_{{S}}} & 2Y_{\mathbf{3}, 3}^{(0)} \\
\end{array}\right).\label{eq:msmmatrx}
\end{align}
In the following section, we show that the neutrino oscillation parameters depend only on the overall scale $k$. Furthermore, parameter values can always be chosen such that the mass hierarchy satisfies  $M_S \ll M_D \ll M_{NS}$. For example, in Model 1 with normal mass ordering in region A, where $k \sim 10^{-4}$, choices of $v_{\Phi}\alpha_{NS} = 10\,\text{TeV}$ and $v_H\alpha_D = 1\,\text{GeV}$ yield an effective scale for $\alpha_S \sim 10\,\text{keV}$.

The matrices in eqs.~\eqref{eq:mdmmatrx}-\eqref{eq:msmmatrx} are described by a total of six dimensionless parameters, which are generally complex. To increase the predictive power of the model, we impose generalized CP symmetry (gCP)~\cite{Novichkov:2019sqv}. The gCP symmetry requires simultaneous transformation $ \tau \to -\tau^* $, $Y^{(k)}_{\mathbf{r}}(\tau) \to X_{\mathbf{r}}  Y^{(k)*}_{\mathbf{r}}(\tau) $, and $ \varphi_i \to X_{\mathbf{r}} \bar{\varphi} $ for fields $ \varphi $, where $ X $ is the $ A_4 $ CP matrix.  This symmetry ensures that, apart from the modulus $\tau$, which remains complex and the sole source of CP violation, the remaining five parameters are constrained to be real. As a result, the model becomes more predictive, with seven real free parameters while maintaining consistency with observed flavor structures and CP-violating phases in the lepton sector.

\subsection{Model 2: triplet-singlet hybrid construction}
 
This model features distinct representation assignments where the lepton doublets $L$ and fermion singlets $S$ transform as $A_4$ triplets, while the right-handed charged leptons $E_R$ and neutrino singlets $N$ are assigned to identical $ A_4 $ singlets $\mathbf{1}$. To prevent lepton mass degeneracy despite identical representations, we implement differentiated modular weights across generations. The field transformations and weight assignments, detailed in table~\ref{tab:fieldcont2}, are chosen to exclude unwanted terms while ensuring modular invariance.
\begin{table}[h!]
   \begin{center}
    \renewcommand{\arraystretch}{1.3}
  \setlength{\tabcolsep}{16pt}
     \begin{tabular}{c||c|c|c|c||c|c}
     \toprule
\text{Fields}&$L$&$e_R\quad\mu_{R}\quad\tau_{R}$&$N$&$S$&$H$&$\Phi$\\
       \hline  
$SU(2)_{L}$&$\mathbf{2}$&$\mathbf{1}$&$\mathbf{1}$&$\mathbf{1}$&$\mathbf{2}$&$\mathbf{1}$\\
    $U(1)_{Y}$&$-\frac{1}{2}$&-1&0&0&$\frac{1}{2}$&$ 0$\\
 $A_4$ & $\mathbf{3}$ &$\mathbf{1\quad1\quad1}$&$\mathbf{1\quad1\quad1}$ &$\mathbf{3}$&$\mathbf{1}$& $\mathbf{1}$\\
       $k$&$\frac{1}{2}$&$-\frac{1}{2}\quad \frac{3}{2} \quad \frac{7}{2}$&$-\frac{1}{2}\quad \frac{3}{2} \quad \frac{7}{2}$&$0$ &$ 0$ & $\frac{1}{2}$\\
       \bottomrule
      \end{tabular}  
   \end{center} 
    \caption{Field content and transformation properties under $SU(2)_L \times U(1)_Y$, modular $A_4$ symmetry, and modular weights $k$ for model 2. The differentiation of modular weights $k_{e_R} = -\frac{1}{2}$, $k_{\mu_R} = \frac{3}{2}$, $k_{\tau_R} = \frac{7}{2}$ across generations prevents mass degeneracy despite identical $A_4$ singlet assignments.}
   \label{tab:fieldcont2}
\end{table} 
The fractional weight assignments $ k_{e_R} = -1/2 $, $ k_{\mu_R} = 3/2 $, $ k_{\tau_R} = 7/2 $ serve dual purposes. They lift generational degeneracy by introducing distinct modular form dependencies for each generation. Also, they forbid unwanted operators like $ \bar{N}N $ through the weight mismatches while allowing the $\bar{S}N \Phi$ term.

The modular-invariant Lagrangian decomposes into charged lepton and neutrino sectors,
\begin{align}
-\mathscr{L}_{\rm CL}= & \alpha_{\rm CL}\left(\bar{L}HY_{\mathbf{3}}^{(0)}\right)_{\mathbf{1}}e_R+\beta_{\rm CL}\left(\bar{L}HY_{\mathbf{3}}^{(-2)}\right)_{\mathbf{1}}\mu_R+\gamma_{\rm CL}\left(\bar{L}HY_{\mathbf{3}}^{(-4)}\right)_{\mathbf{1}}\tau_R+\mathrm{h.c.},\\
-\mathscr{L}_{M_D}=& \alpha_{D}\left(\bar{L}\tilde{H}Y_{\mathbf{3}}^{(0)}\right)_{\mathbf{1}}N_1+\beta_{D}\left(\bar{L}\tilde{H}Y_{\mathbf{3}}^{(-2)}\right)_{\mathbf{1}}N_2+\gamma_{D}\left(\bar{L}\tilde{H}Y_{\mathbf{3}}^{(-4)}\right)_{\mathbf{1}}N_3+\mathrm{h.c.},\\
-\mathscr{L}_{M_{NS}}=& \alpha_{NS}\left(\bar{S}Y_{\mathbf{3}}^{(0)}\right)_{\mathbf{1}}N_1\Phi+\beta_{NS}\left(\bar{S}Y_{\mathbf{3}}^{(-2)}\right)_{\mathbf{1}}N_2\Phi+\gamma_{NS}\left(\bar{S}Y_{\mathbf{3}}^{(-4)}\right)_{\mathbf{1}}N_3\Phi+\mathrm{h.c.}, \\
  -\mathscr{L}_{M_{S}}=& \alpha_{{S}}\left(\bar{S}S\right)_{\mathbf{3}_S}Y_{\mathbf{3}}^{(0)}+\beta_{{S}}\left(\bar{S}S\right)_{\mathbf{1}}Y_{\mathbf{1}}^{(0)}+\mathrm{h.c.}.
\end{align}

Following electroweak symmetry breaking, the dimensionless mass matrices (scaled by $\alpha_{{\rm CL}}$, $\alpha_{D}$, $\alpha_{{NS}}$, and $\alpha_{S}$) emerge with parameter ratios $a = \beta_{{\rm CL}}/\alpha_{{\rm CL}}$, $b = \gamma_{{\rm CL}}/\alpha_{{\rm CL}}$, $c = \beta_{{D}}/\alpha_{{D}}$, $d = \gamma_{{D}}/\alpha_{{D}}$, $e = \beta_{{NS}}/\alpha_{{NS}}$, $f = \gamma_{{NS}}/{\alpha_{{NS}}}$, $g = \beta_{{S}}/\alpha_{{S}}$. The expressions of these dimensionless matrices are as follows
\begin{align}
m_{\rm CL}=& \left(\begin{array}{ccc}
  Y_{\mathbf{3},1}^{(0)}& aY_{\mathbf{3},1}^{(-2)} & bY_{\mathbf{3},1}^{(-4)}\\
  Y_{\mathbf{3},3}^{(0)}&aY_{\mathbf{3},3}^{(-2)}&bY_{\mathbf{3},3}^{(-4)}\\
  Y_{\mathbf{3},2}^{(0)}&aY_{\mathbf{3},2}^{(-2)} &bY_{\mathbf{3},2}^{(-4)}\\
  \end{array}\right),\\
m_{D}=& \left(\begin{array}{ccc}
  Y_{\mathbf{3},1}^{(0)}& cY_{\mathbf{3},1}^{(-2)} & dY_{\mathbf{3},1}^{(-4)}\\
  Y_{\mathbf{3},3}^{(0)}&cY_{\mathbf{3},3}^{(-2)}&dY_{\mathbf{3},3}^{(-4)}\\
  Y_{\mathbf{3},2}^{(0)}&cY_{\mathbf{3},2}^{(-2)} &dY_{\mathbf{3},2}^{(-4)}\\
  \end{array}\right),\\
m_{NS}=& \left(\begin{array}{ccc}
  Y_{\mathbf{3},1}^{(0)}& eY_{\mathbf{3},1}^{(-2)} & fY_{\mathbf{3},1}^{(-4)}\\
  Y_{\mathbf{3},3}^{(0)}&eY_{\mathbf{3},3}^{(-2)}&fY_{\mathbf{3},3}^{(-4)}\\
  Y_{\mathbf{3},2}^{(0)}&eY_{\mathbf{3},2}^{(-2)} &fY_{\mathbf{3},2}^{(-4)}\\
  \end{array}\right),\\
m_{S}=& \left(\begin{array}{ccc}
2Y_{\mathbf{3},1}^{(0)}+g&- Y_{\mathbf{3},3}^{(0)} &- Y_{\mathbf{3},2}^{(0)}\\
-Y_{\mathbf{3},3}^{(0)}  & 2Y_{\mathbf{3},2}^{(0)}& -Y_{\mathbf{3},1}^{(0)}+g \\
-Y_{\mathbf{3},2}^{(0)} & -Y_{\mathbf{3},1}^{(0)}+g & 2Y_{\mathbf{3},3}^{(0)} \\
\end{array}\right).
\end{align}
Each column of the charged lepton mass matrix is composed of $Y_{\mathbf{3}}^{(0)}$, $Y_{\mathbf{3}}^{(-2)}$, and $Y_{\mathbf{3}}^{(-4)}$. The Froggatt-Nielsen mechanisms may apply if they have hierarchical structure, thereby naturally generates the charged lepton mass spectrum $ m_e \ll m_\mu \ll m_\tau $. Knowing the $q$-expansion of these modular forms, $\mathrm{Im}(\tau)$ cannot be very large, otherwise the mass matrix reduces the rank and cannot explain three nonzero charged lepton masses.

In contrast to model 1, the non-diagonal charged lepton mass matrix $m_{\rm CL}$ indicates that the flavor mixing originates from both the charged and the neutral lepton sectors. Imposing gCP symmetry restricts all couplings to real values and fixes $ Y^{(k)}_{\mathbf{3}, i}(-\tau^*) = [Y^{(k)}_{\mathbf{3}, i}(\tau)]^* $, reducing the 16 complex parameters to 9 real degrees of freedom (including $\tau$) for numerical fitting. This framework generates correlated predictions for lepton masses and mixing angles through the $\tau$-dependence of modular forms.

\subsection{Model 3: triplet-dominated construction}

This model represents a distinct configuration where the lepton doublets $L$, right-handed neutrinos $N$, and fermion singlets $S$ all transform as triplets under the $ A_4 $ modular symmetry, while the right-handed charged leptons $ E_R $ are assigned to different singlet representations $ \mathbf{1} $, $ \mathbf{1^\prime} $, $ \mathbf{1^{\prime\prime}} $, respectively. Modular weight assignments, detailed in table~\ref{tab:fieldcont3}, are chosen to maintain consistency with the inverse seesaw mechanism while preventing unwanted mass degeneracies.
\begin{table}[h!]
   \begin{center}
      \renewcommand{\arraystretch}{1.3}
  \setlength{\tabcolsep}{16pt}
     \begin{tabular}{c||c|c|c|c||c|c}
          \toprule
\text{Fields}&$L$&$E_{R}$&$N$&$S$&$H$&$\Phi$\\
       \hline  $SU(2)_{L}$&$\mathbf{2}$&$\mathbf{1}$&$\mathbf{1}$&$\mathbf{1}$&$\mathbf{2}$&$\mathbf{1}$\\
       $U(1)_{Y}$&$-\frac{1}{2}$&-1&0&0&$\frac{1}{2}$&$ 0$\\
       $A_4$&$\mathbf{3}$ &$\mathbf{1\quad1^{\prime\prime}\quad1^{\prime}}$&$\mathbf{3}$ &$\mathbf{3}$&$\mathbf{1}$&$\mathbf{1}$\\
       $k$&$\frac{1}{2}$&$-\frac{1}{2}$&$-\frac{1}{2}$&$0$&$0$&$ \frac{1}{2}$\\\bottomrule
       \end{tabular}
     \caption{Field content and transformation properties under $ SU(2)_L \times U(1)_Y $, modular $ A_4 $ symmetry, and modular weights $ k $ for model 3. The triplet assignments for $ L $, $ N $, and $ S $ lead to richer flavor structures compared to previous models, while the differentiated singlet assignments for $ E_R $ prevent mass degeneracies.
     }
   \label{tab:fieldcont3} 
    \end{center}   
   \end{table} 
The flavor structure emerges from the interplay of $A_4$ triplet representations, with symmetric ${\mathbf{3}_S}$ and antisymmetric  ${\mathbf{3}_A}$ tensor products contributing to the mass matrices. The Lagrangian exhibits a rich flavor structure through distinct $A_4$ tensor decompositions,
\begin{align}
-\mathscr{L}_{\rm CL}=& \alpha_{\rm CL}\left(\bar{L}HY_{\mathbf{3}}^{(0)}\right)_{\mathbf{1}}e_R+\beta_{\rm CL}\left(\bar{L}HY_{\mathbf{3}}^{(0)}\right)_{\mathbf{1^{\prime}}}\mu_R+\gamma_{\rm CL}\left(\bar{L}HY_{\mathbf{3}}^{(0)}\right)_{\mathbf{1^{\prime\prime}}}\tau_R+\mathrm{h.c.}, 
\end{align}
where the charged lepton sector uses different singlet projections. The neutrino sector incorporates both symmetric ${\mathbf{3}_S}$ and antisymmetric ${\mathbf{3}_A}$ triplet contractions,
\begin{align}
-\mathscr{L}_{M_{D}}=& \alpha_{{D}}\left(\bar{L}\tilde{H}N\right)_{\mathbf{1}}Y_{\mathbf{1}}^{(0)}+\beta_{{D}}\left(\bar{L}\tilde{H}N\right)_{\mathbf{3}_S}Y_{\mathbf{3}}^{(0)}+\gamma_{{D}}\left(\bar{L}\tilde{H}N\right)_{\mathbf{3}_A}Y_{\mathbf{3}}^{(0)}+\mathrm{h.c.},\\
-\mathscr{L}_{M_{NS}}=& \alpha_{{NS}}\left(\bar{S}N\right)_{\mathbf{1}}Y_{\mathbf{1}}^{(0)}\Phi+\beta_{{NS}}\left(\bar{S}N\right)_{\mathbf{3}_S}Y_{\mathbf{3}}^{(0)}\Phi+\gamma_{{NS}}\left(\bar{S}N\right)_{\mathbf{3}_A}Y_{\mathbf{3}}^{(0)}\Phi +\mathrm{h.c.},\\
-\mathscr{L}_{M_{S}}=& \alpha_{{S}}\left(\bar{S}S\right)_{\mathbf{3}_S}Y_{\mathbf{3}}^{(0)}+\beta_{{S}}\left(\bar{S}S\right)_{\mathbf{1}}Y_{\mathbf{1}}^{(0)}+\mathrm{h.c.}.
\end{align}

After electroweak breaking, the dimensionless mass matrices take the following particular structures:
\begin{align}
m_{\rm CL}= & \left(\begin{array}{ccc}
Y_{\mathbf{3},1}^{(0)} & aY_{\mathbf{3},2}^{(0)} & bY_{\mathbf{3},3}^{(0)} \\
Y_{\mathbf{3},3}^{(0)}  & aY_{\mathbf{3},1}^{(0)} & bY_{\mathbf{3},2}^{(0)}  \\
Y_{\mathbf{3},2}^{(0)}  & aY_{\mathbf{3},3}^{(0)} & bY_{\mathbf{3},1}^{(0)} \\
\end{array}\right),\\
m_{D}= &
\left(\begin{array}{ccc}
  1+2cY_{\mathbf{3},1}^{(0)}& -cY_{\mathbf{3},3}^{(0)}+dY_{\mathbf{3},3}^{(0)}& -cY_{\mathbf{3},2}^{(0)}-dY_{\mathbf{3},2}^{(0)} \\
  -cY_{\mathbf{3},3}^{(0)}-dY_{\mathbf{3},3}^{(0)} &2cY_{\mathbf{3},2}^{0}&1-cY_{\mathbf{3},1}^{(0)}+dY_{\mathbf{3},1}^{(0)}\\
  -cY_{\mathbf{3},2}^{(0)}+dY_{\mathbf{3},2}^{(0)} & 1-cY_{\mathbf{3},1}^{(0)}-dY_{\mathbf{3},1}^{(0)}&2cY_{\mathbf{3},3}^{0}\\
\end{array}\right),
\end{align}
\begin{align}
m_{NS}=& 
\left(\begin{array}{ccc}
  1+2eY_{\mathbf{3},1}^{(0)}& -eY_{\mathbf{3},3}^{(0)}+fY_{\mathbf{3},3}^{(0)}& -eY_{\mathbf{3},2}^{(0)}-fY_{\mathbf{3},2}^{(0)} \\
  -eY_{\mathbf{3},3}^{(0)}-fY_{\mathbf{3},3}^{(0)} &2eY_{\mathbf{3},2}^{0}&1-eY_{\mathbf{3},1}^{(0)}+fY_{\mathbf{3},1}^{(0)}\\
  -eY_{\mathbf{3},2}^{(0)}+fY_{\mathbf{3},2}^{(0)} & 1-eY_{\mathbf{3},1}^{(0)}-fY_{\mathbf{3},1}^{(0)}&2eY_{\mathbf{3},3}^{0}\\
\end{array}\right),\\
m_{S}=& \left(\begin{array}{ccc}
2Y_{\mathbf{3},1}^{(0)}+g& -Y_{\mathbf{3},3}^{(0)} & -Y_{\mathbf{3},2}^{(0)}\\
-Y_{\mathbf{3},3}^{(0)}  & 2Y_{\mathbf{3},2}^{(0)}& -Y_{\mathbf{3},1}^{(0)}+g \\
-Y_{\mathbf{3},2}^{(0)} & -Y_{\mathbf{3},1}^{(0)}+g & 2Y_{\mathbf{3},3}^{(0)} \\
\end{array}\right),
\end{align}
where the free parameters $a, b,c, d, e, f, g$ defined as coupling ratios analogous to model 2. In particular, the structure of mass matrix $M_S$ remains identical to models 1 and 2 due to its shared representation and weight assignment. The matrix $m_D$ contains antisymmetric contribution, which is proportional to $d$, absent in previous models, and $m_{CL}$ has a cyclic structure revealing non-trivial charged lepton mixing. These characteristics generate testable correlations between $\delta_{\rm CP}$ and charged lepton flavor violation.

The implementation of generalized gCP symmetry in model 3 requires all coupling parameters except the modulus $\tau$ to be real. This approach reduces the parameter space from 16 a priori complex degrees of freedom to 9 real parameters, while automatically satisfying the CP invariance condition for the modular forms. Such construction eliminates unphysical parameter degeneracies without compromising capability of the model to generate CP violation through the vacuum expectation value of $\langle \tau\rangle$.

The non-diagonal structure of the charged lepton mass matrix $M_{\rm CL}$ introduces correlated constraints throughout the lepton sector, with flavor mixing emerging from both the charged lepton $U_e$ and neutrino $U_\nu$ rotations in the PMNS matrix $U_{\rm PMNS} = U_e^\dagger U_\nu$. This constrained parameter space, combined with the predictive power of modular symmetry, will be tested against precision oscillation data in section~\ref{sec:phenom}, with particular focus on the $\text{Im}\,(\tau)$-dependence of CP-violating observables.

\section{Numerical analyses and results}
\label{sec:phenom}

We perform a comprehensive numerical analyses of the three models by scanning the parameter space under gCP symmetry constraints. Dimensionless free parameters are sampled within $[10^{-3}, 10^3]$, while the complex modulus $\tau = \text{Re}\,(\tau) + i\,\text{Im}\,(\tau)$ varies throughout the fundamental domain with $\text{Re}\,(\tau) \in (0, 0.5)$ and $\text{Im}\,(\tau) > \sqrt{3}/2$.  The gCP symmetry allows mapping $\tau \to -\tau^*$ to cover $\text{Re}\,(\tau) \in (-0.5, 0)$ without loss of generality~\cite{Ding:2023htn}. The $\chi^2$ minimization procedure incorporates both neutrino oscillation parameters and charged lepton mass ratios. For the neutrino mixing angles $\theta_{ij}$ and mass-squared differences $\Delta m^2_{ij}$, we use the one-dimensional $\chi^2$ projections from the latest global analysis~ \cite{Esteban:2024eli}, while the charged lepton mass ratios adopt the values from Ref.~\cite{Xing:2007fb}. The total $\chi^2$ function combines these constraints,
\begin{equation}
	\chi^2 = \sum_j \left( \frac{p_j - q_j}{\sigma_j} \right)^2,
\end{equation}
where $p_j$ represents model predictions, and $q_j, \sigma_j$ are, respectively, the experimental central values and their $1\sigma$ uncertainties,  summarized in table~\ref{tab:data}. The minimization is performed using the FlavorPy package~\cite{FlavorPy}, which efficiently explores the high-dimensional parameter space.
\begin{table}[h!]
	\begin{center}
		\renewcommand{\arraystretch}{1.5}
		\setlength{\tabcolsep}{15pt}
		\begin{tabular}{c|c|c}\toprule
			\text{Observable}& $\text{bfp} \pm 1\sigma$&$3\sigma$ range\\
			\hline  
			$\frac{m_e}{m_\mu } $&$0.004737$&$\cdots$\\
			$\frac{m_\mu}{m_\tau }$&$0.05882$&$\cdots$\\
			\hline  
			$\sin^2\theta_{12}$ (NO)&$0.308^{+0.012}_{-0.011}$&0.275$\to$0.345\\
			$\sin^2\theta_{12}$ (IO)&$0.308^{+0.012}_{-0.011}$&0.275$\to$0.345\\
			$\sin^2\theta_{13}$ (NO)&$0.02215^{+0.00056}_{-0.00058}$&0.02030$\to$0.02388\\
			$\sin^2\theta_{13}$ (IO)&$0.02231^{+0.00056}_{-0.00056}$&0.02060$\to$0.02409\\
			$\sin^2\theta_{23}$ (NO)&$0.470^{+0.017}_{-0.013}$&0.435$\to$0.585\\
			$\sin^2\theta_{23}$ (IO)&$0.550^{+0.012}_{-0.015}$&0.440$\to$0.584\\
			
			$\Delta m^2_{21}/10^{-5}~$eV$^2$ &$7.49^{+0.19}_{-0.19}$&6.92$\to$8.05\\
			$\Delta m^2_{31}/10^{-3}~$eV$^2$ (NO)&$2.513^{+0.021}_{-0.019}$&2.451$\to$2.578\\
			$\Delta m^2_{32}/10^{-3}~$eV$^2$ (IO)&$-2.484^{+0.020}_{-0.0.020}$ &-2.547$\to$-2.421\\
			\bottomrule
		\end{tabular}   
	\end{center}   
	\caption{ Current experimental constraints on lepton masses and mixing parameters. Neutrino oscillation parameters show both $ 1\sigma $ and $ 3\sigma $ ranges for normal (NO) and inverted (IO) mass ordering~\cite{Esteban:2024eli}. For the charged lepton mass ratios, we use the $M_Z$-scale values from Ref.~\cite{Xing:2007fb}, and set $1 \sigma$ uncertainty  by the $ 0.1\%$ of their central value, as was used in \cite{Qu:2024rns}. These values serve as benchmarks for testing predictability of the model.}
	\label{tab:data}
\end{table}

The measurement of cosmological parameters by Planck collaboration sets an upper limit of $\sum m_i < 0.12$ eV at $95~\%$ confidence level~\cite{Planck:2018vyg}. Laboratory experiments provide complementary constraints: the KATRIN experiment establishes $m_\beta < 0.45$ eV for the effective electron neutrino mass~\cite{Katrin:2024tvg}, while KamLAND-Zen reports $m_{\beta\beta} < 0.028 - 0.122$ eV for the effective Majorana mass~\cite{KamLAND-Zen:2024eml}. Future experiments such as LEGEND-1000~\cite{LEGEND:2021bnm} and nEXO~\cite{nEXO:2021ujk} aim to reach sensitivities of $m_{\beta\beta} \sim 0.009 - 0.021$ eV and $ 0.0047 - 0.0203$ eV respectively, which will further test the predictions of our models.

\subsection{Phenomenological implications of model 1}

\begin{figure}[h!]
	\centering
	\includegraphics[width=0.6\textwidth]{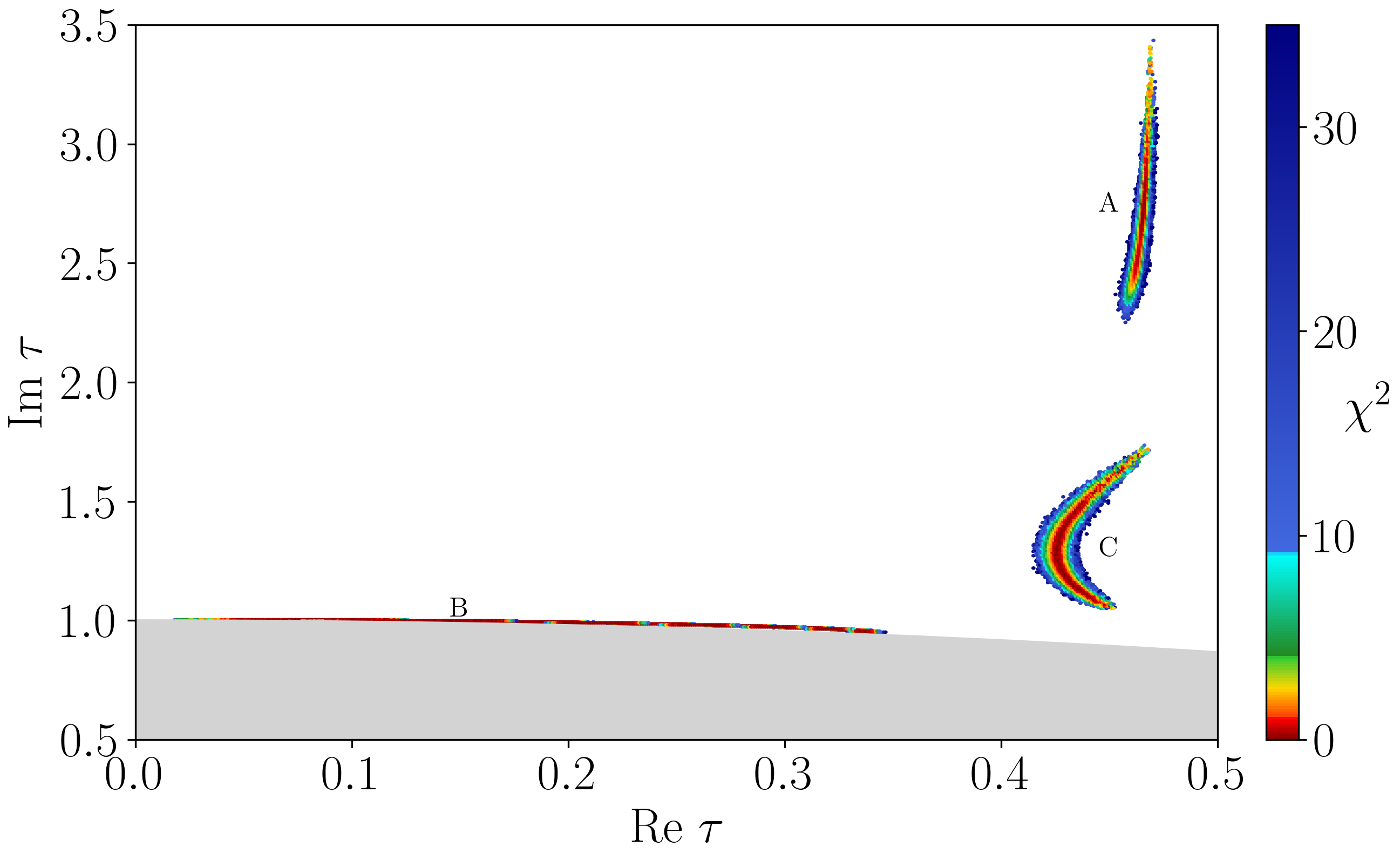}
	\caption{Distribution of viable $\tau$ values for model 1 (NO) in the complex plane. The three distinct regions (A, B, and C) correspond to different solutions that reproduce neutrino oscillation data within $3\sigma$. The color gradient represents the $\chi^2$ values, with brighter regions indicating better agreement with experimental data.}
	\label{fig:m1no_tau}
\end{figure}

Our numerical analysis of model 1 in the NO scenario reveals three phenomenologically viable regions in the complex $\tau$ plane, as shown in figure 
~\ref{fig:m1no_tau}. The best-fit parameters for these regions, presented in table~\ref{tab:m1no_fit}, demonstrate excellent agreement with the neutrino oscillation data. A particularly intriguing feature is the prediction of near-maximal Dirac CP violation across all regions, with $\delta_{\rm CP}=0.496\pi$ for region A, $0.484\pi$ for region B, and $1.495\pi$ for region C. These predictions, which lie within the current experimental uncertainties, will be directly testable by next-generation neutrino oscillation experiments.
On the other hand, the Majorana phases show an interesting pattern, with $\eta_1$ clustering around $1.5\pi$ in all three regions. While $\eta_2$ follows this trend in regions A and C, region B exhibits a distinct value of $\eta_2 = 0.485\pi$. This variation in Majorana phases leads to different predictions for neutrinoless double beta decay rates between the regions, providing potential targets for experimental discrimination. The consistency of these results across multiple viable $\tau$ regions demonstrates the robustness of predictions of model 1 while maintaining sufficient flexibility to accommodate the current experimental uncertainties.
\begin{table}[h!]
	\begin{center}
		\renewcommand{\arraystretch}{1.3}
		\setlength{\tabcolsep}{13pt}
		\begin{tabular}{c|c|c|c}\toprule
			\text{Model 1 (NO)}& \text{region A} &\text{region B}&\text{region C}\\
			\hline  
			$\tau $&$0.4663+2.7215i$&$0.2057+0.9912i$&$0.4276+1.3342i$\\
			$\frac{\beta_{D}}{\alpha_{D} }$&44.6847&826.7185&363.5671\\ 
			$\frac{\gamma_{D}}{\alpha_{D} }$&635.2967&706.4313&828.0536\\
			$\frac{\beta_{{NS}}}{\alpha_{{NS}}}$&208.3283&469.3657&38.2725\\
			$\frac{\gamma_{{NS}}}{\alpha_{{NS}} }$&127.5405&370.4232&83.1532\\
			$\frac{\beta_{S}}{\alpha_{S} }$&577.7306&1.3131&0.4242\\
			\hline  
			$k$ (eV)&0.0006219&0.00005028&0.0005630\\
			\hline  
			$\sin^2\theta_{12}$&0.3055&0.3101&0.3050\\
			$\sin^2\theta_{13}$&0.02215&0.02209&0.02223\\
			$\sin^2\theta_{23}$&0.4703&0.4686&0.4686\\
			$\delta_{\rm CP}/\pi$&0.4960&0.4842&1.495\\
			$\eta_1/\pi$&1.504&1.494&1.489\\
			$\eta_2/\pi$&1.493&0.4850&1.497\\
			\hline 
			$\Delta m^2_{21}/10^{-5}$ (eV$^2$)& 7.500&7.501&7.483\\
			$\Delta m^2_{31}/10^{-3}$ (eV$^2$)& 2.510&2.509&2.515\\
			$m_1$ (eV)&0.09626&0.0003536&0.002457\\
			$m_2$ (eV)&0.09665&0.008668&0.008993\\
			$m_3$ (eV)&0.1085&0.05009&0.05021\\
			$\sum m_i$ (eV)&0.3014&0.05912&0.06167\\
			$m_{\beta}$ (eV)&0.09667&0.008941&0.009270\\
			$m_{\beta\beta}$ (eV)&0.09660&0.003975&0.005481\\
			\hline 
			$\chi^2_{\rm min}$&0.02113&0.02159&0.03007\\
			\bottomrule
		\end{tabular}   
	\end{center}   
	\caption{Best-fit parameters and predictions for model 1 (NO) in three $\tau$ regions. Neutrino masses and mixing parameters reproduce experimental data within $1\sigma$ while predicting distinct signatures for CP violation and effective neutrino mass to be tested in the neutrinoless double beta decay experiments.}
	\label{tab:m1no_fit}
\end{table}

The mass spectra show important variations between regions. Region A produces a quasi-degenerate spectrum with $\sum m_i = 0.3014$ eV, which exceeds the upper bound of 0.12 eV, while regions B and C predict more hierarchical patterns with $\sum m_i \approx 0.06$ eV. These differences have significant implications for cosmological observations and direct neutrino mass measurements. It is worth noting that the mass $m_1 \approx 0.00035$ eV in region B is extremely light, which represents an especially strong hierarchy case that could be distinguished by future precision measurements of the absolute neutrino mass scale.

The predicted effective neutrino masses exhibit distinct patterns across the three regions, with important implications for current and future experiments. For the effective electron neutrino mass $ m_\beta $, all regions yield values within experimentally accessible ranges, spanning from $0.0089$ eV to $0.0967$ eV. The effective Majorana mass $ m_{\beta\beta} $ shows a particularly interesting variation. Region A predicts $ m_{\beta\beta} = 0.0966 $ eV, which lies within the current sensitivity window of KamLAND-Zen, while regions B and C predict substantially smaller values of $0.00398$ eV and $0.00548$ eV respectively. In particular, the prediction of $ m_{\beta\beta} = 0.00548$ eV from region C falls within the projected sensitivity range of the next-generation nEXO experiment, offering a clear experimental signature for this scenario. In contrast,  the result $0.00398$ eV from region B lies even below this future sensitivity threshold, representing a particularly challenging case for experimental verification. These differences in $ m_{\beta\beta} $ predictions provide a potential pathway for discriminating between viable regions through precision neutrinoless double beta decay searches.

\begin{figure}[h!]
	\centering
	\begin{subfigure}[b]{0.23\textwidth}
		\includegraphics[width=\textwidth]{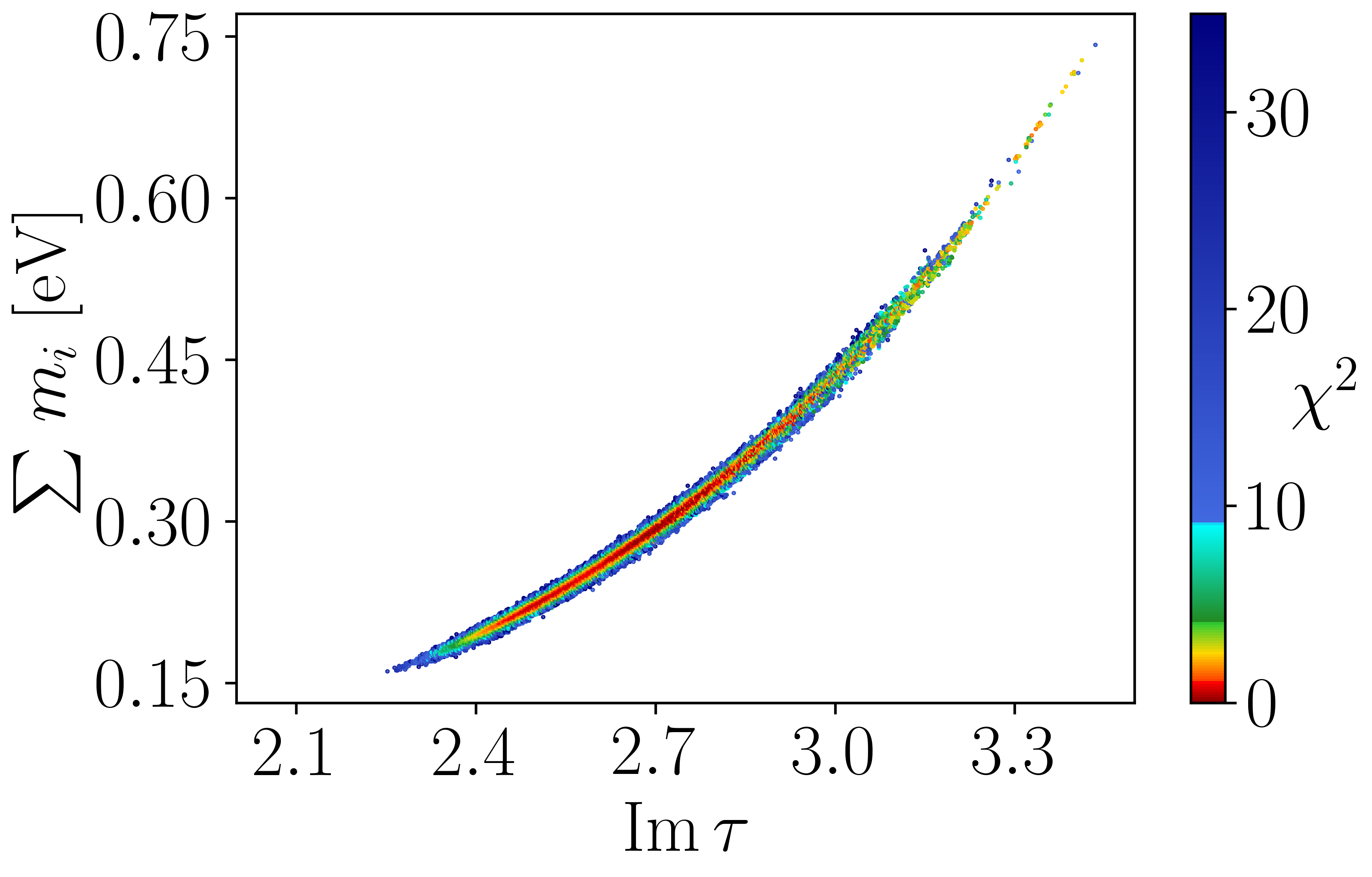}
		\caption*{A(1)}
		\label{fig:m1no_sub1}
	\end{subfigure}
	\hfill
	\begin{subfigure}[b]{0.23\textwidth}
		\includegraphics[width=\textwidth]{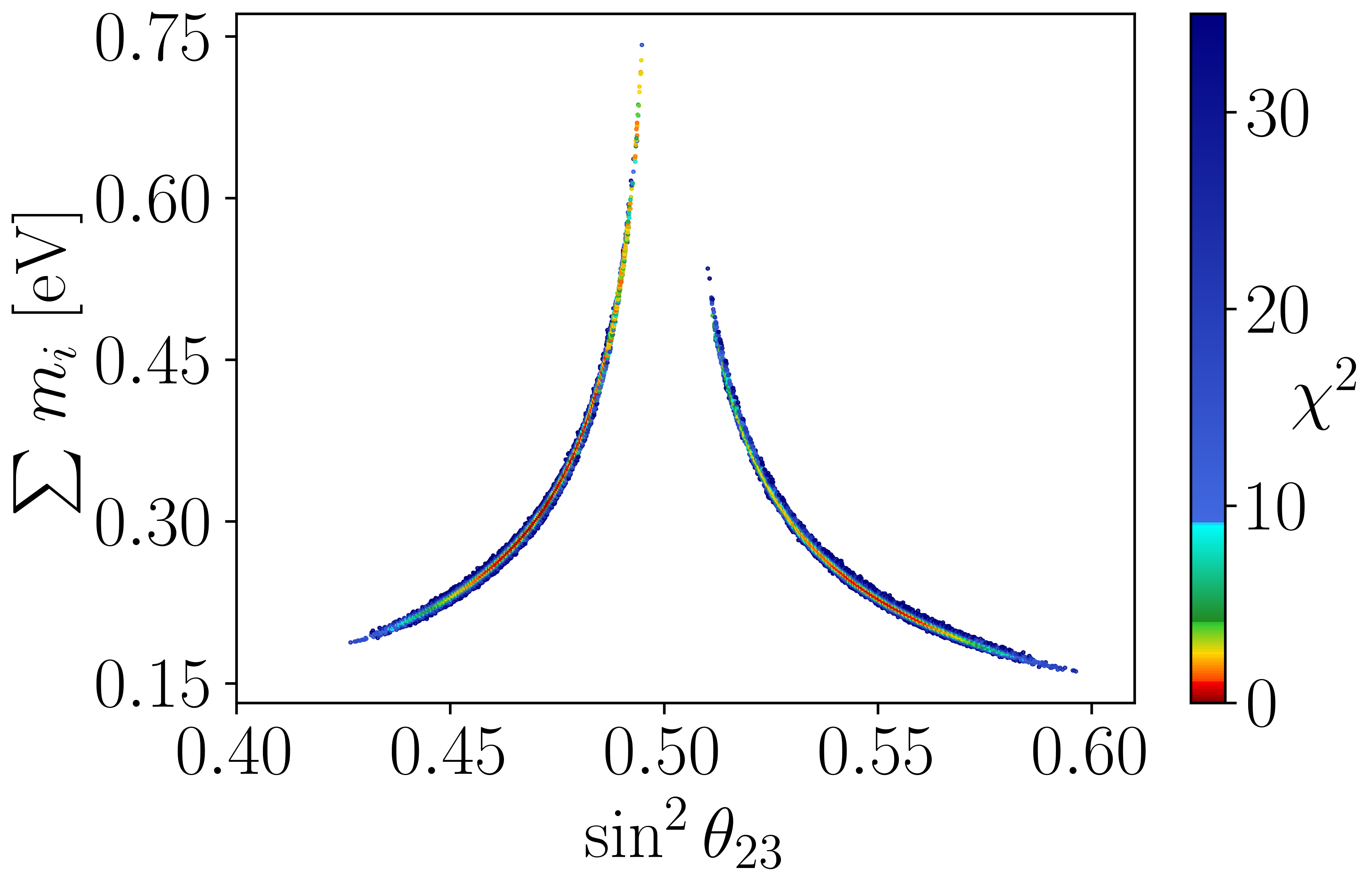} 
		\caption*{A(2)}
		\label{fig:m1no_sub2}
	\end{subfigure}
	\hfill
	\begin{subfigure}[b]{0.23\textwidth}
		\includegraphics[width=\textwidth]{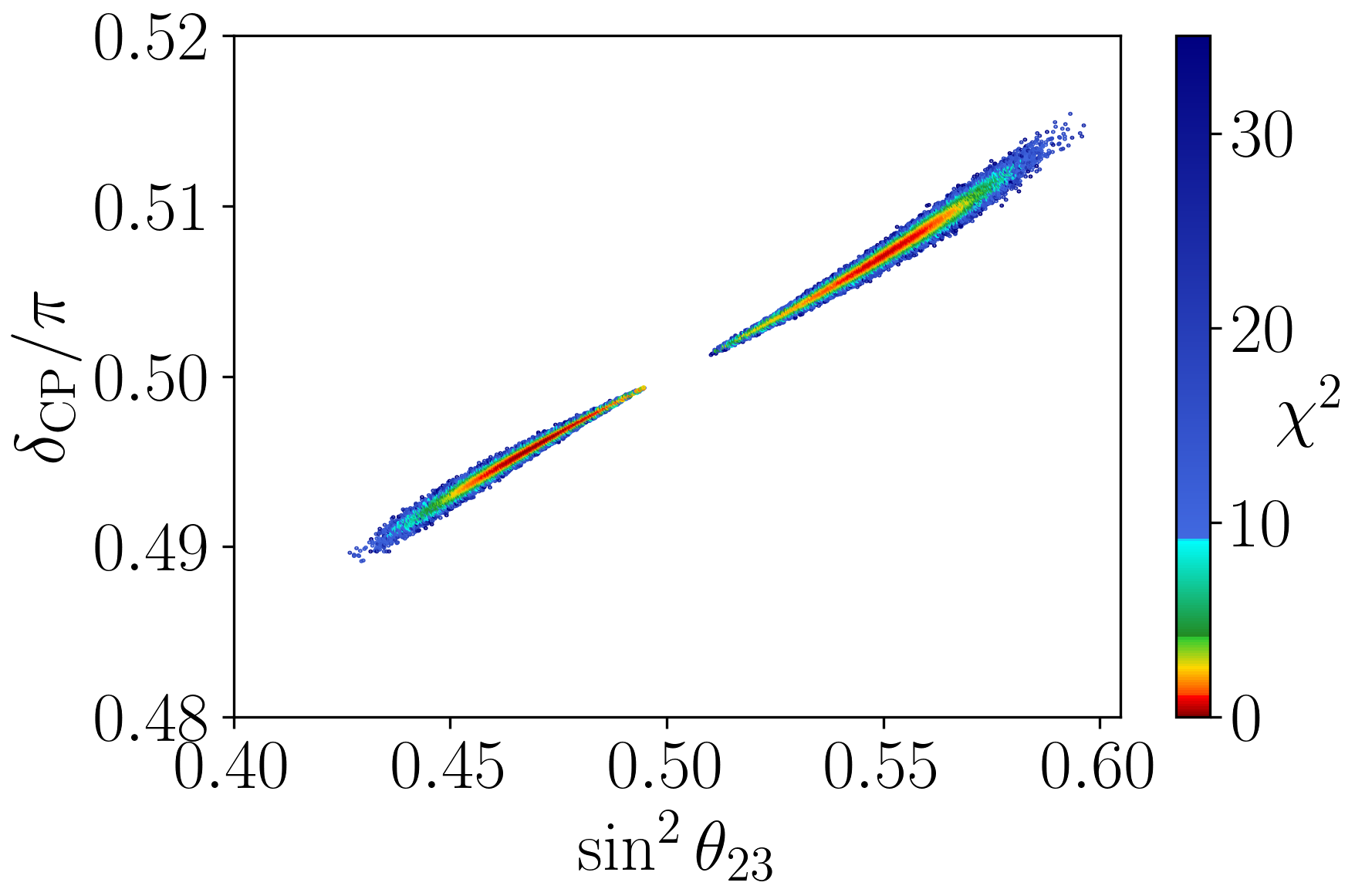} 
		\caption*{A(3)}
		\label{fig:m1no_sub3}
	\end{subfigure}
	\hfill
	\begin{subfigure}[b]{0.23\textwidth}
		\includegraphics[width=\textwidth]{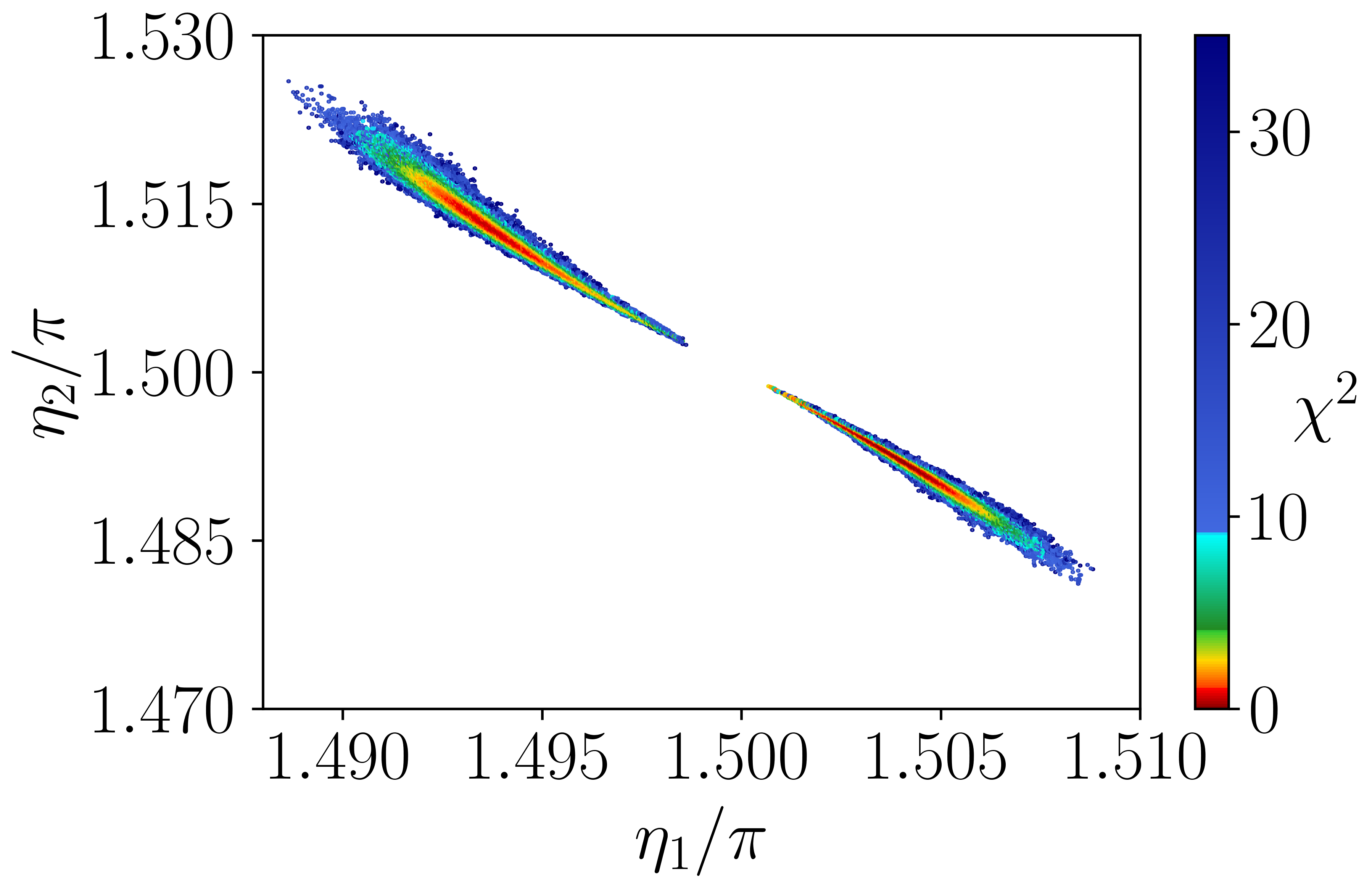} 
		\caption*{A(4)}
		\label{fig:m1no_sub4}
	\end{subfigure}
	\vspace{0.5cm} 
	\begin{subfigure}[b]{0.23\textwidth}
		\includegraphics[width=\textwidth]{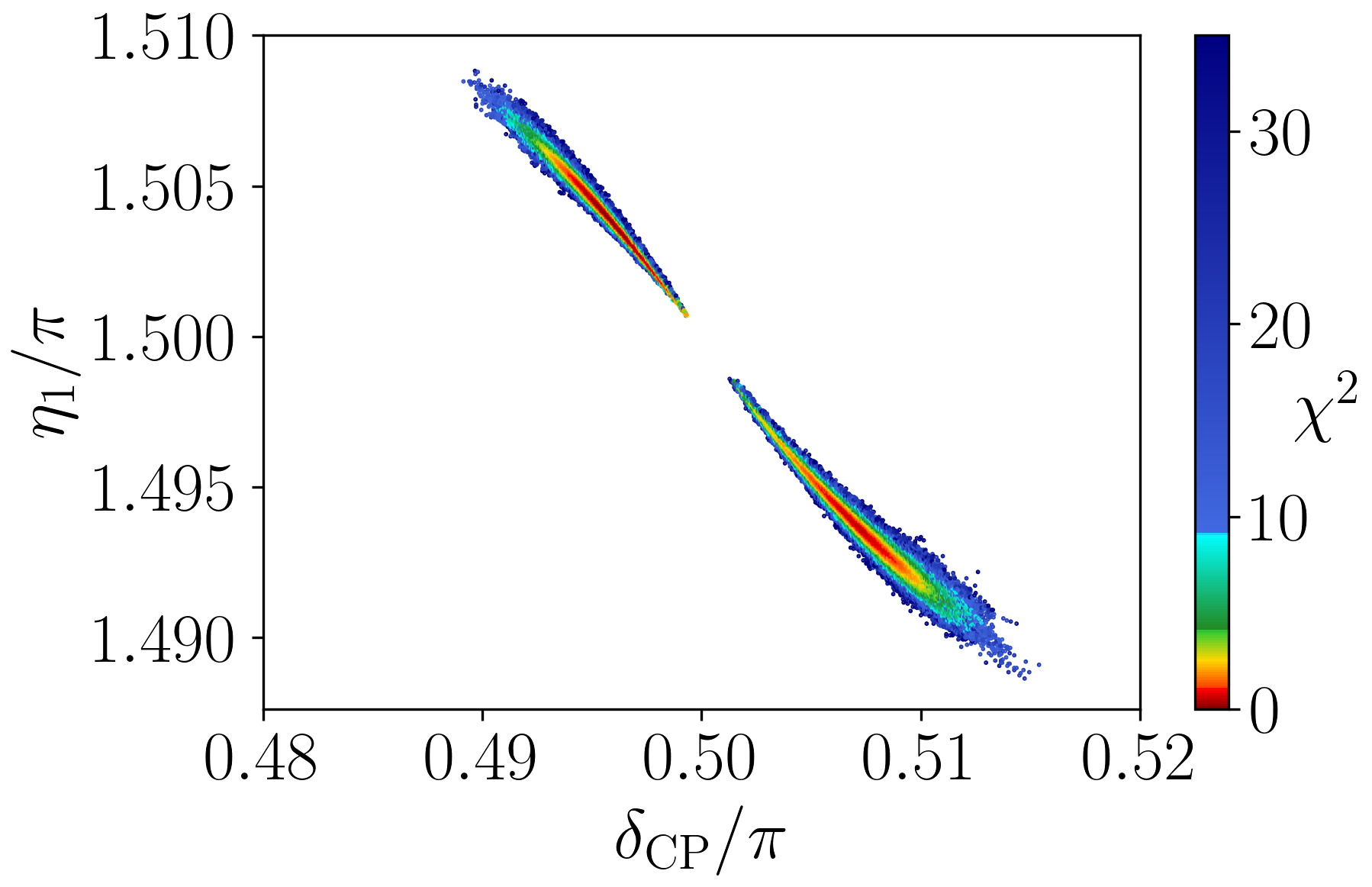}
		\caption*{A(5)}
		\label{fig:m1no_sub5}
	\end{subfigure}
	\hfill
	\begin{subfigure}[b]{0.23\textwidth}
		\includegraphics[width=\textwidth]{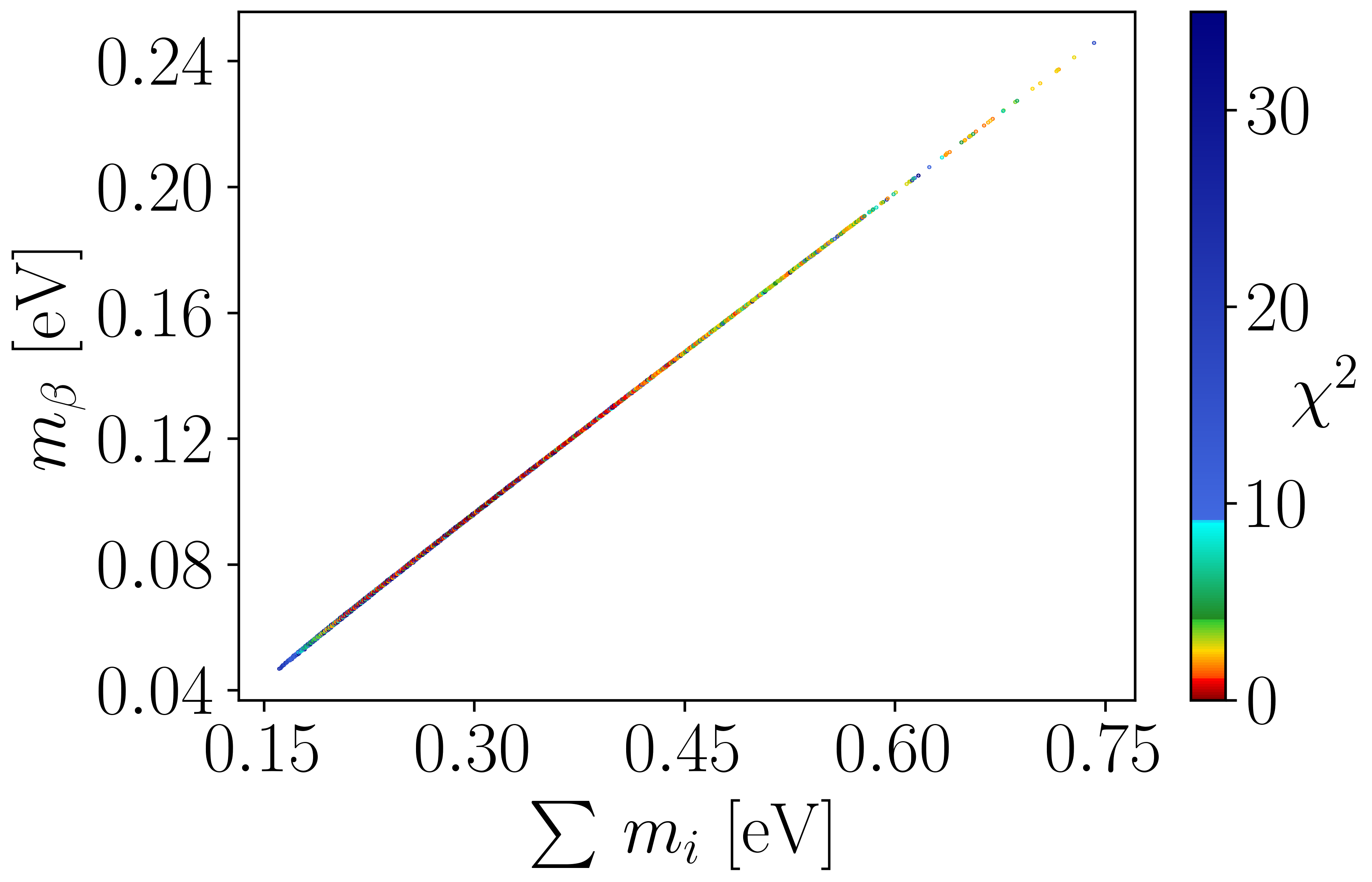} 
		\caption*{A(6)}
		\label{fig:m1no_sub6}
	\end{subfigure}
	\hfill 
	\begin{subfigure}[b]{0.23\textwidth}
		\includegraphics[width=\textwidth]{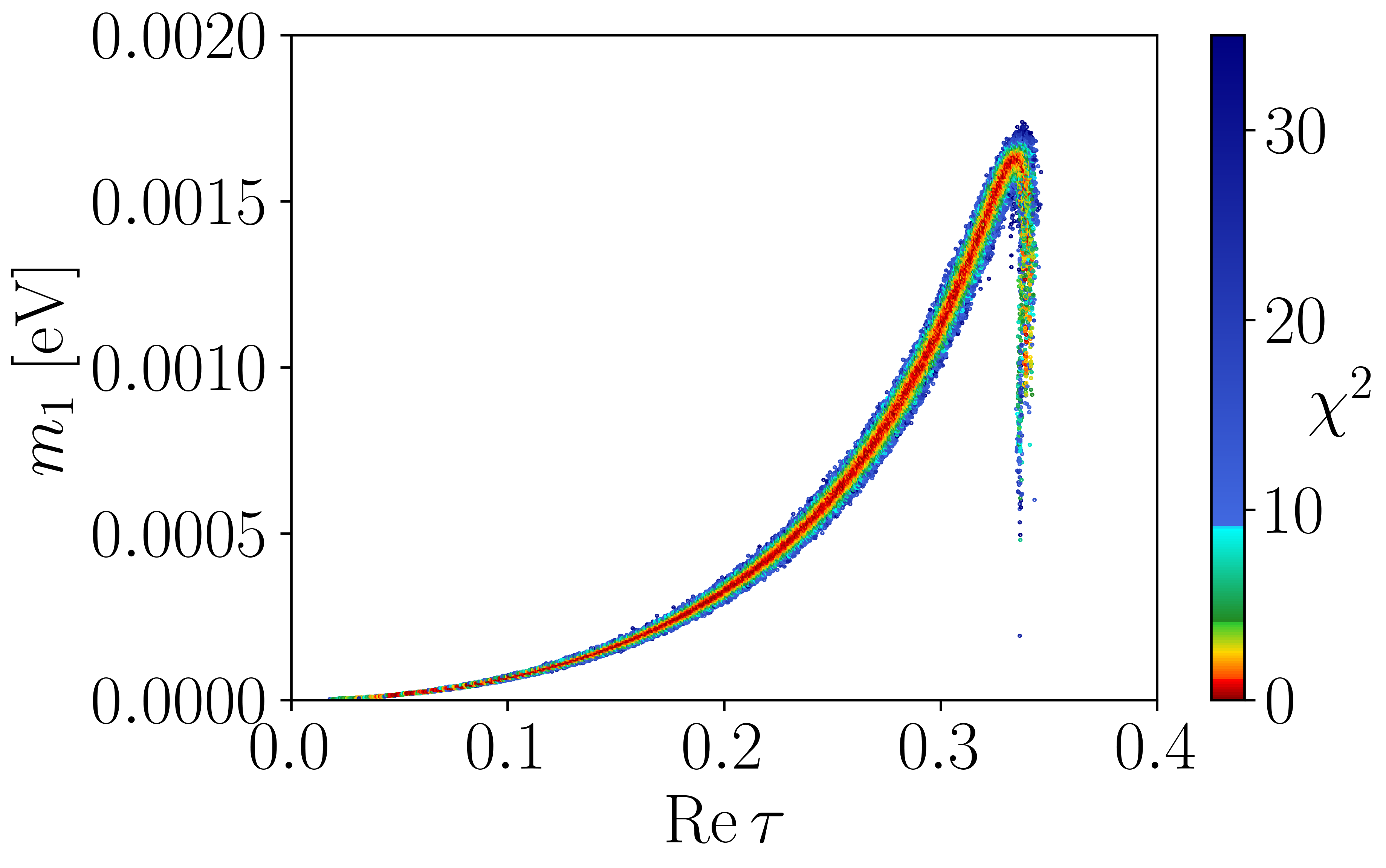}
		\caption*{B(1)}
		\label{fig:m1no_sub7}
	\end{subfigure}
	\hfill
	\begin{subfigure}[b]{0.23\textwidth}
		\includegraphics[width=\textwidth]{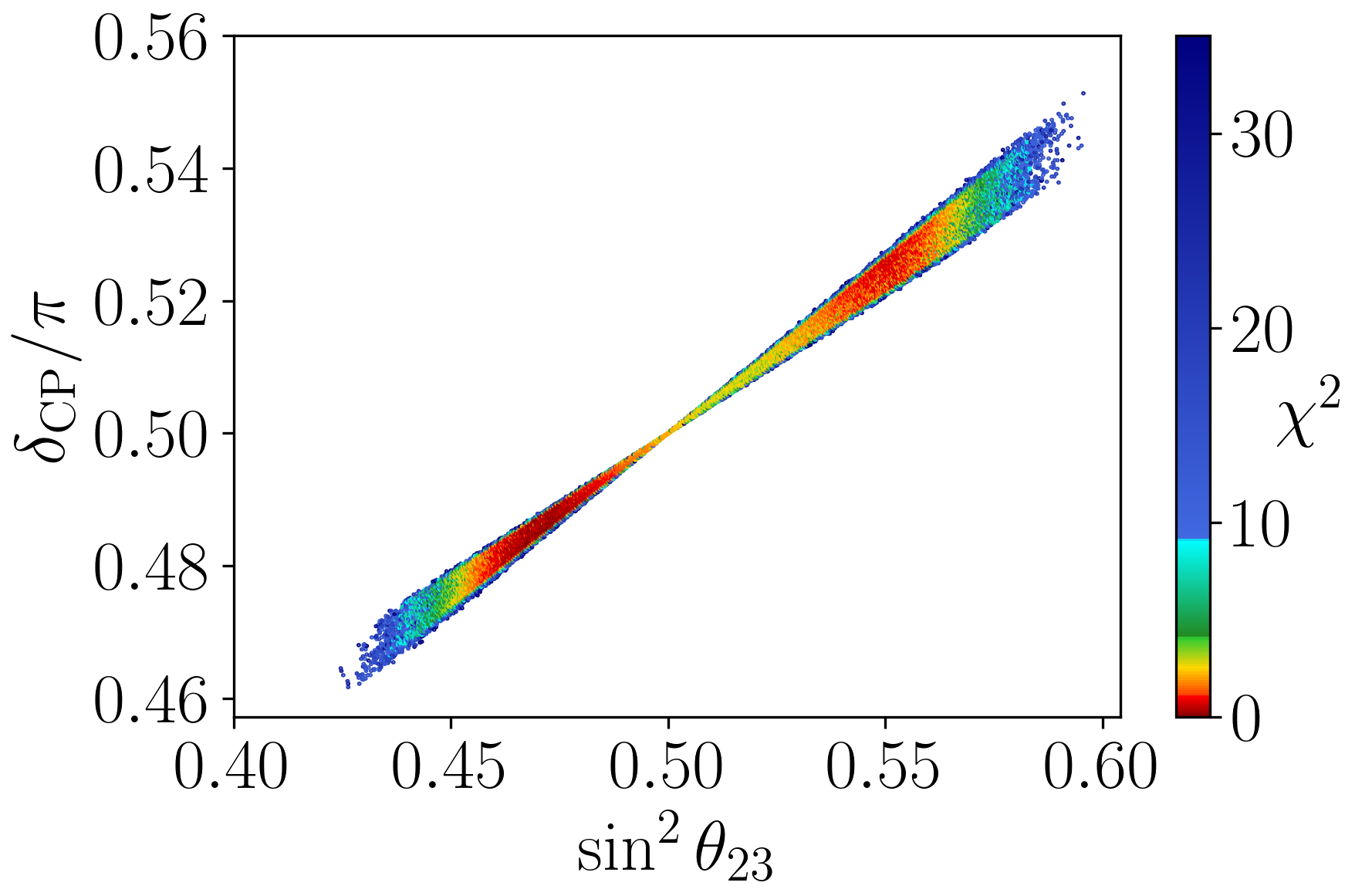} 
		\caption*{B(2)}
		\label{fig:m1no_sub8}
	\end{subfigure}  
	\vspace{0.5cm} 
	\begin{subfigure}[b]{0.23\textwidth}
		\includegraphics[width=\textwidth]{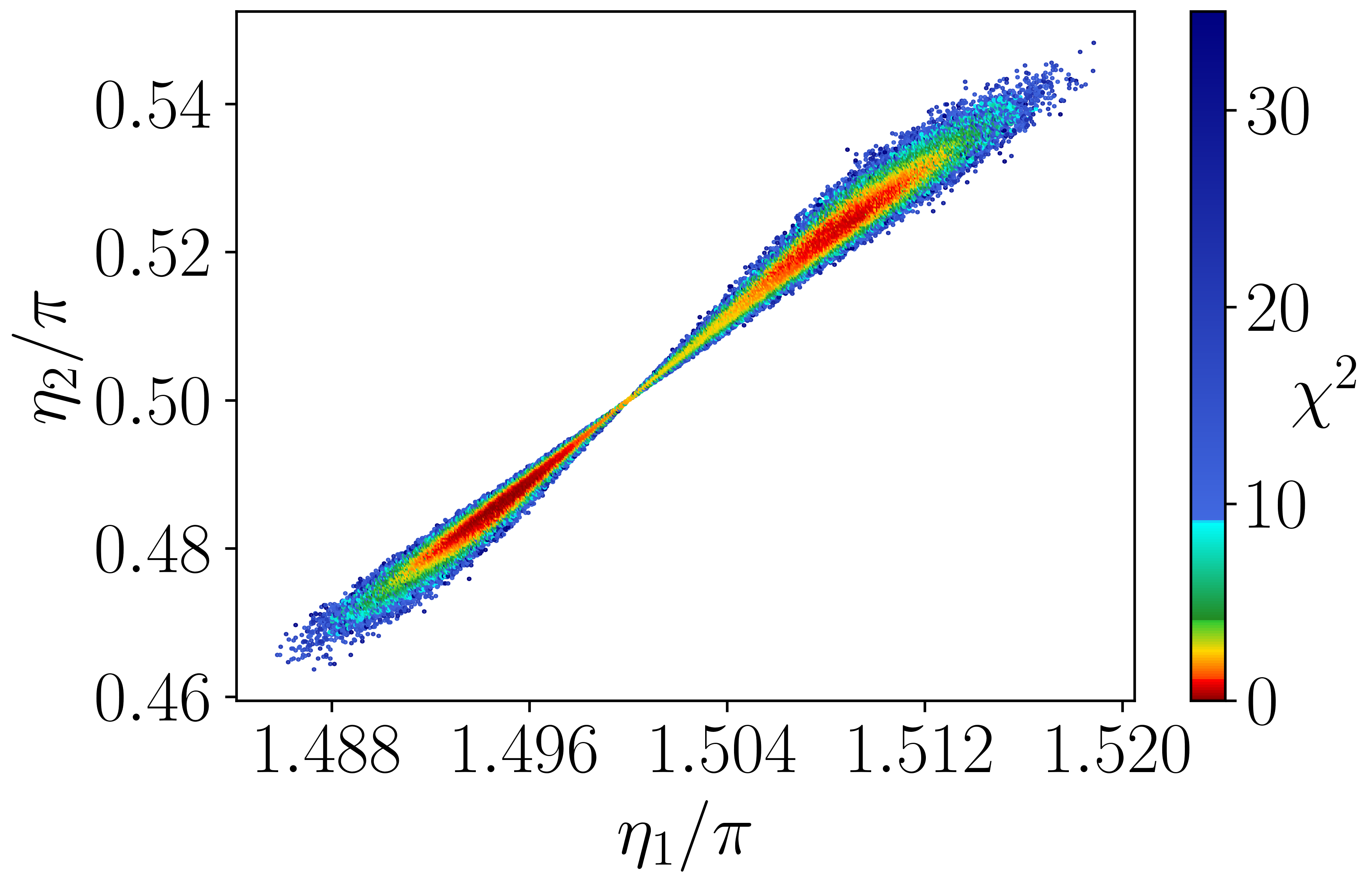} 
		\caption*{B(3)}
		\label{fig:m1no_sub9}
	\end{subfigure}
	\hfill
	\begin{subfigure}[b]{0.23\textwidth}
		\includegraphics[width=\textwidth]{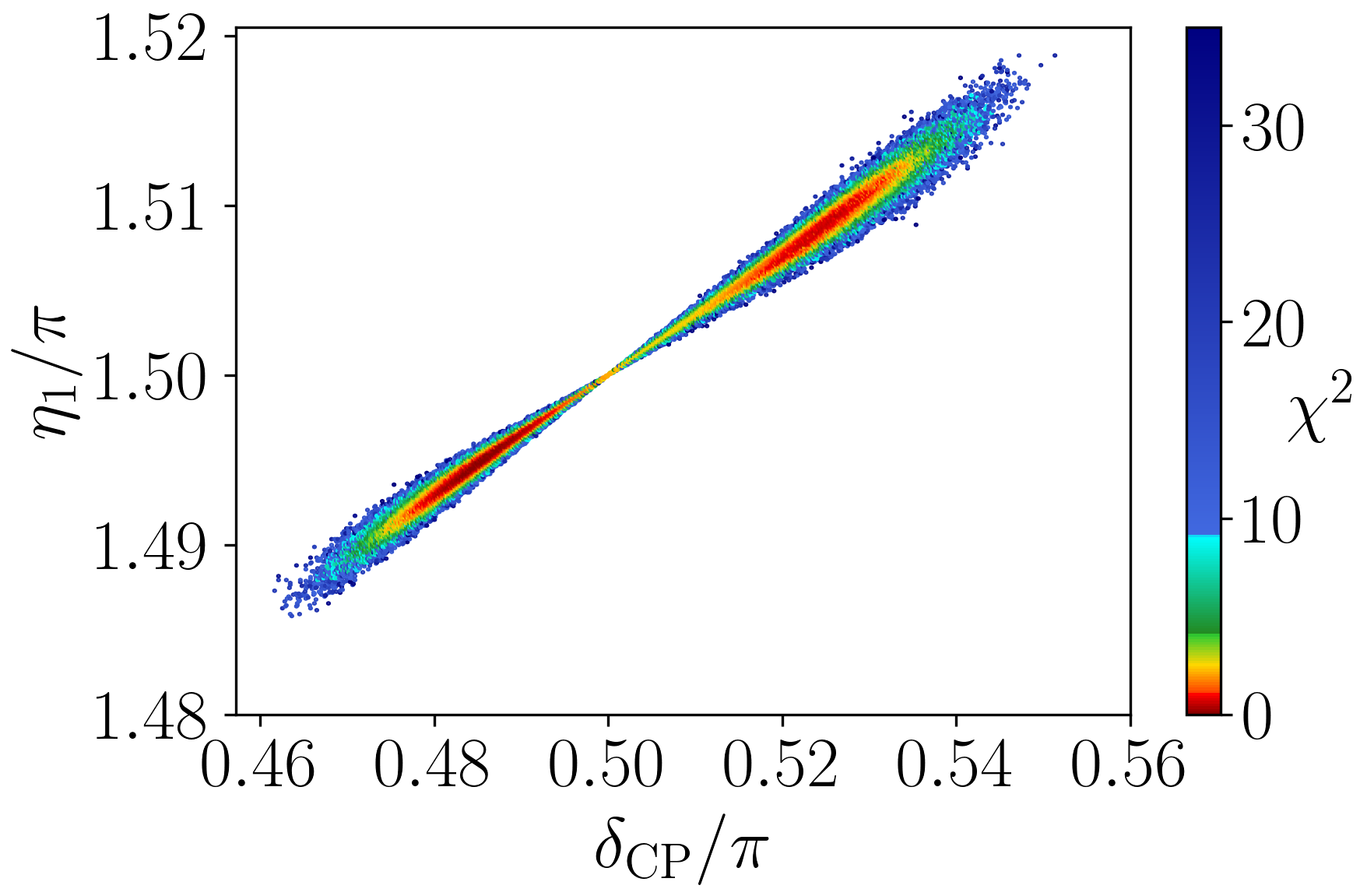} 
		\caption*{B(4)}
		\label{fig:m1no_sub10}
	\end{subfigure}
	\hfill
	\begin{subfigure}[b]{0.23\textwidth}
		\includegraphics[width=\textwidth]{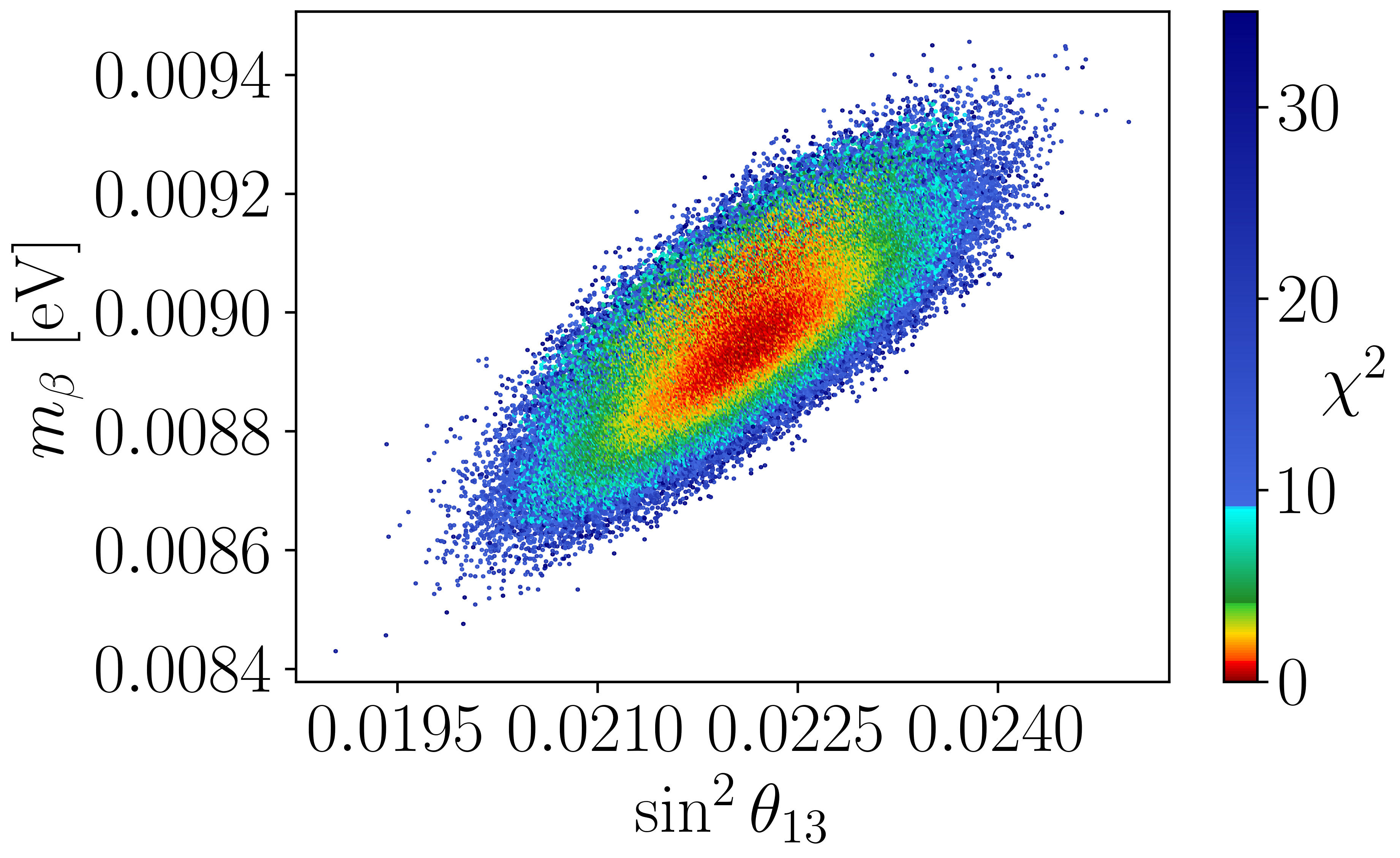} 
		\caption*{B(5)}
		\label{fig:m1no_sub11}
	\end{subfigure}
	\hfill
	\begin{subfigure}[b]{0.23\textwidth}
		\includegraphics[width=\textwidth]{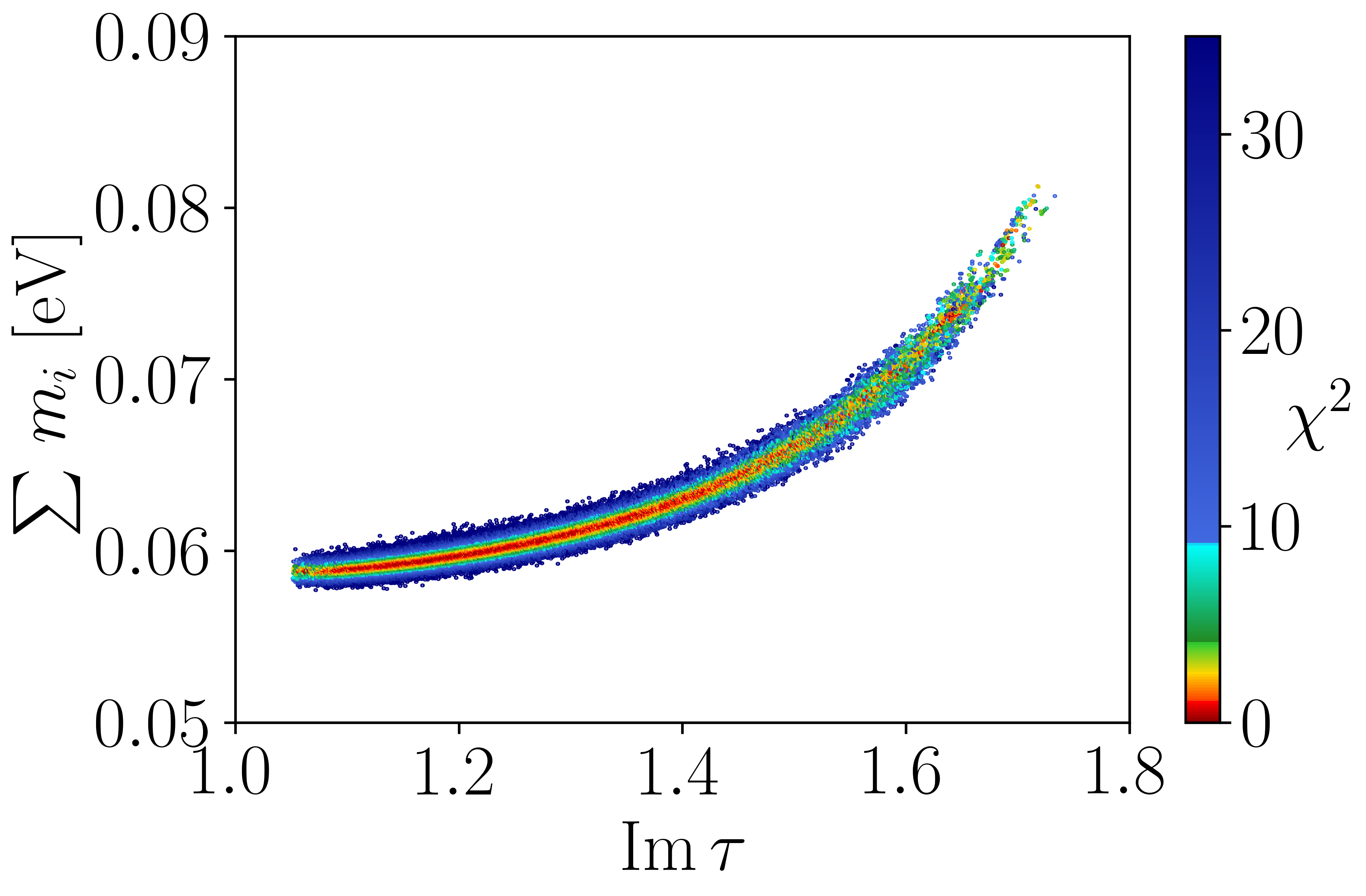}
		\caption*{B(6)}
		\label{fig:m1no_sub12}
	\end{subfigure}  
	\vspace{0.5cm} 
	\begin{subfigure}[b]{0.23\textwidth}
		\includegraphics[width=\textwidth]{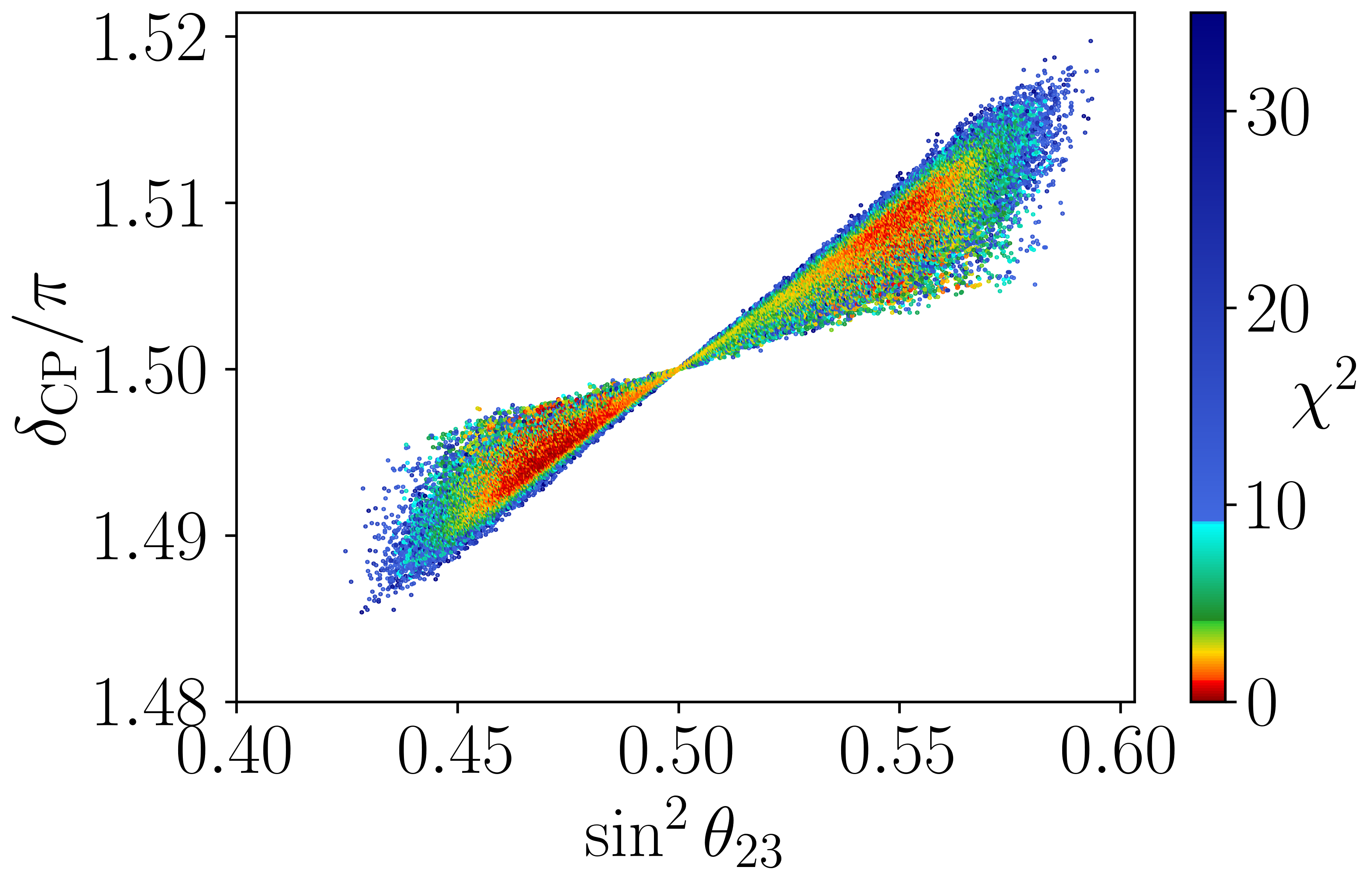} 
		\caption*{C(1)}
		\label{fig:m1no_sub13}
	\end{subfigure}
	\hfill
	\begin{subfigure}[b]{0.23\textwidth}
		\includegraphics[width=\textwidth]{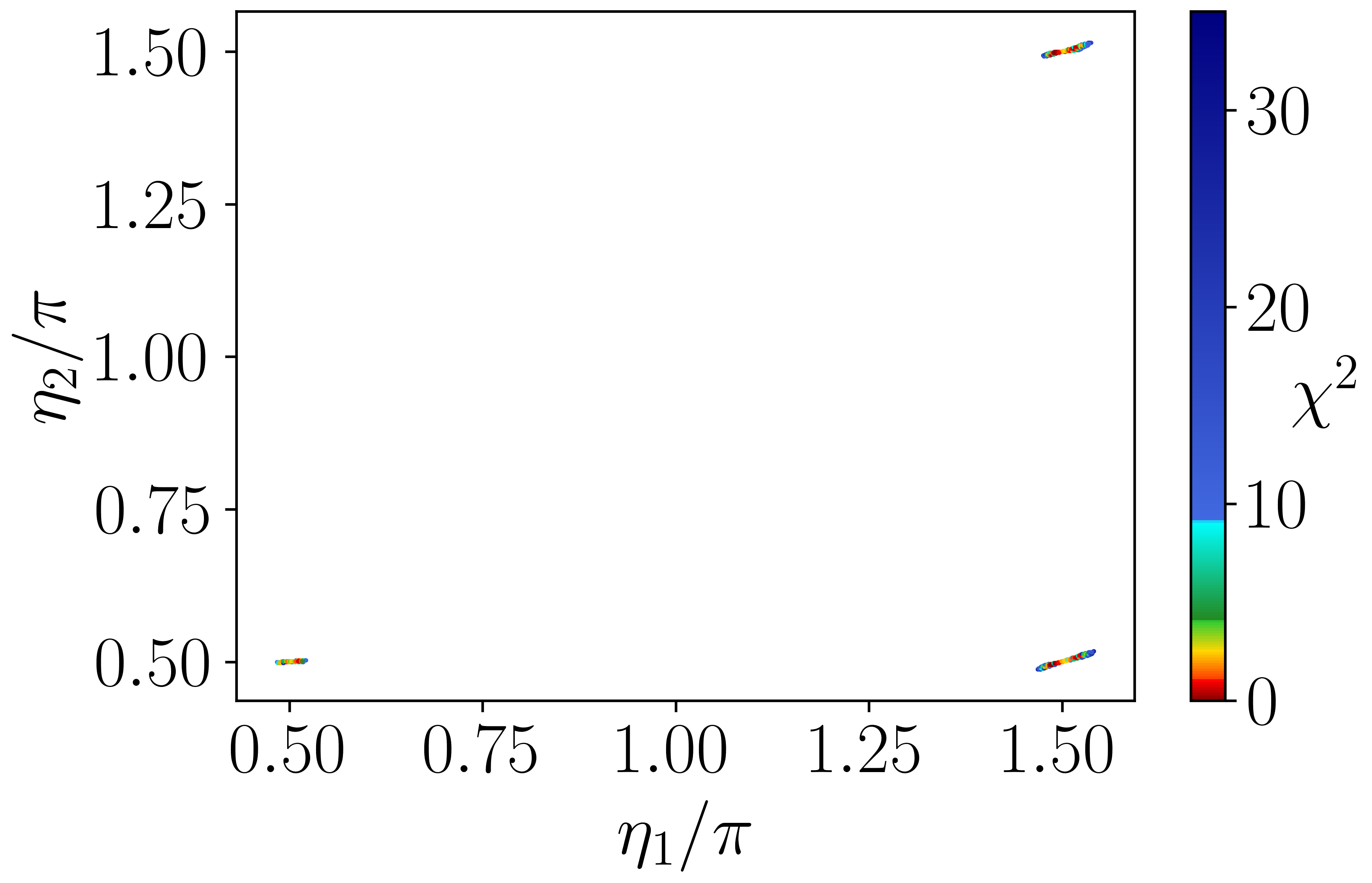}
		\caption*{C(2)}
		\label{fig:m1no_sub14}
	\end{subfigure}
	\hfill
	\begin{subfigure}[b]{0.23\textwidth}
		\includegraphics[width=\textwidth]{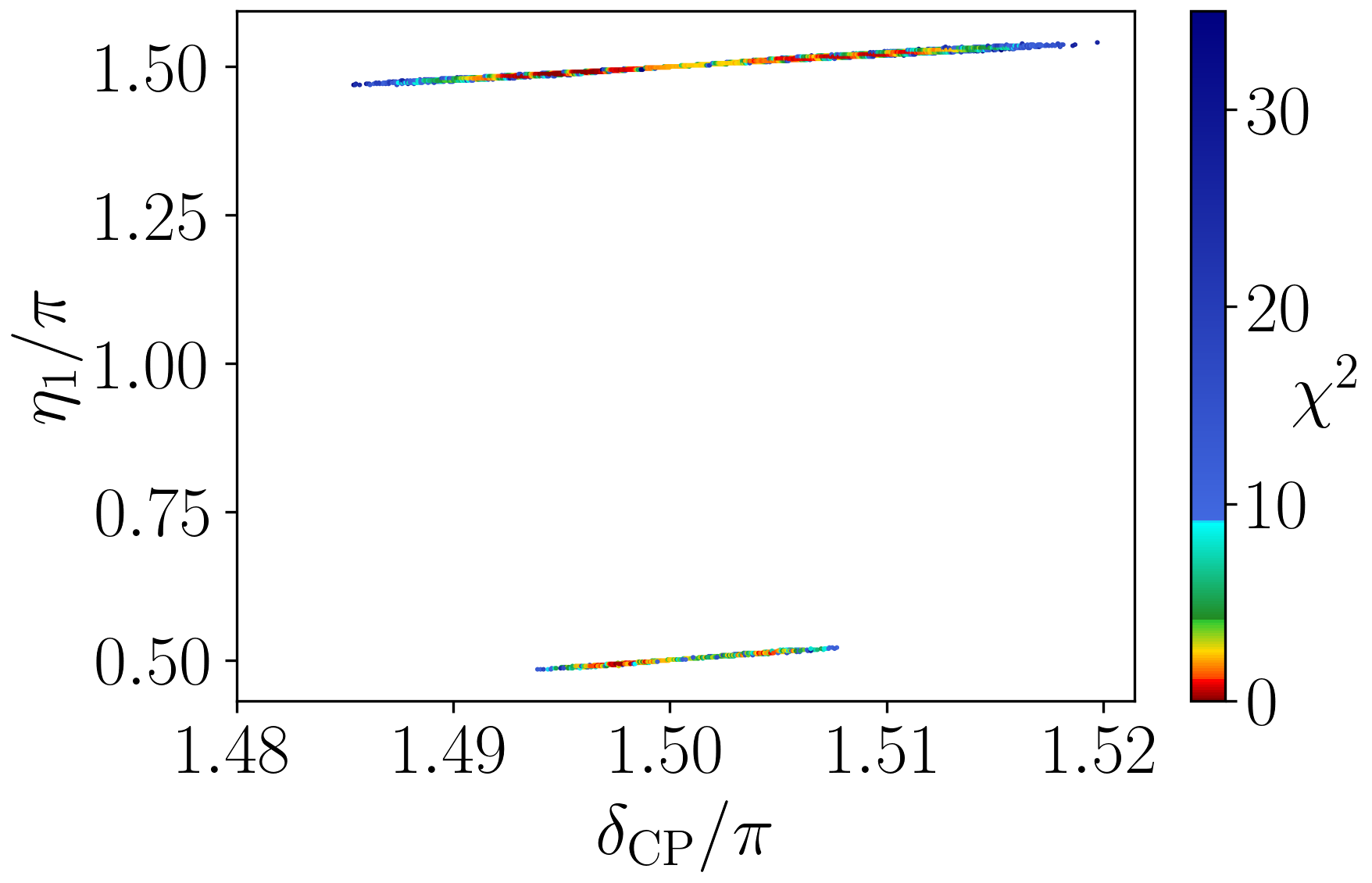} 
		\caption*{C(3)}
		\label{fig:m1no_sub15}
	\end{subfigure}
	\hfill
	\begin{subfigure}[b]{0.23\textwidth}
		\includegraphics[width=\textwidth]{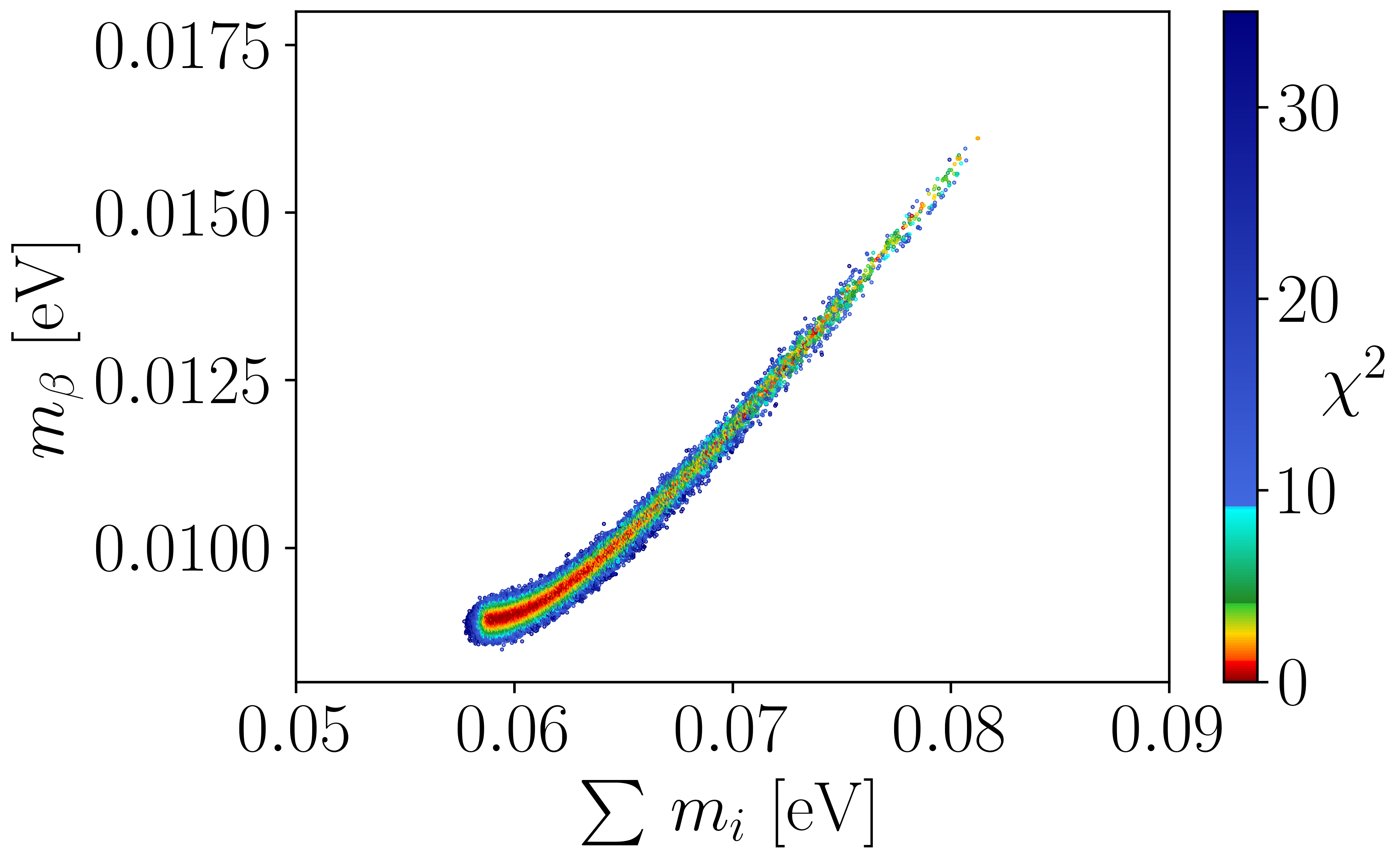} 
		\caption*{C(4)}
		\label{fig:m1no_sub16}
	\end{subfigure}
	\caption{Parameter correlations for model 1 NO in the three $\tau$ regions A, B, and C. The top row panels A(1)-A(6) correspond to region A, the middle row B(1)-B(6) to region B, and the bottom row C(1)-C(4) to region C. The panels display two-dimensional correlations of neutrino parameters including sum of neutrino masses $\sum m_i$, mixing angles $\sin^2\theta_{23}$ and $\sin^2\theta_{13}$, CP-violating phases $\delta_{\rm CP}$, $\eta_1$, and $\eta_2$, absolute neutrino mass $m_1$, the effective Majorana mass $m_{\beta\beta}$, and the modulus $\tau$. The color gradient represents the $\chi^2$ value.}
	\label{fig:m1no}
\end{figure}

Figure \ref{fig:m1no} illustrates the fundamental correlations between neutrino parameters in the three viable regions of $\tau$. The Dirac CP phase $\delta_{\rm CP}$ clusters near maximal values ($\left| \delta_{\rm CP}\right| \approx  \pi/2, 3\pi/2$) in all regions, demonstrating a robust prediction of near-maximal CP violation. A pronounced correlation emerges between $\sin^2\theta_{23}$ and $\delta_{\rm CP}$, where sine square of the atmospheric mixing angle spans the experimentally allowed range ($0.43–0.58$) while maintaining $\delta_{\rm CP}$ within $10\%$ of maximum values.

Absolute neutrino masses exhibit distinct $\tau$-dependencies. Regions A and B show positive correlations between $\sum m_i $ and $\text{Im}\,(\tau)$, with the total mass increasing with the imaginary modulus component. Region B shows a unique linear relationship between $m_1$ and $\text{Re}\,(\tau)$, where the lightest neutrino mass scales proportionally to $\text{Re}\,(\tau)$. This dependence is absent in other regions. Majorana phases show region-specific patterns, with regions A and C exhibiting anti-correlation between $\eta_1$ and $\eta_2$, while region B clusters $\eta_2$ around $0.5\pi$. These differences impact neutrinoless double beta decay predictions. Region A yields $m_{\beta\beta} \approx 0.1$ eV within the current sensitivity, while predictions ($3.4–10.6$) meV of region C fall within next-generation sensitivity ranges. Region B produces bifurcated solutions with both detectable and sub-threshold $m_{\beta\beta}$ values. These correlation structures provide clear experimental targets for distinguishing between $\tau$ regions through precision measurements of CP violation, absolute mass scales, and neutrinoless double beta decay.

Figure 3 shows the prediction of  model 1 for the effective Majorana neutrino mass $m_{\beta\beta}$ as a function of the lightest neutrino mass $m_{1}$ under NO. Theoretical points delineate three distinct regimes. Regions A and B form a plateau at intermediate masses $m_{1} \sim 10^{-4}-10^{-2}$ eV. Region C (on top-right) exhibits a linear rise where $m_{\beta\beta} \propto m_{1}$ for $m_{1} \gtrsim 0.04$ eV. Noticeably, the entire predicted curve—spanning regions A, B, and C traces the boundary of the gray dashed area, indicating that model 1 saturates the NO-favored parameter space while remaining consistent with this mass ordering preference.  
\begin{figure}[h!]
	\centering
	\includegraphics[width=0.7\textwidth]{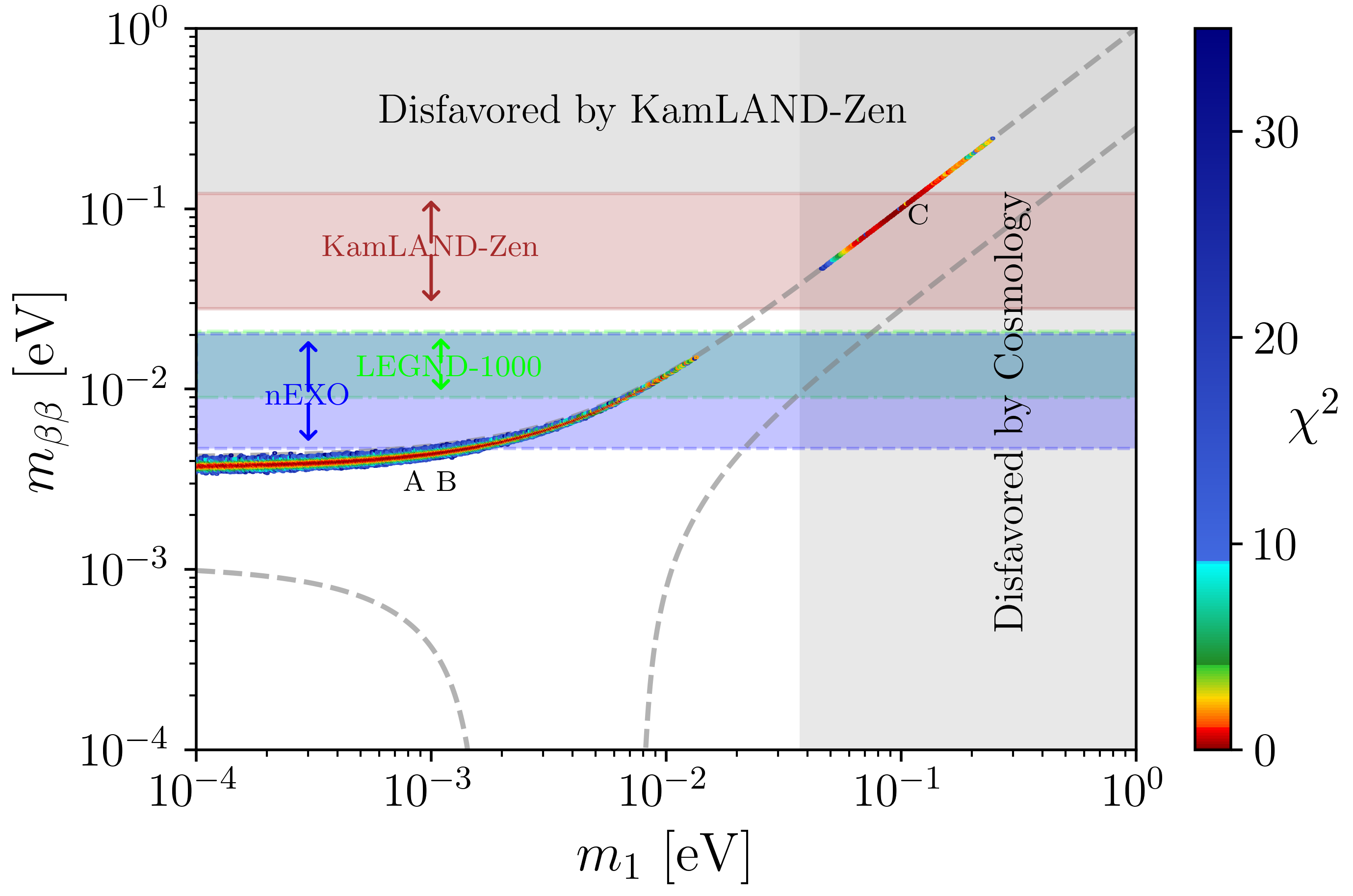}
	\caption{Prediction of model 1 for the effective Majorana neutrino mass $m_{\beta\beta}$ as a function of the lightest neutrino mass $m_{1}$ under NO. Regions A, B (some overlap with A), and C denote distinct phenomenological regimes. The gray dashed area indicates parameter space favored by NO. Experimental constraints include KamLAND-Zen exclusion limit (brown band, $m_{\beta\beta} < 0.028$–$0.122~\mathrm{eV}$~\cite{KamLAND-Zen:2024eml}), target sensitivities of next-generation $0\nu\beta\beta$ experiments such as LEGEND-1000 (green band, $0.009$–$0.021~\mathrm{eV}$~\cite{LEGEND:2021bnm}) and nEXO (blue band, $0.0047$–$0.0203~\mathrm{eV}$~\cite{nEXO:2021ujk}), as well as cosmological exclusion of $m_{1} \gtrsim 0.037~\mathrm{eV}$  (vertical gray band~\cite{GAMBITCosmologyWorkgroup:2020rmf}).}
	\label{fig:m1no_0vbb}
\end{figure}

Experimental constraints further contextualize these predictions: KamLAND-Zen (brown band) excludes the upper regime from region C, while next-generation experiments nEXO (blue) and LEGEND-1000 (green) will probe the  regions (A/B) where the curve intersects their sensitivity bands. The vertical gray band (cosmological bounds) disfavors $m_{1} \gtrsim 0.037$ eV, excluding the rightmost portion of Region C. This positioning demonstrates that the phenomenology of model 1 is testable across the viable NO parameter space, with regions A/B particularly accessible to imminent $0\nu\beta\beta$ searches.

\begin{figure}[h!]
	\centering
	\includegraphics[width=0.6\textwidth]{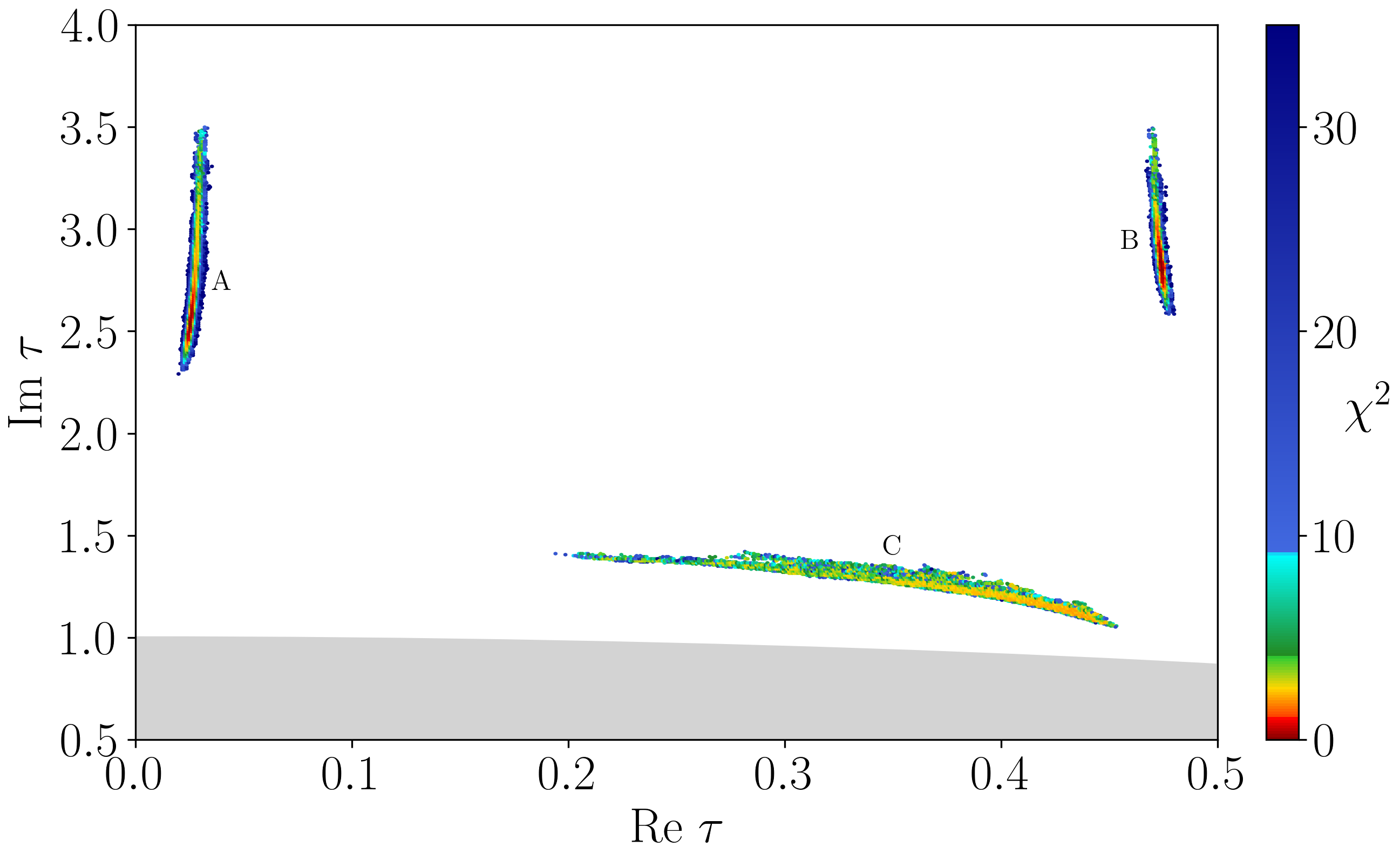}
	\caption{Distribution of the modular parameter $\tau$ in model 1 for IO case. There exist three distinct viable regions A, B, and C that reproduce observed oscillation parameters. Regions A  and B occupy the high-imaginary regime ($\text{Im}(\tau) > 2$), while region C  appears at intermediate values.}
	\label{fig:m1io_tau}
\end{figure}

Having discussed the case of NO, we now analyze its implications for the IO case. The modular parameter $\tau$ distributes in three disconnected regions A, B, and C, as shown in figure~ \ref{fig:m1io_tau}. These regions correspond to local minima in the $\chi^2$ landscape, each yielding compatible neutrino oscillation parameters. Regions A and B occupy the upper half-plane with $\text{Im}(\tau) > 2$, suggesting a preference for larger imaginary components. Their position appear near the boundary of the fundamental domain. Region C, with $\text{Im}(\tau) \approx 1.2$, lies at lower imaginary values. The best-fit parameters for these regions are summarized in table~\ref{tab:m1io_fit}.  Region C gives a larger $\chi^2$ value (1.9821) compared to Regions A (0.01233) and B (0.02723), suggesting a poorer fit to the current oscillation data. This region predicts a neutrino mass sum $\sum m_i= 0.09989$ eV, which lies below the cosmological upper limit of $0.12$ eV. In contrast, regions A and B yield larger mass sums $0.2427$ eV and $0.2384$ eV respectively, potentially in tension with cosmological constraints.

\begin{figure}[h!]
	\centering
	\begin{subfigure}[b]{0.3\textwidth}
		\includegraphics[width=\textwidth]{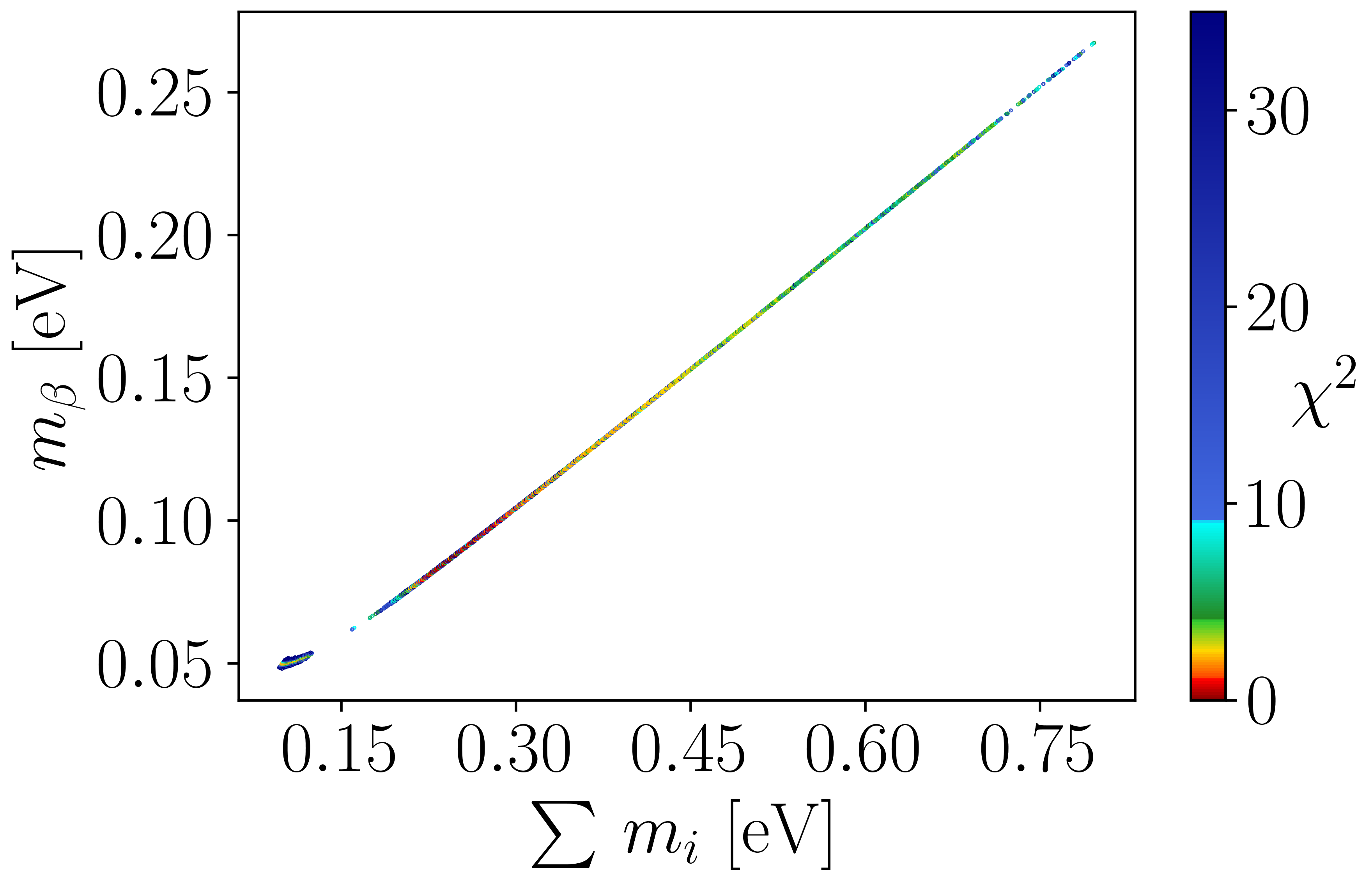}
		\caption{}
	\end{subfigure}
	\hfill
	\begin{subfigure}[b]{0.3\textwidth}
		\includegraphics[width=\textwidth]{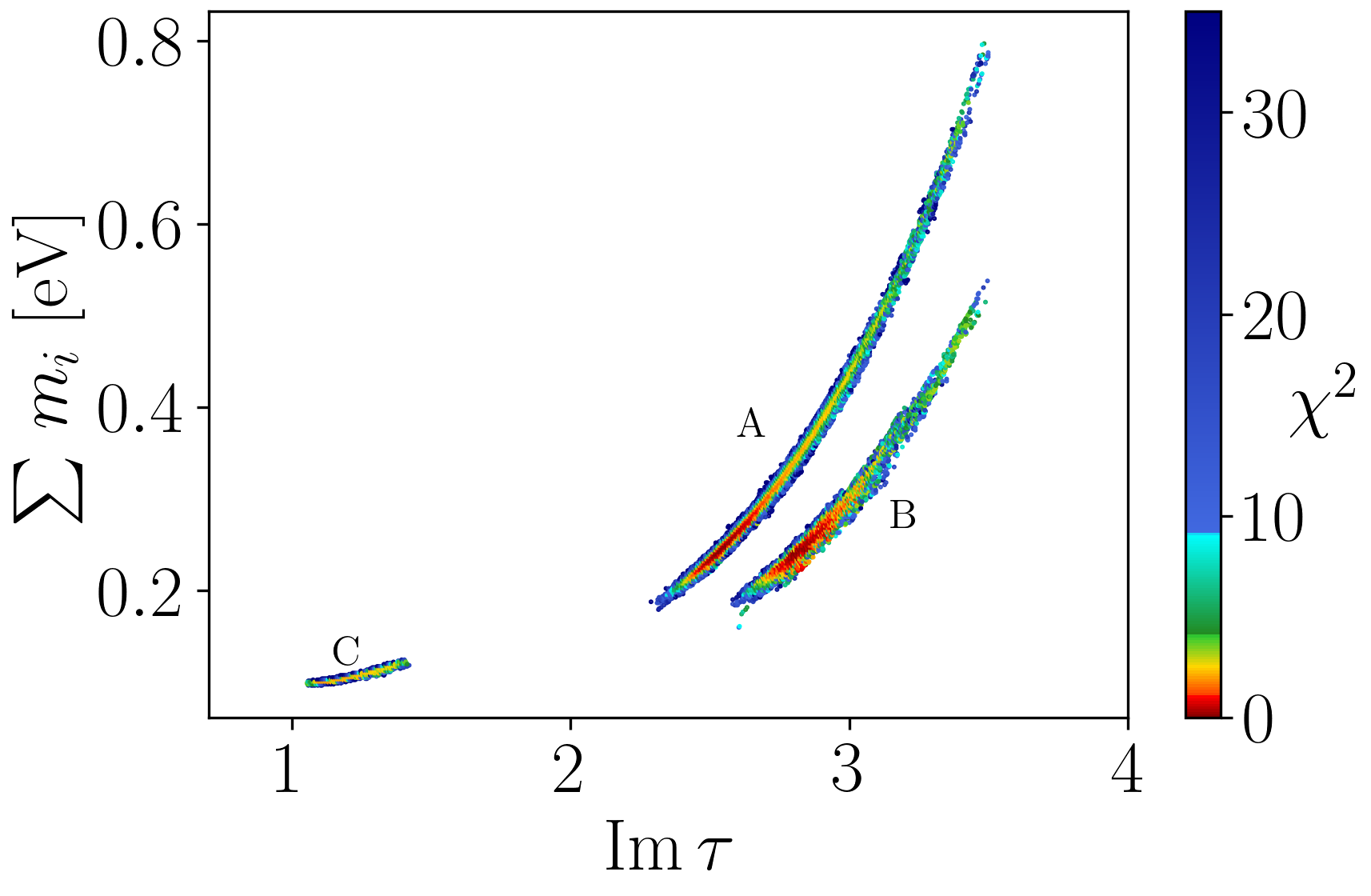}
		\caption{}
	\end{subfigure}
	\hfill
	\begin{subfigure}[b]{0.3\textwidth}
		\includegraphics[width=\textwidth]{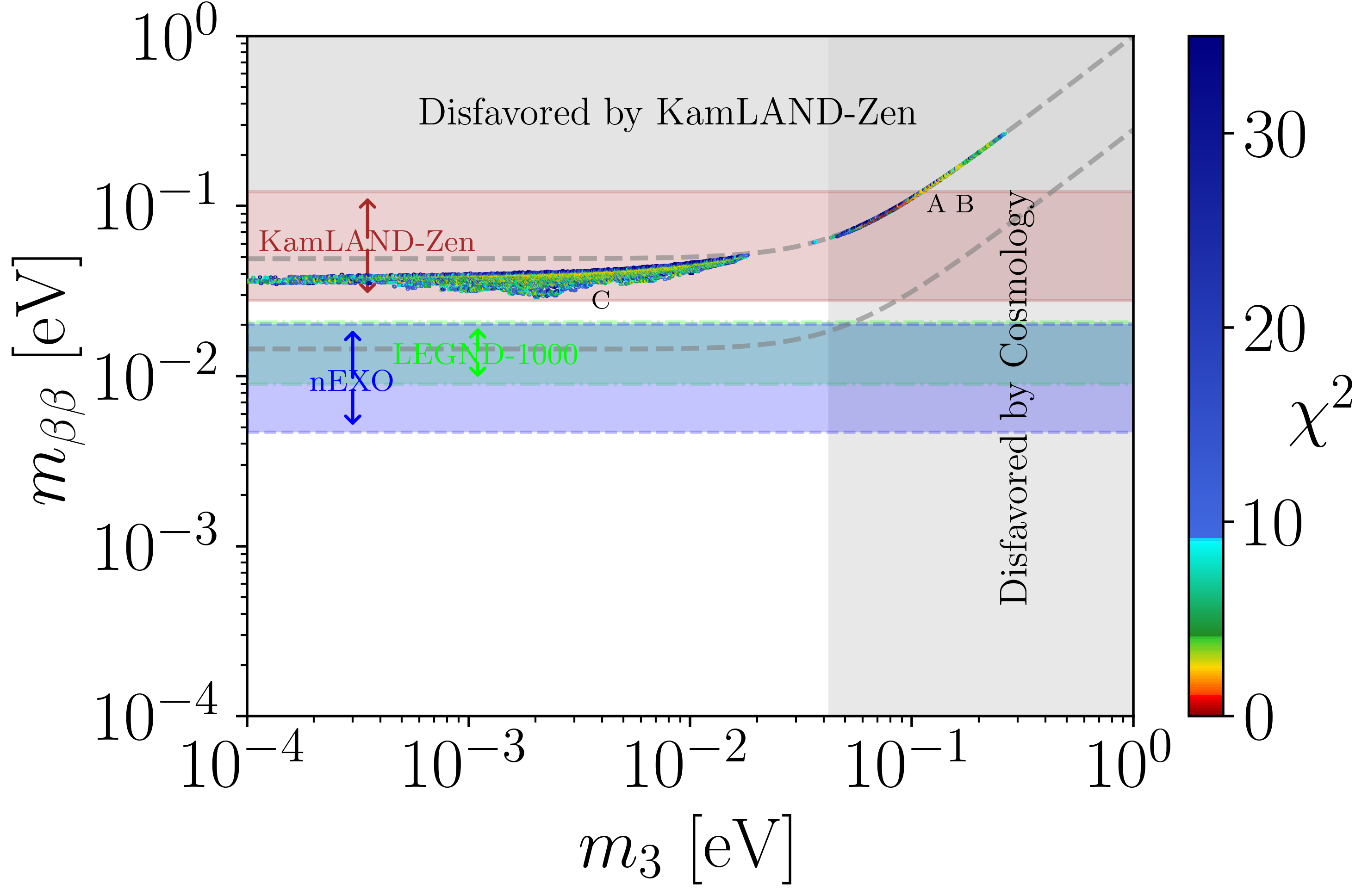}
		\caption{}
	\end{subfigure}
	\vspace{0.5cm}	
	\begin{subfigure}[b]{0.3\textwidth}
		\includegraphics[width=\textwidth]{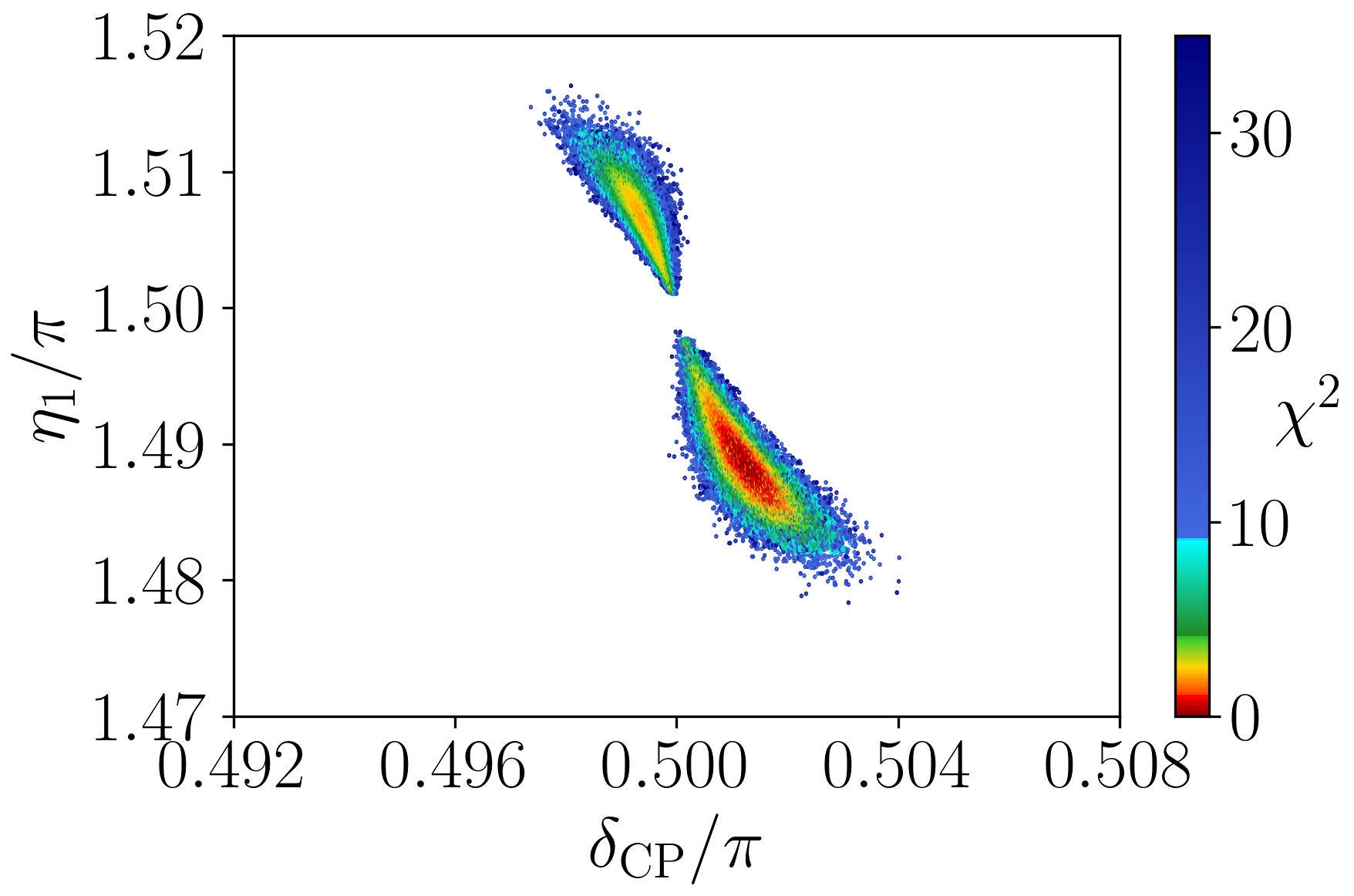}
		\caption*{A(1)}
	\end{subfigure}
	\hfill
	\begin{subfigure}[b]{0.3\textwidth}
		\includegraphics[width=\textwidth]{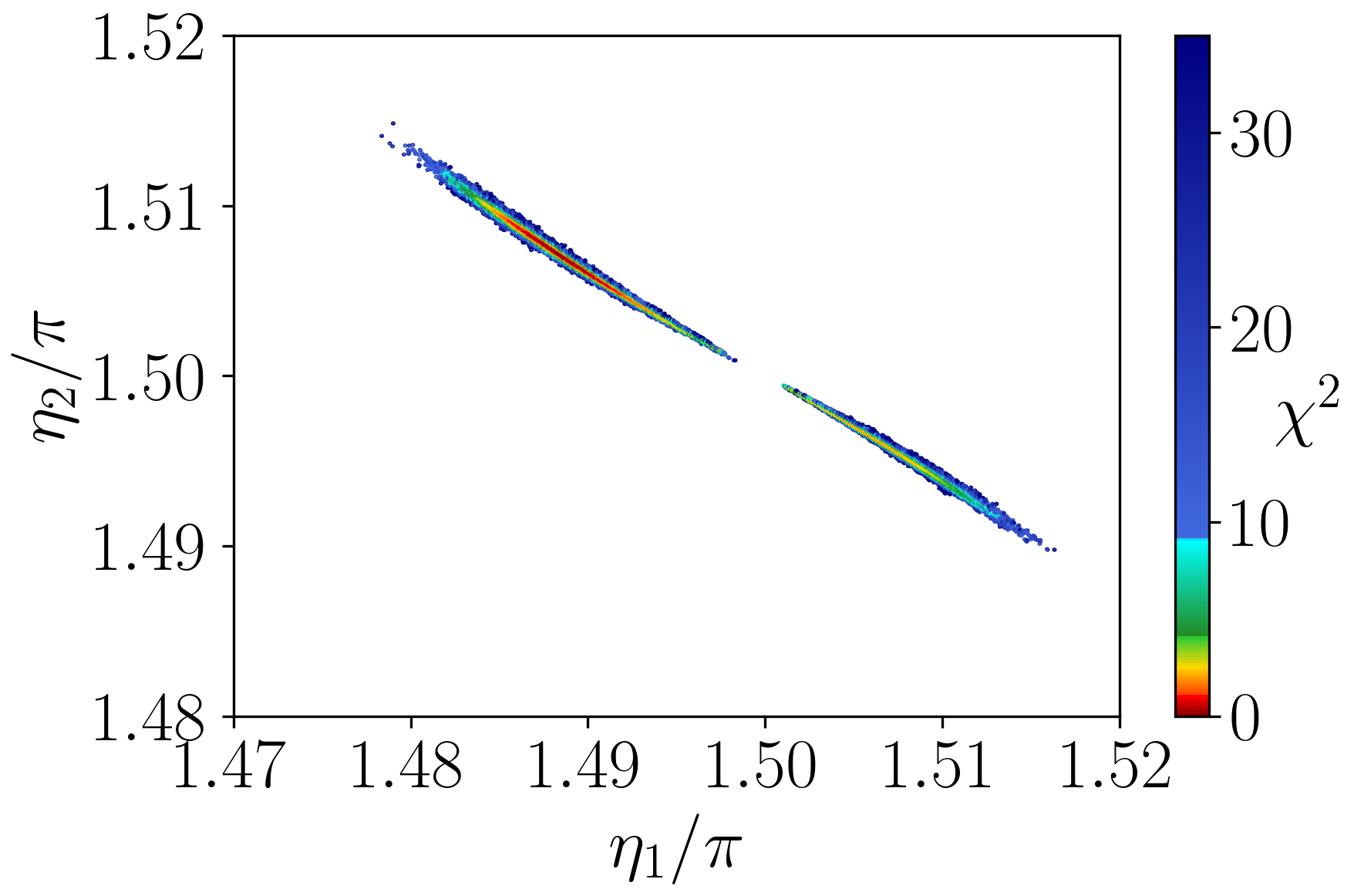}
		\caption*{A(2)}
	\end{subfigure}
	\hfill
	\begin{subfigure}[b]{0.3\textwidth}
		\includegraphics[width=\textwidth]{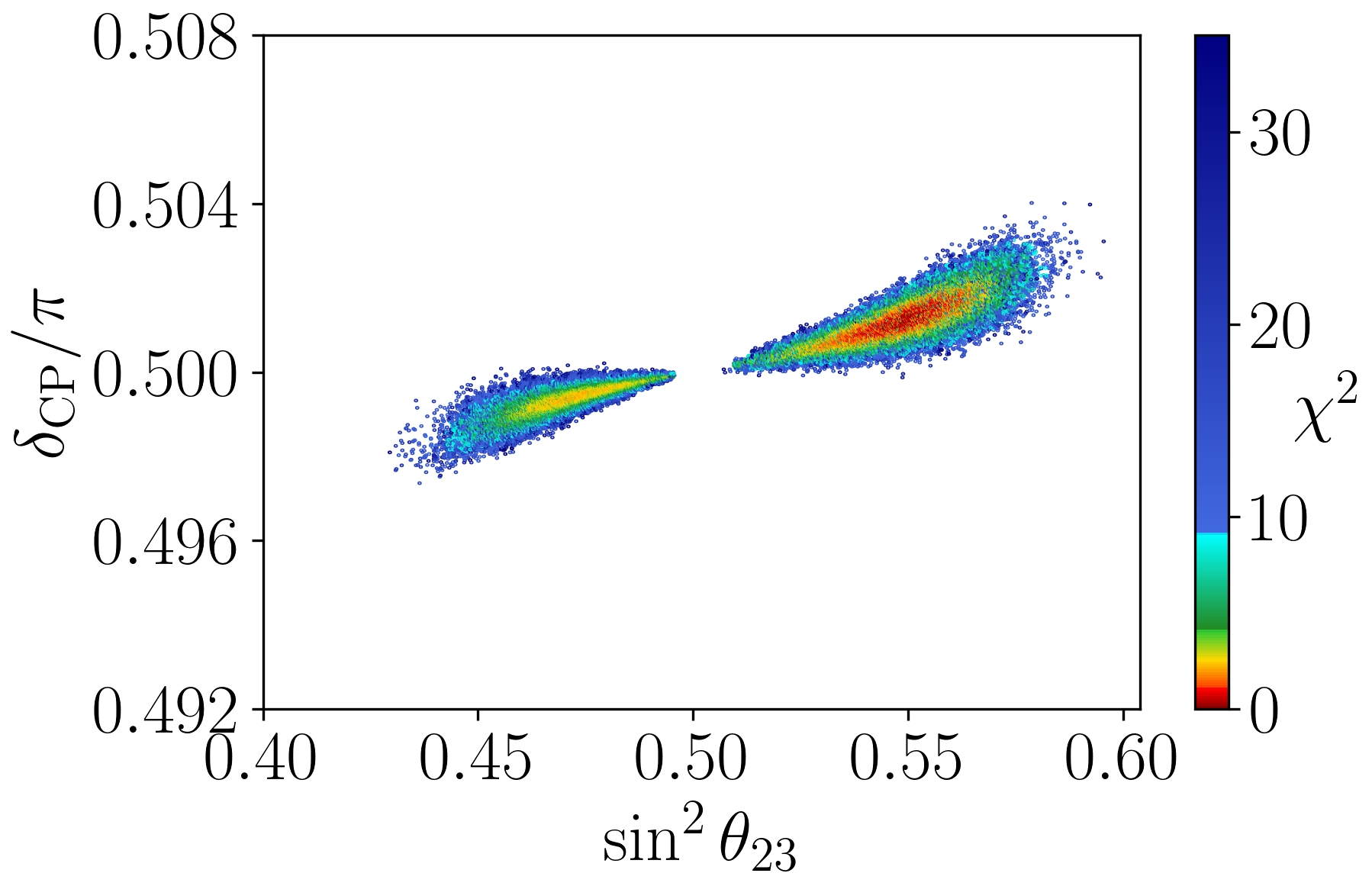}
		\caption*{A(3)}
	\end{subfigure}
	\vspace{0.5cm}
	\begin{subfigure}[b]{0.3\textwidth}
		\includegraphics[width=\textwidth]{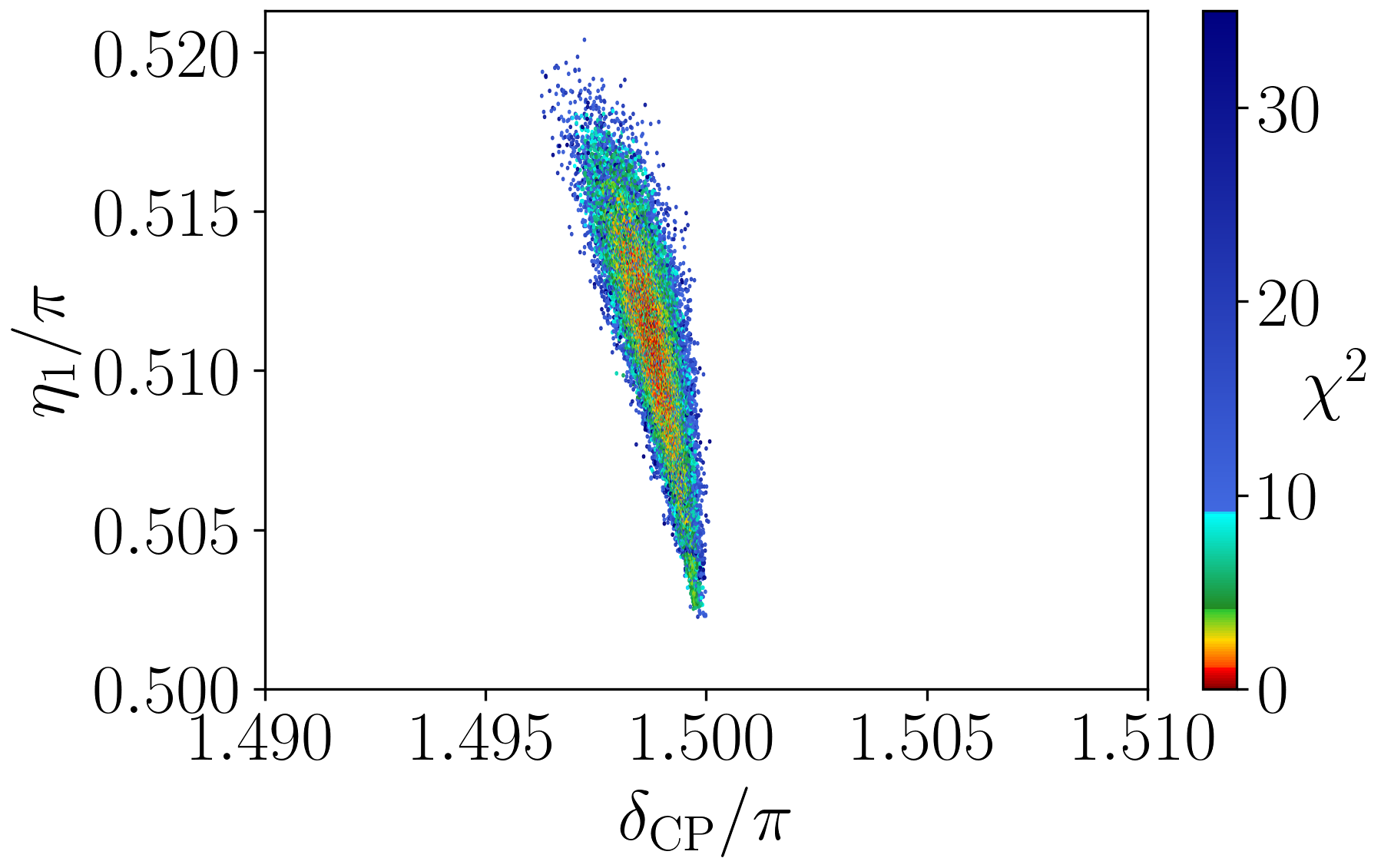}
		\caption*{B(1)}
	\end{subfigure}
	\hfill
	\begin{subfigure}[b]{0.3\textwidth}
		\includegraphics[width=\textwidth]{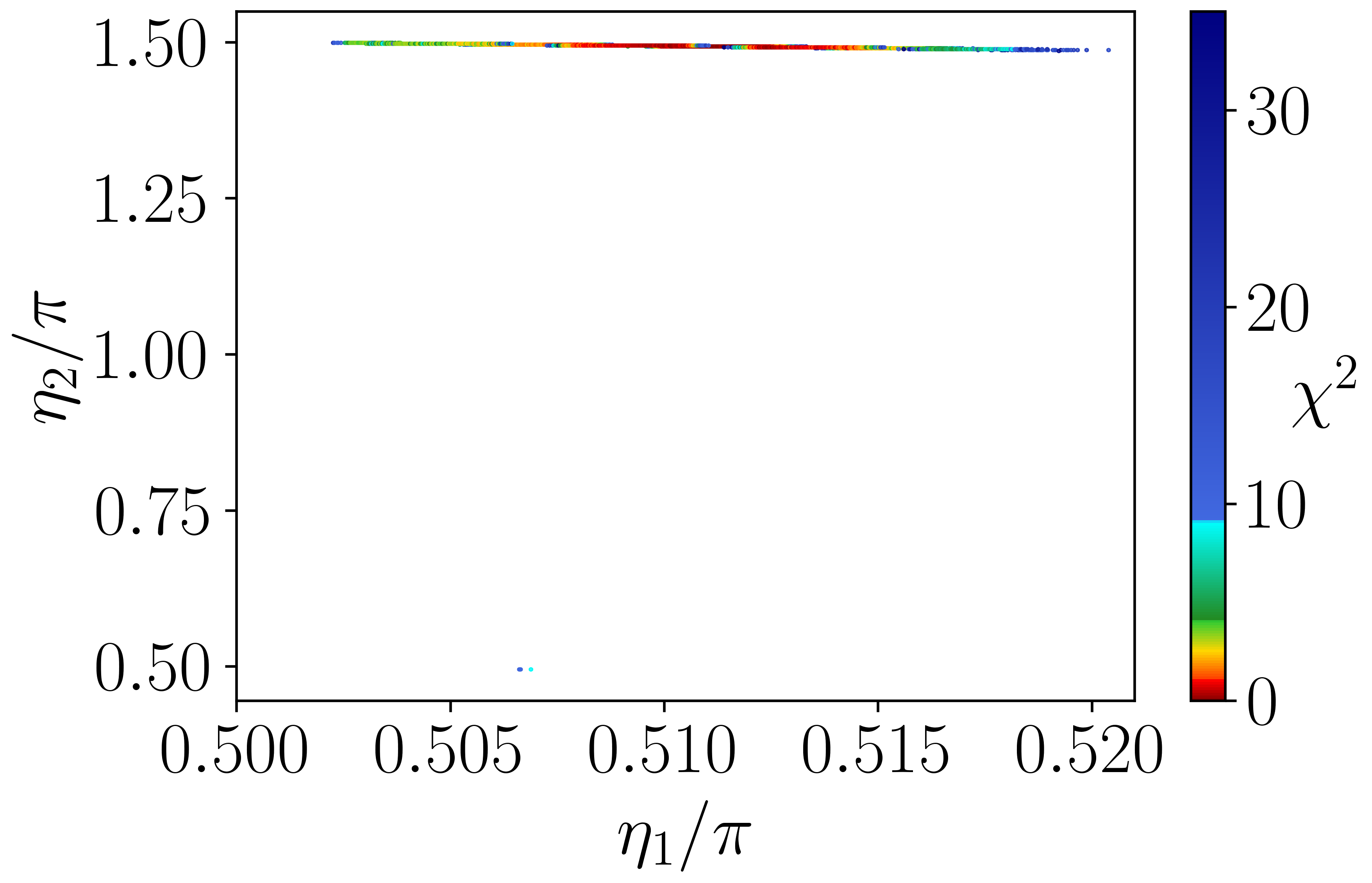}
		\caption*{B(2)}
	\end{subfigure}
	\hfill
	\begin{subfigure}[b]{0.3\textwidth}
		\includegraphics[width=\textwidth]{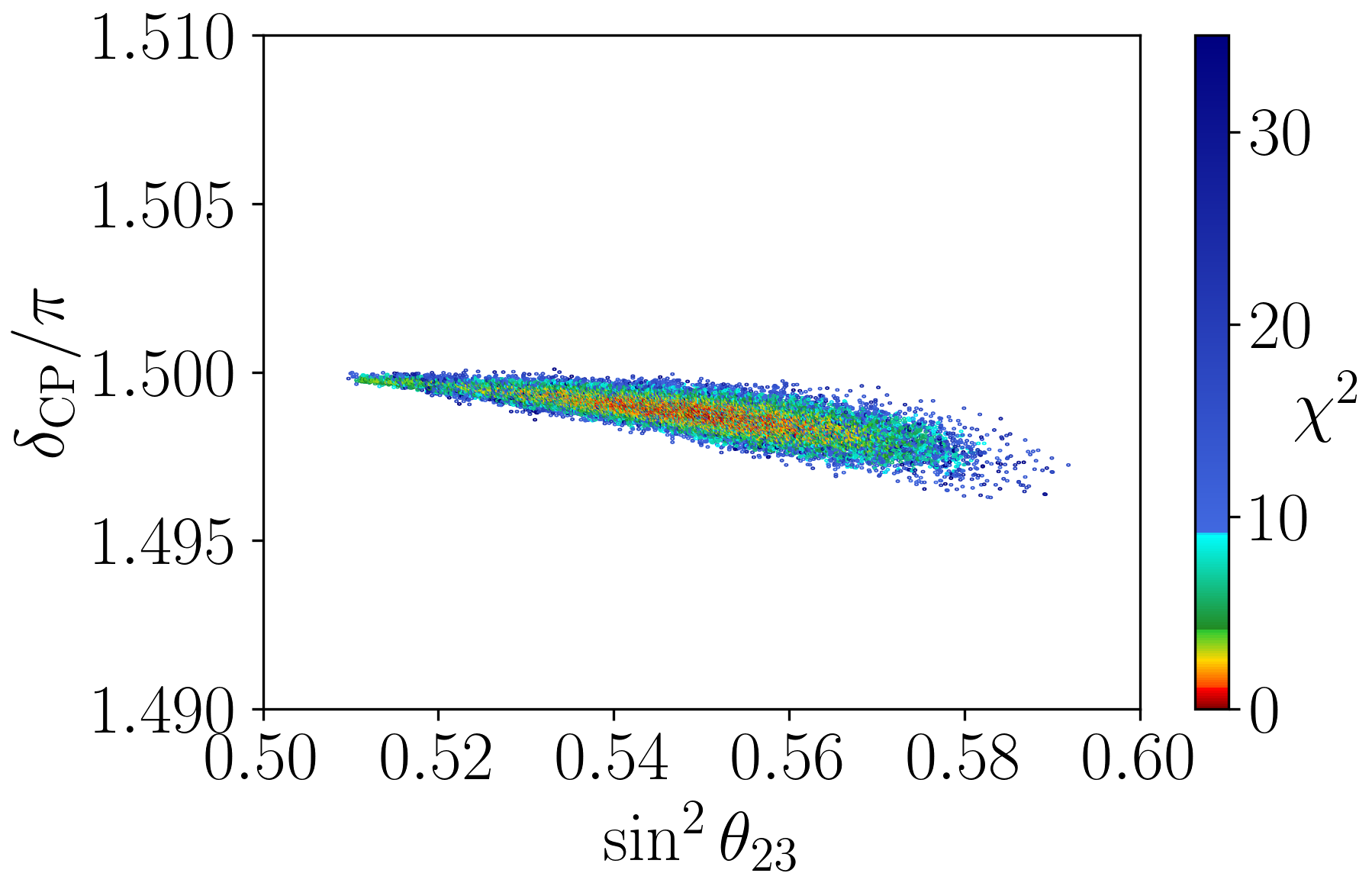}
		\caption*{B(3)}
	\end{subfigure}
	\vspace{0.5cm}
	\begin{subfigure}[b]{0.3\textwidth}
		\includegraphics[width=\textwidth]{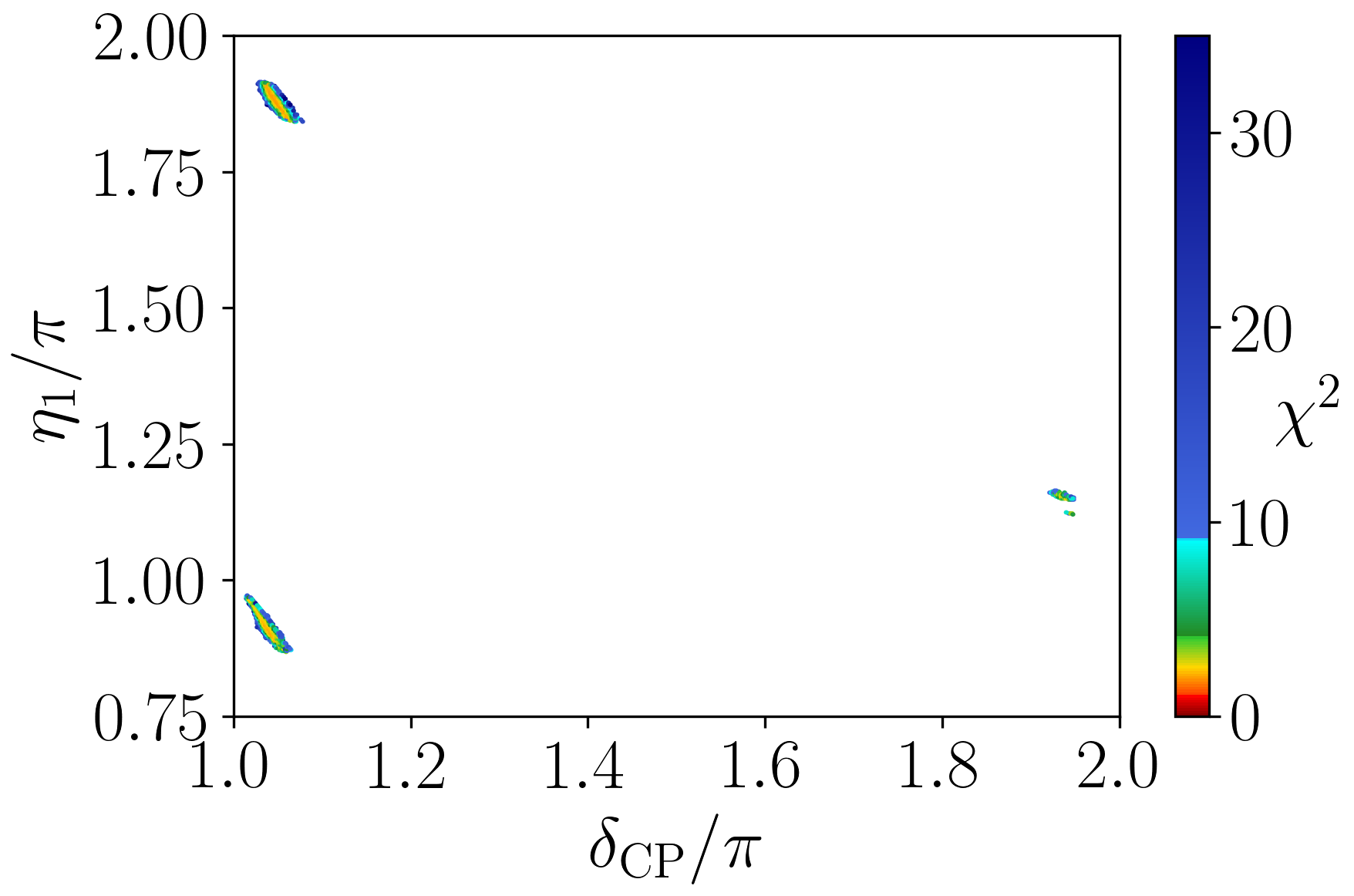}
		\caption*{C(1)}
	\end{subfigure}
	\hfill
	\begin{subfigure}[b]{0.3\textwidth}
		\includegraphics[width=\textwidth]{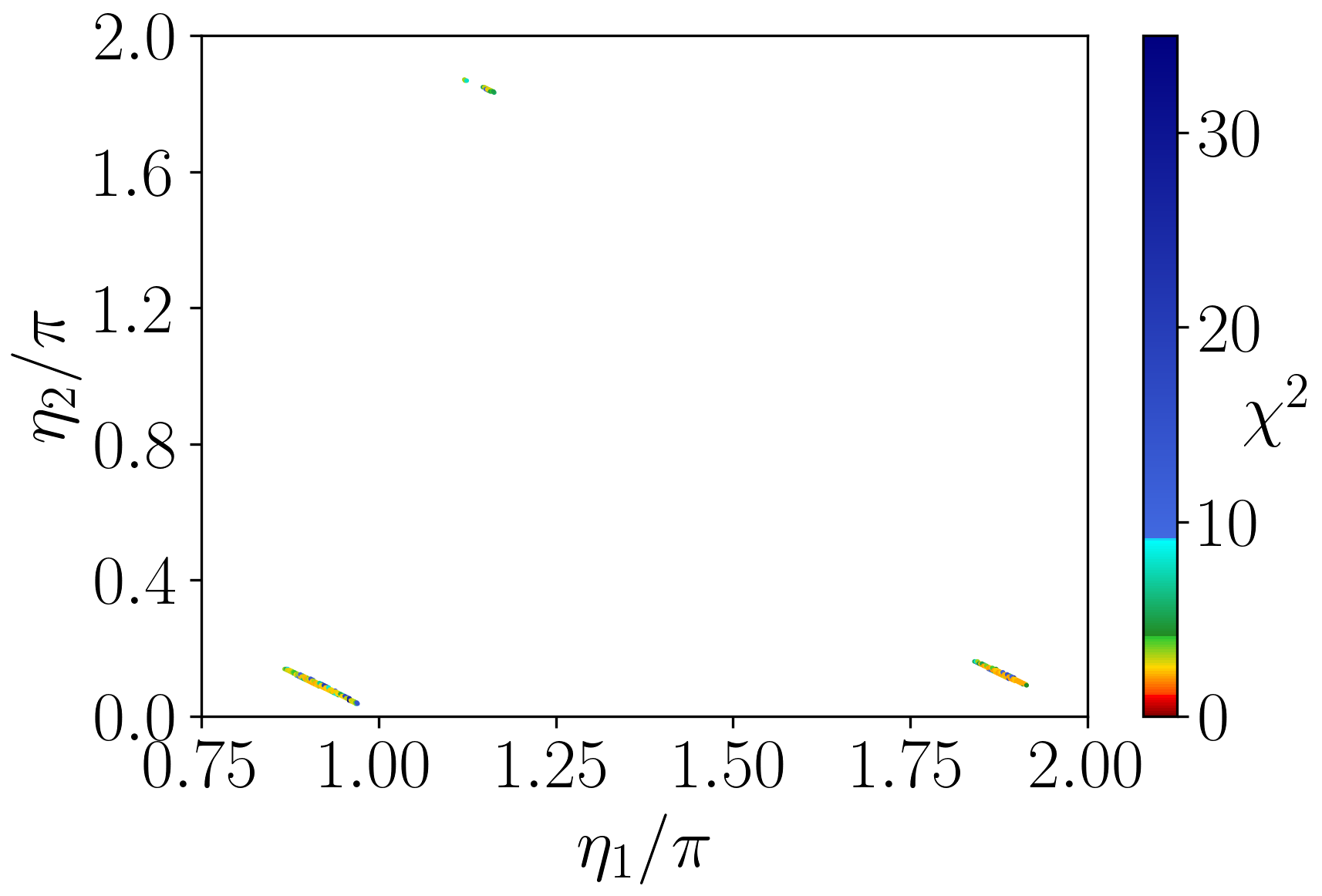}
		\caption*{C(2)}
	\end{subfigure}
	\hfill
	\begin{subfigure}[b]{0.3\textwidth}
		\includegraphics[width=\textwidth]{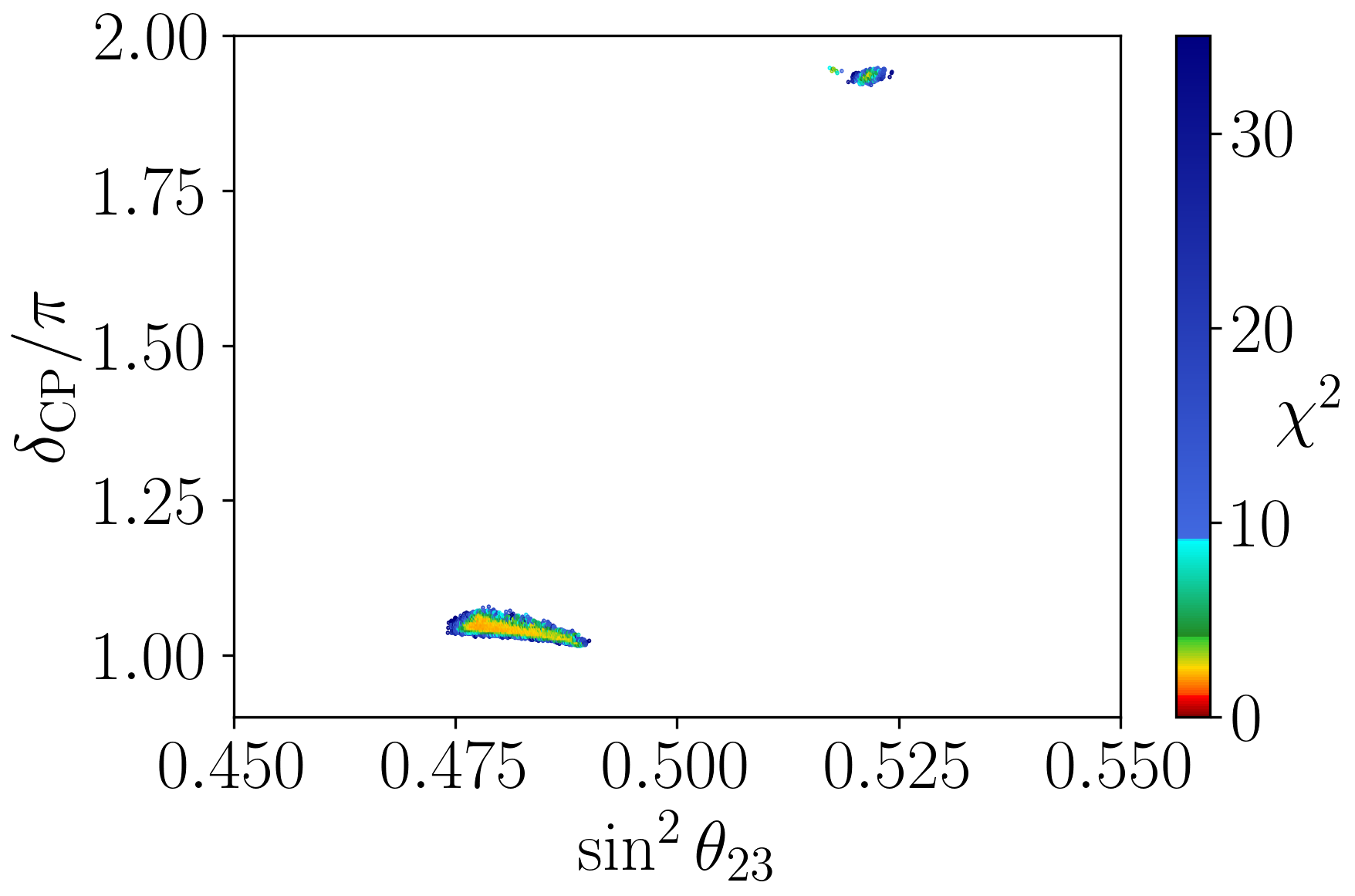}
		\caption*{C(3)}
	\end{subfigure}	
	\caption{Parameter correlations for model 1 under  IO case, organized by  three regions A, B, and C. Top three panels exhibit the following:  (a) $m_\beta$ vs. $\sum m_i$, (b) $\sum m_i$ vs. $\text{Im}(\tau)$, (c) $m_{\beta\beta}$ vs. Lightest mass $m_3$. Lower panels show region-specific correlations for regions A (A(1)-A(3)), B (B(1)-B(3)), and C (C(1)-C(3)), with columns representing CP phases and $\theta_{23}$ octant dependence relationships respectively.}
	\label{fig:m1io}
\end{figure}

In the IO scenario of model 1, similar to the NO case, we observe that $\tau$ is distributed across three disconnected regions, as shown in figure \ref{fig:m1io_tau}. The best-fit values for these regions are summarized in table~\ref{tab:m1io_fit}.
\begin{table}[h!]
	\begin{center}
		\renewcommand{\arraystretch}{1.3}
		\setlength{\tabcolsep}{13pt}
		\begin{tabular}{c|c|c|c}\toprule
			\text{Model 1 (NO)}& \text{region A} &\text{region B}&\text{region C}\\
			\hline  
			$\tau $&$0.02594+2.5360i$&$0.4747+2.7989i$&$0.4392+1.1027i$\\
			$\frac{\beta_{D}}{\alpha_{D} }$&185.5795&779.9672&739.9807\\ 
			$\frac{\gamma_{D}}{\alpha_{D} }$&898.1528&737.3879&934.7742\\
			$\frac{\beta_{{NS}}}{\alpha_{{NS}}}$&950.2446&830.5287&689.3298\\
			$\frac{\gamma_{{NS}}}{\alpha_{{NS}} }$&189.4644&53.2445&912.4368\\
			$\frac{\beta_{S}}{\alpha_{S} }$&354.7910&2.4657&0.006796\\
			\hline  
			$k$ (eV)&0.0007310&0.05260&0.002476\\
			\hline  
			$\sin^2\theta_{12}$&0.3098&0.3099&0.3052\\
			$\sin^2\theta_{13}$&0.02229&0.02229&0.02233\\
			$\sin^2\theta_{23}$&0.5508&0.5524&0.4775\\
			$\delta_{\rm CP}/\pi$&0.5012&1.499&1.048\\
			$\eta_1/\pi$&1.488&0.5115&1.882\\
			$\eta_2/\pi$&1.507&1.493&0.1193\\
			\hline 
			$\Delta m^2_{21}/10^{-5}$ (eV$^2$)&7.497&7.490&7.487\\
			$\Delta m^2_{32}/10^{-5}$ (eV$^2$)&-2.482&-2.484&-2.485\\
			$m_1$ (eV)&0.08589&0.08456&0.04910\\
			$m_2$ (eV)&0.08633&0.08500&0.04986\\
			$m_3$ (eV)&0.07050&0.06886&0.0009332\\
			$\sum m_i$ (eV)&0.2427&0.2384&0.09989\\
			$m_{\beta}$ (eV)&0.08630&0.08498&0.04981\\
			$m_{\beta\beta}$ (eV)&0.08555&0.08423&0.03760\\
			\hline 
			$\chi^2_{\rm min}$&0.01233&0.02723&1.9821\\
			\bottomrule
		\end{tabular}   
	\end{center}   
	\caption{Best-fit parameters for model 1 in IO across the three viable $\tau$ regions A, B, and C. Regions A and B exhibit $\sum m_i > 0.12$ eV (cosmologically disfavored) with maximal CP violation, while region C shows $\sum m_i = 0.09989$ eV (cosmologically allowed) with suppressed CP violation. The $\chi^2_{\rm min}$ values indicate regions A/B provide superior fits to current oscillation data compared to region C.}
	\label{tab:m1io_fit}
\end{table}
In particular, region C yields a minimum $\chi^2$ value of $1.9821$, which is greater than those of regions A (0.1233) and B (0.02723). Despite this, the neutrino mass $\sum m_i$ in region C remains below 0.12 eV. In contrast, the best-fit points for regions A and B predict neutrino masses greater than this limit, with values clustering around 0.24 eV. The Dirac and Majorana phases predictions for regions A and B closely resemble those observed in the NO scenario, as illustrated in figure \ref{fig:m1io}. 
The parameter correlations and ranges predicted by these two regions exhibit similarities to those of region A in the NO case. The upper row displays global relationships. Panel (a) shows the linear correlation between the neutrino mass sum and the effective beta-decay mass, confirming consistent mass scaling of the model. Panel (b) shows a proportionality between $\text{Im}(\tau)$ and $\sum m_i$, which demonstrates how the modular parameter governs the absolute neutrino mass scale. Panel (c) plots $m_{\beta\beta}$ versus the lightest neutrino mass $m_3$, which exhibits the overlapping results of observed in the Majorana mass predictions in these three regions. 

The Dirac CP Phase $\delta_{\rm CP}$ of regions A and B show it is clustered near maximum values $\pi/2$ or $3\pi/2$, while region C allows moderate CP violation, $\delta_{\rm CP} \approx  1.05\pi$. Regions A and B favor Majorana CP phase configurations $\eta_1 \approx \pi/2$ or $\pi$, $\eta_2 \approx \pi$, this can lead to a strong interference in $0\nu\beta\beta$ decay that suppresses $m_{\beta\beta}$ (0.084–0.085 eV) despite large $\sum m_i > 0.12$ eV, while region C allows CP-violating alignment $\eta_1 \approx 1.88\pi$, $\eta_2 \approx 0.12\pi$, yielding $m_{\beta\beta} \approx 0.038$ eV. These phase distributions arise from modular symmetry breaking—high $\text{Im}(\tau)$ stabilizes phases at fixed points, whereas lower $\text{Im}(\tau)$ permits deviation—and are correlated with the Dirac CP violation. The maximum $\delta_{\rm CP}$ in region A/B coexists with the CP-conserving Majorana phases, while the moderate $\delta_{\rm CP} \approx 1.05\pi$ in region C accompanies full leptonic CP violation. The phase-driven $m_{\beta\beta}$ suppression in regions A/B and enhancement in region C provide testable signatures for next-generation $0\nu\beta\beta$ experiments, with the potential to distinguish between these symmetry regimes.  

The octant-dependent $\theta_{23}$ behavior manifests itself most in the $\theta_{23}$- $\delta_{\rm CP}$ plane  in figures A(3), B(3), and C(3). For $\theta_{23}>45^\circ$ in regions A and B, $\delta_{\rm CP}$ correlates with $\theta_{23}$ in region A but anticorrelates in region B. For $\theta_{23}<45^\circ$ in regions A and C, $\delta_{\rm CP}$ positively correlates with $\theta_{23}$ in region A but slightly anticorrelates in region C. This reversal across the octant boundary provides a useful signature to test the $\theta_{23}$ octant problem.

In general, model 1 exhibits a rich phenomenology that encompasses both maximal Dirac CP violation and scenarios with minimal CP-violating effects. In particular, in parameter space regions where the imaginary component of the modular parameter $\tau$ satisfies $\text{Im}(\tau)>2$, predictions for NO and IO scenarios converge. It demonstrates a positive correlation between $\text{Im}(\tau)$ and neutrino mass. The distinctive behavior of the atmospheric mixing angle $\theta_{23}$ around $45^{\circ}$ is particularly striking. The model predicts a fundamental change in the $\theta_{23}$-mass correlation depending on which side of this critical value the mixing angle is located. For $\theta_{23}>45^\circ$, the correlation between $\theta_{23}$ and neutrino mass parameters inverts relative to the $\theta_{23}<45^\circ$ regime, representing a characteristic signature of the underlying modular symmetry structure.

\subsection{Phenomenological implications of model 2}

\begin{figure}[h!]
	\centering
	\includegraphics[width=0.6\textwidth]{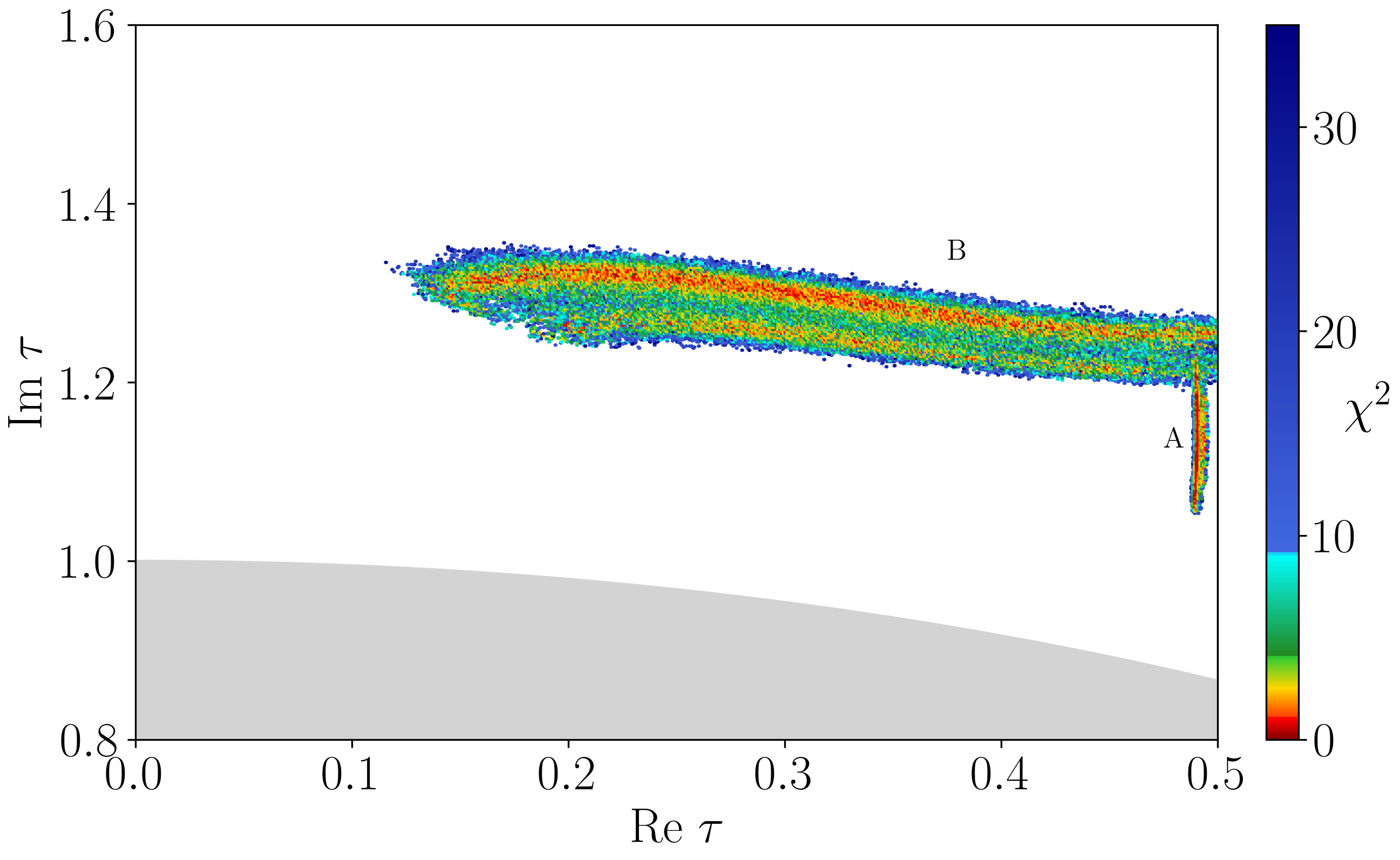}
	\caption{Viable $\tau$ parameter space for model 2 under NO. Two regions A and  B have partial overlap, which indicates shared symmetry constraints despite phenomenological differences.}
	\label{fig:m2no_tau}
\end{figure}

Having explored the phenomenology of model 1 under both mass orderings, we now turn to model 2, which features distinct modular symmetry assignments and Yukawa structures. This model with NO exhibits broader parametric freedom in the complex $\tau$ plane, as evidenced by figure~\ref{fig:m2no_tau}, where two viable regions A and B emerge with a partially overlapping parameter space.
Despite of the overlap, we perform separate discussions, due to their differing predictions. For region A, the best-fit point corresponds to $\tau = 0.4906 + 1.0965i$, with $\chi^2_{\rm min} = 0.09846$, and the relevant physical parameters are calculated as
\begin{equation}
  \begin{aligned}
    &m_e/m_{\mu}=0.004738,\quad m_{\mu}/m_{\tau}=0.05882,\\
    &\sin^2\theta_{12}=0.3056,\quad \sin^2\theta_{13}=0.02205,\quad \sin^2\theta_{23}=0.4716,\\
    &\delta_{\rm CP}=0.2374\pi,\quad \eta_1=1.031\pi,\quad \eta_2=1.776\pi,\\
    &m_1=0.03105~\mathrm{eV},\quad m_2=0.03223~\mathrm{eV},\quad m_3=0.05895 ~\mathrm{eV},\\
    &\sum m_i=0.1222~\mathrm{eV},\quad m_{\beta}=0.03230 ~\mathrm{eV},\quad m_{\beta\beta}=0.02334~\mathrm{eV}.
    \end{aligned}
    \label{eq:mod2noregA}
  \end{equation}
This accurately reproduces the data of neutrino oscillation experiments. As shown, the charged lepton mass ratios, the neutrino masses, the effective mass in the $\beta$ decay, as well as the $0\nu\beta\beta$ are within the current experimentally allowed ranges. 

For region B, the corresponding modulus at $\tau = 0.1690+1.3152i $ results in $\chi^2_{\rm min} = 0.1066$. Computed best-fit values for the physical observables are the following:
\begin{equation}
  \begin{aligned}
    &m_e/m_{\mu}=0.004737,\quad m_{\mu}/m_{\tau}=0.05883,\\
    &\sin^2\theta_{12}=0.3068,\quad \sin^2\theta_{13}=0.02227,\quad \sin^2\theta_{23}=0.4723,\\
    &\delta_{\rm CP}=1.203\pi,\quad  \eta_1=1.194\pi,\quad  \eta_2=0.2554\pi,\\
    &m_1=0.0009878~\mathrm{eV},\quad  m_2=0.008706~ \mathrm{eV},\quad  m_3=0.05017 ~\mathrm{eV},\\
    &\sum m_i=0.05986~\mathrm{eV},\quad  m_{\beta}=0.009005~ \mathrm{eV},\quad  m_{\beta\beta}=0.002207~\mathrm{eV}
    \end{aligned}
  \end{equation}

\begin{figure}[h!]
	\centering
	\includegraphics[width=0.7\textwidth]{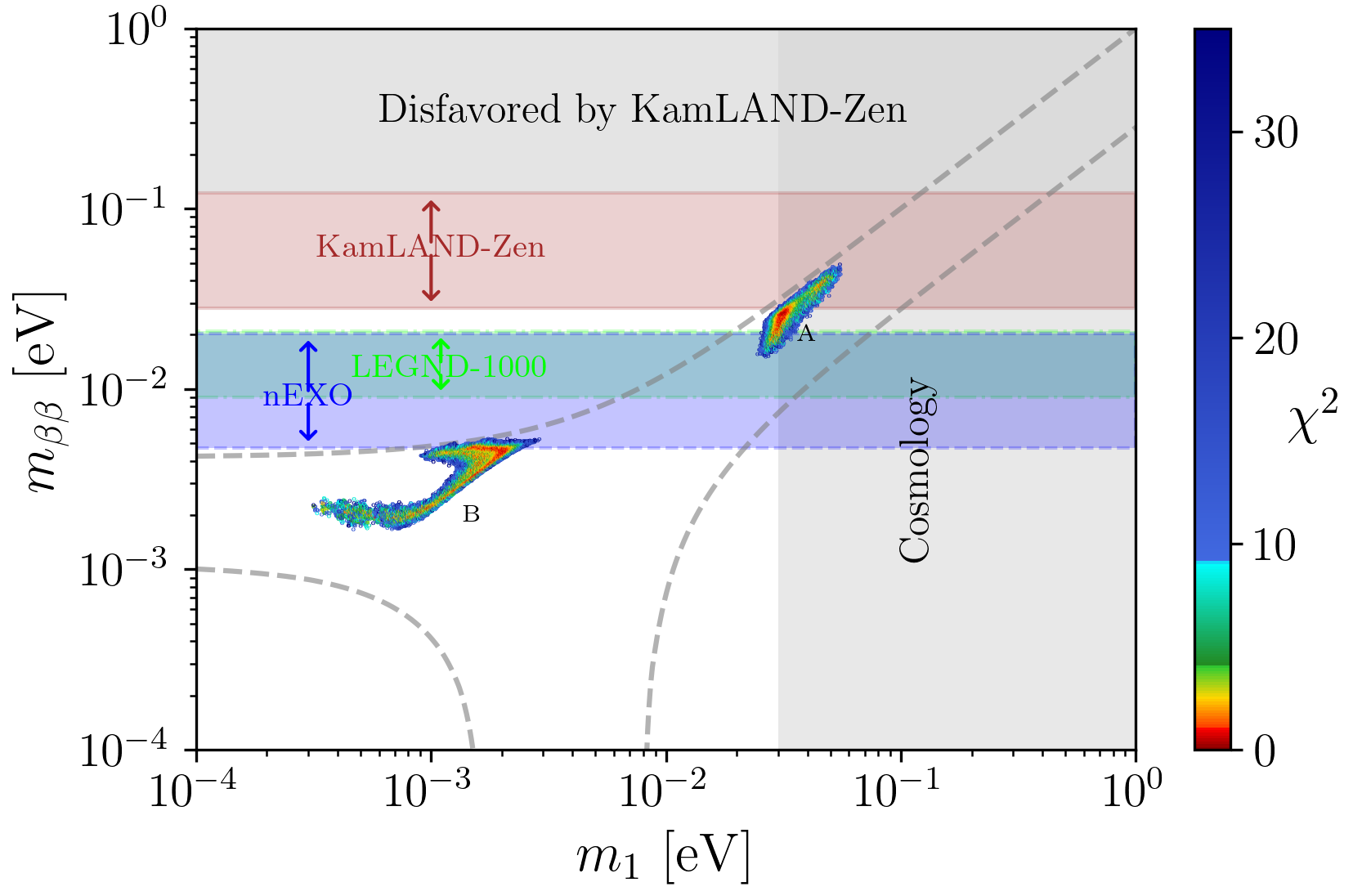}
	\caption{Viable regions for $m_{\beta\beta}$ versus lightest neutrino mass $m_{1}$ in model 2 with NO. Region A predicts $m_{\beta\beta} \approx 0.023$ eV within nEXO/LEGEND sensitivity. Region B shows strong suppression ($m_{\beta\beta} \lesssim 0.003$ eV) due to Majorana phase cancellation, evading next-generation experiments. }
	\label{fig:m2no_0vbb}
\end{figure}

Figure~\ref{fig:m2no_0vbb} displays the prediction of model 2 for the effective Majorana neutrino mass \(m_{\beta\beta}\) as a function of the lightest neutrino mass $m_{1}$ in the NO case. The plot includes two distinct phenomenological regimes corresponding to regions A and B identified in figure~\ref{fig:m2no_tau}.
Region A covers the domain $m_{1} \gtrsim 0.03$ eV, where $m_{\beta\beta}$ clusters near $0.023$ eV, within the sensitivity range of next-generation $0\nu\beta\beta$ experiments such as nEXO and LEGEND-1000. This reflects constructive interference between mass eigenstates, consistent with Majorana phases $\eta_1 \approx 1.03\pi$ and $\eta_2 \approx 1.78\pi$ from eq.~\eqref{eq:mod2noregA}.  Region B exhibits strong suppression at low masses when $m_{1} \lesssim 0.001$ eV, with $m_{\beta\beta} \approx 0.002$ eV falling below current experimental detection thresholds. This suppression stems from destructive interference between Majorana phases $\eta_1 \approx 1.19\pi$, $\eta_2 \approx 0.26\pi$, resulting in a cancellation minimum. In particular, the absence of a plateau regime, unlike model 1, highlights a distinct phase structure of model 2. The current constraint from KamLAND-Zen excludes the upper parameter space but leaves both regions viable. The sharp transition at $m_{1} \approx 0.01$ eV marks the boundary between regions A and B, without an intermediate cancellation phase. This dichotomy arises from the implementation of modular symmetry in model 2, where residual symmetries might prevent partial phase alignment.

\begin{figure}[h!]
	\centering
	\begin{subfigure}[b]{0.3\textwidth}
		\includegraphics[width=\textwidth]{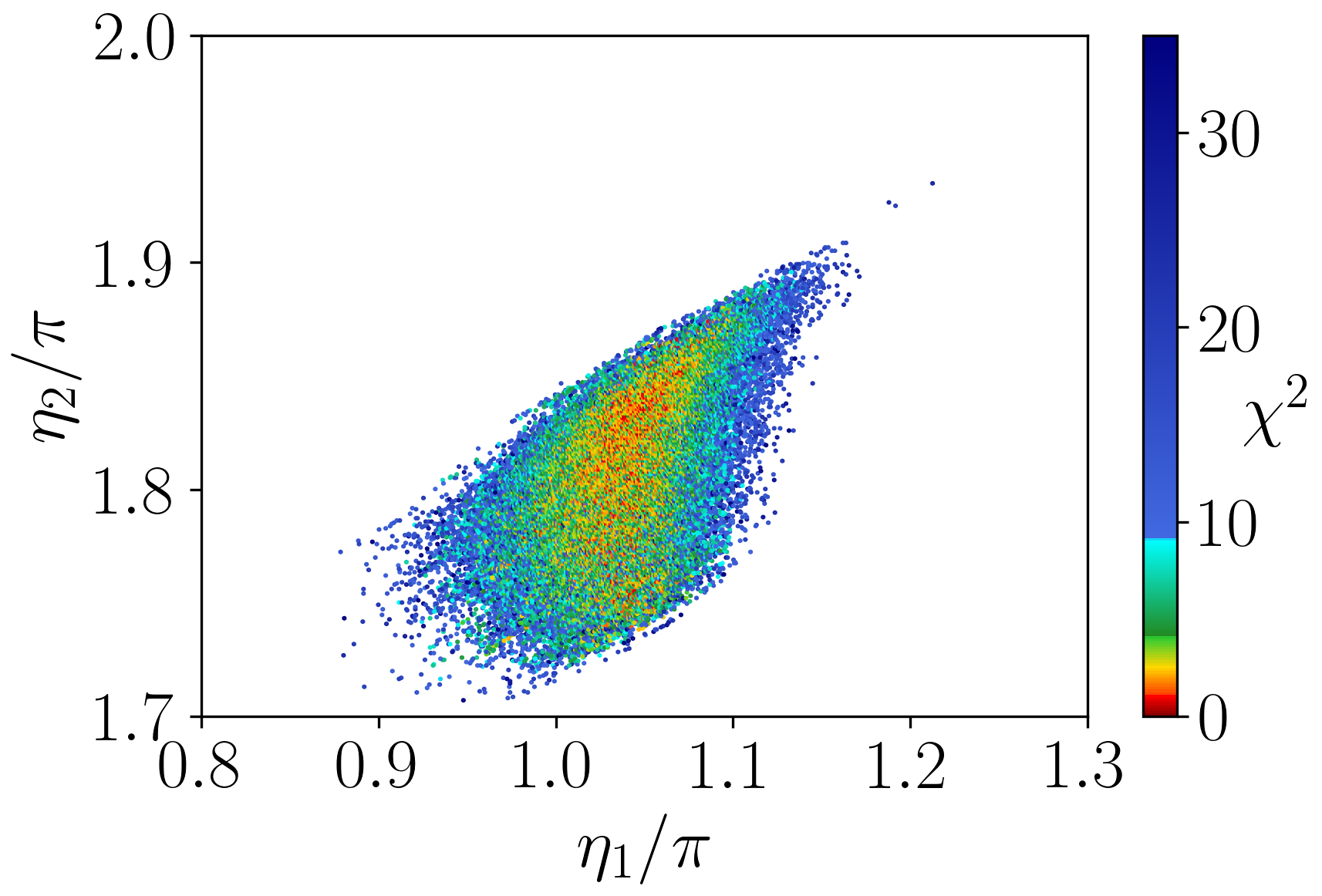}
		\caption*{A(1)}
	\end{subfigure}
	\hfill
	\begin{subfigure}[b]{0.3\textwidth}
		\includegraphics[width=\textwidth]{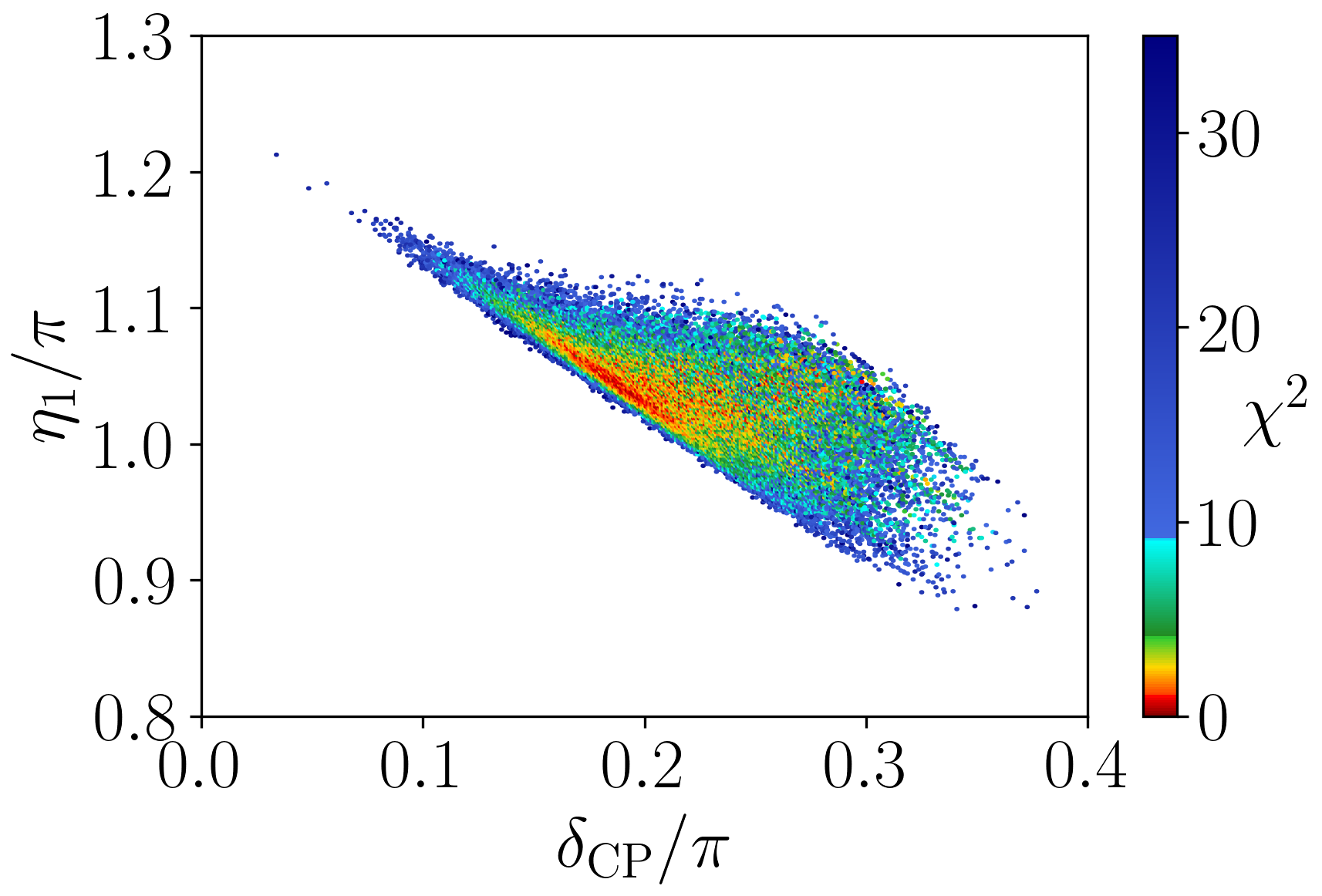}
		\caption*{A(2)}
	\end{subfigure}
	\hfill
	\begin{subfigure}[b]{0.3\textwidth}
		\includegraphics[width=\textwidth]{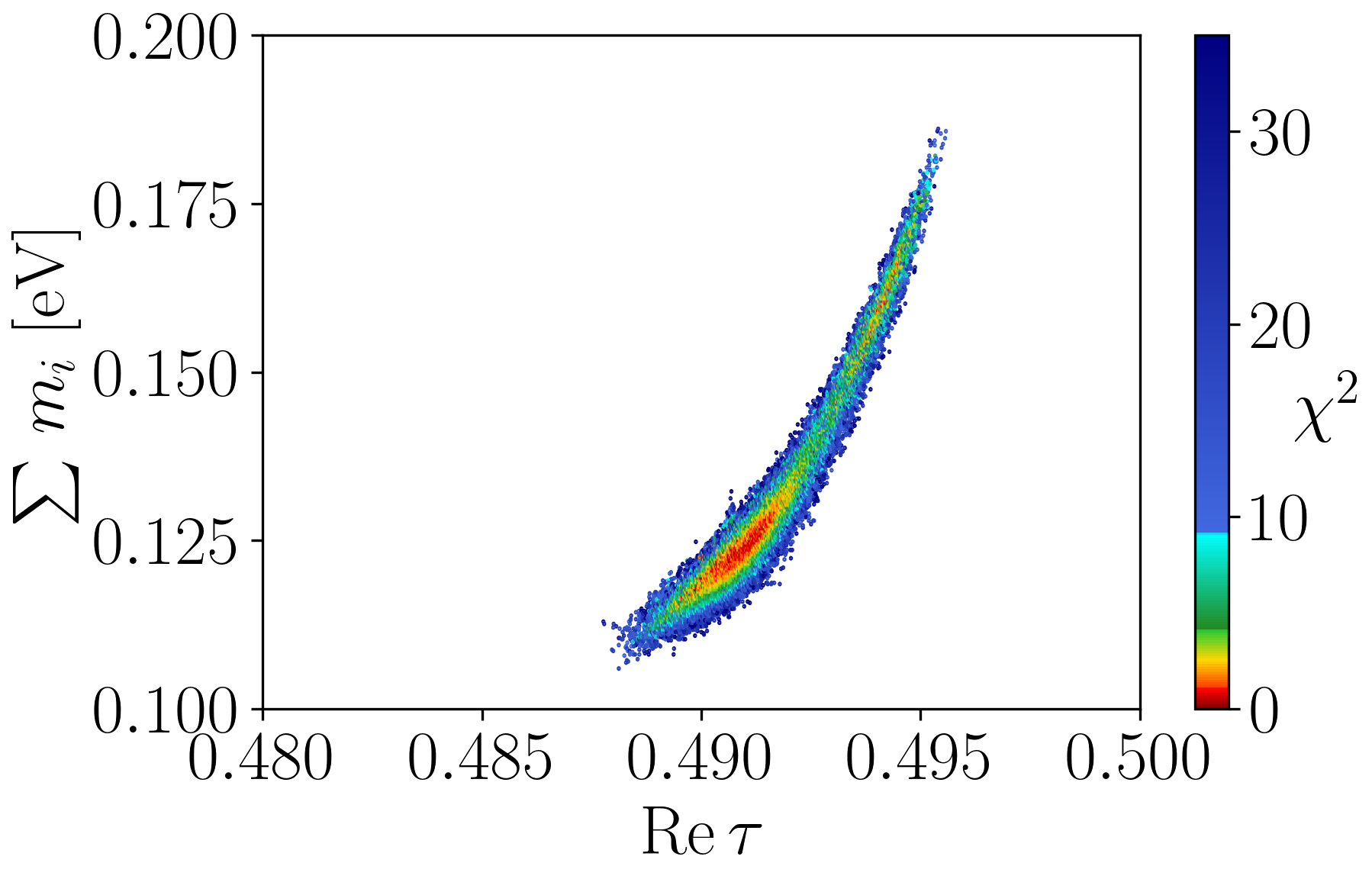}
		\caption*{A(3)}
	\end{subfigure}
	\vspace{0.5cm}	
	\begin{subfigure}[b]{0.3\textwidth}
		\includegraphics[width=\textwidth]{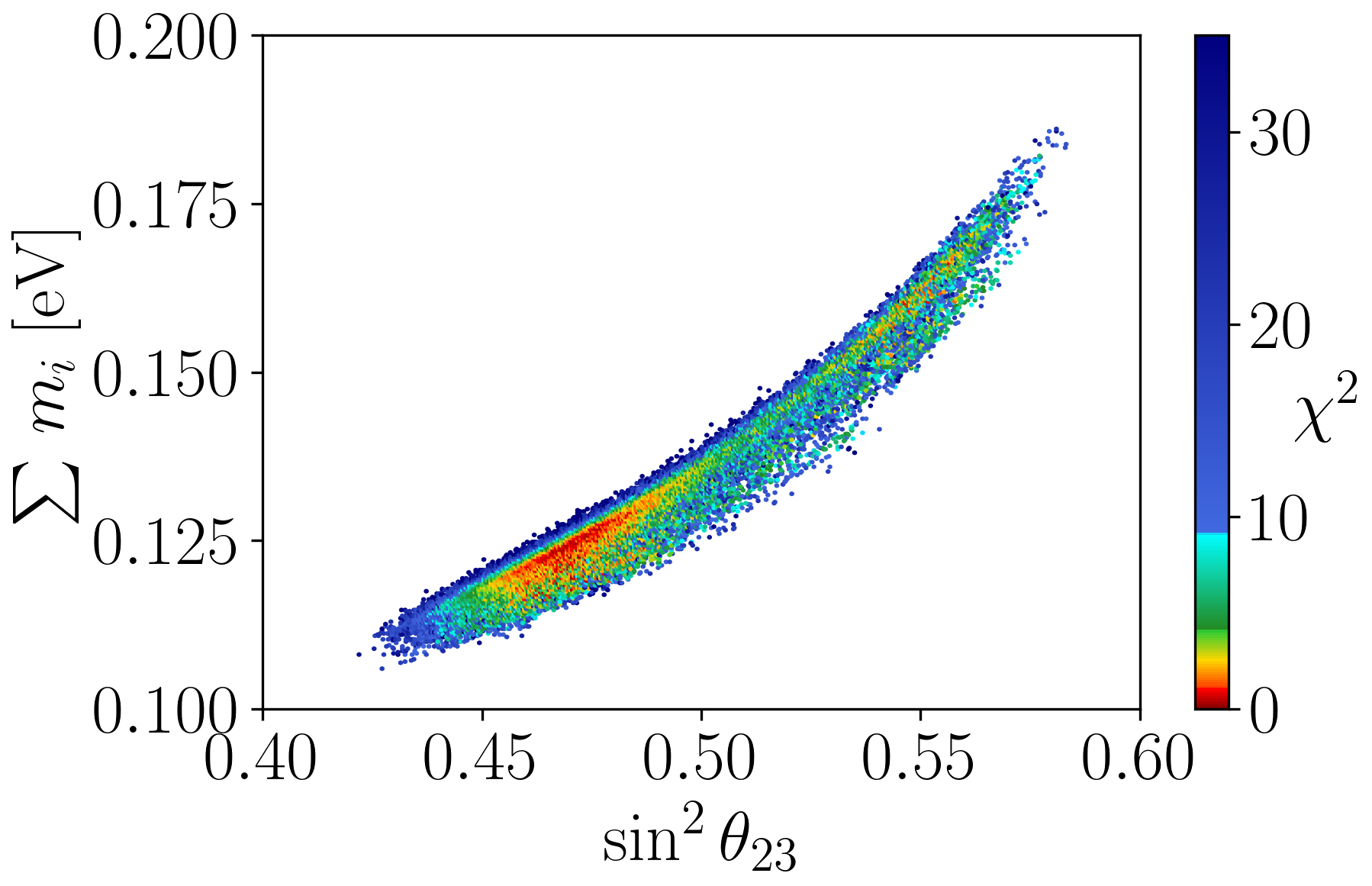}
		\caption*{A(4)}
	\end{subfigure}
	\hfill
	\begin{subfigure}[b]{0.3\textwidth}
		\includegraphics[width=\textwidth]{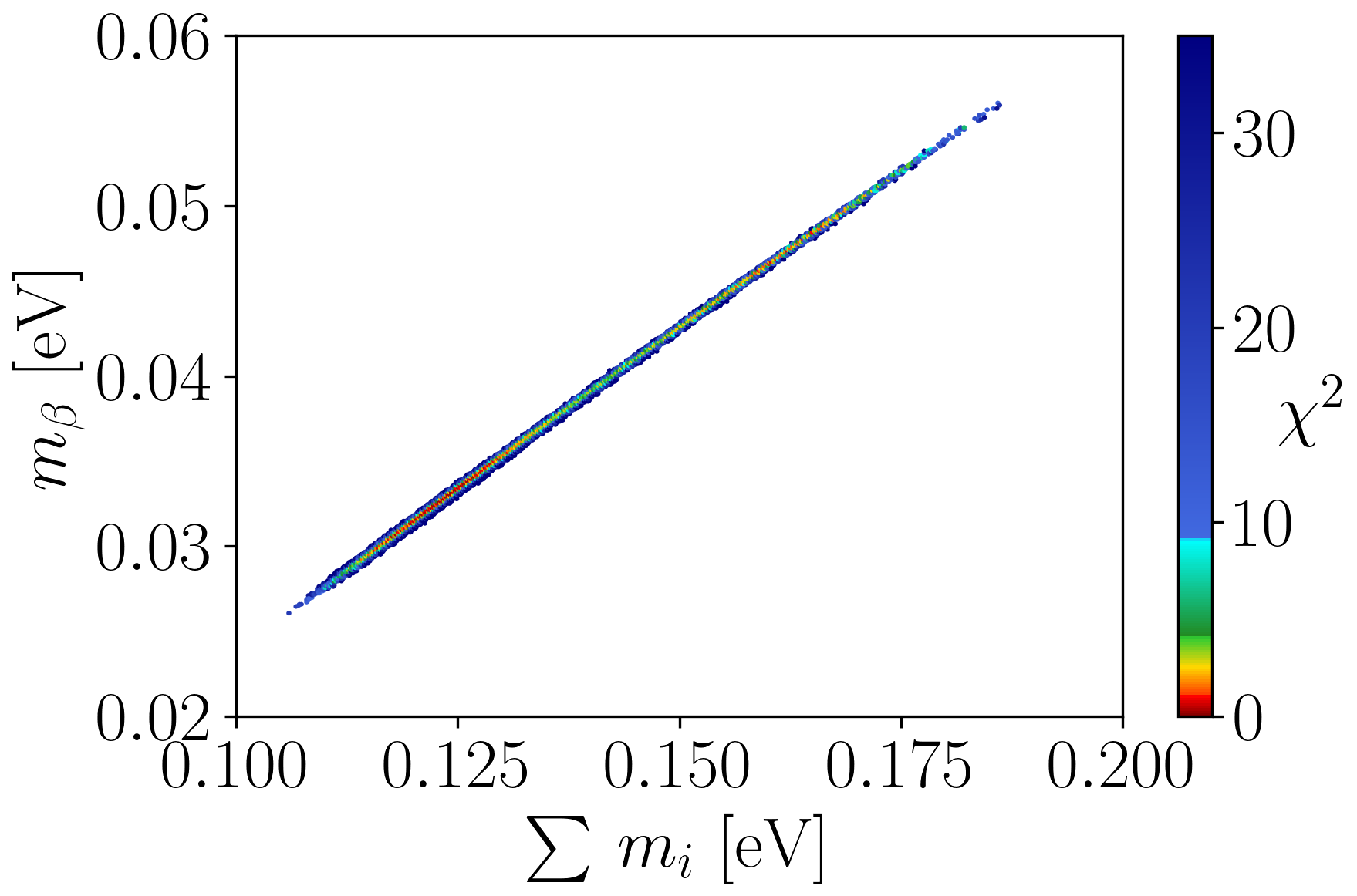}
		\caption*{A(5)}
	\end{subfigure}
	\hfill
	\begin{subfigure}[b]{0.3\textwidth}
		\includegraphics[width=\textwidth]{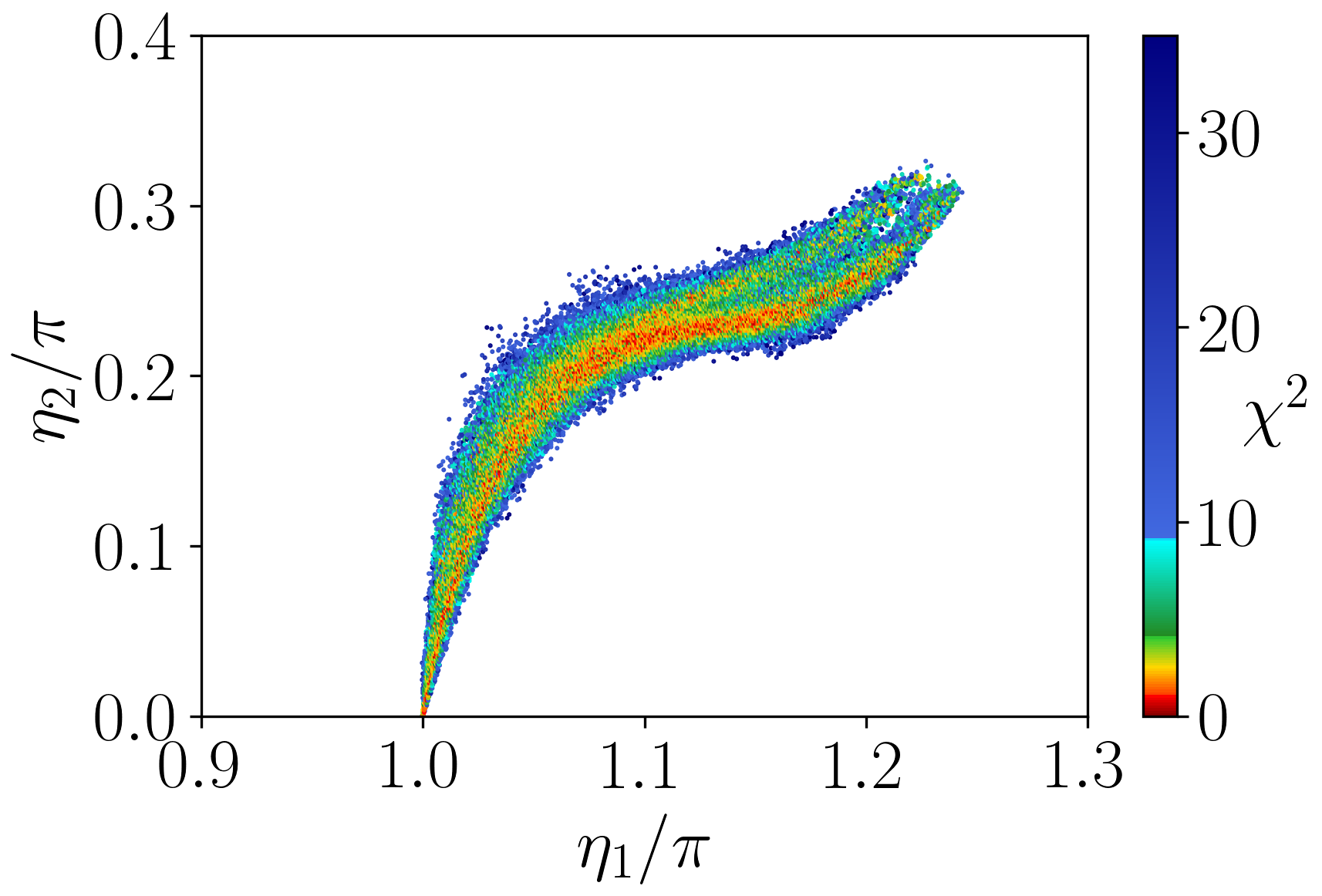}
		\caption*{B(1)}
	\end{subfigure}
	\vspace{0.5cm}
	\begin{subfigure}[b]{0.3\textwidth}
		\includegraphics[width=\textwidth]{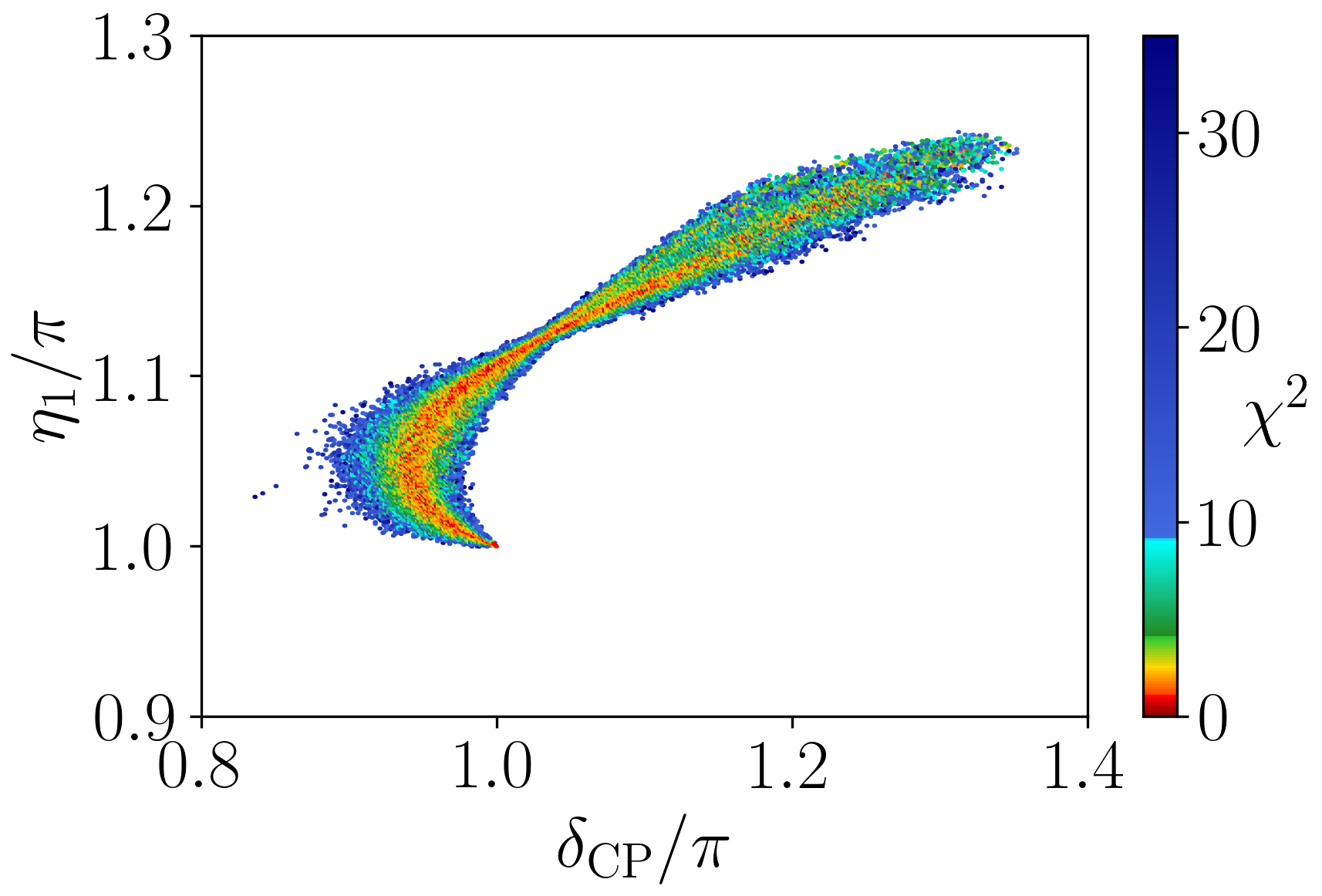}
		\caption*{B(2)}
	\end{subfigure}
	\hfill
	\begin{subfigure}[b]{0.3\textwidth}
		\includegraphics[width=\textwidth]{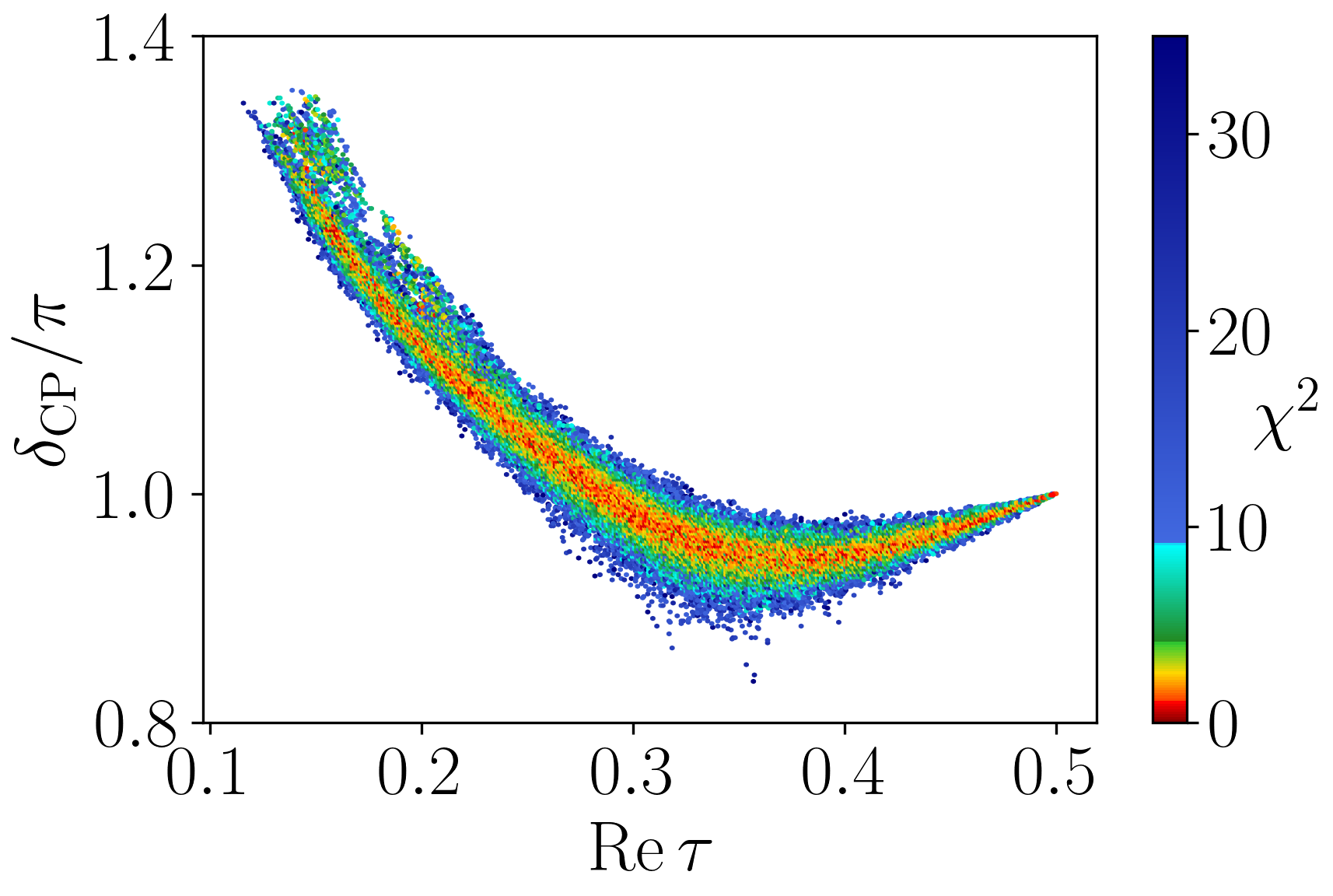}
		\caption*{B(3)}
	\end{subfigure}
	\hfill
	\begin{subfigure}[b]{0.3\textwidth}
		\includegraphics[width=\textwidth]{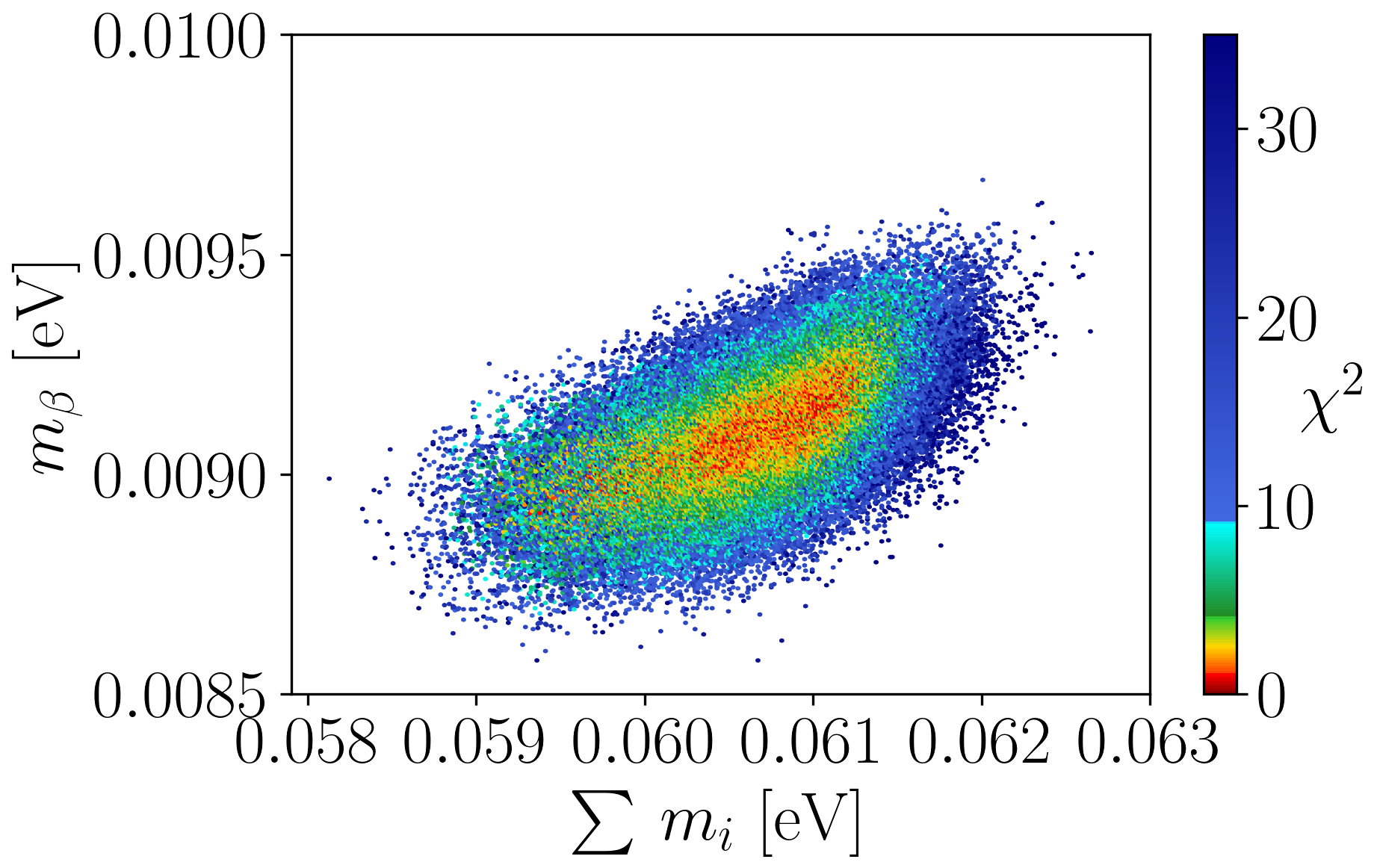}
		\caption*{B(4)}
	\end{subfigure}
	\caption{Parameter correlations for model 2 with NO. Region A: (1) $\eta_2$ vs. $\eta_1$, (2) $\eta_1$ vs. $\delta_{\rm CP}$, (3) $\sum m_i$ vs. $\text{Re}(\tau)$, (4) $\sum m_i$ vs. $\sin^2\theta_{23}$, (5) $m_\beta$ vs. $\sum m_i$. Region B: (1) $\eta_2$ vs. $\eta_1$, (2) $\eta_1$ vs. $\delta_{\rm CP}$, (3) $\delta_{\rm CP}$ vs. $\text{Re}(\tau)$, (4) $\sum m_i$ vs. $m_\beta$. }
	\label{fig:m2no}
\end{figure}

Furthermore, the figure~\ref{fig:m2no} presents an investigation of the relationships between physical parameters in model 2 with NO, and reveals connections between the modulus $\tau$ and neutrino observables. The analysis demonstrates how distinct regions in the $\tau$ plane give rise to different phenomenological regimes through their influence on the mass and mixing parameters. In region A, the Majorana phases $\eta_1$ and $\eta_2$ demonstrate a rather broad correlation, with best-fit values of $\eta_1 \approx 1.03\pi$ and $\eta_2 \approx 1.78\pi$. This particular phase combination leads to partial constructive interference in $0\nu\beta\beta$ decay, resulting in $m_{\beta\beta} \approx 0.023$ eV. The Dirac CP phase $\delta_{\rm CP}$ clusters around $0.24\pi$, indicating moderate CP violation that remains stable throughout the parameter space. The total neutrino mass $\sum m_i$ shows a strong dependence on $\text{Re}(\tau)$, the best fitting appear near $\text{Re}(\tau) \approx 0.491$ where $\sum m_i \approx 0.122$ eV. This arises from residual symmetry enhancement at the fundamental domain boundary (see Refs.~\cite{Novichkov:2019sqv,Ding:2023htn} for relevant discussions). The atmospheric mixing angle spans the range $0.42 \leq \sin^2\theta_{23} \leq 0.58$, crossing the maximal mixing boundary at $\sin^2\theta_{23} = 0.5$. This behavior indicates that the model does not predict a definitive octant preference for $\theta_{23}$. The effective neutrino mass $m_\beta$ and the neutrino mass sum $\sum m_i$ have almost linear correlations, spanning $0.105 ~\text{eV}\le \sum m_i \le 0.190$ eV and $0.026~\text{eV}\le m_\beta \le 0.056$ eV.
The parameter correlations in region B show a qualitatively different behavior compared to that in region A. The Majorana phases in this region take positive correlations, with the best-fit points $\eta_1 \approx 1.19\pi$ and $\eta_2 \approx 0.26\pi$, creating a strong suppression of $m_{\beta\beta}$ to be approximately $0.002$ eV. Dirac CP pahse $\delta_{\rm CP}$ spans $0.88\pi-1.35\pi$, and the best-fit value of  $\delta_{\rm CP} = 1.20\pi$ represents a mild CP violation. The correlation between the Majorana phase $\eta_1$ and the Dirac CP phase $\delta_{\rm CP}$ exhibits a positive relationship. The alignment of $\eta_1 \approx 1.194\pi$ and $\delta_{\rm CP} \approx 1.203\pi$ at the best fit indicates moderate CP violation in both the oscillation and $0\nu\beta\beta$ sectors. Their distributions show that $\eta_1$ and $\delta_{\rm CP}$ vary respectively across $(1.00-1.24)\pi$ and $(0.88-1.35)\pi$, suggesting that their behavior is governed by common symmetry constraints in parameter space of model 2. There are anti-correlations between $\delta_{\rm CP}$ and $\text{Re}(\tau)$. In particular, at $\text{Re}(\tau) = 0.5$, $\delta_{\rm CP} = \pi$, a CP conserving value, occurs due to unbroken $\mathbb{Z}_2$ symmetry, which is a hallmark of modular-invariant theories. The last panel illustrates the correlated area between effective beta-decay mass $ m_\beta $ and the sum of neutrino masses $ \sum m_i $. The best-fit point $ m_\beta \approx 0.0090$ eV with $ \sum m_i \approx 0.060$ eV is within the reach of next-generation beta decay experiments, providing a critical test of neutrino mass structure.  

\begin{figure}[h!]
	\centering
	\includegraphics[width=0.6\textwidth]{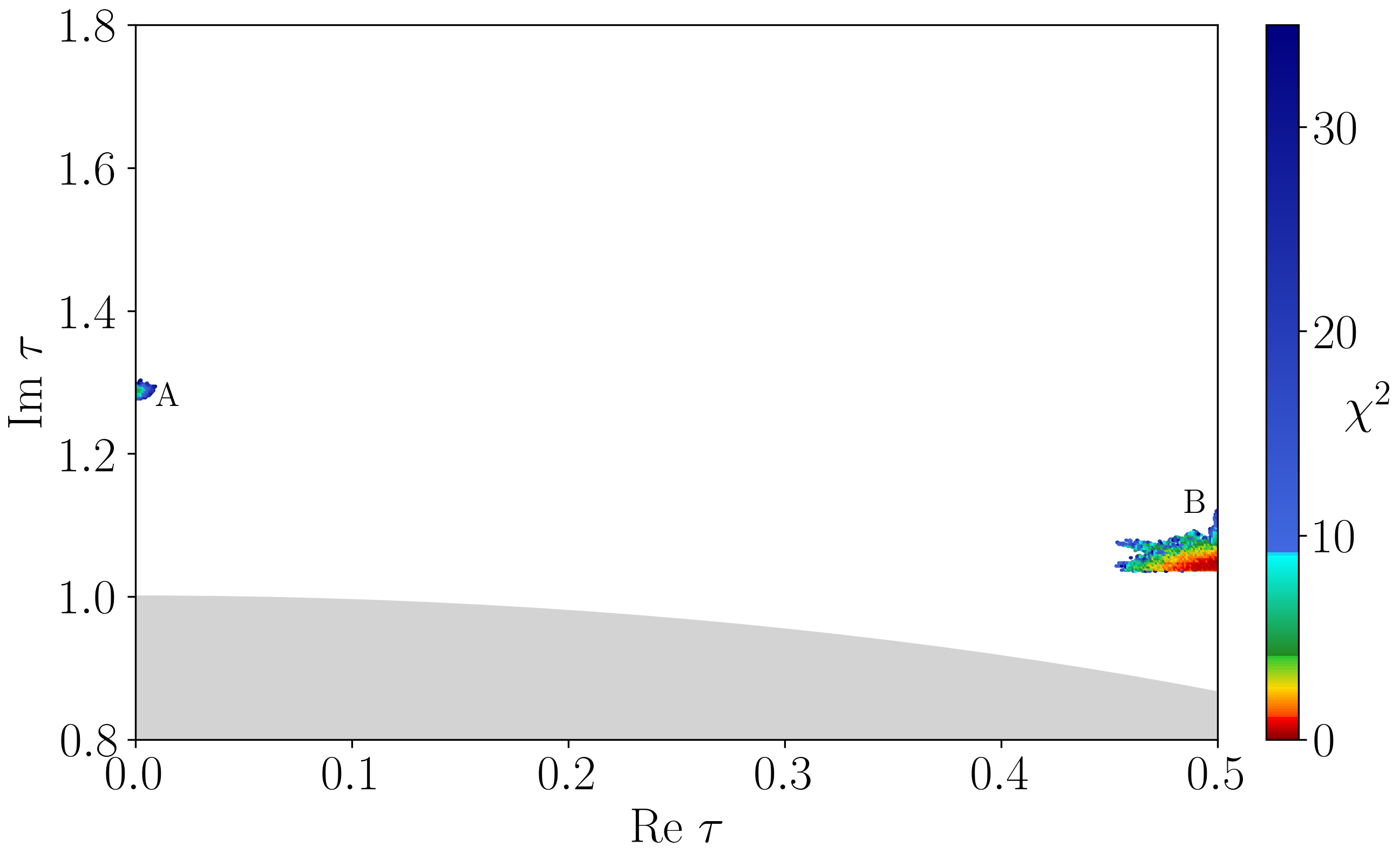}
	\caption{Viable parameter space of the modulus $\tau$ for model 2 with IO case. There are two distinct regions, marked A and B, that reproduce observed neutrino oscillation parameters. Region A is localized near the imaginary axis, $\text{Re}(\tau) \approx 0$, while region B sets closed to the another boundary of the fundamental domain, $\text{Re}(\tau) \approx 0.5$. The color gradient indicates the quality of $\chi^2$ fit.}
	\label{fig:m2io_tau}
\end{figure}

Similarly to the NO case, in the IO scenario of model 2, we identify two viable regions in the complex $\tau$ plane, as depicted in Figure \ref{fig:m2io_tau}. Both are located near the two boundaries, having the best point of $\tau$ in region A is $\tau=1.7094 \times 10^{-6} + 1.2885i$ with $\chi^2_{\text{min}} = 4.838$ and that of region B is $\tau=0.4999 + 1.0430i$ with $\chi^2_{\text{min}} =0.1029 $. In what follows, we analyze the physical parameters and their correlations in these two regions separately.   
Region A, centered at $\tau=1.7094 \times 10^{-6} + 1.2885i$, yields the corresponding best-fit values of mass, mixing, and CP-violating parameter are the following:
\begin{equation}
	\begin{aligned}
		&m_e/m_{\mu}=0.004736,\quad m_{\mu}/m_{\tau}=0.05883,\\
		&\sin^2\theta_{12}=0.3277,\quad  \sin^2\theta_{13}=0.02244,\quad  \sin^2\theta_{23}=0.4707,\\
		&\delta_{\rm CP}=0.9999\pi,\quad  \eta_1=\pi,\quad  \eta_2=\pi,\\
		&m_1=0.06231~ \mathrm{eV},\quad  m_2=0.06291~ \mathrm{eV},\quad  m_3=0.03836 ~\mathrm{eV},\\
		&\sum m_i=0.1636~\mathrm{eV},\quad  m_{\beta}=0.06286~\mathrm{eV},\quad  m_{\beta\beta}=0.06197~\mathrm{eV}.
	\end{aligned}
\end{equation}
The best-fit point predicts near-exact CP conservation in both the Dirac $\delta_{\rm CP} = 0.9999\pi$ and Majorana $\eta_1 = \eta_2 = \pi$ sectors, which indicates a CP symmetry-preserving configuration. The neutrino mixing parameters fall within the experimental $3\sigma$ bounds $\sin^2\theta_{12}$, $\sin^2\theta_{13}$, $\sin^2\theta_{23}$, which confirm the consistency with the oscillation data. The effective Majorana mass $m_{\beta\beta} = 0.06197$ eV lies within the sensitivity ranges of next-generation $0\nu\beta\beta$ experiments such as nEXO and LEGEND. However, significant tension arises in the mass sector. The total neutrino mass is $\sum m_i = 0.1636$ eV, which exceeds the cosmological upper limit $ 0.12$ eV. This suggests that region A may be cosmologically disfavored despite its compatibility with oscillation constraints.  

\begin{figure}[!h]
	\centering
	\begin{subfigure}[b]{0.3\textwidth}
		\includegraphics[width=\textwidth]{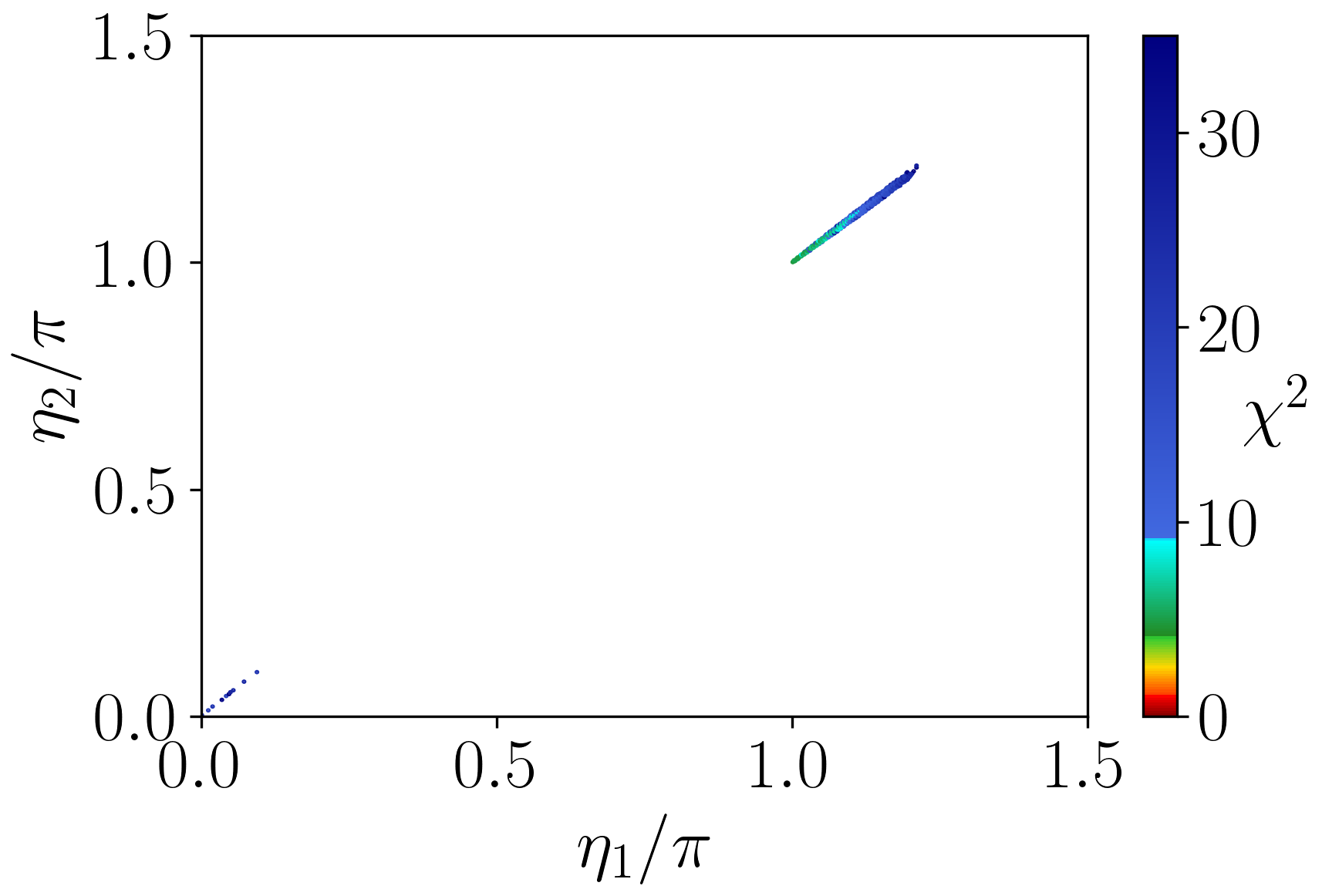}
		\caption*{A(1)}
	\end{subfigure}
	\hfill
	\begin{subfigure}[b]{0.3\textwidth}
		\includegraphics[width=\textwidth]{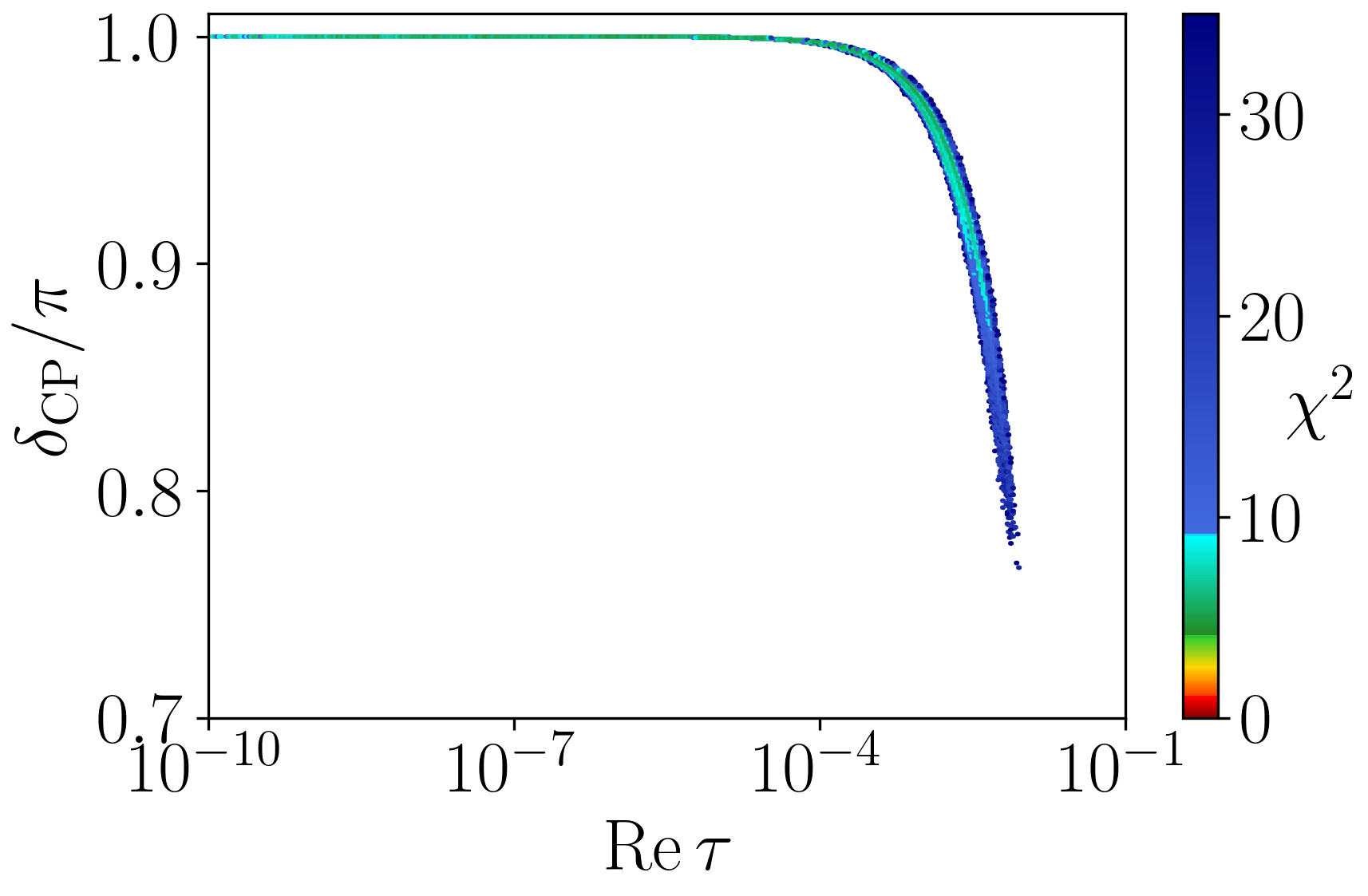}
		\caption*{A(2)}
	\end{subfigure}
	\hfill
	\begin{subfigure}[b]{0.3\textwidth}
		\includegraphics[width=\textwidth]{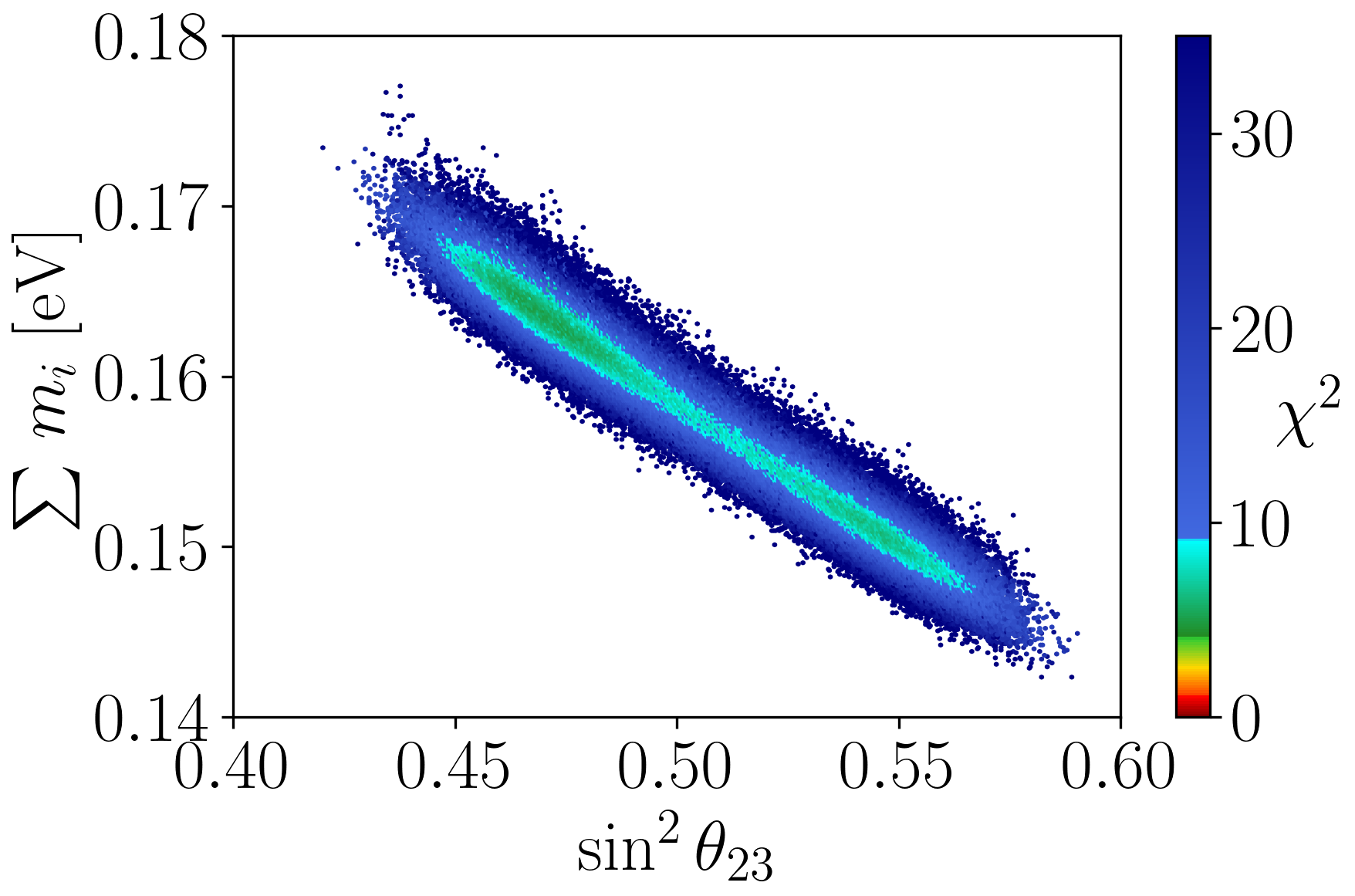}
		\caption*{A(3)}
	\end{subfigure}
	\vspace{0.5cm}
	\begin{subfigure}[b]{0.3\textwidth}
		\includegraphics[width=\textwidth]{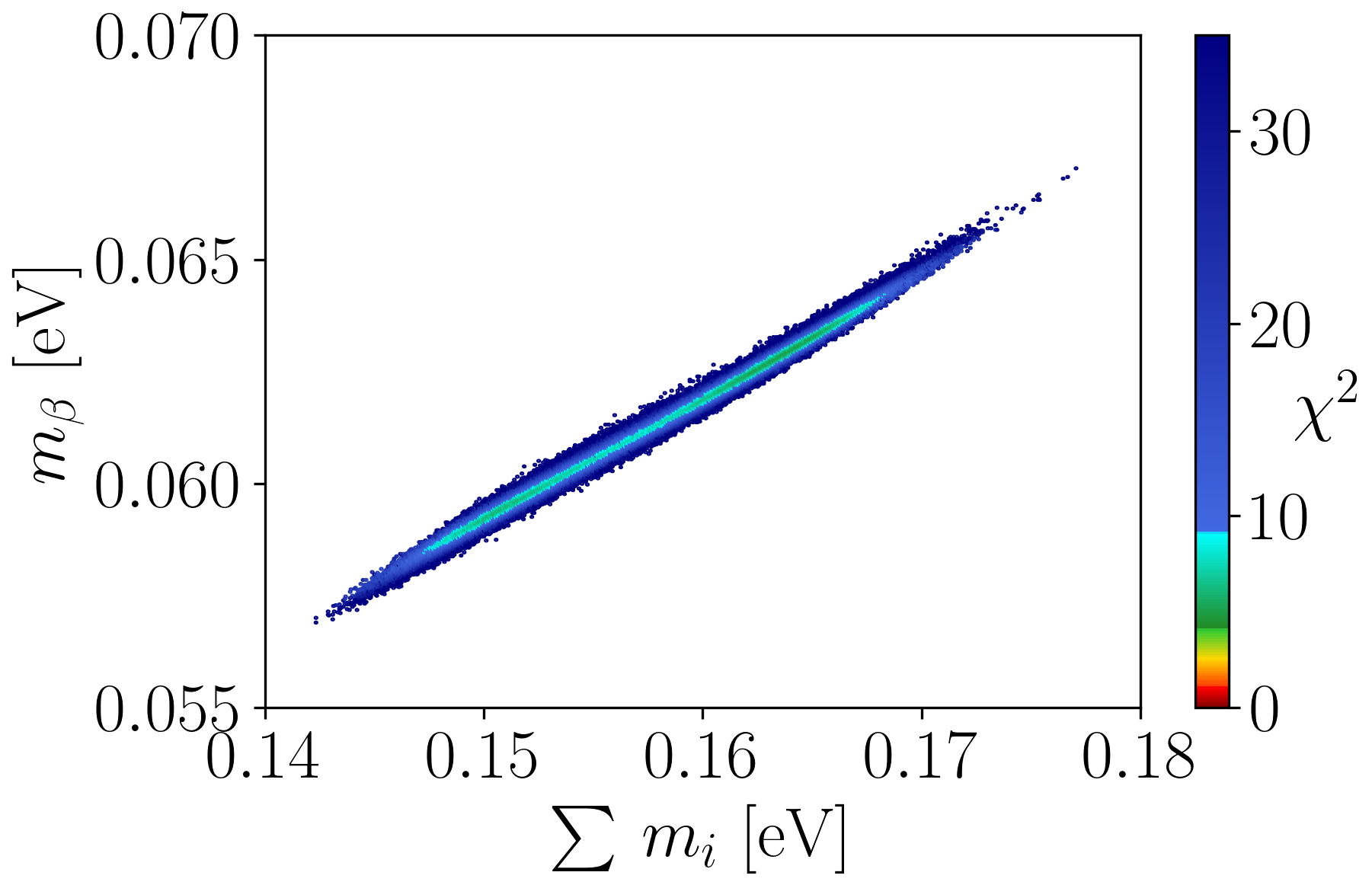}
		\caption*{A(4)}
	\end{subfigure}
	\hfill
	\begin{subfigure}[b]{0.3\textwidth}
		\includegraphics[width=\textwidth]{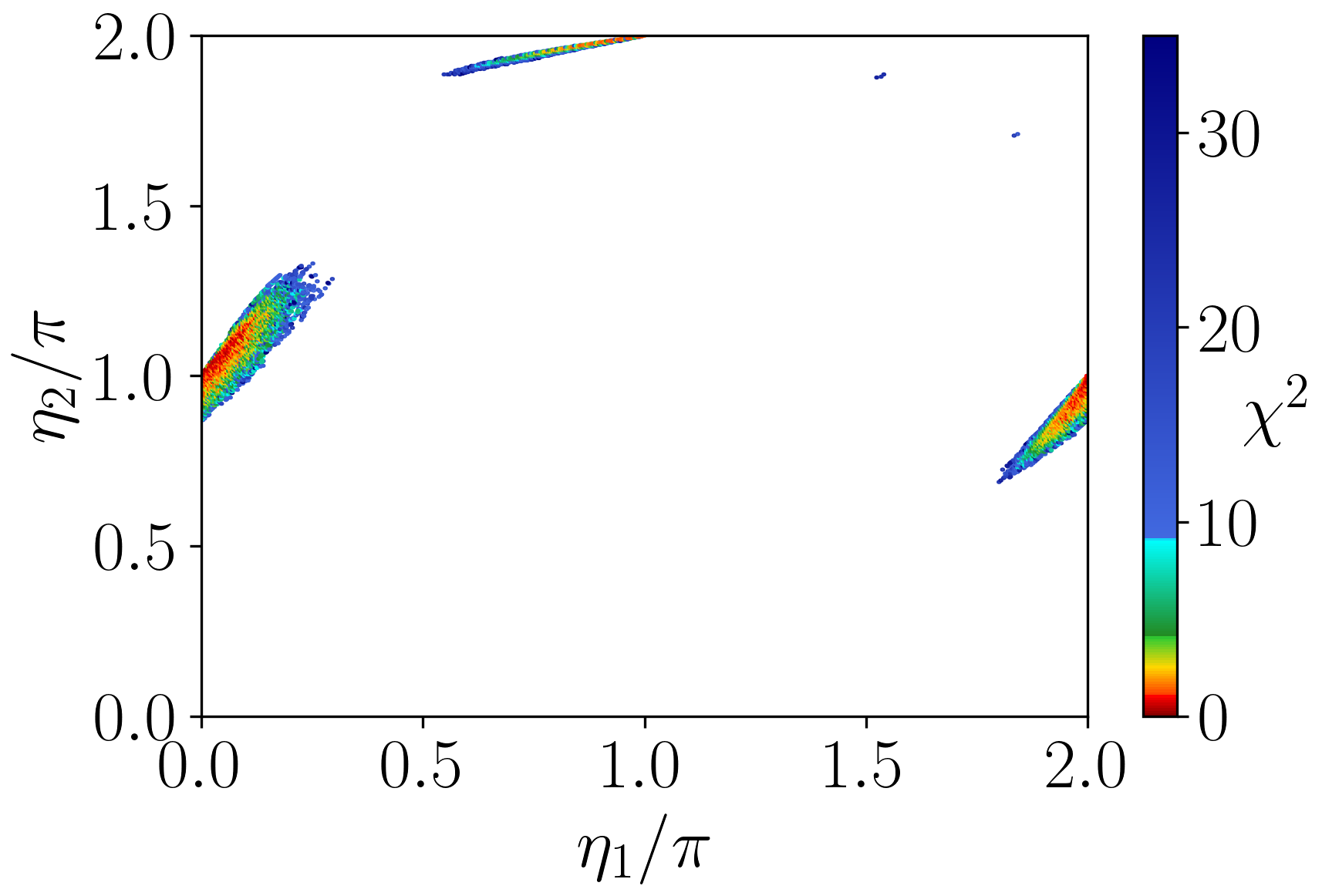}
		\caption*{B(1)}
	\end{subfigure}
	\hfill
	\begin{subfigure}[b]{0.3\textwidth}
		\includegraphics[width=\textwidth]{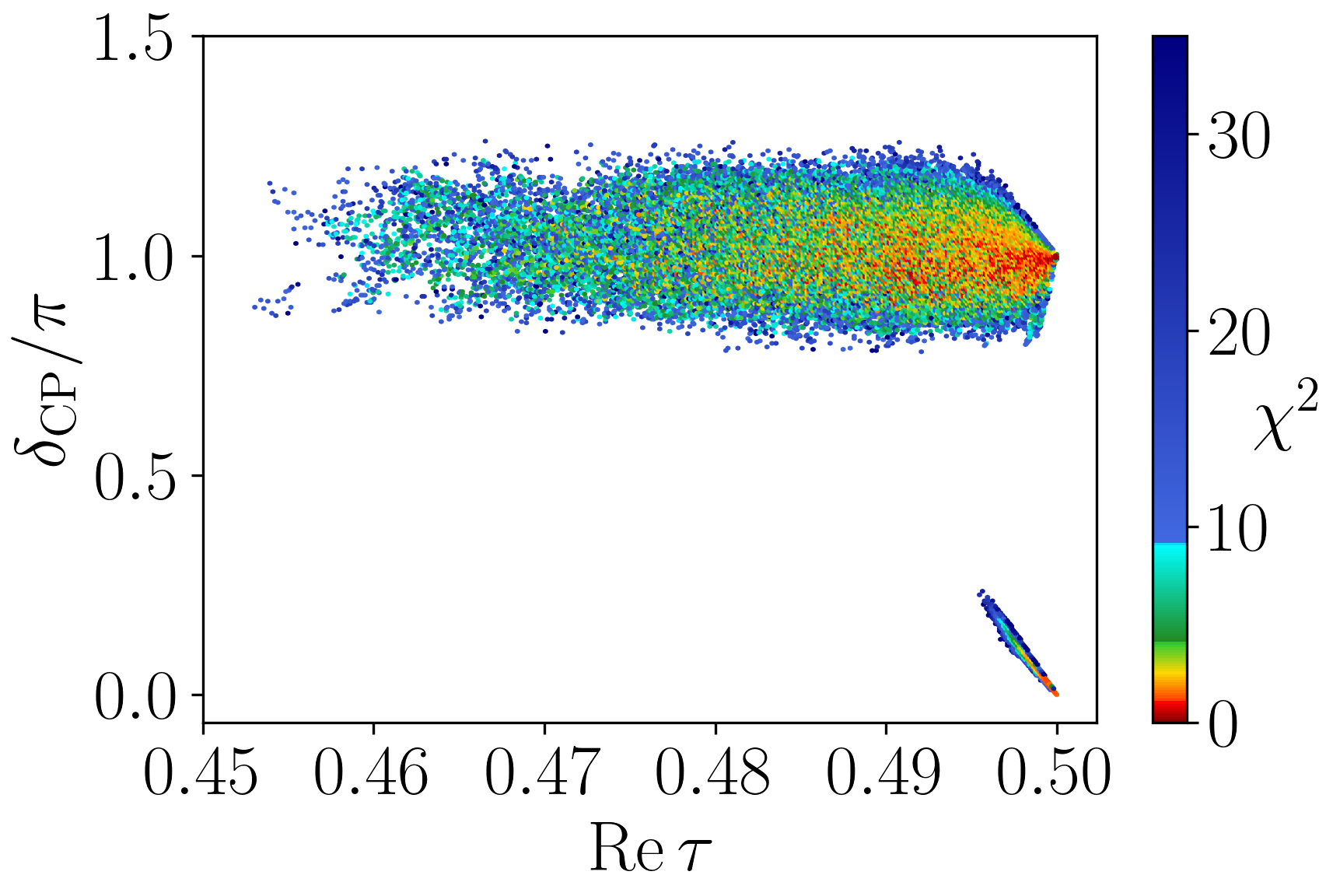}
		\caption*{B(2)}
	\end{subfigure}
	\vspace{0.5cm}
	\begin{subfigure}[b]{0.3\textwidth}
		\includegraphics[width=\textwidth]{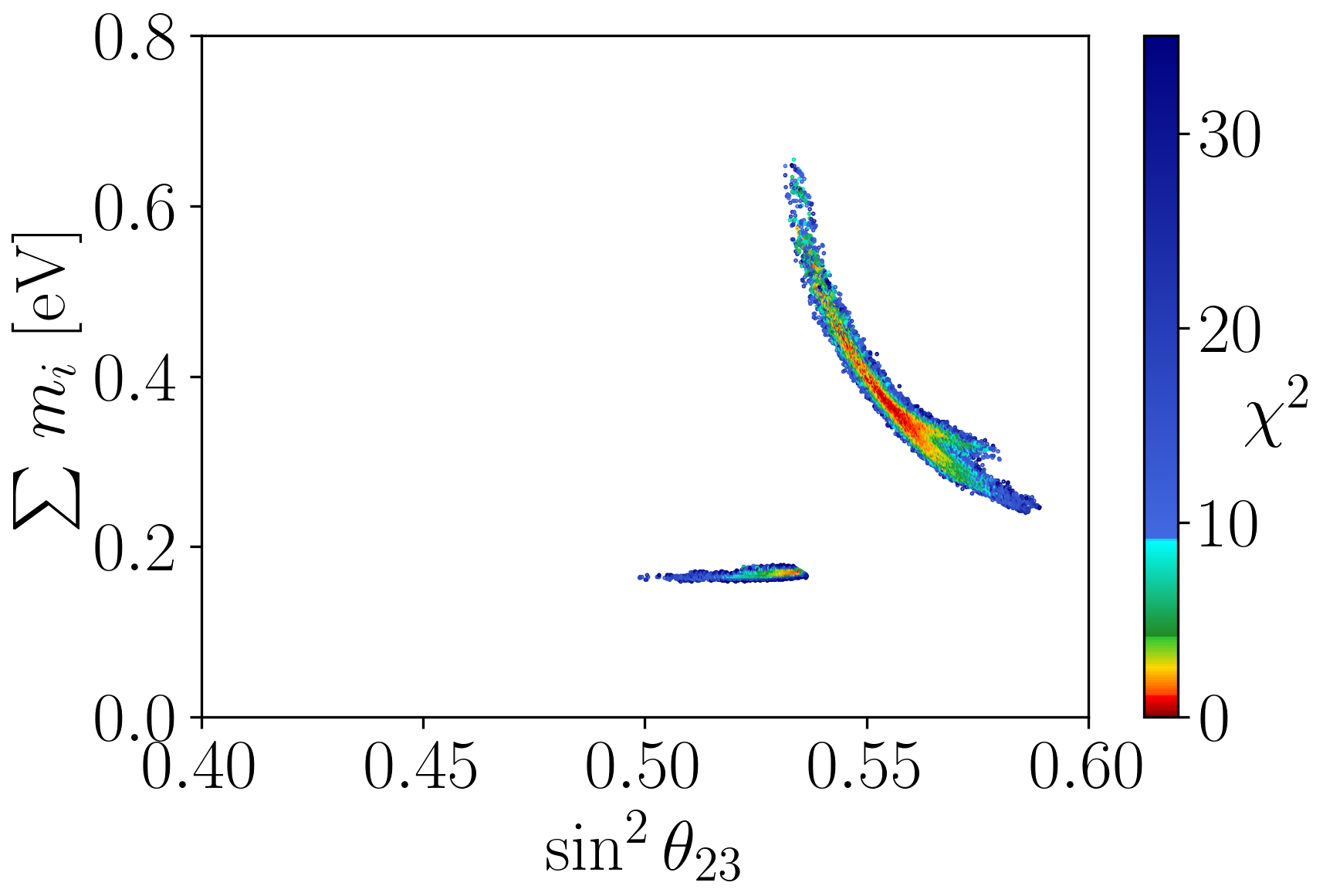}
		\caption*{B(3)}
	\end{subfigure}
	\hfill
	\begin{subfigure}[b]{0.3\textwidth}
		\includegraphics[width=\textwidth]{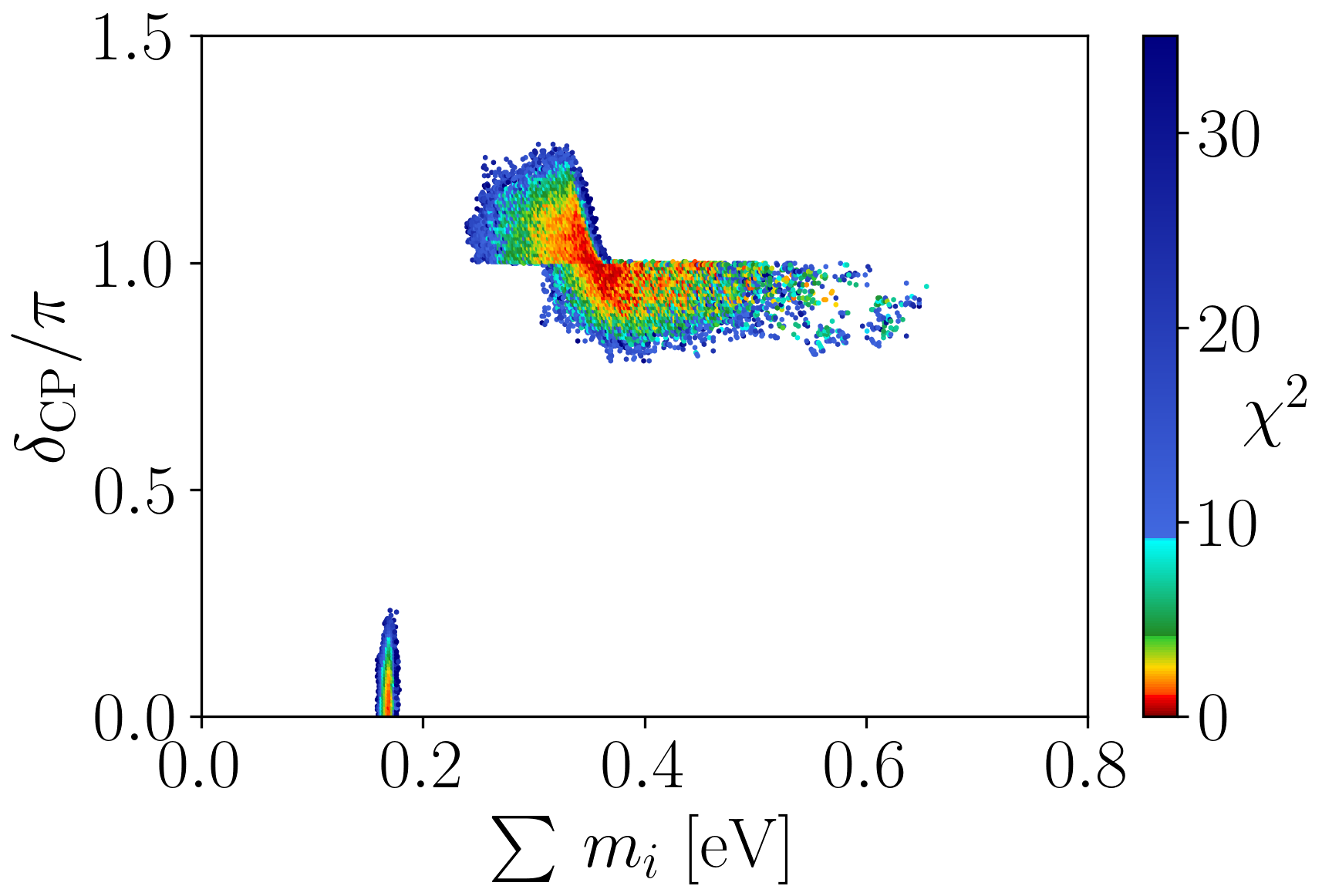}
		\caption*{B(4)}
	\end{subfigure}
	\hfill
	\begin{subfigure}[b]{0.3\textwidth}
		\includegraphics[width=\textwidth]{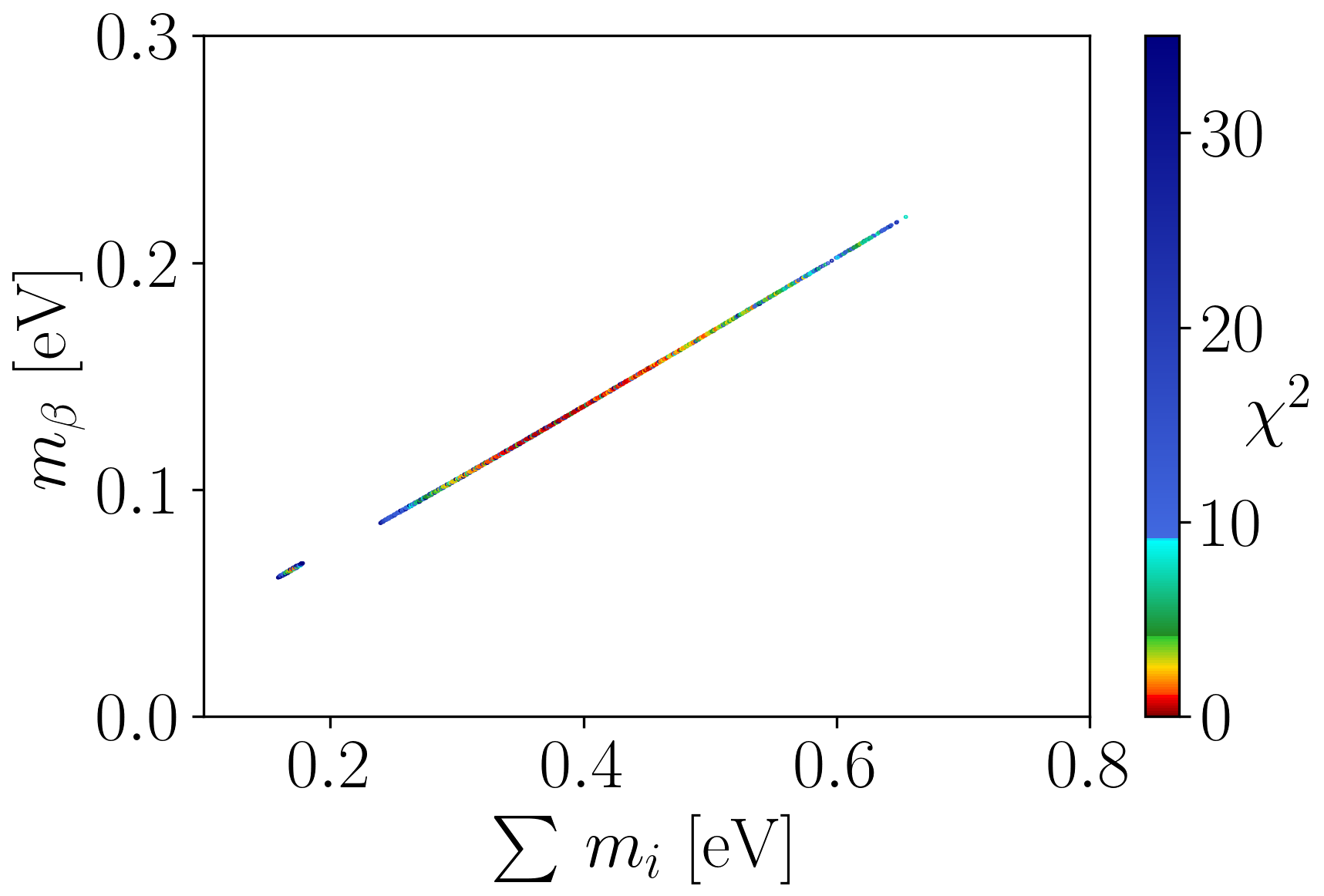}
		\caption*{B(5)}
	\end{subfigure}
	
	\caption{Parameter correlations for model 2 in IO case. Region A: Majorana phases $\eta_2$ vs. $\eta_1$, $\delta_{\rm CP}$ vs. $\text{Re}(\tau)$, $\sum m_i$ vs. $\sin^2\theta_{23}$, and $m_\beta$ vs. $\sum m_i$; Region B: $\eta_2$ vs. $\eta_1$, $\delta_{\rm CP}$ vs. $\text{Re}(\tau)$, $\sum m_i$ vs. $\sin^2\theta_{23}$, $\delta_{\rm CP}$ vs. $\sum m_i$, and $m_\beta$ vs. $\sum m_i$.}
	\label{fig:m2io}
\end{figure}

\begin{figure}[h!]
	\centering
	\includegraphics[width=0.7\textwidth]{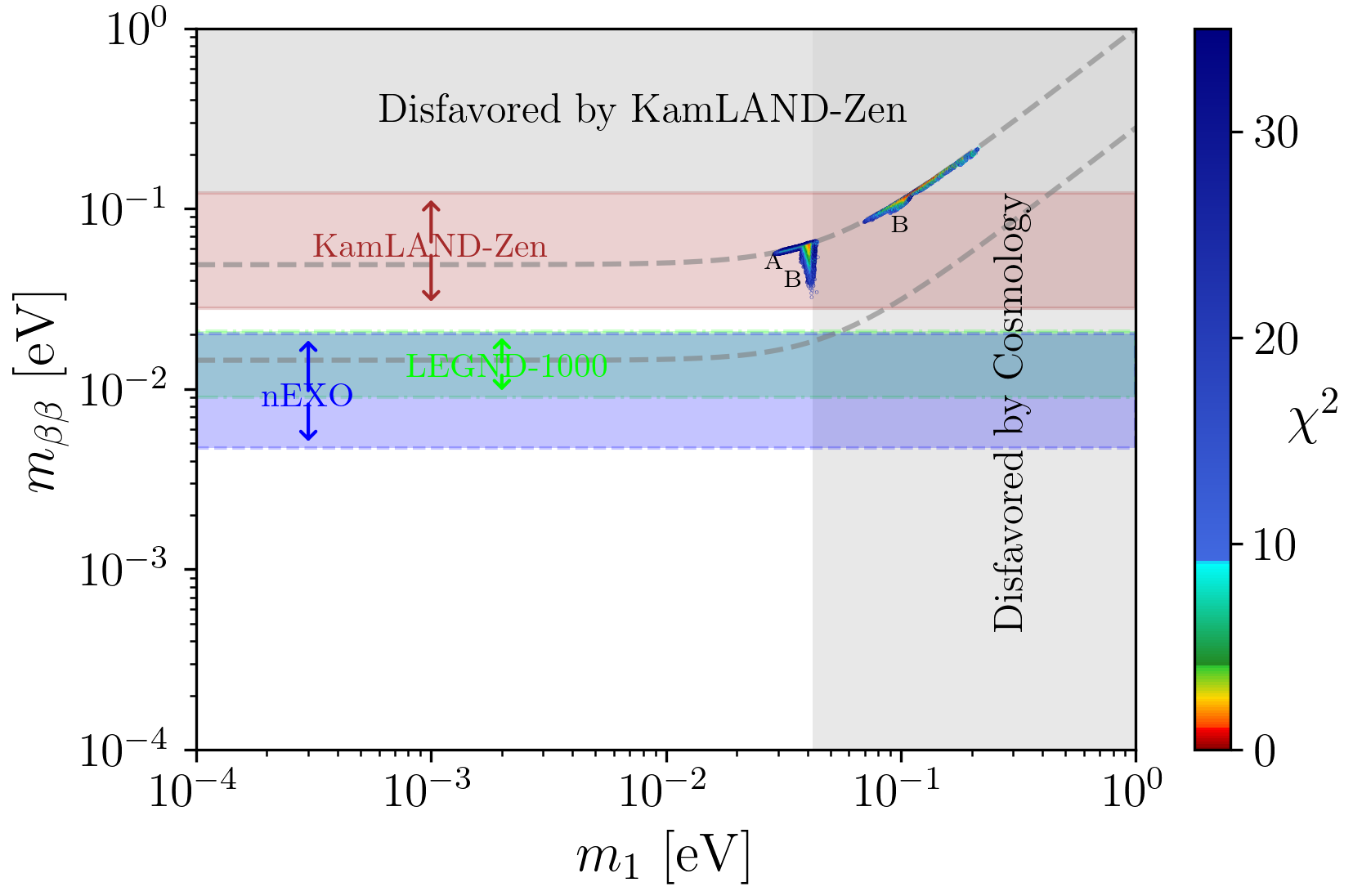}
	\caption{Effective Majorana neutrino mass $m_{\beta\beta}$ versus the lightest neutrino mass $m_3$ for model 2 in the IO case. Region A yields $m_{\beta\beta} \approx 0.056$–$0.066$ eV, stay below the current experimental and cosmological limits. The right part of region B is excluded by cosmology, while the left part remain viable below the both experimental and cosmological limits. Future sensitivities of KamLAND-Zen (brown), nEXO (blue), LEGEND-1000 (green) exclusion bands, and cosmological limit (gray) are overlaid.}
	\label{fig:m2io_0vbb}
\end{figure}

Although region B demonstrates relatively good agreement with the oscillation data, with the best-fit point of $\tau$ yielding $\chi^2_{\rm min}= 0.1029$, whose prediction for the effective Majorana mass $m_{\beta\beta}=0.1411~\mathrm{eV}$ exceeds the current experimental upper limit of KamLAND-Zen ($m_{\beta\beta} < 0.122$ eV). This tension with neutrinoless double beta decay constraints prevents a detailed phenomenological analysis of this region as a viable scenario. Nevertheless, we include its parameter correlations with region A in figures~\ref{fig:m2io} and \ref{fig:m2io_0vbb} to illustrate the broader behavior of the parameter space of model 2. We note that a subset of points in region B remains experimentally allowed, suggesting possible parameter configurations where the predictions might satisfy all current constraints.

Given the parameter correlations in figure~\ref{fig:m2io}, there are different CP properties between the two regions. In region A, we observe complete CP conservation in both the Dirac and Majorana sectors. The Majorana phases $\eta_1$ and $\eta_2$ remain fixed at $\pi$ throughout the parameter space, while the Dirac CP phase  $\delta_{\rm CP}$  maintains $\delta_{\rm CP}\approx \pi$ when $\tau$ lies close to the imaginary axis $\text{Re}(\tau)<10^{-4}$. As $\text{Re}(\tau)$ increases from $10^{-4}$ to $10^{-2}$, $\delta_{\rm CP}$ gradually deviates from $\pi$, consistent with the expected CP violation when $\tau$ moves away from boundary of the fundamental domain. The atmospheric mixing angle $\theta_{23}$ spans both octants $0.42\le \sin^2\theta_{23} \le 0.59$, and shows an anti-correlation with the total neutrino mass $\sum m_i$. However, this region is cosmologically excluded as $\sum m_i$ exceeds the $0.12$ eV limit, despite the clear linear correlation between $m_\beta$ and $\sum m_i$ evident in panel A(4).

Region B exhibits different behavior. The Majorana phases $\eta_1$ and $\eta_2$ cluster into three distinct configurations: $\eta_1\approx 1.2\pi$ with $\eta_2\approx 0.25\pi$, $\eta_1\approx 0.8\pi$ with $\eta_2\approx 1.9\pi$, and $\eta_1\approx 1.9\pi$ with $\eta_2\approx 0.9\pi$. The Dirac CP phase $\delta_{\rm CP}$ demonstrates a discontinuous transition from CP violation point $\delta_{\rm CP}\sim 1.2\pi$ to conservation $\delta_{\rm CP}\sim \pi$ at the symmetry point $\text{Re}(\tau) = 0.5$, reflecting the underlying modular geometry. Unlike region A, $\theta_{23}$  remains in the upper octant (as $\sin^2\theta_{23} >0.5$) throughout this region. While $\delta_{\rm CP}$ clusters near CP-conserving values, $m_\beta$ and $\sum m_i$ maintain a linear relationship, the predicted $\sum m_i$ values still violate the cosmological upper limit of $0.12$ eV.

For the predictions of $0\nu\beta\beta$ experiments, region A gives $m_{\beta\beta}\approx 0.056 –0.066$ eV within the horizontal band, resulting from interference between Majorana phases fixed at $\eta_1 = \eta_2 \approx \pi$. While these values lie below the KamLAND-Zen exclusion limit $m_{\beta\beta}<0.122$ eV, they violate cosmological limit $\sum m_i > 0.12$ eV. Region B exhibits two distinct separate correlations, in the first $m_{\beta\beta}$ stays near the cosmological bound, while the other falls inside the excluded region. A subset of points near $m_3\approx 0.04$ eV yields $m_{\beta\beta}\approx 0.056 - 0.066$ eV, satisfying both the KamLAND-Zen constraints and cosmological limits while falling within the sensitivity ranges of next-generation  $0\nu\beta\beta$ experiments. This suppression arises from interference between $\eta_1\approx 1.2\pi$ and $\eta_2\approx 0.25\pi$, demonstrating how Majorana phase configurations enable experimentally viable parameter space.

\subsection{Phenomenological implications of model 3}

\begin{figure}[h!]
	\centering
	\includegraphics[width=0.6\textwidth]{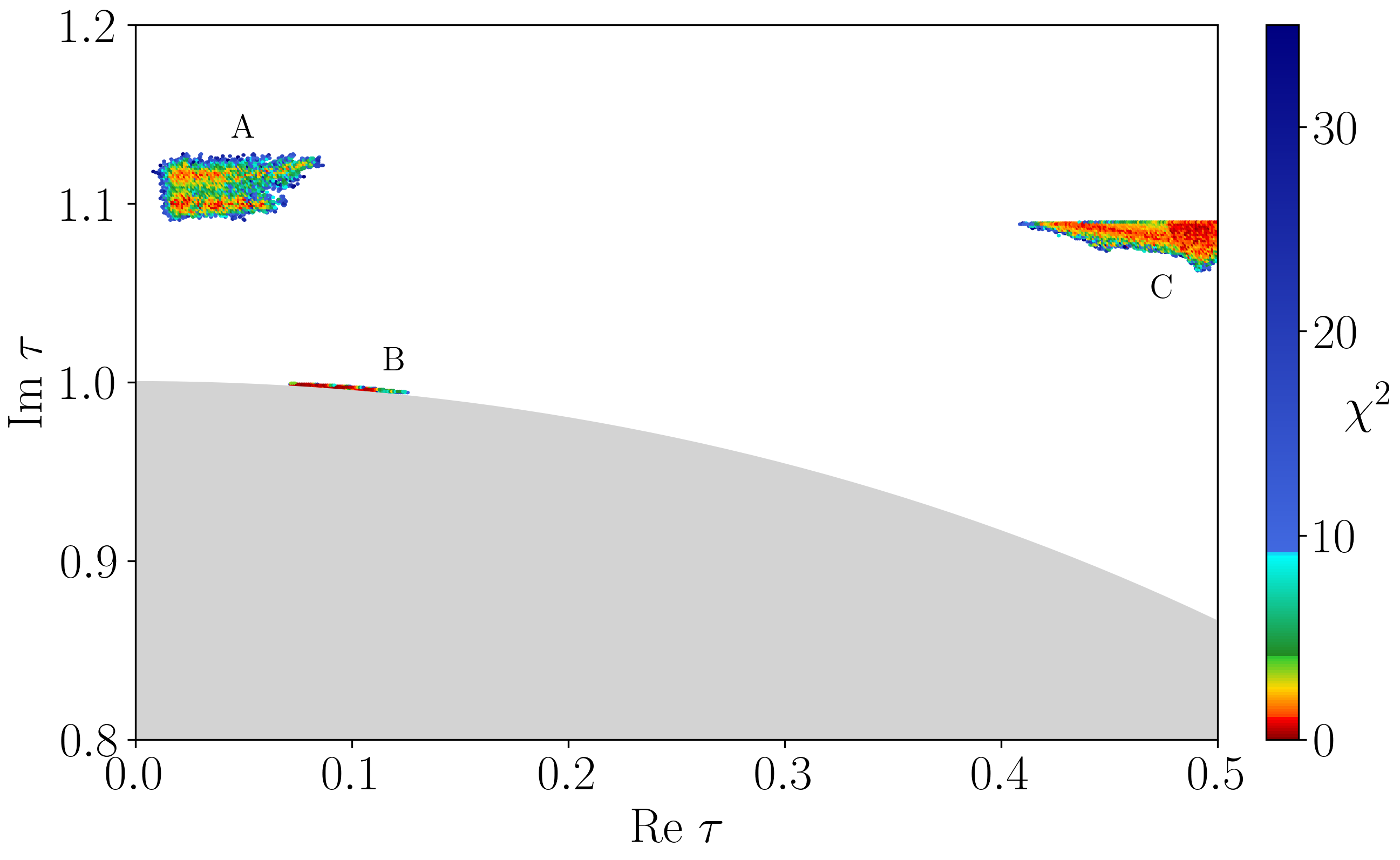}
	\caption{Viable $\tau$ parameter space for model 3 under NO case. Region A yields quasi-degenerate neutrino masses with $\sum m_i = 0.5036$ eV, region B produces hierarchical spectra consistent with cosmological bound ($\sum m_i < 0.12$ eV), and region C exhibits intermediate masses with $\sum m_i = 0.1270$ eV. Color intensity denotes $\chi^2$ fit quality, with brighter regions indicating better agreement with oscillation data. }
	\label{fig:m3no_tau}
\end{figure}

Completing our analysis for the phenomenology of model 2, we now examine model 3 in the NO case. This model reveals three parameter space regions marked by A, B, and C in the complex $\tau$ plane, shown in figure~\ref{fig:m3no_tau}. The best-fit values for each region are presented in table~\ref{tab:m3no_fit}. Each of these regions exhibits different mass scales and CP violation patterns that highlight the diversity of modular symmetry implementations.  
Region A, at the best-fit point, indicates the value of total neutrino mass $\sum m_i = 0.5036$ eV that exceeds the cosmological upper bound of $0.12$ eV. It predicts near-maximal Dirac CP violation $\delta_{\rm CP}\approx 1.518\pi$ and Majorana phases $\eta_1\approx 1.745\pi$, $\eta_2\approx 1.125\pi$ that generate $m_{\beta\beta} = 0.08045$ eV, potentially detectable in next-generation experiments but cosmologically excluded. The atmospheric angle resides in the lower octant $\sin^2\theta_{23}\approx  0.469$. Region B emerges as the only cosmologically viable sector $\sum m_i = 0.07767$ eV $< 0.12$ eV. Its hierarchical mass spectrum $m_1\approx 0.0117$ eV, $m_3\approx 0.0515$ eV combines with moderate Dirac CP violation $\delta_{\rm CP}\approx 0.864\pi$ and Majorana phases $\eta_1\approx 0.338\pi$, $\eta_2\approx 1.404\pi$ that induce partial destructive interference, thus resulting in $m_{\beta\beta} = 0.01227$ eV. This value falls within the sensitivity windows of the nEXO and LEGEND-1000 experiments, making region B a prime target for experimental verification. Region C shows a marginal cosmological tension $\sum m_i = 0.1270$ eV $> 0.12$ eV and CP violation  with $\delta_{\rm CP}\approx 1.834\pi$. The Majorana phases configuration $\eta_1\approx 1.130\pi$, $\eta_2\approx 0.162\pi$ produces $m_{\beta\beta} = 0.03375$ eV, which is detectable by future $0\nu\beta\beta$ experiments if cosmological constraints relax. Like other regions, it predicts the lower octant for $\theta_{23}$, with $\sin^2\theta_{23}\approx  0.472$.  

\begin{figure}[!h]
	\centering
	\begin{subfigure}[b]{0.3\textwidth}
		\includegraphics[width=\textwidth]{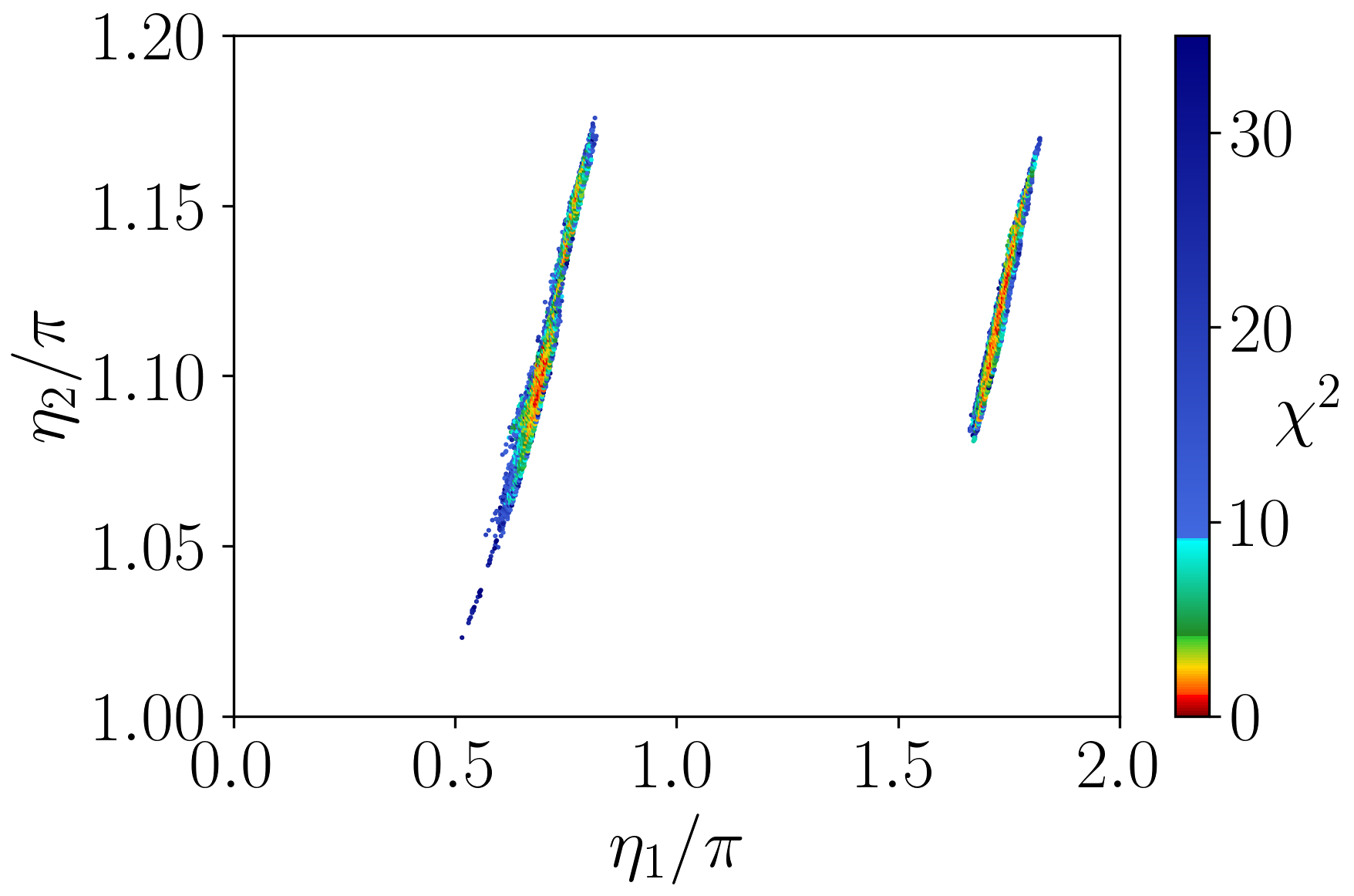}
		\caption*{A(1)}
	\end{subfigure}
	\vspace{0.5cm}
	\hfill
	\begin{subfigure}[b]{0.3\textwidth}
		\includegraphics[width=\textwidth]{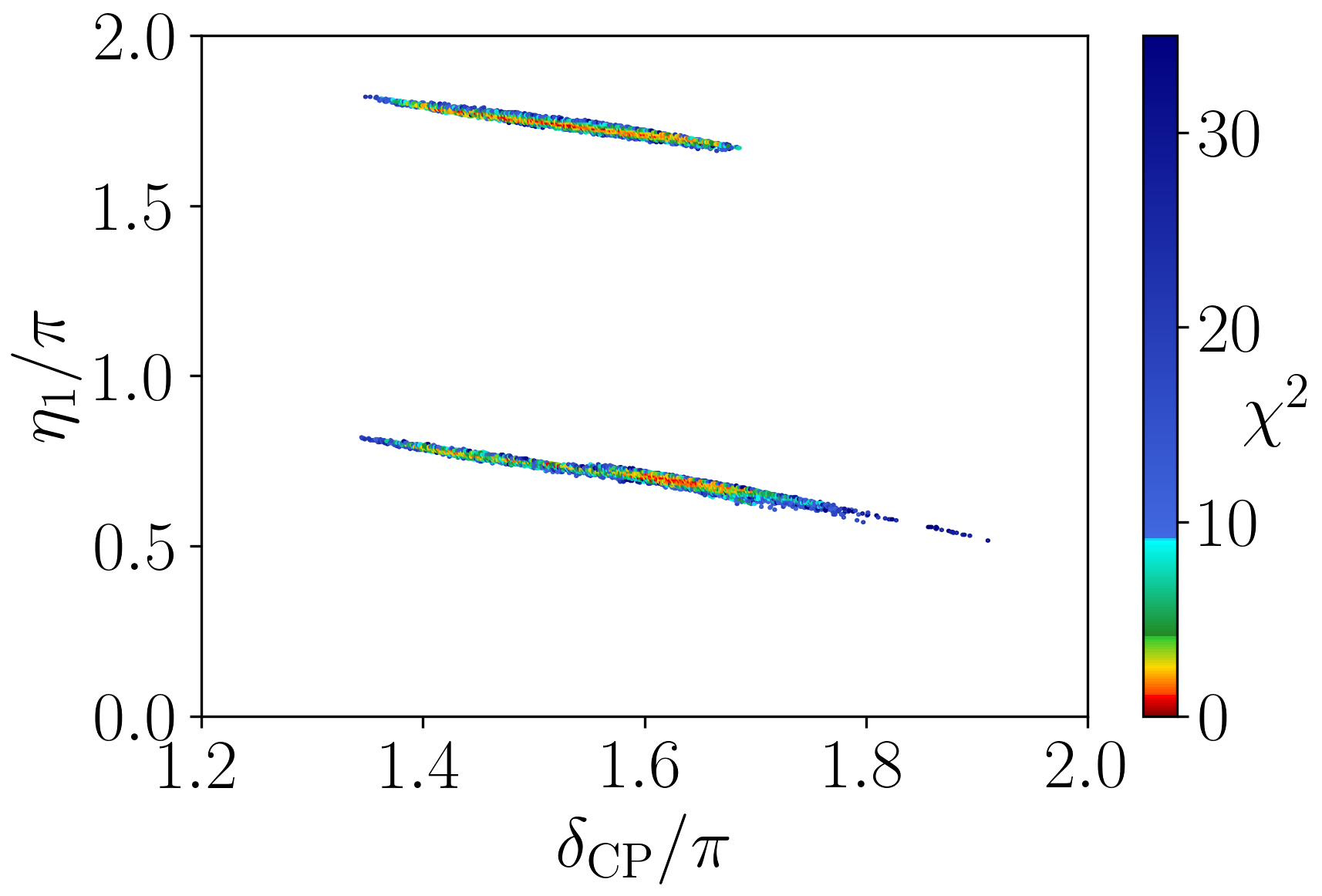}
		\caption*{A(2)}
	\end{subfigure}
	\hfill
	\begin{subfigure}[b]{0.3\textwidth}
		\includegraphics[width=\textwidth]{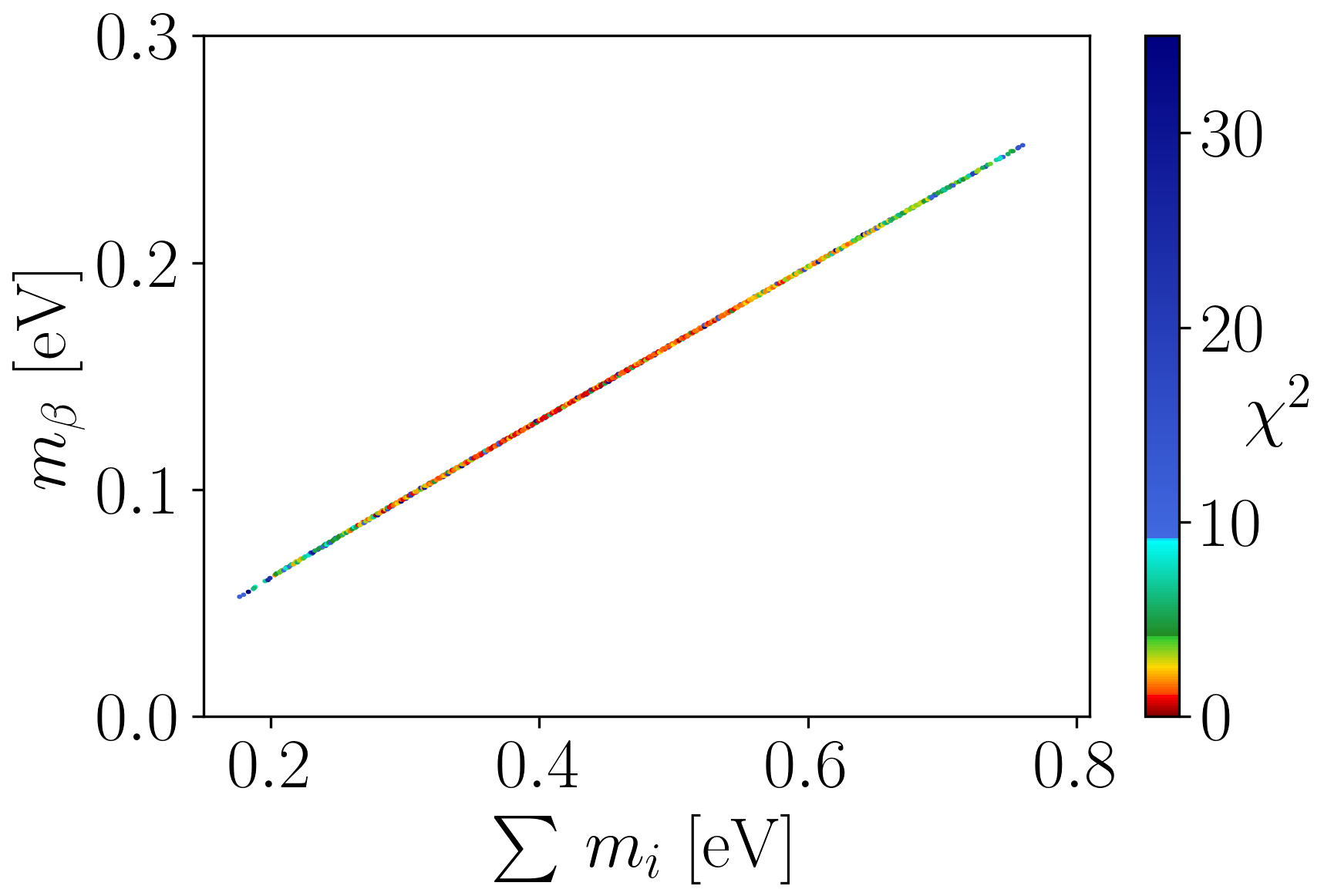}
		\caption*{A(3)}
	\end{subfigure} 
	\vspace{0.5cm}
	\begin{subfigure}[b]{0.3\textwidth}
		\includegraphics[width=\textwidth]{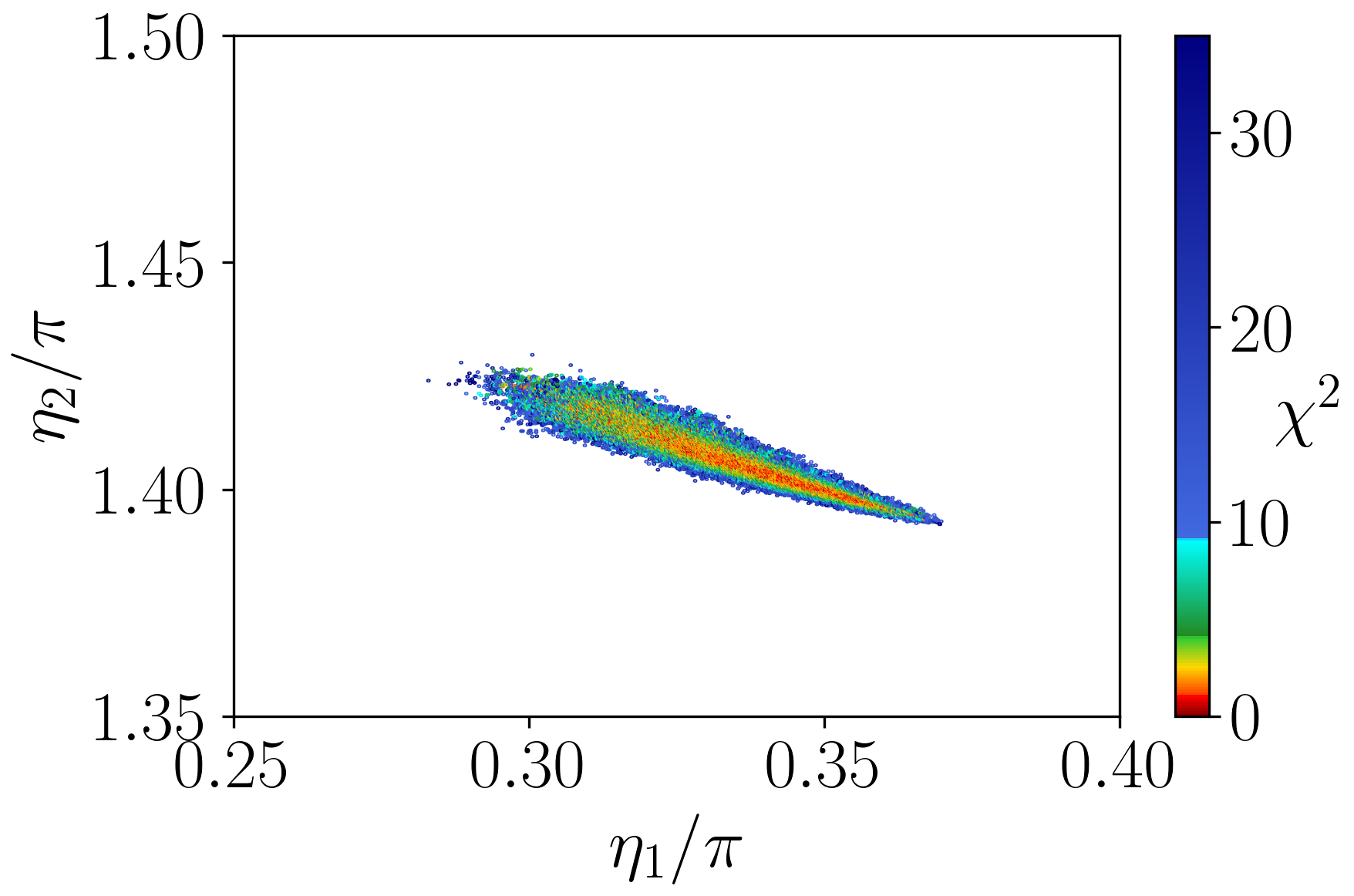}
		\caption*{B(1)}
	\end{subfigure}
	\hfill
	\begin{subfigure}[b]{0.3\textwidth}
		\includegraphics[width=\textwidth]{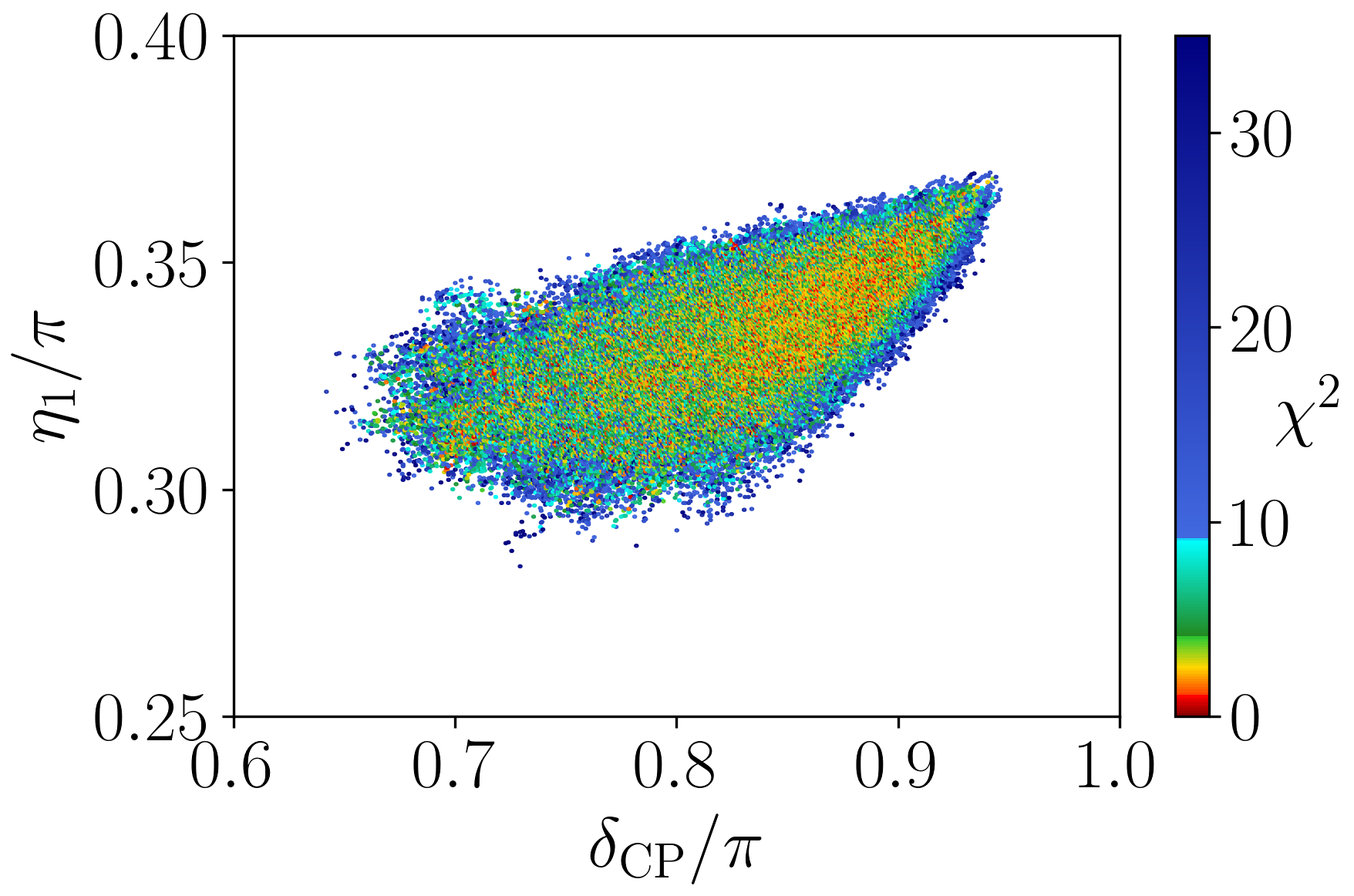}
		\caption*{B(2)}
	\end{subfigure}
	\hfill
	\begin{subfigure}[b]{0.3\textwidth}
		\includegraphics[width=\textwidth]{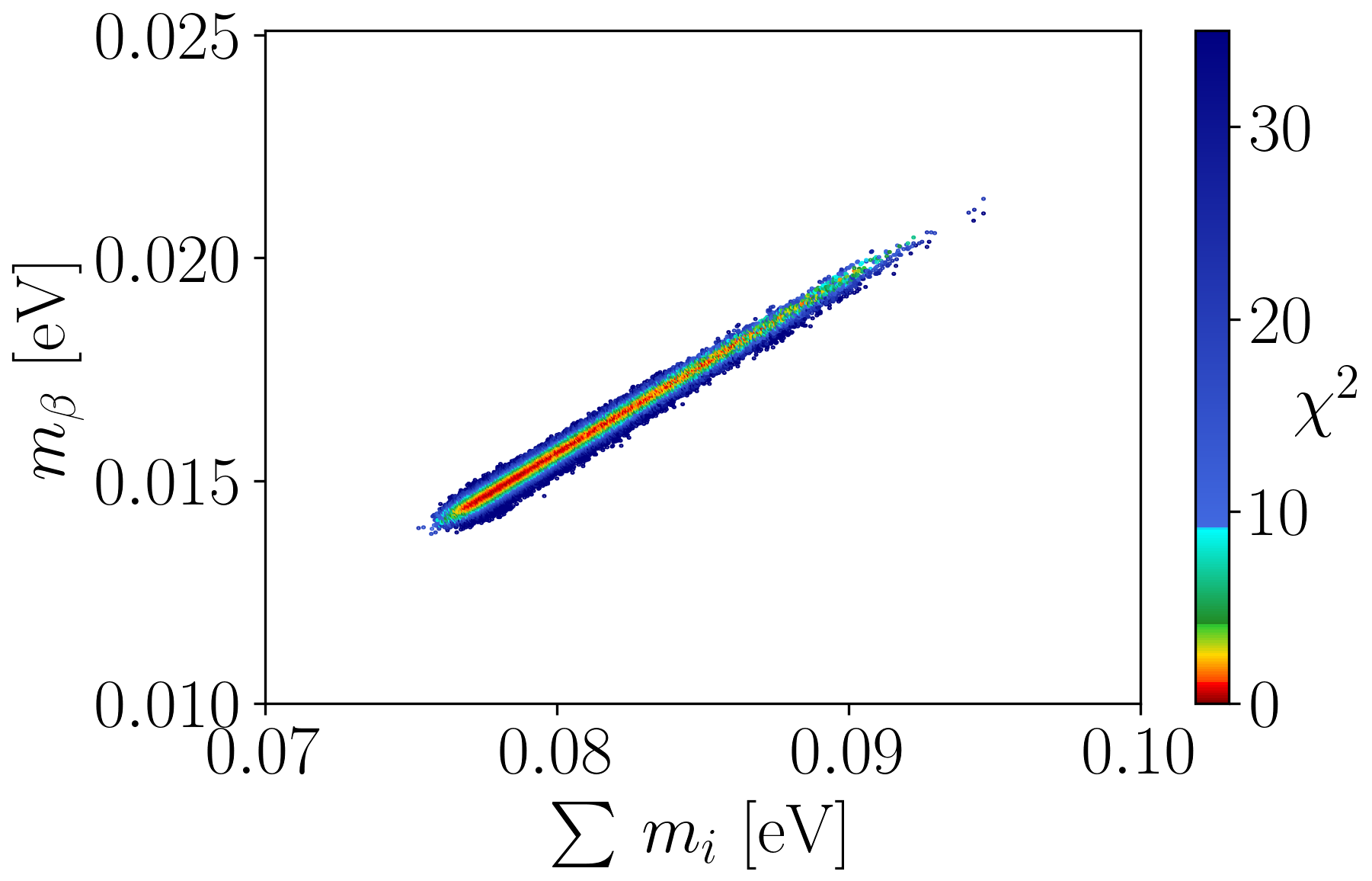}
		\caption*{B(3)}
	\end{subfigure}
	\vspace{0.5cm}
	\begin{subfigure}[b]{0.3\textwidth}
		\includegraphics[width=\textwidth]{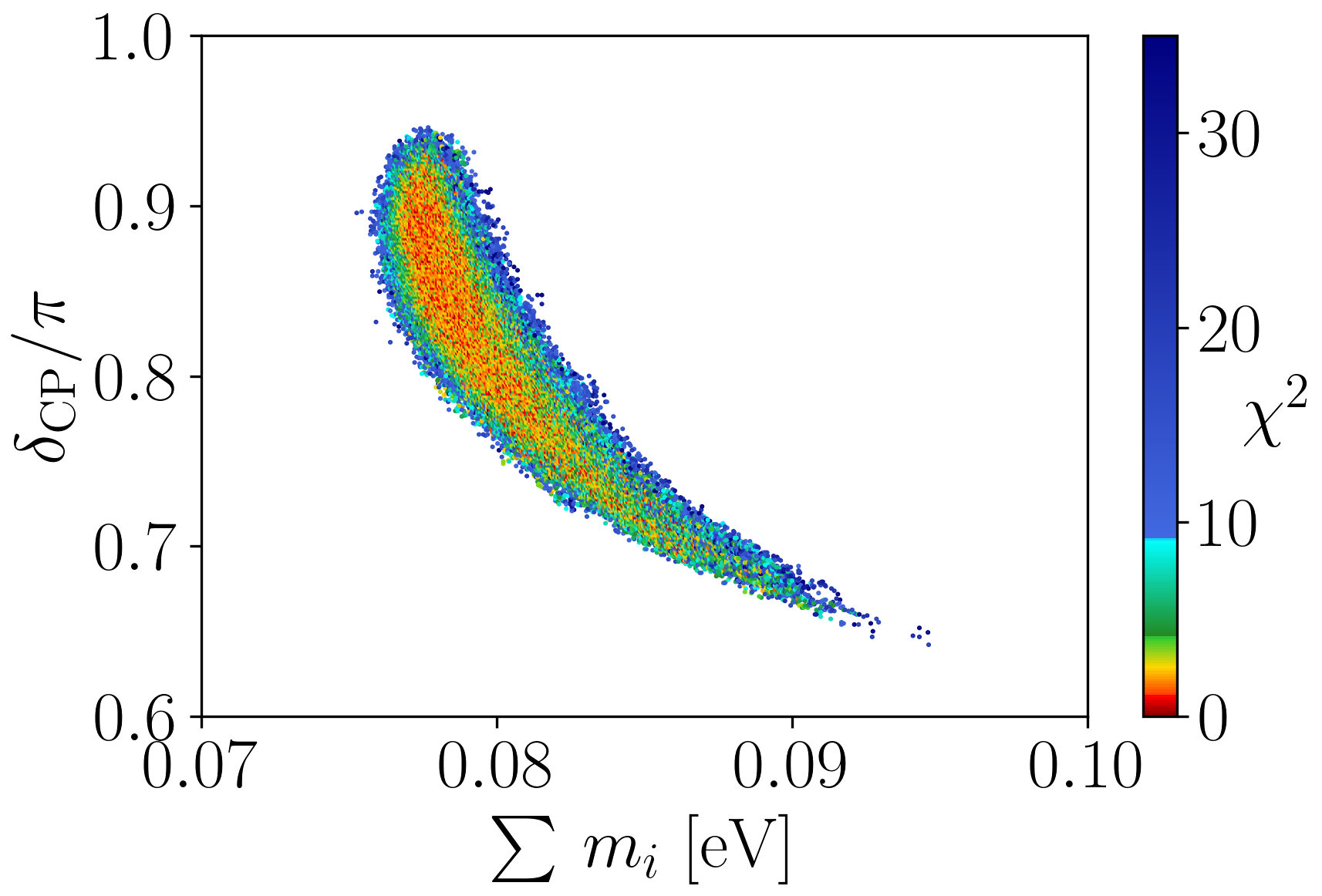}
		\caption*{B(4)}
	\end{subfigure}
	\hfill
	\begin{subfigure}[b]{0.3\textwidth}
		\includegraphics[width=\textwidth]{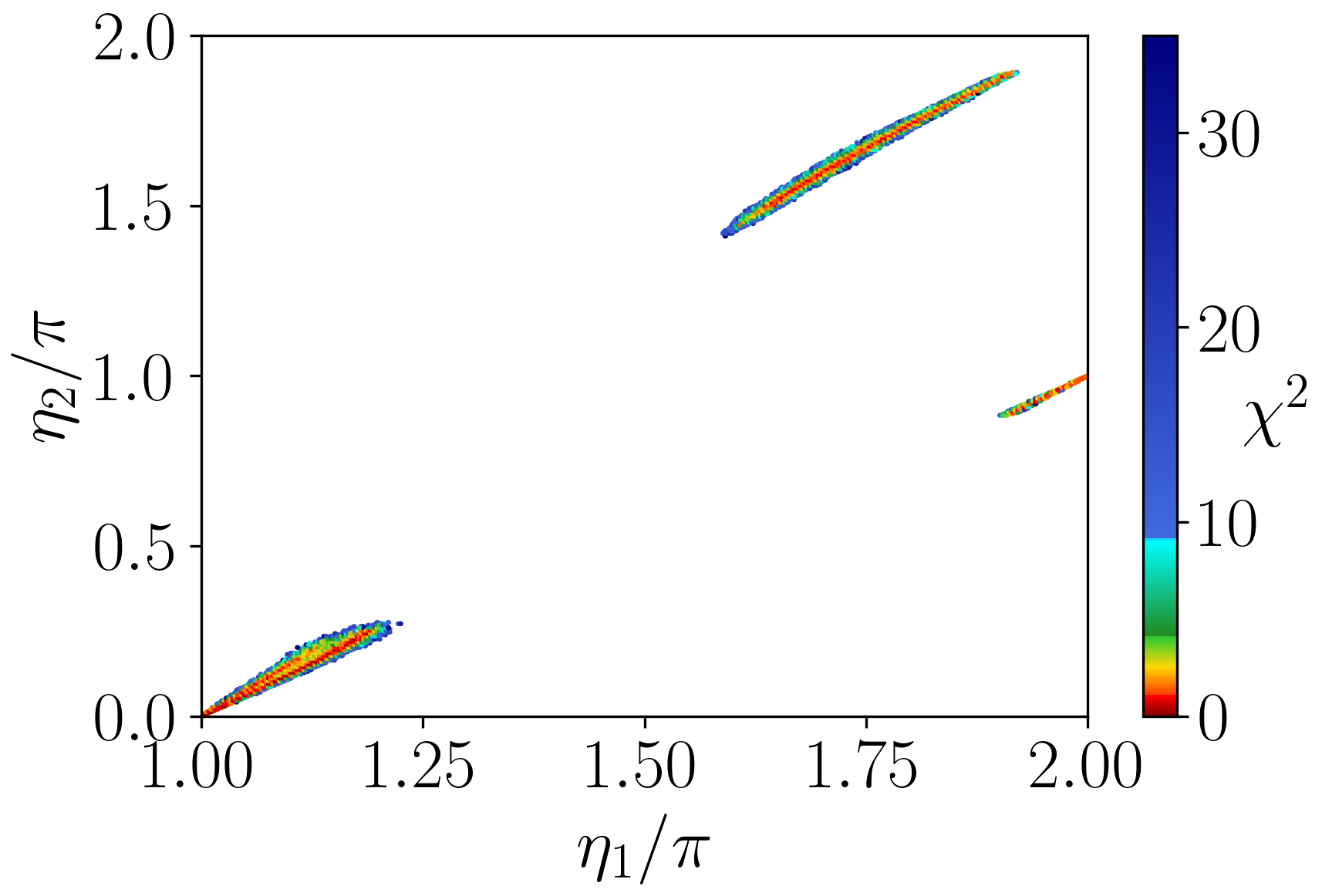}
		\caption*{C(1)}
	\end{subfigure}
	\hfill
	\begin{subfigure}[b]{0.3\textwidth}
		\includegraphics[width=\textwidth]{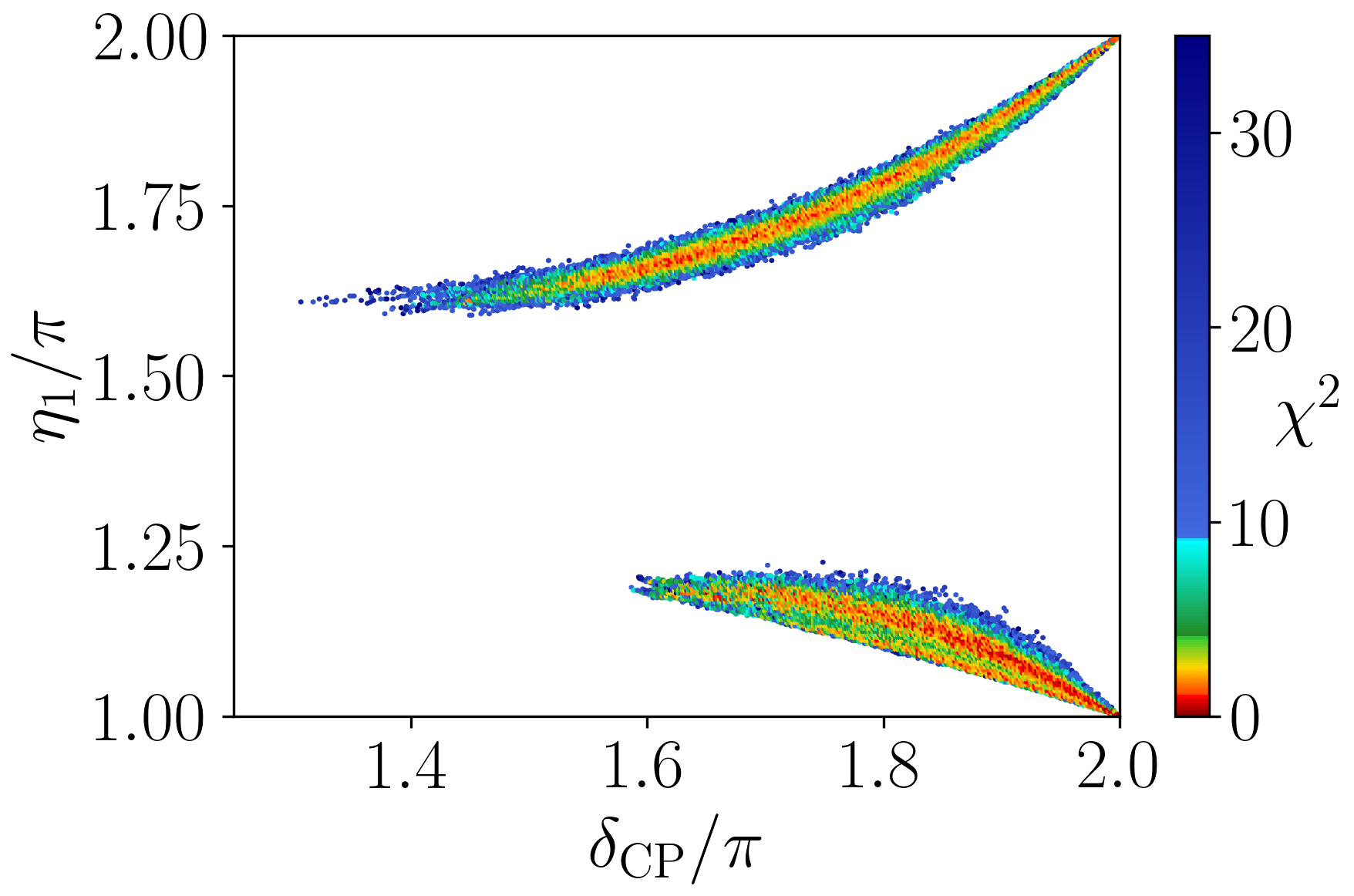}
		\caption*{C(2)}
	\end{subfigure}
	\vspace{0.5cm}
	\begin{subfigure}[b]{0.3\textwidth}
		\includegraphics[width=\textwidth]{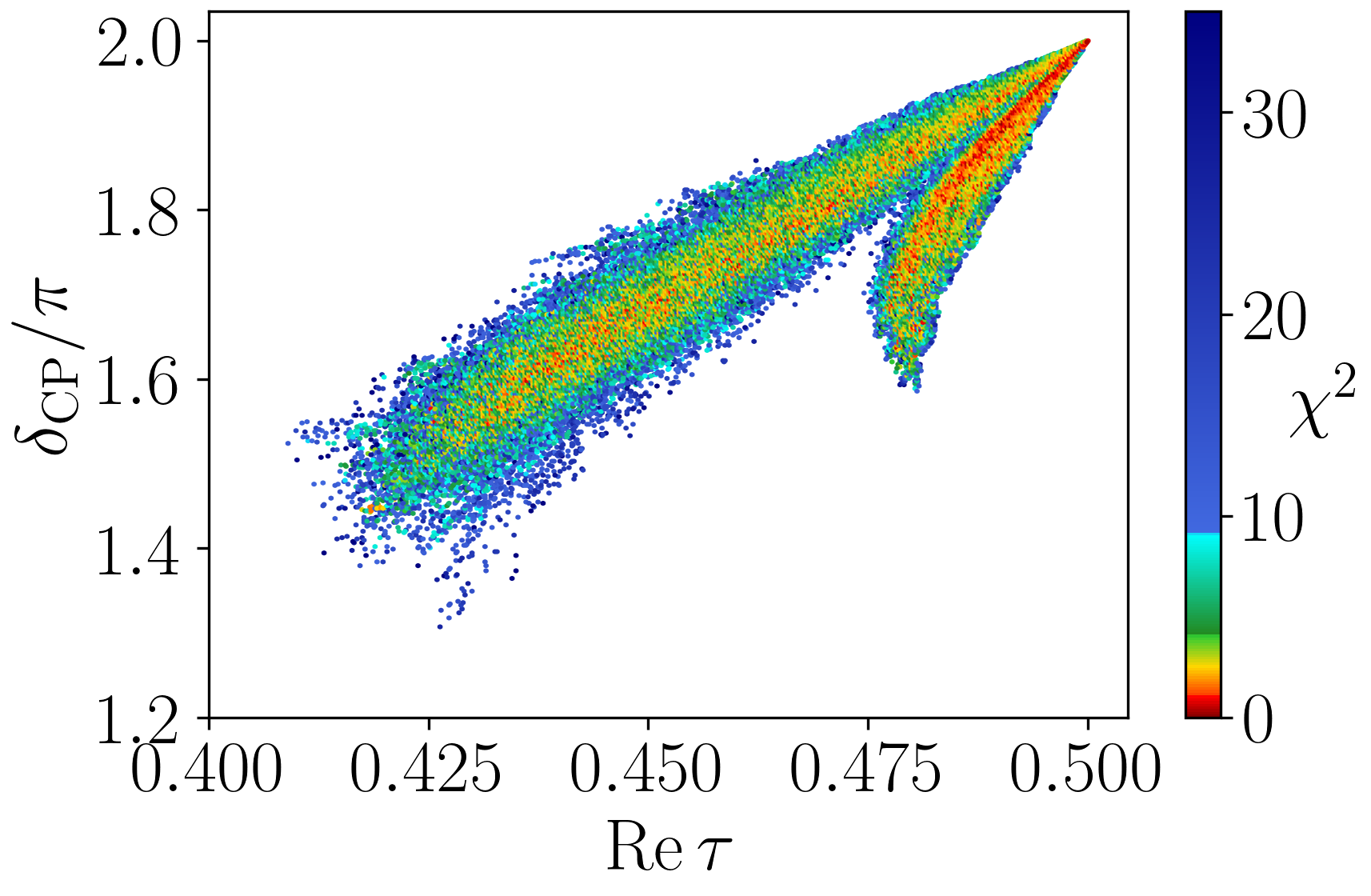}
		\caption*{C(3)}
	\end{subfigure}
	\hfill
	\begin{subfigure}[b]{0.3\textwidth}
		\includegraphics[width=\textwidth]{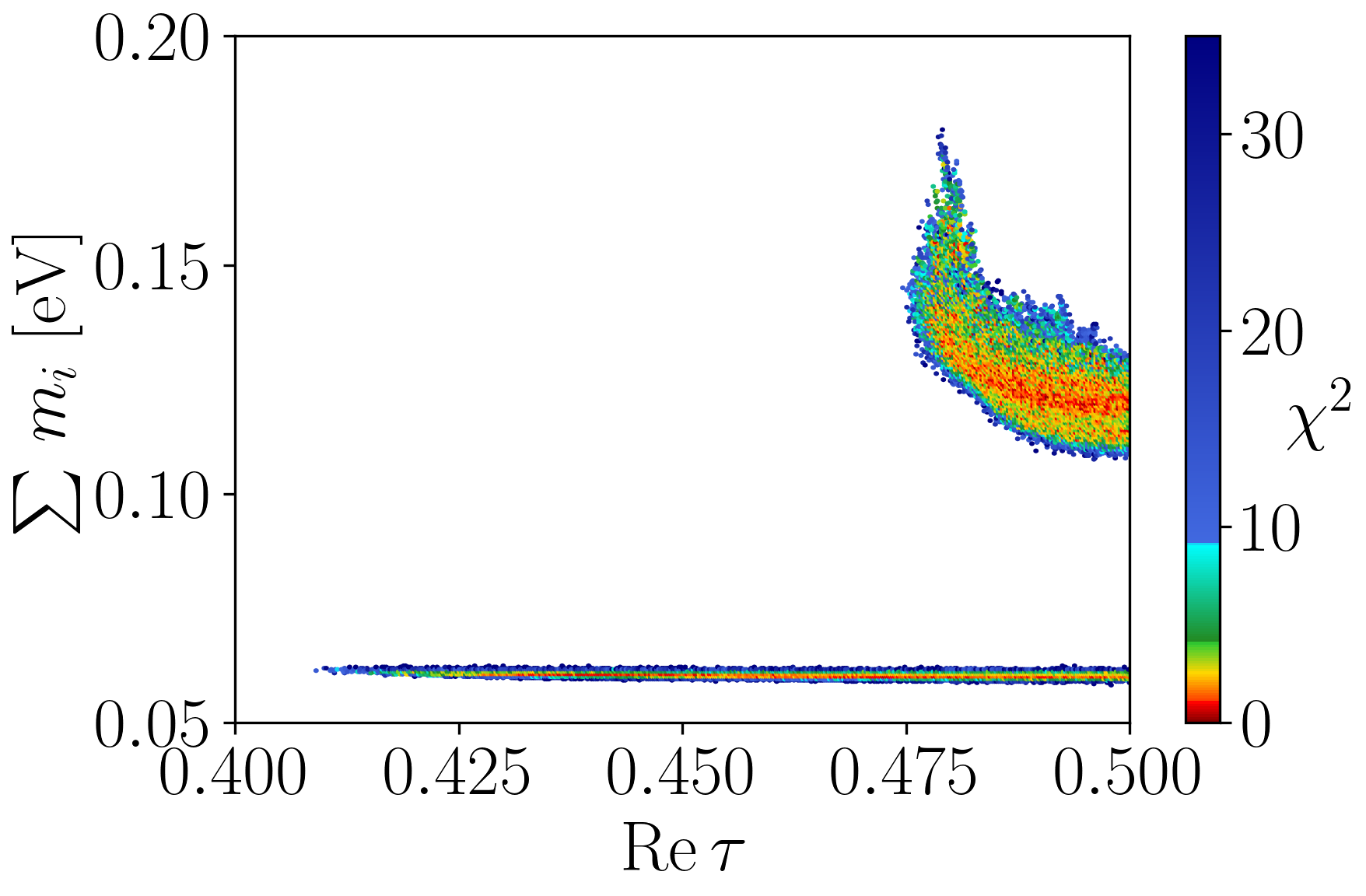}
		\caption*{C(4)}
	\end{subfigure}
	\hfill
	\begin{subfigure}[b]{0.3\textwidth}
		\includegraphics[width=\textwidth]{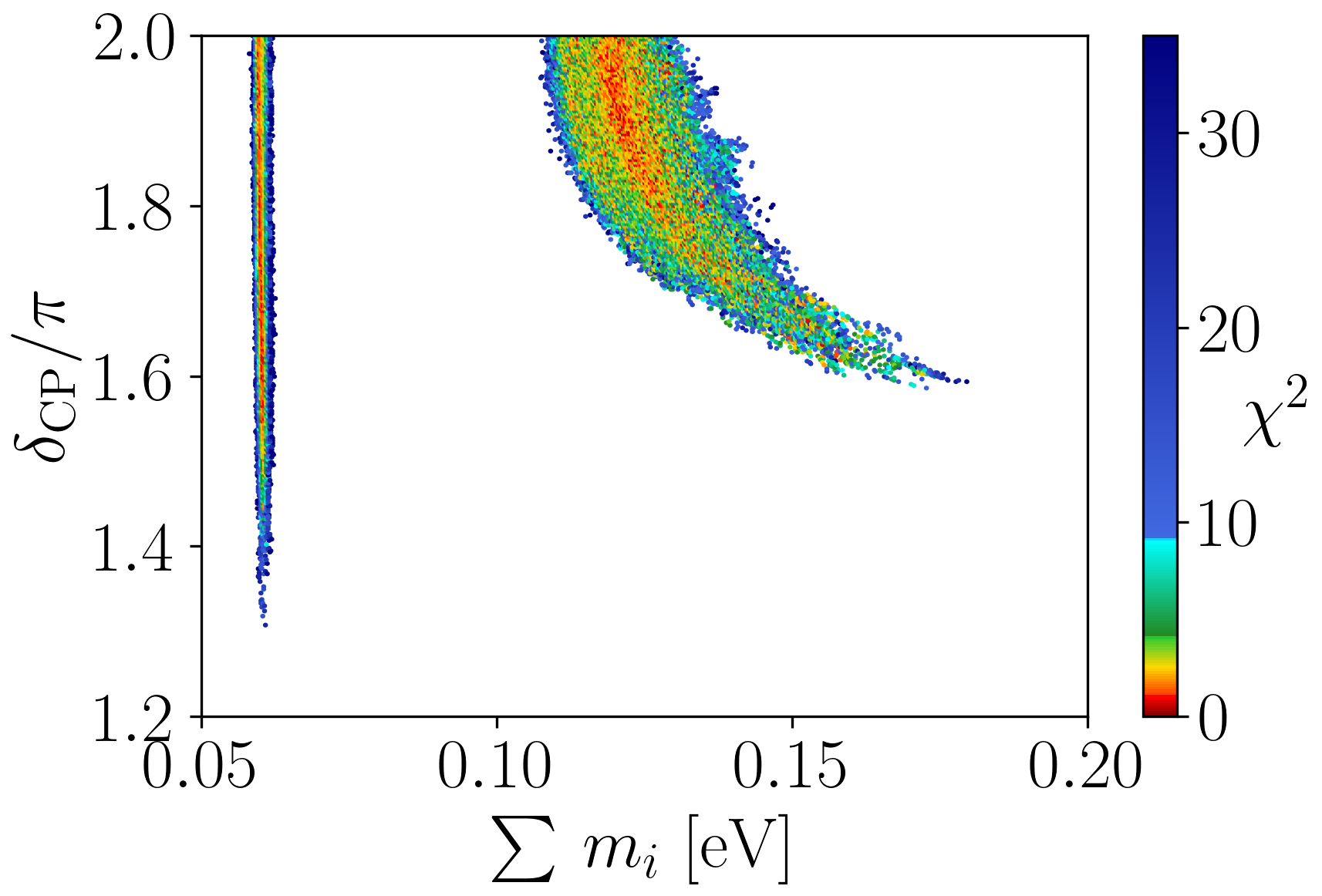}
		\caption*{C(5)}
	\end{subfigure}
	\caption{Correlations between physical parameters in model 3 NO, partitioned into phenomenological regions A (with subfigures A(1)–A(3)), B (with B(1)–B(4)), and C (with C(1)–C(5)). Region A exhibits minimal linear correlations. Region B shows anticorrelation of $\eta_1$ – $\eta_2$, positive correlations of $\eta_1$ – $\delta_{\rm CP}$ and of $m_\beta$ - $\sum m_i$, and neutrino mass–$\delta_{\rm CP}$ couplings. Region C features CP violating phase correlations $\eta_1$ - $\eta_2$, $\eta_1$ - $\delta_{\rm CP}$, ${\rm Re}(\tau)$-driven constraints, and a bimodal distribution of $\sum m_i$ with distinct experimental implications. }
	\label{fig:m3no}
\end{figure}

Figure~\ref{fig:m3no} illustrates the correlations among the neutrino parameters in three distinct regions A, B, and C for the NO case of model 3. 
\begin{table}[h!]
	\begin{center}
		\renewcommand{\arraystretch}{1.2}
		\setlength{\tabcolsep}{10pt}
		\begin{tabular}{c|c|c|c}\toprule
			\text{Model 3 (NO)}&\text{region A}&\text{region B}&\text{region C}\\
			\hline  
			$\tau $&$0.03807+1.0982i$&$0.08864+0.9978i$&$0.4856+1.0828i$\\
			$\frac{\beta_{\rm CL}}{\alpha_{\rm CL} }$&0.005410&203.2314&57.5523\\ 
			$\frac{\gamma_{\rm CL}}{\alpha_{\rm CL} }$&8.6155&396.8323&851.0561\\
			$\frac{\beta_{D}}{\alpha_{D} }$&167.7448&919.9664&0.5228\\ 
			$\frac{\gamma_{D}}{\alpha_{D} }$&631.6569&163.1855&4.3394\\
			$\frac{\beta_{{NS}}}{\alpha_{{NS}}}$&250.7566&444.7744&1.4974\\
			$\frac{\gamma_{{NS}}}{\alpha_{{NS}} }$&949.8954&5.7388&4.5820\\
			$\frac{\beta_{S}}{\alpha_{S} }$&420.0563&1.2148&355.3579\\
			\hline  
			$k$ (eV)&0.0008922&0.005075&0.0001720\\
			\hline 
			$\frac{m_e}{m_{\mu} }$&0.004737&0.004736&0.004736\\
			$\frac{m_{\mu}}{m_{\tau}}$&0.05880&0.05882&0.05881\\
			\hline  
			$\sin^2 \theta_{12}$&0.3117&0.3059&0.3069\\
			$\sin^2 \theta_{13}$&0.02224&0.02203&0.02217\\
			$\sin^2 \theta_{23}$&0.4692&0.4724&0.4719\\
			$\delta_{\rm CP}/\pi$&1.518&0.8638&1.834\\
			$\eta_1/\pi$&1.745&0.3382&1.130\\
			$\eta_2/\pi$&1.125&1.404&0.1621\\
			\hline 
			$\Delta m_{21}$ (eV)&7.491&7.498&7.497\\
			$\Delta m_{31}$ (eV)&2.513&2.5105&2.5106\\
			$m_1$ (eV)&0.1653&0.01168&0.03296\\
			$m_2$ (eV)&0.1656&0.01454&0.03408\\
			$m_3$ (eV)&0.1728&0.05145&0.05997\\
			$\sum m_i$ (eV)&0.5036&0.07767&0.1270\\
			$m_{\beta}$ (eV)&0.1656&0.01469&0.03415\\
			$m_{\beta\beta}$ (eV)&0.08045&0.01227&0.03375\\
			\hline 
			$\chi^2_{\rm min}$&0.2087&0.1039&0.1017\\
			\bottomrule
		\end{tabular}   
	\end{center}   
	\caption{Best-fit parameters for model 3 in NO across three distinct $\tau$ regions A, B, and C. Listed quantities include the modular parameter $\tau$, free parameters ratios, charged lepton mass ratios, neutrino mixing parameters, CP-violating phases, neutrino masses, mass-squared differences, effective $\beta$-decay mass, $0\nu\beta\beta$ decay mass, and the minimum $\chi^2$ values. Region B provides the only cosmologically viable scenario with $\sum m_i = 0.07767$ eV.}
	\label{tab:m3no_fit}
\end{table}
Region A includes three subfigures that explore fundamental parameter relationships. In subfigure A(1) , the Majorana phases $\eta_1$ and $\eta_2$ exhibit a two separately distributed relationship, with $\eta_1$ aggregates in two intervals $0.52\pi\le \eta_1 \le 0.81\pi$ and $1.66\pi\le \eta_1 \le 1.82\pi$, $\eta_2$ correspondingly changes in the intervals $(1.02, 1.17)\pi$ and $(1.07, 1.17)\pi$. These indicate a preference for CP violation in the Majorana sector. There are also two separate correlations between $\eta_1$ and $\delta_{\rm CP}$, $\eta_1 \in (0.52, 0.81)\pi$ for $\delta_{\rm CP}\in (1.35, 1.91)\pi$ as well as $\eta_1 \in (1.66, 1.82)\pi$ for $\delta_{\rm CP}\in (1.35,1.69)\pi$. Both of them align with global fit hints for near-maximal CP violation. The effective neutrino mass $m_\beta$ displays a tight linear correlation with the total neutrino mass $\sum m_i$, constrained to be $\sum m_{i} \approx (0.18$–$0.76)$ eV, which contradicts with existing upper bound. 

Region B extends the analysis through four subfigures. The anticorrelation between $\eta_2$ and $\eta_1$ suggests modular symmetry-driven constraints, with $\eta_1$ preferentially in $(0.28, 0.37)\pi$ and $\eta_2$ in $(1.39, 1.43)\pi$. The Majorana phase $\eta_1$ adopts a rather broad distribution over the Dirac phase $\delta_{\rm CP}$, favoring  $\eta_1 \in (0.28, 0.37)\pi$ for  $\delta_{\rm CP} \in (0.64, 0.95)\pi$. This indicates strict CP violations in both the Dirac and Majorana sectors. Unlike the situation in region A, $\sum m_i$ spans a range ($0.075$–$0.095$) eV, with $m_{\beta}$ changing in $(0.013, 0.022)$ eV. This region also reveals an anti-correlation of $\delta_{\rm CP}$ on $\sum m_i$. Higher mass scales correlate with $\delta_{\rm CP} \approx 0.6\pi$, while lower masses favor $\delta_{\rm CP} \approx 0.95\pi$, highlighting a direct interplay between absolute neutrino mass and CP-violating phases. These predictions partially overlap with the sensitivity of next generation $0\nu\beta\beta$ experiments, positioning region B as a testable frontier.  

Region C contains five subfigures and exhibits complex behavior tied to the modular parameter $\tau$. The Majorana phases $\eta_1$ and $\eta_2$ display trifurcated clustering, reflecting weaker constraints from the model. The large portion of the parameter space reveals the trend of CP violation, although small fragments near $\eta_1 \approx \pi, \eta_2 \approx 0$ and $\eta_1 \approx 2\pi, \eta_2 \approx \pi$ maintain a characteristic of CP-conservation. The correlation between the Majorana phase $\eta_1$ and the Dirac phase $\delta_{\rm CP}$ show two fragmented regions, covered with points in $3\pi/2 \le \eta_1\le 2\pi$ and $1.3\pi \le \delta_{\rm CP} \le 2.0\pi$, the other in $\pi\le \eta_1\le 5\pi/4$ and $1.57\pi \le \delta_{\rm CP} \le 2.0\pi$. Both parts indicate CP-violation, except for the right-ends that skew toward $\delta_{\rm CP} = 2\pi$. The correlation of the Dirac phase $\delta_{\rm CP}$ and $\text{Re}(\tau)$ divides into two branches. Most of the points lead Dirac CP-violation, while the upper-right edge has CP-conserving. In particular, at $\text{Re}(\tau) = 0.5$, a symmetry-enhanced point in the fundamental domain of $\tau$, $\delta_{\rm CP}$ collapses to $2\pi$. This signals exact CP conservation at the fixed point. The total neutrino mass $\sum m_i$ fragments into two regimes with continuous change of $\text{Re}(\tau)$. The lower $\sum m_i \approx 0.06$ eV stays almost constant over the change $\text{Re}(\tau) \approx 0.410$–$0.500$ and higher $\sum m_i\approx 0.10$–$0.18$ eV within the change $\text{Re}(\tau) \approx 0.475$–$0.500$. It is clear that most of the higher mass range is excluded. In the last panel, the correlation between the total neutrino mass $\sum m_i$ and the Dirac CP phase $\delta_{\rm CP}$ emerges as a bimodal distribution, with two distinct regimes governed by the interplay of modular symmetry and symmetry breaking effects. The primary cluster at $\sum m_i \approx 0.11-0.18$ eV (centered near 0.13 eV) shows a negative correlation with $\delta_{\rm CP}$, whose parameter values span $(1.6–2.0)\pi$. The secondary cluster at $\sum m_i \approx 0.06-0.07$ eV forms a near-vertical band of $\delta_{\rm CP}$ spanning $(1.3–2.0)\pi$.  The higher $\sum m_i$ values surpass the upper cosmological bound ($\sum m_i < 0.12$ eV), making it less favorable. This bifurcation arises from the  $\tau$-dependent Yukawa couplings that simultaneously modulate the absolute mass scale and the CP phases. The correlations in this figure offer testable targets for next-generation neutrino experiments and cosmological surveys.

\begin{figure}[h!]
	\centering
	\includegraphics[width=0.7\textwidth]{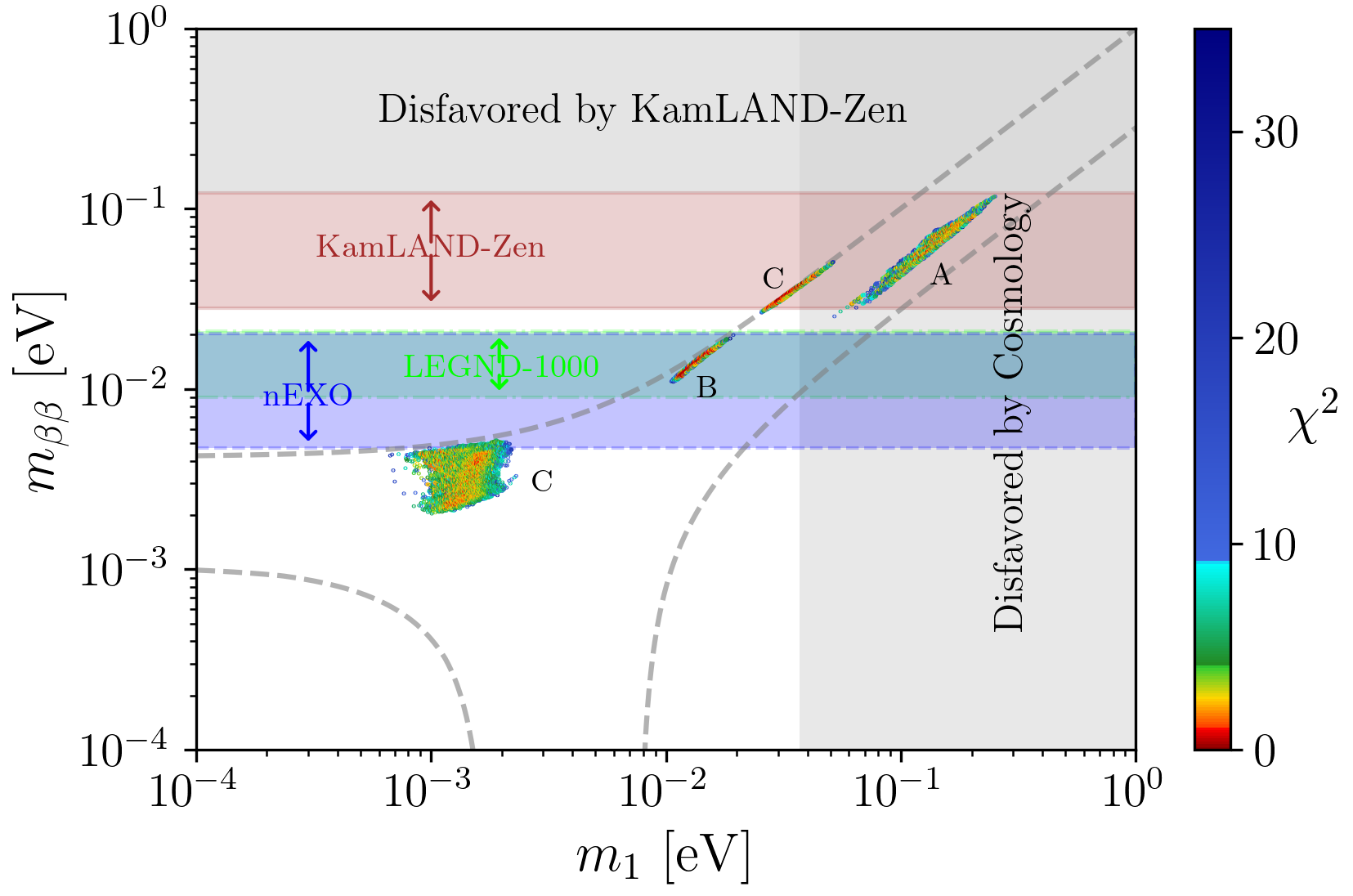}
	\caption{Predicted effective Majorana mass $m_{\beta\beta}$ as a function of the lightest neutrino mass in model 3 wth NO case. There are three distinct regions A, B, and C with different experimental implications.}
	\label{fig:m3no_0vbb}
\end{figure}

Figure~\ref{fig:m3no_0vbb} displays the effective Majorana mass $m_{\beta\beta}$ versus the lightest neutrino mass for the NO case of model 3, with regions A, B, and C delimitated, and overlaid with current and projected experimental sensitivities for $0\nu\beta\beta$ decay. Region A is constrained to $m_1 \sim 0.05-0.4$ eV where $m_{\beta\beta}\sim 2.5\times 10^{-2}-10^{-1}$ eV. This concentration suggests phase constraints that shape the lightest mass while allowing moderate variation in $m_{\beta\beta}$. These predicted points are entirely within the reach of next-generation experiments, however, they fall inside the cosmologically disfavored area.  Region B occupies $m_1\sim 0.01-0.02$ eV and $m_{\beta\beta} \sim 10^{-2}-10^{-3}$ eV, which overlaps the projected sensitivities of next generation $0\nu\beta\beta$ experiments (LEGEND-1000, nEXO).  While the prediction of region C has two regions, one of which is mostly within the upper limit given by the current KamLAND-Zen experiment, the prediction of another region covers a small part within the sensitivity of that next generation $0\nu\beta\beta$ experiment. The majority of prediction in the region C is below the sensitivities of future experiments. This suppresion is due to the role of phase cancellations in the effective mass expression.

\begin{figure}[h!]
	\centering
	\includegraphics[width=0.6\textwidth]{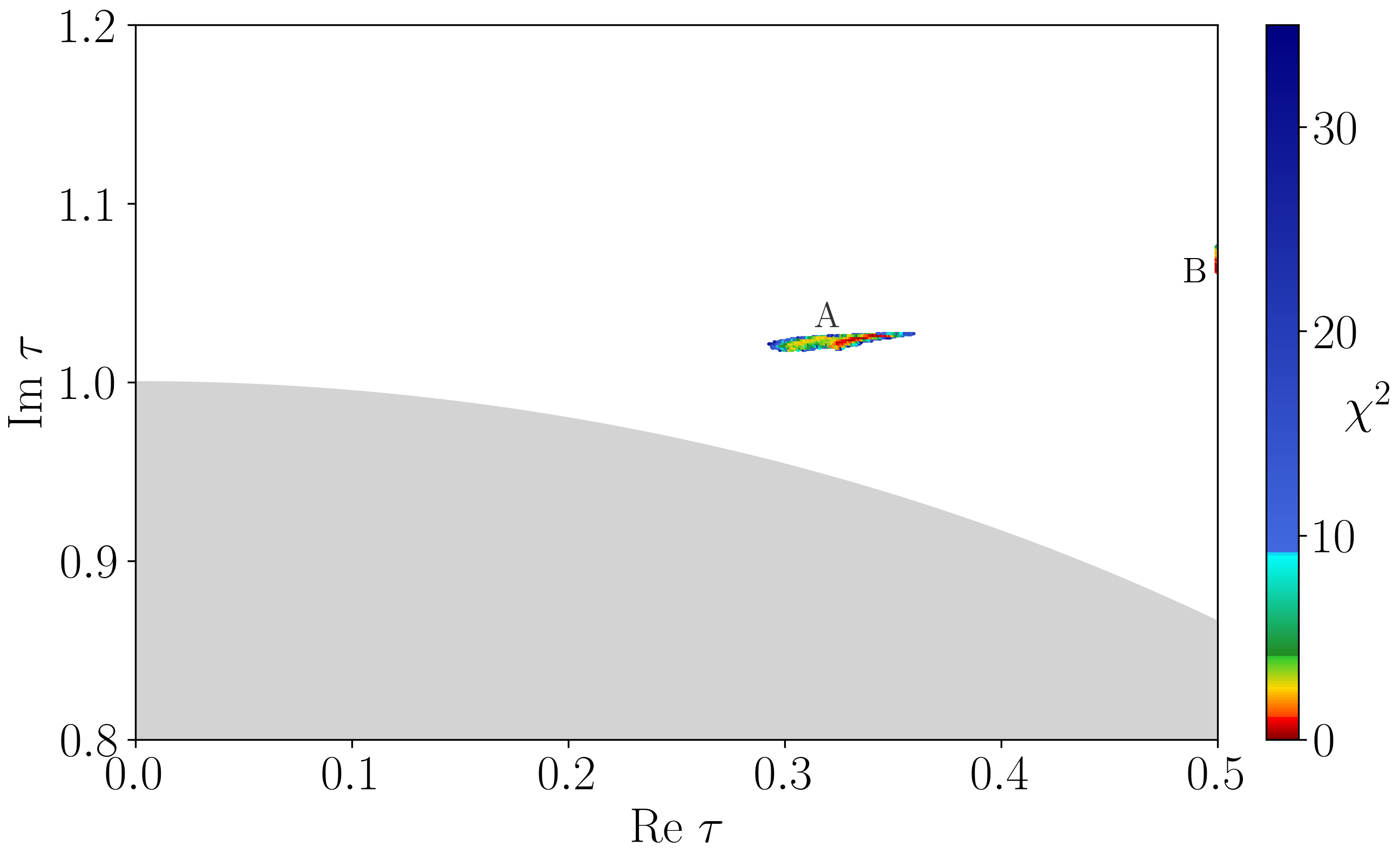}
	\caption{Feasible regions in the $\mathrm{Re}(\tau)$–$\mathrm{Im}(\tau)$ plane for model 3 with IO. Region A has $\chi^2_{\mathrm{min}} = 0.1037$ at $\tau = 0.3327 + 1.0242i$, and exhibits significant CP violation, while region B, with $\chi^2_{\mathrm{min}} = 0.2056$ at $\tau = 0.4999 + 1.0628i$, shows CP suppression near $\mathrm{Re}(\tau) = 0.5$. Contours denote boundary of fundamental domain, and color intensity indicates $\chi^2$ values.}
	\label{fig:m3io_tau}
\end{figure}

Figure~\ref{fig:m3io_tau} delineates the feasible regions of the modulus parameter $\tau$ for model 3 in the case of IO. There are two phenomenologically distinct regions, marked A and B. Region A converges at $\tau = 0.3327 + 1.0242i$ with $\chi^2_{\mathrm{min}} = 0.1037$, while region B lies near the boundary $\mathrm{Re}(\tau) \sim 0.5$ at $\tau = 0.4999 + 1.0628i$ with $\chi^2_{\mathrm{min}} = 0.2056$. Both regions predict neutrino masses below current experimental limits, featuring an inverted hierarchy where the lightest mass $m_3 \sim 10^{-5}$ eV is negligible compared to $m_1 \approx m_2 \sim 0.049$ eV, resulting in $\sum m_i \approx 0.099$ eV, which is consistent with cosmological bounds.  
Specifically, the predicted values of neutrino parameters from the best-fit point of region A are 
  \begin{equation}
	\begin{aligned}
		&m_e/m_{\mu}=0.004737,\quad m_{\mu}/m_{\tau}=0.05882,\\
		&\sin^2\theta_{12}=0.3054,\quad \sin^2\theta_{13}=0.02221,\quad \sin^2\theta_{23}=0.5538,\\
		&\delta_{\rm CP}=1.848\pi,\quad \eta_1=0.2271\pi,\quad \eta_2=1.605\pi,\\
		&m_1=0.04909~ \mathrm{eV},\quad m_2=0.04985~ \mathrm{eV},\quad m_3=0.00002079~\mathrm{eV},\\
		&\sum m_i=0.09896~\mathrm{eV},\quad m_{\beta}=0.04981~ \mathrm{eV},\quad m_{\beta\beta}=0.02486~\mathrm{eV}
	\end{aligned}
\end{equation}
and from region B are 
\begin{equation}
	\begin{aligned}
		&m_e/m_{\mu}=0.004737,\quad m_{\mu}/m_{\tau}=0.05883,\\
		&\sin^2\theta_{12}=0.3103,\quad \sin^2\theta_{13}=0.02216,\quad \sin^2\theta_{23}=0.5545,\\
		&\delta_{\rm CP}=0.05518\pi,\quad \eta_1=0.004608\pi,\quad \eta_2=0.9985\pi,\\
		&m_1=0.04909~ \mathrm{eV},\quad m_2=0.04985~ \mathrm{eV},\quad m_3=0.00007811~\mathrm{eV},\\
		&\sum m_i=0.09901~\mathrm{eV},\quad m_{\beta}=0.04980 ~\mathrm{eV},\quad m_{\beta\beta}=0.04822~\mathrm{eV}
	\end{aligned}
\end{equation}
From these expressions, we can see that region A exhibits near-maximal Dirac CP violation, $\delta_{\mathrm{CP}} = 1.848\pi$, and non-trivial Majorana phases $\eta_1 = 0.2271\pi$, $\eta_2 = 1.605\pi$. These phases induce negative interference in neutrinoless double beta decay, yielding $m_{\beta\beta} = 0.02486$ eV, a value challenging for current experiments but potentially accessible to next-generation detectors. In contrast, region B demonstrates suppressed CP violation $\delta_{\mathrm{CP}} = 0.05518\pi$ as $\mathrm{Re}(\tau)$ approaches the symmetry-enforced boundary at 0.5, with Majorana phases favoring positive interference $\eta_1 \approx 0$, $\eta_2 \approx \pi$. This alignment produces $m_{\beta\beta} = 0.04822$ eV, placing it within the sensitivity range of upcoming experiments like nEXO.

\begin{figure}[h!]
	\centering
	\begin{subfigure}[b]{0.3\textwidth}
		\includegraphics[width=\textwidth]{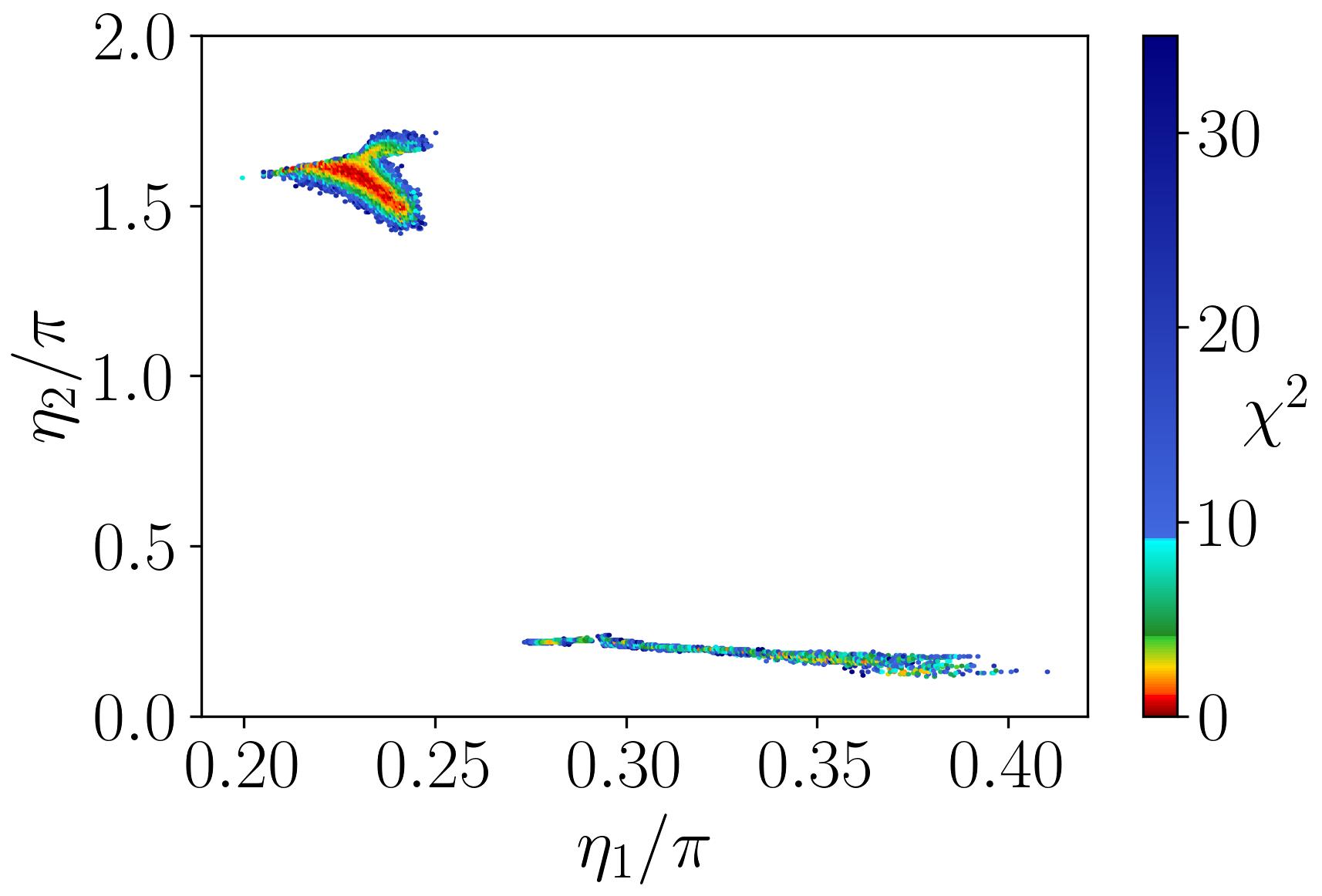}
		\caption*{A(1)}
	\end{subfigure}
	\hfill
	\begin{subfigure}[b]{0.3\textwidth}
		\includegraphics[width=\textwidth]{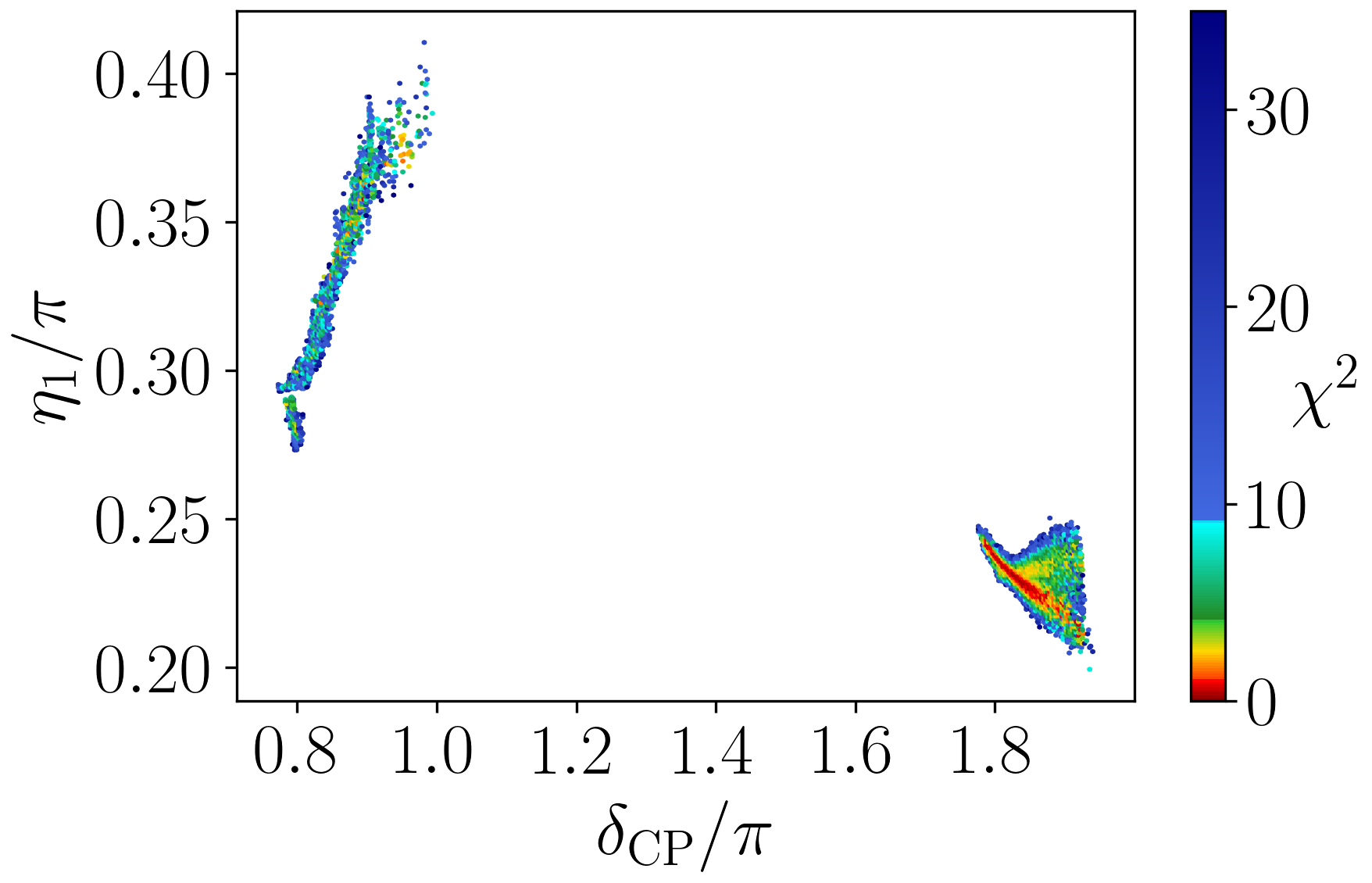}
		\caption*{A(2)}
	\end{subfigure}
	\hfill
	\begin{subfigure}[b]{0.3\textwidth}
		\includegraphics[width=\textwidth]{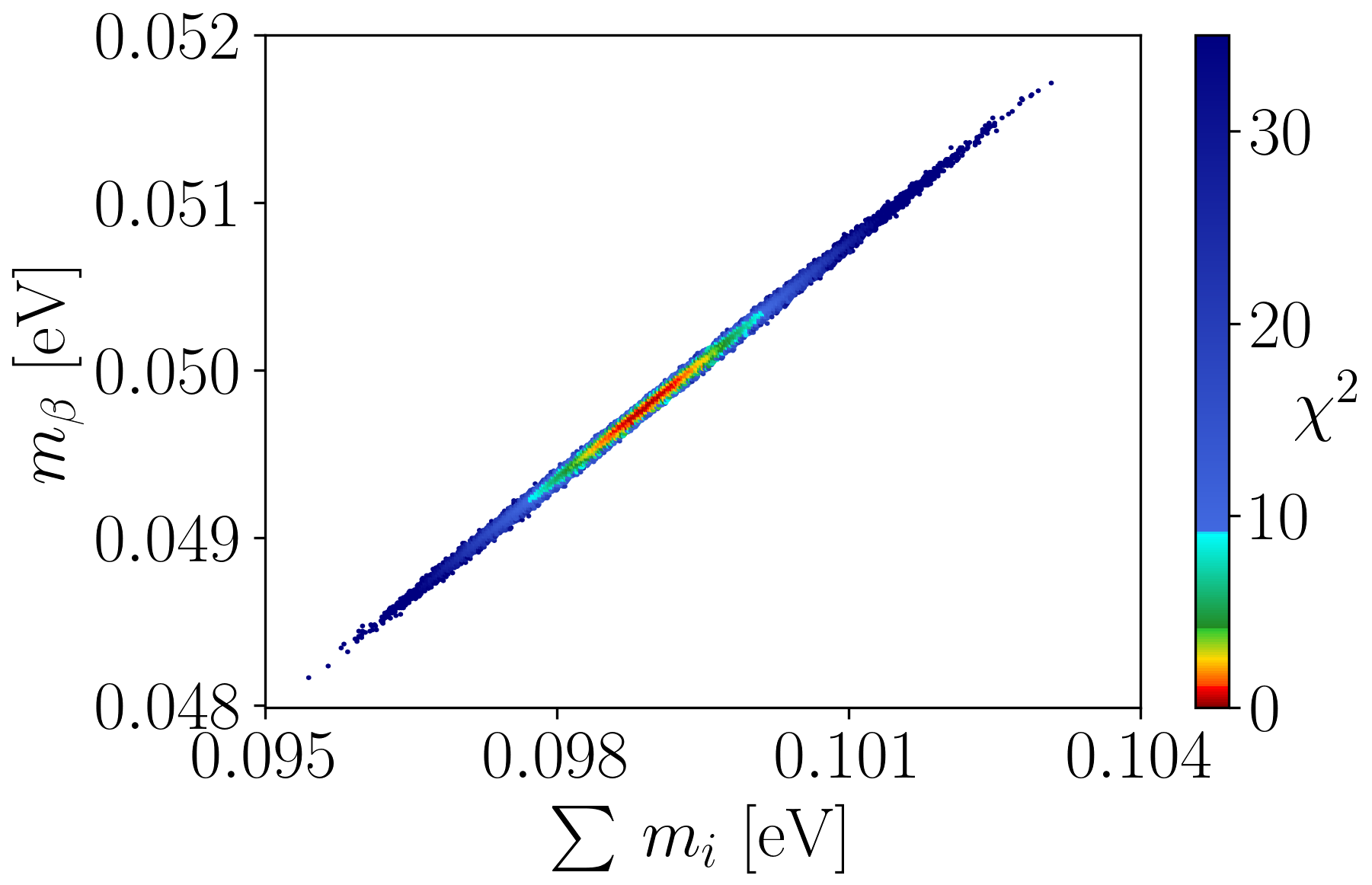}
		\caption*{A(3)}
	\end{subfigure}
	\vspace{0.5cm}
	\begin{subfigure}[b]{0.3\textwidth}
		\includegraphics[width=\textwidth]{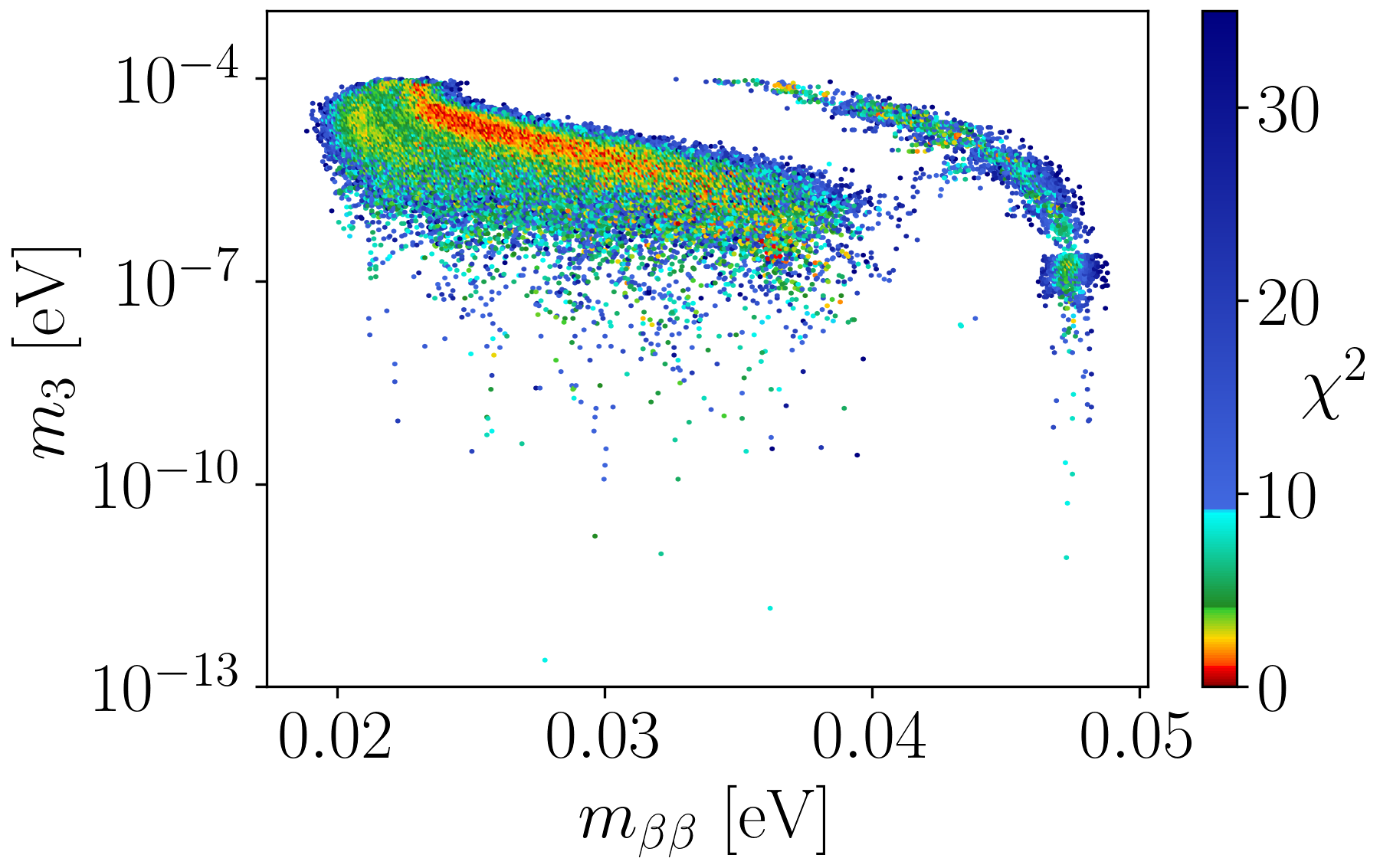}
		\caption*{A(4)}
	\end{subfigure}
	\hfill
	\begin{subfigure}[b]{0.3\textwidth}
		\includegraphics[width=\textwidth]{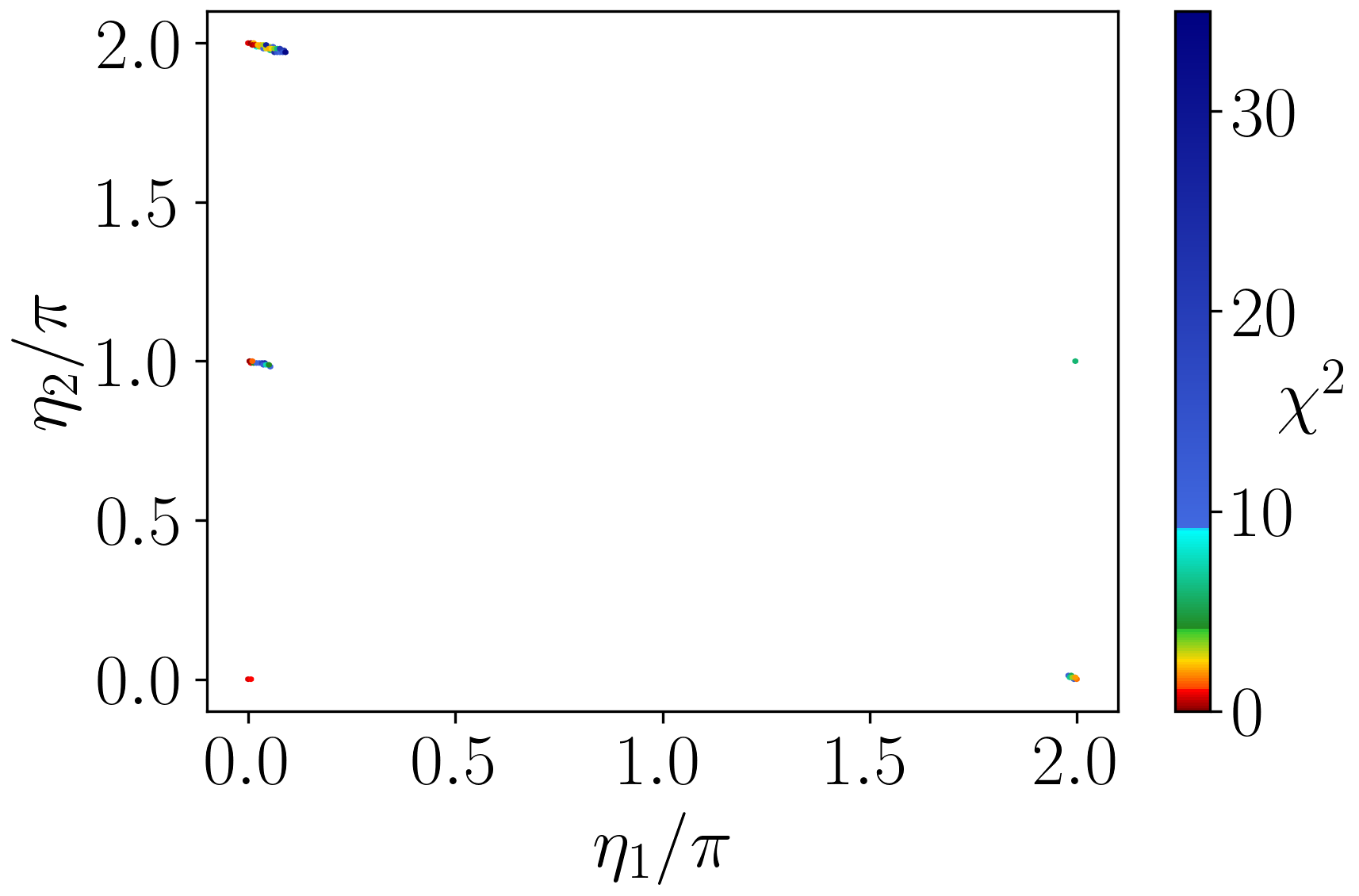}
		\caption*{B(1)}
	\end{subfigure}
	\hfill
	\begin{subfigure}[b]{0.3\textwidth}
		\includegraphics[width=\textwidth]{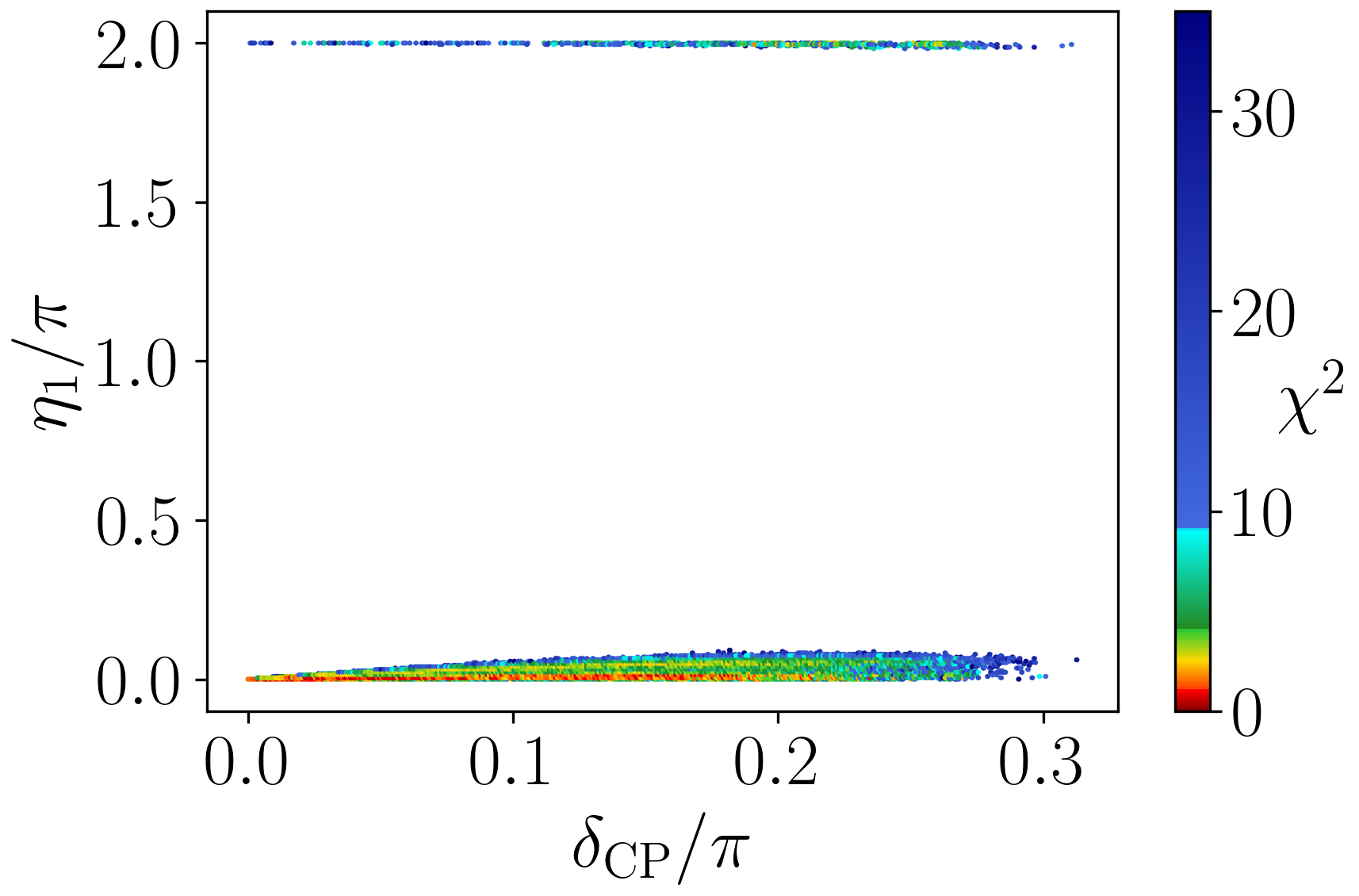}
		\caption*{B(2)}
	\end{subfigure}
	\vspace{0.5cm}
	\begin{subfigure}[b]{0.3\textwidth}
		\includegraphics[width=\textwidth]{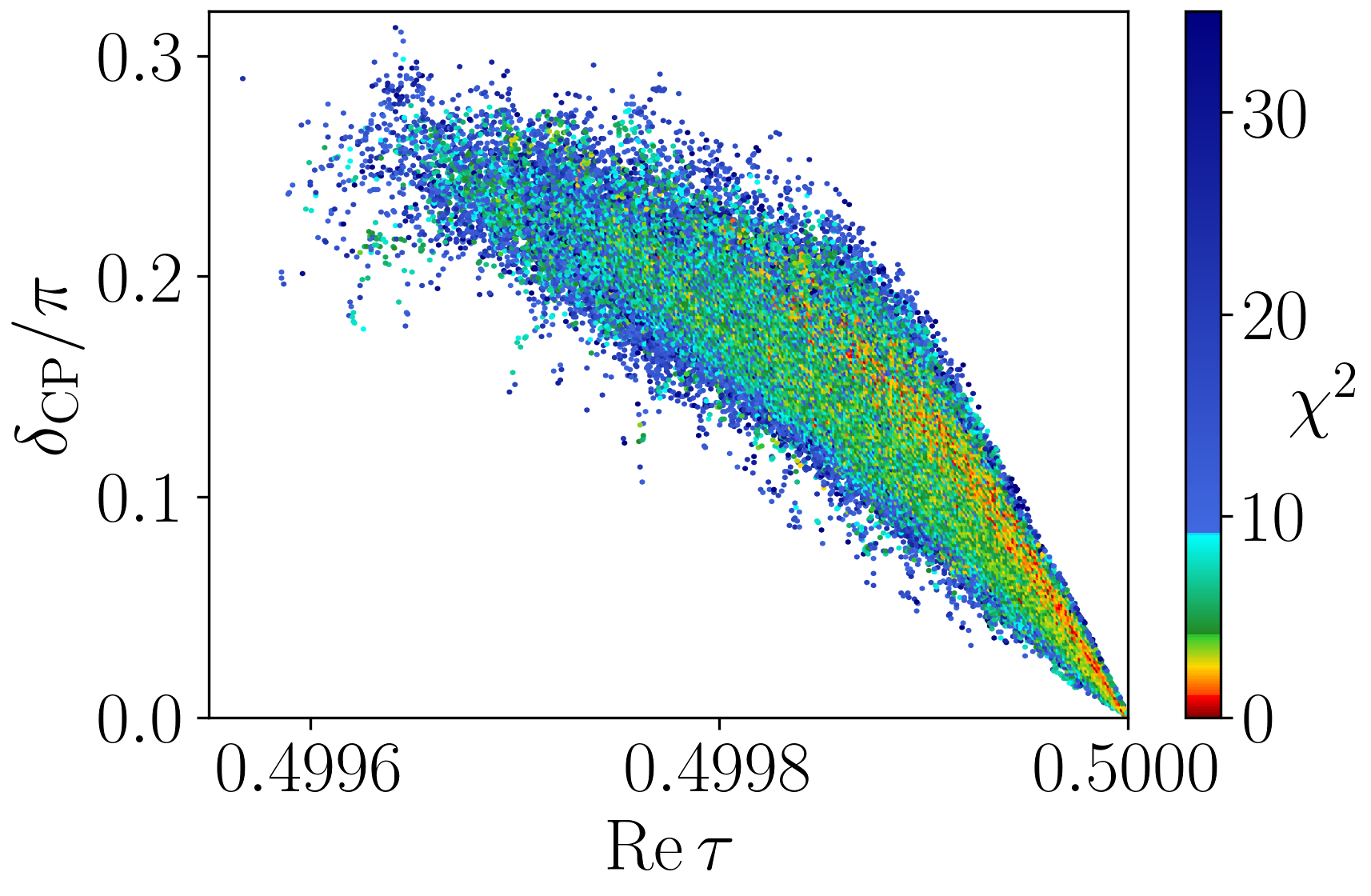}
		\caption*{B(3)}
	\end{subfigure}
	\hfill
	\begin{subfigure}[b]{0.3\textwidth}
		\includegraphics[width=\textwidth]{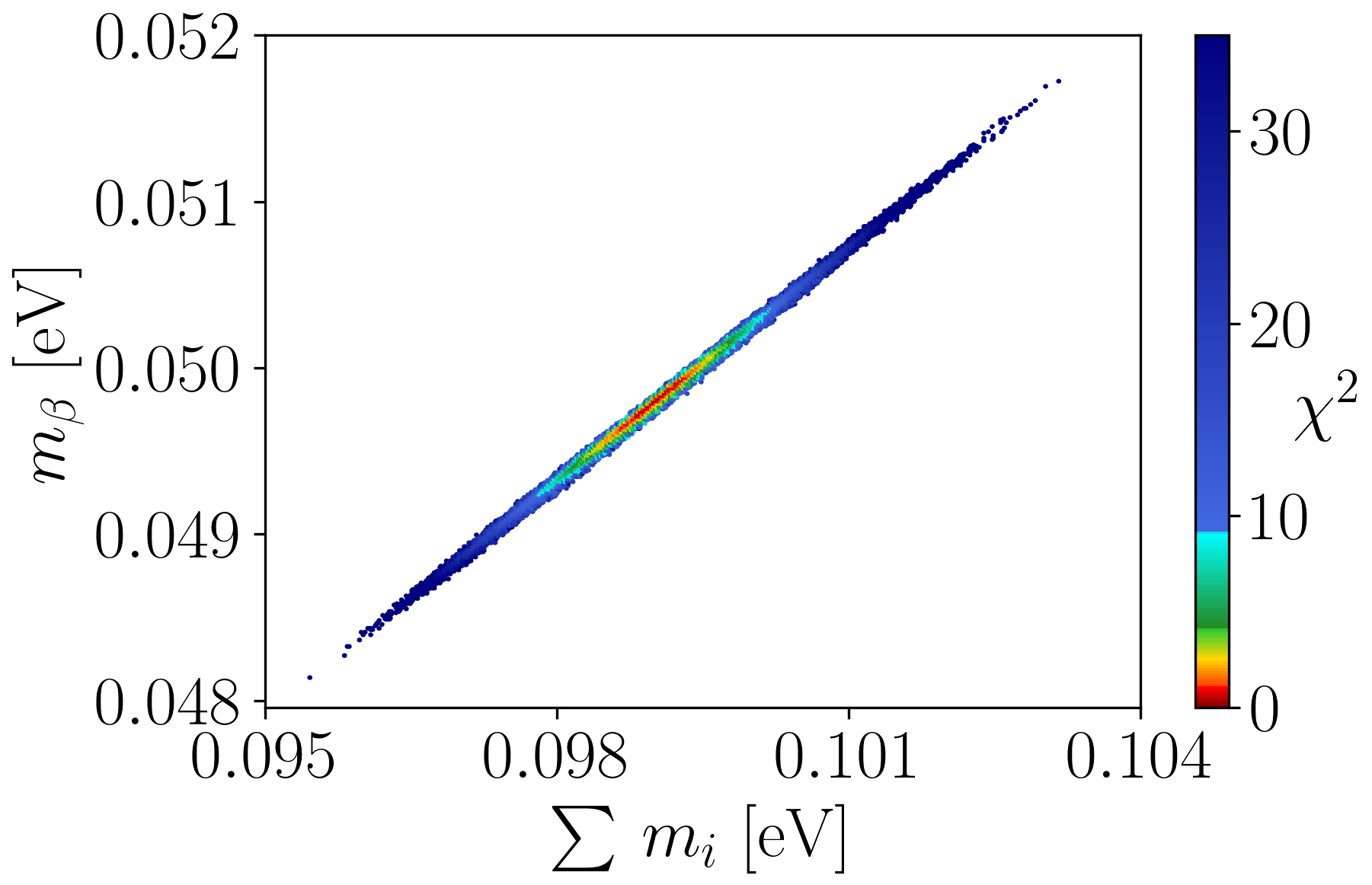}
		\caption*{B(4)}
	\end{subfigure}
	\hfill
	\begin{subfigure}[b]{0.3\textwidth}
		\includegraphics[width=\textwidth]{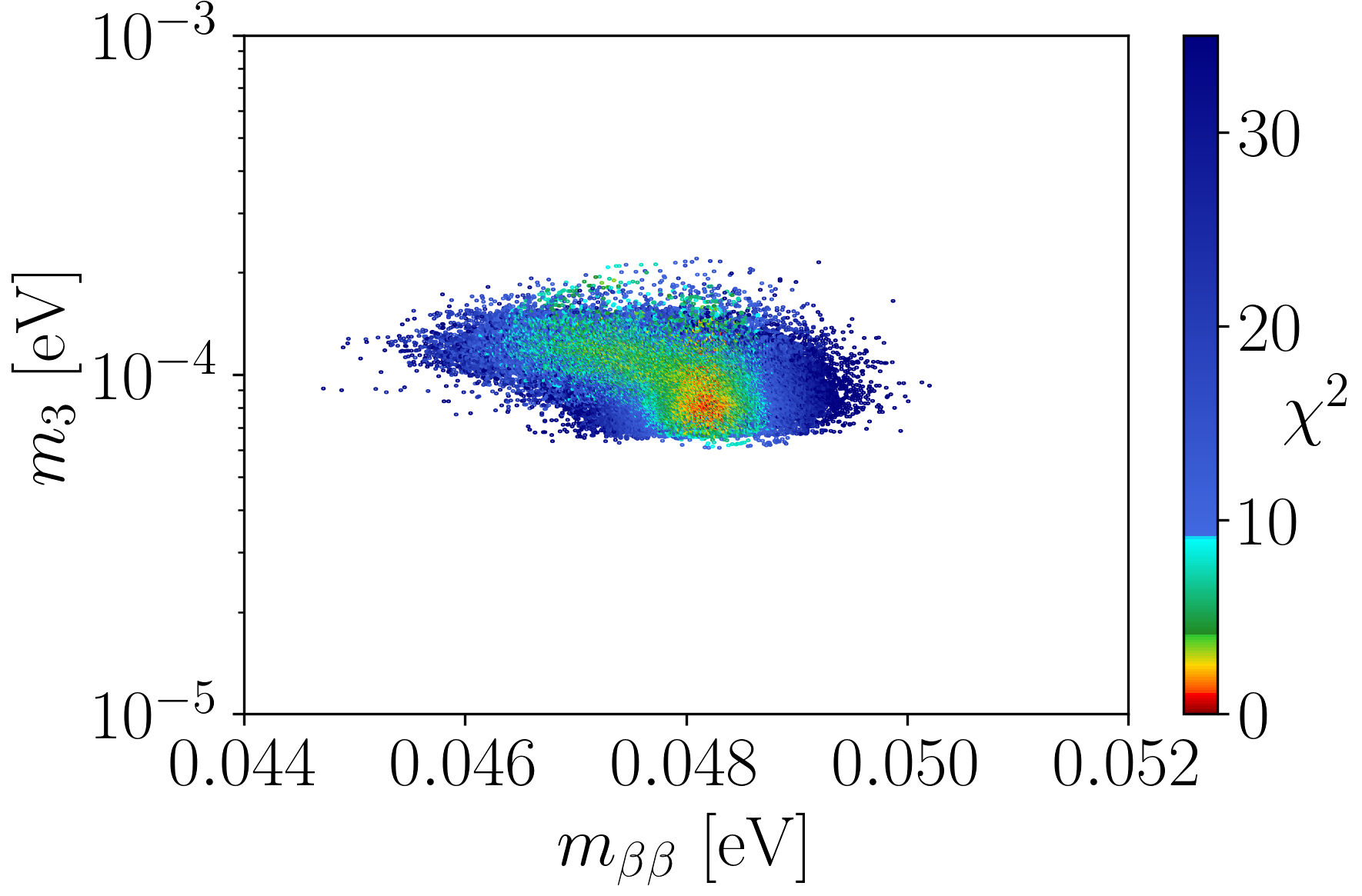}
		\caption*{B(5)}
	\end{subfigure}
	
	\caption{Parameter correlations in model 3 for IO case. Region A: Majorana phase correlation A(1), $\eta_1$-$\delta_{\rm CP}$ relation A(2), $m_\beta$ vs. $\sum m_i$ A(3), $m_3$ vs. $m_{\beta\beta}$ A(4). Region B: $\eta_2$-$\eta_1$ correlation B(1), Majorana and Dirac phase relation B(2), $\delta_{\rm CP}$ -  $\mathrm{Re}(\tau)$ boundary effect B(3), $m_3$ vs. $\sum m_i$ B(4), $m_{\beta\beta}$ vs. $m_3$ B(5).}
	\label{fig:m3io}
\end{figure}

Figure~\ref{fig:m3io} details the correlations between the physical parameters in model 3 in the IO scenario, divided into regions A and B. The Majorana phases $\eta_1$ and $\eta_2$ exhibit a quasi-uniform distribution across $\eta_2\approx 0.2\pi $ when $\eta_1 \approx (0.27-0.41)\pi$, with secondary clusters at $\eta_2 \approx 1.6\pi$ when $\eta_1\approx (0.20-0.25)\pi$. The clustering aligns with the best-fit point ($\eta_1 = 0.2271\pi$, $\eta_2 = 1.605\pi$) and reflects phase configurations that partially suppress $m_{\beta\beta}$ through negative interference. The relationship between Majorana phase $\eta_1$ and Dirac CP phase $\delta_{\rm CP}$ shows two partitioned distribution. There is a pronounced correlation; $\eta_1$ increases from $0.27\pi$ to $0.41\pi$ as $\delta_{\rm CP}$ increases from $0.8\pi$ to $1.1\pi$. In addition, $\eta_1$ clusters in $(0.20-0.25)\pi$ when $\delta_{\rm CP}$ varies in $(1.77-1.88)\pi$. This region covers the best-fit point $\delta_{\rm CP} = 1.848\pi, \eta_1 = 0.2271\pi$, confirming the interplay between the Majorana and Dirac phases. The third subplot illustrates the $\beta$-decay effective neutrino mass $m_\beta$ correlations with the sum of neutrino masses $\sum m_i$. There is a strong linear positive correlation among them, and both of them lie below the current experimental upper bounds. The plot of the effective Majorana mass $m_{\beta\beta}$ versus the lightest neutrino mass $m_3$ reveals a tight clustering around $m_{\beta\beta} \approx 0.025$ eV  for $m_3 < 10^{-4}$ eV, confirming phase-mediated suppression. This matches the best-fit value $m_{\beta\beta} = 0.02486$ eV and reflects destructive interference from Majorana phase alignments.  The effective Majorana mass  $m_{\beta\beta}\approx 18–49$ meV lies partially within the sensitivity of future experiments. A subset of points below 20 meV evades detection even in next-generation searches.  

Parallel with the analysis of region A, the corresponding plots in region B exhibit very different patterns. Majorana phases $\eta_1$-$\eta_2$ correlation with extreme clustering at three points $\eta_1 \approx 0, \eta_2 \approx 1.0\pi$ or $2.0\pi$, and $\eta_1 \approx \eta_2 \approx 2.0\pi$. This is due to the special position of region B placed very near the boundary of the fundamental domain. These contain the best fit values $\eta_1 \approx 0, \eta_2 \approx \pi$ and indicate extremized $m_{\beta\beta}$. The correlation of $\eta_1$ and $\delta_{\rm CP}$ reveals uniform high-density concentration of $\eta_1 \approx 0$ or $2.0\pi$ over the change of $\delta_{\rm CP} \approx (0-0.3)\pi$. This CP phase distribution in region B contrasts sharply with the results from region A. $\delta_{\rm CP}$ versus $\mathrm{Re}(\tau)$ correlation demonstrates the critical boundary effect. As $\mathrm{Re}(\tau) \rightarrow 0.5$, $\delta_{\rm CP} \rightarrow 0$ with a steep trajectory. The best fit point ($\mathrm{Re}(\tau) = 0.4999, \delta_{\rm CP} = 0.05518\pi$) is at the convergence, which shows that modular symmetry suppresses CP violation near $\mathrm{Re}(\tau) = 0.5$. The points for $\sum m_\beta$ versus $\sum m_i$ mirror the distribution of Region A. Both $\sum m_i \approx (0.095 -0.103)$ eV and $m_\beta \approx ( 0.048-0.052)$ eV are viable from the point of view of terrestrial and space-based experiments. In the last plot, very tight correlation between $m_{\beta\beta}$ versus $m_3$ is illustrated. A dense cluster of $m_{\beta\beta} \approx 0.048$ eV at $m_3 \sim 10^{-4}$ eV lies very left-edge of the conventional $0\nu\beta\beta$ plot. Since it is viable and almost invisible when plotting with experimental contours, here we magnify the predicted areas without showing these constraints.  

As a comment for results from these three model analyzed, in comparison with previous works on inverse seesaw model with holomorphic $A_4$ symmetry~\cite{Nomura:2019xsb,Nomura:2020cog,Gogoi:2022jwf,Kobayashi:2023qzt,Pathak:2024sei,Pathak:2025zdp}, we observe that the phenomenological predictions of different models depend on the specific model construction, such as the range of $\tau$ values, the choice of different weights, and the representations of matter fields. The results do not show particularly strong preferences. This suggests that we should pay more attention to the modulus stabilization problem. If we can determine the specific range or values of the modulus $\tau$, the differences and preferences among various models may become more pronounced, since the Yukawa couplings are controlled by the VEV of $\tau$. We refer appendix~\ref{app:modstabls} to some remarks on the modulus stabilization.

\FloatBarrier

\section{Conclusions}
\label{sec:concluds}

In this paper, we have presented a comprehensive study of lepton flavor structure within the framework of non-supersymmetric modular symmetry, specifically focusing on the finite modular group $\Gamma _3\cong A_4$. Our model employs the inverse seesaw mechanism for neutrino mass generation while maintaining modular invariance through the use of polyharmonic Maa{\ss} forms as Yukawa couplings. The incorporation of gCP symmetry serves to reduce the number of free parameters, leading to a more constrained and predictive framework. We have developed three distinct models based on polyharmonic Maa{\ss} forms with weights $k = -4, -2, 0$, building upon established mathematical foundations.

The phenomenological implications of our models have been thoroughly investigated through detailed numerical analyses. Model 1 features a diagonal charged lepton sector with modular symmetry constraints applied exclusively to neutrinos, whereas models 2 and 3 introduce more complex structures affecting both charged leptons and neutrinos. All three models successfully accommodate both normal and inverted mass hierarchies within current experimental bounds, providing predictions for key observables including the Dirac CP-violating phase, Majorana phases, and absolute neutrino mass scales. The effective neutrino masses $m_\beta$ and $m_{\beta\beta}$ have been calculated, offering concrete targets for ongoing and future experimental searches.

Our analyses show several interesting correlations between physical parameters, particularly in specific regions of the modulus $\tau$ space. Within the current theoretical framework and experimental constraints, the constructed parameter space of our model demonstrates significant physical plausibility. All three models are capable of accommodating both the normal and inverted hierarchies of the latest neutrino oscillation data. Within the $3\sigma$ confidence interval of the neutrino oscillation parameters, these models provide predictions for the three CP-violating phases, which are still poorly understood. Specific values of the absolute neutrino mass scale, the effective neutrino masses $m_{\beta}$ and $m_{\beta\beta}$ are predicted or constrained in very tight regions. Strong relationships emerge between the Dirac CP phase and atmospheric mixing angle in certain parameter regimes, while other observables show more moderate dependencies. These patterns are systematically presented through comprehensive tables and figures, highlighting the predictive power of the modular symmetry approach. The viable parameter spaces we identify maintain consistency with existing experimental constraints while making testable predictions that can be probed by next-generation facilities, such as next-generation long-baseline experiments DUNE~\cite{DUNE:2020ypp} and Hyper-Kamiokande~\cite{Hyper-Kamiokande:2018ofw}, as well as precision determinations of neutrino mass by KATRIN~\cite{KATRIN:2022ayy} or Project 8~\cite{Project8:2022wqh}. Such high-precision experiments will not only rigorously test existing theoretical models but also may reveal some new features beyond our understanding.

The results demonstrate that non-holomorphic modular flavor symmetry represents a promising direction for understanding the lepton flavor structure. Although alternative theoretical approaches exist, the mathematical elegance and phenomenological viability of this framework warrant continued investigation. Future work could explore extensions to the quark sector, incorporation into a unified framework, or deeper connections to fundamental theory. As experimental precision continues to improve, particularly in neutrino oscillation studies and neutrinoless double beta decay searches, the predictions of modular symmetry models will face increasingly stringent tests. Our findings contribute to this growing field by establishing concrete phenomenological consequences of this theoretical approach and identifying specific signatures for experimental verification.

\section*{Acknowledgements}
This work was supported by the National Natural Science Foundation of China under Grant No. 12565016, and the Natural Science Foundation of Xinjiang Uygur Autonomous Region of China under Grant No. 2022D01C52.


\appendix

\section{The modular symmetry and modular forms}
\label{app:modelularsymandform}

The special linear group $SL(2, \mathbb{R})$ acts on the upper half-plane $\mathcal{H} = \left\{\tau \in \mathbb{C} \, | \, \text{Im}(\tau) > 0\right\}$ through linear fractional transformations, which are conformal mappings that preserve angles on the Riemann sphere. A key discrete subgroup of $SL(2, \mathbb{R})$ is the modular group $\Gamma \equiv SL(2, \mathbb{Z})$, consisting of $2 \times 2$ matrices with integer entries and unit determinant. For $\gamma = \begin{pmatrix} a & b \\ c & d \end{pmatrix} \in \Gamma$, the modular transformation of $\tau \in \mathcal{H}$ is given by  
\begin{equation}
	\tau \to \gamma\tau = \frac{a\tau + b}{c\tau + d}.
	\label{eq:modgrp}
\end{equation}
Since $\gamma$ and $-\gamma$ induce identical transformations, the effective symmetry group is the projective modular group $\overline{\Gamma} = \Gamma / \{\pm \mathbf{I}\}$. The modular group is generated by the elements $S$ and $T$, represented by the following matrices:  
\begin{equation}
	S = \begin{pmatrix}
		0 & 1 \\
		-1 & 0
	\end{pmatrix}, \quad T = \begin{pmatrix}
		1 & 1 \\
		0 & 1
	\end{pmatrix},
\end{equation}
which satisfy the relations $S^2 = (ST)^3 = \mathbf{I}$.  

An important class of normal subgroups of $\Gamma$ are the principal congruence subgroups $\Gamma(N)$, defined for $N \in \mathbb{Z}^+$ as  
\begin{equation}
	\Gamma(N) = \left\{ \begin{pmatrix}
		a & b \\
		c & d
	\end{pmatrix} \in SL(2, \mathbb{Z}) \;\Bigg|\; \begin{pmatrix}
		a & b \\
		c & d
	\end{pmatrix} \equiv \begin{pmatrix}
		1 & 0 \\
		0 & 1
	\end{pmatrix} \pmod{N} \right\}.
\end{equation}
For $N > 2$, $\Gamma(N)$ does not contain $-\mathbf{I}$, and thus $\overline{\Gamma}(N) \cong \Gamma(N)$. Finite modular groups arise as quotients of $\Gamma$ and $\overline{\Gamma}$ 
\begin{equation}
	\Gamma_N \equiv \overline{\Gamma} / \overline{\Gamma}(N), \quad \Gamma'_N \equiv \Gamma / \Gamma(N),
\end{equation}
where $\Gamma_N$ is the homogeneous finite modular group and $\Gamma'_N$ is its inhomogeneous counterpart. For $N > 2$, $\Gamma'_N \cong \Gamma_N / \{\pm \mathbf{I}\}$. At low levels, these groups coincide with well-known permutation and alternating groups: $\Gamma_2 \cong S_3$, $\Gamma_3 \cong A_4$, $\Gamma_4 \cong S_4$, and $\Gamma_5 \cong A_5$~\cite{deAdelhartToorop:2011re}, while $\Gamma'_N$ correspond to their double covers. 

Modular forms are holomorphic functions $f(\tau)$ on $\mathcal{H}$ that transform under $\Gamma(N)$ with a specific weight $k$ and a representation structure~\cite{2008The}. These forms are central to string theory and have recently been employed in construction of flavor models for particle physics. A weight-$k$ modular form satisfies the transformation law  
\begin{equation}
	f(\gamma \tau) = (c\tau + d)^k f(\tau), \quad \forall \gamma \in \Gamma(N).
\end{equation}
The space of such forms, denoted $M_k(\Gamma(N))$, is finite-dimensional and furnishes representations of the modular group. Modular multiplets $Y_{\mathbf{r}}(\tau)$ transform under irreducible representations $\rho_{\mathbf{r}}$ of the finite modular group as~\cite{Ding:2023htn}  
\begin{equation}
	Y_{\mathbf{r}}(\gamma \tau) = (c\tau + d)^k \rho_{\mathbf{r}}(\gamma) Y_{\mathbf{r}}(\tau),
\end{equation}
where $k > 0$ is the modular weight and $N$ is the level.

In $N=1$ supersymmetric theories, the matter superfields $\Phi$ transform under modular symmetry as~\cite{Feruglio:2017spp}  
\begin{equation}
	\Phi \mapsto (c\tau + d)^{-k_\Phi} \rho_{\mathbf{r}}(\gamma) \Phi,
\end{equation}
where $k_\Phi$ is the so-called modular weight (in the formal sense) of $\Phi$. This ensures the modular invariance of the superpotential while distinguishing matter fields from modular forms.  

The framework can be extended to non-holomorphic modular forms, where Yukawa couplings are polyharmonic Maa{\ss} forms satisfying the Laplace condition~\cite{Qu:2024rns},  
\begin{equation}
	\Delta_k Y(\tau) = 0,
\end{equation}
with the weight-$k$ Laplace operator defined as  
\begin{equation}
	\Delta_k = -y^2\left(\frac{\partial^2}{\partial x^2} + \frac{\partial^2}{\partial y^2}\right) + iky\left(\frac{\partial}{\partial x} + i\frac{\partial}{\partial y}\right) = -4y^2\frac{\partial}{\partial \tau}\frac{\partial}{\partial \bar{\tau}} + 2iky\frac{\partial}{\partial \bar{\tau}},
\end{equation}
where $\tau = x + iy$. This generalizes the holomorphic case while preserving modular covariance.

\section{$A_4$ symmetry essentials}
\label{app:a4modsym}

The alternating group $ A_4 $, isomorphic to the finite modular group $ \Gamma_3 $, is a non-Abelian discrete group of order 12, consisting of all even permutations of four objects. Geometrically, it describes the rotational symmetry group of a regular tetrahedron. In particle physics, $ A_4 $ is particularly significant as the smallest non-Abelian group with a three-dimensional irreducible representation, making it a useful flavor symmetry for model building.  

The group $ A_4 $ is generated by two elements, $ S $ and $ T $, which satisfy the defining relations  
\begin{equation}
S^2 = T^3 = (ST)^3 = \mathbf{I}.
\end{equation}  
These generators correspond to specific rotational symmetries of the tetrahedron: $ S $ represents a $ \pi $-rotation about a vertex, while $ T $ corresponds to a $ 2\pi/3 $-rotation about a face center. $ A_4 $ has four irreducible representations, including three one-dimensional representations $ \mathbf{1} $, $ \mathbf{1^\prime} $, and $ \mathbf{1^{\prime\prime}} $, and one three-dimensional representation $ \mathbf{3} $. The Clebsch-Gordan decompositions for tensor products involving singlets are given by  
\begin{equation}
	\begin{aligned}
		& \mathbf{1^{\prime}}\otimes \mathbf{1^{\prime\prime}} =  \mathbf{1},\\
		& \mathbf{3} \otimes \mathbf{3}= \mathbf{1}\oplus \mathbf{1^{\prime}}\oplus \mathbf{1^{\prime\prime}}\oplus\mathbf{3}_A\oplus\mathbf{3}_S,
	\end{aligned}
\end{equation} 
where the subscripts $ S $ and $ A $ denote symmetric and antisymmetric combinations, respectively.  

For two $ A_4 $-triplets $ \mathbf{a} = (a_1, a_2, a_3) $ and $ \mathbf{b} = (b_1, b_2, b_3) $, the tensor product $ \mathbf{a} \otimes \mathbf{b} $ decomposes explicitly as  
\begin{equation}
\begin{aligned}
a \otimes b = &(a_1b_1 + a_2b_3 + a_3b_2)_{\mathbf{1}} \oplus (a_3b_3 + a_1b_2 + a_2b_1)_{\mathbf{1^\prime}} \oplus (a_2b_2 + a_1b_3 + a_3b_1)_{\mathbf{1^{\prime\prime}}} \\
& \oplus \begin{pmatrix}
a_2b_3 - a_3b_2 \\
a_1b_2 - a_2b_1 \\
a_3b_1 - a_1b_3
\end{pmatrix}_{\mathbf{3}_A} \oplus \begin{pmatrix}
2a_1b_1 - a_3b_2 - a_2b_3 \\
2a_3b_3 - a_1b_2 - a_2b_1 \\
2a_2b_2 - a_3b_1 - a_1b_3
\end{pmatrix}_{\mathbf{3}_S}.
\end{aligned}
\end{equation}

\section{Polyharmonic Maa{\ss} forms for $A_4$ modular symmetry }
\label{app:polmaasa4}

We present explicit expressions for polyharmonic Maa{\ss} forms of level $N=3$ with weights $k = -4$, $-2$, and $0$, which transform under the $A_4$ finite modular group. These non-holomorphic modular forms extend the traditional holomorphic framework and are essential for constructing modular-invariant theories~\cite{Qu:2024rns,Qu:2025ddz}. 

For $k=0$, there exists a trivial modular singlet $Y^{(0)}_{\mathbf{1}} = 1$. The non-trivial $A_4$-triplet components are given by
\begin{align}
  Y^{(0)}_{\mathbf{3},1}(\tau)=& y-\frac{3e^{-4\pi y}}{\pi q}-\frac{9e^{-8\pi y}}{2\pi q^2}-\frac{e^{-12\pi y}}{\pi q^3}-\frac{21e^{-16\pi y}}{4\pi q^4}-\frac{18e^{-20\pi y}}{5\pi q^5}+\cdots    \nonumber \\
  & -\frac{9\log3}{4\pi}-\frac{3q}{\pi}-\frac{9q^2}{2\pi}-\frac{q^3}{\pi}-\frac{21q^4}{4\pi}-\frac{18q^5}{5\pi}+\cdots  ,  
  \end{align}
\begin{align}
 Y^{(0)}_{\mathbf{3},2}(\tau)=& \frac{27q^{\frac{1}{3} }e^{\frac{\pi  y}{3}}}{\pi  }\left(\frac{e^{-3\pi y}}{4 q}+\frac{e^{-7\pi y}}{5q^2}+\frac{5e^{-11\pi y}}{16 q^3}+\frac{2e^{-15\pi y}}{11q^4}+\frac{2e^{-15\pi y}}{7q^5}+\cdots \right) \nonumber  \\
  & +\frac{9q^{\frac{1}{3}} }{2\pi} \left(1+\frac{7q}{4}+\frac{8q^2}{7}+\frac{9q^3}{5}+\frac{14q^4}{13}+\frac{31q^5}{16}+\cdots \right),  
  \end{align}
\begin{align}
 Y^{(0)}_{\mathbf{3},3}(\tau)=& \frac{9q^{\frac{2}{3} }e^{\frac{2\pi y}{3}}}{\pi  }\left(\frac{e^{-2\pi y}}{q}+\frac{7e^{-6\pi y}}{4q^2}+\frac{8e^{-10\pi y}}{7q^3}+\frac{9e^{-14\pi y}}{5q^4}+\frac{14e^{-18\pi y}}{13q^5}+\cdots \right)  \nonumber  \\
  & +\frac{27q^{\frac{2}{3}} }{ \pi} \left(\frac{1}{4}+\frac{q}{5}+\frac{5q^2}{16}+\frac{2q^3}{5}+\frac{2q^4}{7}+\frac{9q^5}{17}+\cdots \right) . 
\end{align}

The weight $k=-2$ triplet components take the form
\begin{align}
  Y^{(-2)}_{\mathbf{3},1}(\tau )=& \frac{y^3}{3} +\frac{21\Gamma (3,4\pi y)}{16\pi^3 q} +\frac{189\Gamma (3,8\pi y)}{128\pi^3 q^2} +\frac{169\Gamma (3,12\pi y)}{144\pi^3 q^3} +\frac{1533\Gamma (3,16\pi y)}{1024\pi^3 q^4}+\cdots   \nonumber \\
  & +\frac{\pi \zeta(3)}{40\zeta(4)}+\frac{21q}{8\pi^3}+\frac{189q^2}{64\pi^3}+\frac{169q^3}{72\pi^3}+\frac{1533q^4}{512\pi^3}+\frac{1323q^5}{500\pi^3}+\cdots  , 
\end{align}
\begin{align}
  Y^{(-2)}_{\mathbf{3},2}(\tau )=& -\frac{729q^{\frac{1}{3}} }{16\pi^3 } \left(\frac{\Gamma (3,\frac{8\pi y}{3} )}{16 q} +\frac{7\Gamma (3,\frac{20\pi y}{3} )}{125q^2} +\frac{65\Gamma (3,\frac{32\pi y}{3} )}{1024q^3} +\frac{74\Gamma (3,\frac{44\pi y}{3} )}{1331q^4}+\cdots\right)    \nonumber \\
  & -\frac{81q\frac{1}{3} }{16\pi^3}\left(1+\frac{73q}{64}+\frac{344q^2}{343}+\frac{567q^3}{500}+\frac{20198q^4}{2197}+\frac{4681q^5}{4096}+\cdots \right),  
  \end{align}
\begin{align}
    Y^{(-2)}_{\mathbf{3},3}(\tau )=& -\frac{81q^{\frac{2}{3}} }{32\pi^3 } \left(\frac{\Gamma (3,\frac{4\pi y}{3} )}{q} +\frac{73\Gamma (3,\frac{16\pi y}{3} )}{64q^2} +\frac{344\Gamma (3,\frac{28\pi y}{3} )}{343q^3} +\frac{567\Gamma (3,\frac{40\pi y}{3} )}{500q^4}+\cdots\right)  \nonumber \\
  & -\frac{729q\frac{2}{3} }{8\pi^3}\left(\frac{1}{16} +\frac{7q}{125}+\frac{65q^2}{1024}+\frac{74q^3}{1331}+\cdots \right) .
\end{align}

The weight $k=-4$ forms have the following structure:
\begin{align}
  Y_{3,1}^{(-4)}(\tau)= & \frac{y^{5}}{5}-\frac{549}{3328 \pi^{5}}\left(\frac{\Gamma(5,4 \pi y)}{q}+\frac{33 \Gamma(5,8 \pi y)}{32 q^{2}}+\frac{14641 \Gamma(5,12 \pi y)}{14823 q^{3}}+\frac{1057 \Gamma(5,16 \pi y)}{1024 q^{4}}+\cdots\right) \nonumber \\
  & -\frac{3 \pi}{728} \frac{\zeta(5)}{\zeta(6)}-\frac{1647}{416 \pi^{5}}\left(q+\frac{33 q^{2}}{32}+\frac{14641 q^{3}}{14823}+\frac{1057 q^{4}}{1024}+\frac{3126 q^{5}}{3125}+\cdots\right), 
  \end{align}
\begin{align}
  Y_{3,2}^{(-4)}(\tau)= & \frac{72171 q^{\frac{1 }{3} }}{212992 \pi^{5}}\left(\frac{\Gamma(5,\frac{8 \pi y }{3} )}{q}+\frac{33344 \Gamma(5,\frac{20 \pi y }{3})}{34375 q^{2}}+\frac{1025 \Gamma(5,\frac{32 \pi y }{3})}{1024 q^{3}}+\cdots\right) \nonumber \\
  & +\frac{6561 q^{\frac{1 }{3} }}{832 \pi^{5}}\left(1+\frac{1057 q}{1024}+\frac{16808 q^{2}}{16807}+\frac{51579 q^{3}}{50000}+\frac{371294 q^{4}}{371293}+\cdots\right),
  \end{align}
\begin{align}
  Y_{\mathbf{3}, 3}^{(-4)}(\tau)= & \frac{2187 q^{\frac{2}{3}}}{6656 \pi^{5}}\left(\frac{\Gamma(5,\frac{4 \pi y }{3})}{q}+\frac{1057 \Gamma(5,\frac{16\pi y }{3})}{1024 q^{2}}+\frac{16808 \Gamma(5,\frac{28 \pi y }{3})}{16807 q^{3}}+\cdots\right) \nonumber \\
  & +\frac{216513 q^{\frac{2}{3} }}{26624 \pi^{5}}\left(1+\frac{33344 q}{34375}+\frac{1025 q^{2}}{1024}+\frac{1717888 q^{3}}{1771561}+\frac{16808 q^{4}}{16807}+\cdots\right) .
\end{align}

These polyharmonic Maa{\ss} forms, first presented in~\cite{Qu:2024rns,Qu:2025ddz}, provide a complete basis for $A_4$-symmetric modular constructions beyond the holomorphic limit.

\section{Remarks on the modulus stabilization }
\label{app:modstabls}

While this work  employs a bottom-up approach to model-building with $A_4$ modular flavor symmetry, a complete understanding requires top-down considerations. The modulus $\tau$ is not merely a free parameter but arises naturally in string theory from the geometry of compactified extra dimensions. As a coordinate in moduli space, $\tau$ can, a priori, take any value within the fundamental domain. Its value is critical, as it determines the Yukawa couplings and thereby plays a decisive role in fermion masses and mixing patterns. Consequently, the stabilization of the modulus, the dynamical mechanism that fixes its VEV, becomes an essential theoretical problem. The challenge in string theory is to find potentials that stabilize all moduli at global minima corresponding to de Sitter (dS) vacua, consistent with the observed small positive cosmological constant. Supergravity provides the effective field theory framework in which this issue is typically addressed.
 
Several mechanisms for modulus stabilization have been explored in the literature. For instance, the inclusion of 3-form fluxes in Type IIB string theory compactified on a $T^6/(\mathbb{Z}_2\times \mathbb{Z}_2^{\prime})$ orbifold can generate a potential for complex structure moduli, with minima found to cluster near the left cusp of the fundamental domain, $\tau = \omega$~\cite{Ishiguro:2020tmo}. Other analyses, involving modular symmetries acting on K\"ahler moduli, also yield stabilization at symmetric points like $\tau = \omega$, though often in anti-de Sitter (AdS) vacua~\cite{Gonzalo:2018guu,Novichkov:2022wvg}. Frameworks such as Kachru-Kallosh-Linde-Trivedi~\cite{Kachru:2003aw} can subsequently uplift these AdS vacua to dS vacua, for example, through the inclusion of matter effects in the superpotential~\cite{Knapp-Perez:2023nty}. An alternative approach utilizes non-perturbative corrections to the dilaton K\"ahler potential, which can generate metastable dS vacua directly at the fixed points $\tau=i$ and $\tau=\omega$ without requiring a matter superpotential~\cite{Leedom:2022zdm,King:2023snq}.

A notable pattern emerging from these diverse top-down approaches is the tendency for the modulus $\tau$ to be stabilized at or near the symmetric points of the modular group. This convergence suggests a link between the fundamental physics of compactification and the observed structures in flavor. Although bottom-up phenomenology is a powerful tool, identifying a self-consistent ultraviolet-complete theory that naturally explains the stabilization of $\tau$ remains a challenge for particle physics.

%
%

\bibliographystyle{RAN}
\bibliography{biblio}

\begin{thebibliography}{10}
\providecommand{\url}[1]{\texttt{#1}}
\providecommand{\urlprefix}{URL }
\providecommand{\eprint}[2][]{\url{#2}}

\bibitem{Mohapatra:1986aw}
R.~N. Mohapatra, \emph{{Mechanism for Understanding Small Neutrino Mass in
  Superstring Theories}},
  \MYhref[journalLinks]{http://dx.doi.org/10.1103/PhysRevLett.56.561}{Phys.
  Rev. Lett.
  }\MYhref[journalLinks]{http://dx.doi.org/10.1103/PhysRevLett.56.561}{\textbf{56}
  (1986) 561--563}.

\bibitem{Mohapatra:1986bd}
R.~N. Mohapatra and J.~W.~F. Valle, \emph{{Neutrino Mass and Baryon Number
  Nonconservation in Superstring Models}},
  \MYhref[journalLinks]{http://dx.doi.org/10.1103/PhysRevD.34.1642}{Phys. Rev.
  D
  }\MYhref[journalLinks]{http://dx.doi.org/10.1103/PhysRevD.34.1642}{\textbf{34}
  (1986) 1642}.

\bibitem{Deppisch:2004fa}
F.~Deppisch and J.~W.~F. Valle, \emph{{Enhanced lepton flavor violation in the
  supersymmetric inverse seesaw model}},
  \MYhref[journalLinks]{http://dx.doi.org/10.1103/PhysRevD.72.036001}{Phys.
  Rev. D
  }\MYhref[journalLinks]{http://dx.doi.org/10.1103/PhysRevD.72.036001}{\textbf{72}
  (2005) 036001},
  \MYhref[eprintLinks]{http://arxiv.org/abs/hep-ph/0406040}{{\ttfamily
  arXiv:hep-ph/0406040}}.

\bibitem{Dev:2009aw}
P.~S.~B. Dev and R.~N. Mohapatra, \emph{{TeV Scale Inverse Seesaw in $SO(10)$
  and Leptonic Non-Unitarity Effects}},
  \MYhref[journalLinks]{http://dx.doi.org/10.1103/PhysRevD.81.013001}{Phys.
  Rev. D
  }\MYhref[journalLinks]{http://dx.doi.org/10.1103/PhysRevD.81.013001}{\textbf{81}
  (2010) 013001},
  \MYhref[eprintLinks]{http://arxiv.org/abs/0910.3924}{{\ttfamily
  arXiv:0910.3924 [hep-ph]}}.

\bibitem{CentellesChulia:2020dfh}
S.~Centelles~Chuli\'a, R.~Srivastava and A.~Vicente, \emph{{The inverse seesaw
  family: Dirac and Majorana}},
  \MYhref[journalLinks]{http://dx.doi.org/10.1007/JHEP03(2021)248}{JHEP
  }\MYhref[journalLinks]{http://dx.doi.org/10.1007/JHEP03(2021)248}{\textbf{03}
  (2021) 248}, \MYhref[eprintLinks]{http://arxiv.org/abs/2011.06609}{{\ttfamily
  arXiv:2011.06609 [hep-ph]}}.

\bibitem{Feruglio:2017spp}
F.~Feruglio, \emph{{Are neutrino masses modular forms?}}, pages 227--266 (2019)
  \MYhref[eprintLinks]{http://arxiv.org/abs/1706.08749}{{\ttfamily
  arXiv:1706.08749 [hep-ph]}}.

\bibitem{Feruglio:2019ybq}
F.~Feruglio and A.~Romanino, \emph{{Lepton flavor symmetries}},
  \MYhref[journalLinks]{http://dx.doi.org/10.1103/RevModPhys.93.015007}{Rev.
  Mod. Phys.
  }\MYhref[journalLinks]{http://dx.doi.org/10.1103/RevModPhys.93.015007}{\textbf{93}
  (2021) 1 015007},
  \MYhref[eprintLinks]{http://arxiv.org/abs/1912.06028}{{\ttfamily
  arXiv:1912.06028 [hep-ph]}}.

\bibitem{Ding:2023htn}
G.-J. Ding and S.~F. King, \emph{{Neutrino mass and mixing with modular
  symmetry}},
  \MYhref[journalLinks]{http://dx.doi.org/10.1088/1361-6633/ad52a3}{Rept. Prog.
  Phys.
  }\MYhref[journalLinks]{http://dx.doi.org/10.1088/1361-6633/ad52a3}{\textbf{87}
  (2024) 8 084201},
  \MYhref[eprintLinks]{http://arxiv.org/abs/2311.09282}{{\ttfamily
  arXiv:2311.09282 [hep-ph]}}.

\bibitem{Kobayashi:2023zzc}
T.~Kobayashi and M.~Tanimoto, \emph{{Modular flavor symmetric models}},
  \MYhref[journalLinks]{http://dx.doi.org/10.1142/S0217751X24410124}{Int. J.
  Mod. Phys. A
  }\MYhref[journalLinks]{http://dx.doi.org/10.1142/S0217751X24410124}{\textbf{39}
  (2024) 09n10 2441012},
  \MYhref[eprintLinks]{http://arxiv.org/abs/2307.03384}{{\ttfamily
  arXiv:2307.03384 [hep-ph]}}.

\bibitem{Kobayashi:2018vbk}
T.~Kobayashi, K.~Tanaka and T.~H. Tatsuishi, \emph{{Neutrino mixing from finite
  modular groups}},
  \MYhref[journalLinks]{http://dx.doi.org/10.1103/PhysRevD.98.016004}{Phys.
  Rev. D
  }\MYhref[journalLinks]{http://dx.doi.org/10.1103/PhysRevD.98.016004}{\textbf{98}
  (2018) 1 016004},
  \MYhref[eprintLinks]{http://arxiv.org/abs/1803.10391}{{\ttfamily
  arXiv:1803.10391 [hep-ph]}}.

\bibitem{Kobayashi:2018scp}
T.~Kobayashi et~al., \emph{{Modular A$_{4}$ invariance and neutrino mixing}},
  \MYhref[journalLinks]{http://dx.doi.org/10.1007/JHEP11(2018)196}{JHEP
  }\MYhref[journalLinks]{http://dx.doi.org/10.1007/JHEP11(2018)196}{\textbf{11}
  (2018) 196}, \MYhref[eprintLinks]{http://arxiv.org/abs/1808.03012}{{\ttfamily
  arXiv:1808.03012 [hep-ph]}}.

\bibitem{Novichkov:2018yse}
P.~P. Novichkov, S.~T. Petcov and M.~Tanimoto, \emph{{Trimaximal Neutrino
  Mixing from Modular $A_4$ Invariance with Residual Symmetries}},
  \MYhref[journalLinks]{http://dx.doi.org/10.1016/j.physletb.2019.04.043}{Phys.
  Lett. B
  }\MYhref[journalLinks]{http://dx.doi.org/10.1016/j.physletb.2019.04.043}{\textbf{793}
  (2019) 247--258},
  \MYhref[eprintLinks]{http://arxiv.org/abs/1812.11289}{{\ttfamily
  arXiv:1812.11289 [hep-ph]}}.

\bibitem{Nomura:2019yft}
T.~Nomura and H.~Okada, \emph{{A two loop induced neutrino mass model with
  modular $A_4$ symmetry}},
  \MYhref[journalLinks]{http://dx.doi.org/10.1016/j.nuclphysb.2021.115372}{Nucl.
  Phys. B
  }\MYhref[journalLinks]{http://dx.doi.org/10.1016/j.nuclphysb.2021.115372}{\textbf{966}
  (2021) 115372},
  \MYhref[eprintLinks]{http://arxiv.org/abs/1906.03927}{{\ttfamily
  arXiv:1906.03927 [hep-ph]}}.

\bibitem{Ding:2019zxk}
G.-J. Ding, S.~F. King and X.-G. Liu, \emph{{Modular A$_{4}$ symmetry models of
  neutrinos and charged leptons}},
  \MYhref[journalLinks]{http://dx.doi.org/10.1007/JHEP09(2019)074}{JHEP
  }\MYhref[journalLinks]{http://dx.doi.org/10.1007/JHEP09(2019)074}{\textbf{09}
  (2019) 074}, \MYhref[eprintLinks]{http://arxiv.org/abs/1907.11714}{{\ttfamily
  arXiv:1907.11714 [hep-ph]}}.

\bibitem{Nomura:2019lnr}
T.~Nomura, H.~Okada and O.~Popov, \emph{{A modular $A_4$ symmetric scotogenic
  model}},
  \MYhref[journalLinks]{http://dx.doi.org/10.1016/j.physletb.2020.135294}{Phys.
  Lett. B
  }\MYhref[journalLinks]{http://dx.doi.org/10.1016/j.physletb.2020.135294}{\textbf{803}
  (2020) 135294},
  \MYhref[eprintLinks]{http://arxiv.org/abs/1908.07457}{{\ttfamily
  arXiv:1908.07457 [hep-ph]}}.

\bibitem{Ding:2019gof}
G.-J. Ding, S.~F. King, X.-G. Liu and J.-N. Lu, \emph{{Modular S$_{4}$ and
  A$_{4}$ symmetries and their fixed points: new predictive examples of lepton
  mixing}},
  \MYhref[journalLinks]{http://dx.doi.org/10.1007/JHEP12(2019)030}{JHEP
  }\MYhref[journalLinks]{http://dx.doi.org/10.1007/JHEP12(2019)030}{\textbf{12}
  (2019) 030}, \MYhref[eprintLinks]{http://arxiv.org/abs/1910.03460}{{\ttfamily
  arXiv:1910.03460 [hep-ph]}}.

\bibitem{Zhang:2019ngf}
D.~Zhang, \emph{{A modular $A_4$ symmetry realization of two-zero textures of
  the Majorana neutrino mass matrix}},
  \MYhref[journalLinks]{http://dx.doi.org/10.1016/j.nuclphysb.2020.114935}{Nucl.
  Phys. B
  }\MYhref[journalLinks]{http://dx.doi.org/10.1016/j.nuclphysb.2020.114935}{\textbf{952}
  (2020) 114935},
  \MYhref[eprintLinks]{http://arxiv.org/abs/1910.07869}{{\ttfamily
  arXiv:1910.07869 [hep-ph]}}.

\bibitem{Nomura:2019xsb}
T.~Nomura, H.~Okada and S.~Patra, \emph{{An inverse seesaw model with $A_4$
  -modular symmetry}},
  \MYhref[journalLinks]{http://dx.doi.org/10.1016/j.nuclphysb.2021.115395}{Nucl.
  Phys. B
  }\MYhref[journalLinks]{http://dx.doi.org/10.1016/j.nuclphysb.2021.115395}{\textbf{967}
  (2021) 115395},
  \MYhref[eprintLinks]{http://arxiv.org/abs/1912.00379}{{\ttfamily
  arXiv:1912.00379 [hep-ph]}}.

\bibitem{Kobayashi:2019gtp}
T.~Kobayashi, T.~Nomura and T.~Shimomura, \emph{{Type II seesaw models with
  modular $A_4$ symmetry}},
  \MYhref[journalLinks]{http://dx.doi.org/10.1103/PhysRevD.102.035019}{Phys.
  Rev. D
  }\MYhref[journalLinks]{http://dx.doi.org/10.1103/PhysRevD.102.035019}{\textbf{102}
  (2020) 3 035019},
  \MYhref[eprintLinks]{http://arxiv.org/abs/1912.00637}{{\ttfamily
  arXiv:1912.00637 [hep-ph]}}.

\bibitem{Wang:2019xbo}
X.~Wang, \emph{{Lepton flavor mixing and CP violation in the minimal
  type-(I+II) seesaw model with a modular $A_4$ symmetry}},
  \MYhref[journalLinks]{http://dx.doi.org/10.1016/j.nuclphysb.2020.115105}{Nucl.
  Phys. B
  }\MYhref[journalLinks]{http://dx.doi.org/10.1016/j.nuclphysb.2020.115105}{\textbf{957}
  (2020) 115105},
  \MYhref[eprintLinks]{http://arxiv.org/abs/1912.13284}{{\ttfamily
  arXiv:1912.13284 [hep-ph]}}.

\bibitem{Ding:2020yen}
G.-J. Ding and F.~Feruglio, \emph{{Testing Moduli and Flavon Dynamics with
  Neutrino Oscillations}},
  \MYhref[journalLinks]{http://dx.doi.org/10.1007/JHEP06(2020)134}{JHEP
  }\MYhref[journalLinks]{http://dx.doi.org/10.1007/JHEP06(2020)134}{\textbf{06}
  (2020) 134}, \MYhref[eprintLinks]{http://arxiv.org/abs/2003.13448}{{\ttfamily
  arXiv:2003.13448 [hep-ph]}}.

\bibitem{Nomura:2020cog}
T.~Nomura and H.~Okada, \emph{{Modular $A_4$ symmetric inverse seesaw model
  with $SU(2)_L$ multiplet fields}}  (2020),
  \MYhref[eprintLinks]{http://arxiv.org/abs/2007.15459}{{\ttfamily
  arXiv:2007.15459 [hep-ph]}}.

\bibitem{Aoki:2020eqf}
M.~Aoki and D.~Kaneko, \emph{{A hybrid seesaw model and hierarchical neutrino
  flavor structures based on $A_{4}$ symmetry}},
  \MYhref[journalLinks]{http://dx.doi.org/10.1093/ptep/ptab008}{PTEP
  }\MYhref[journalLinks]{http://dx.doi.org/10.1093/ptep/ptab008}{\textbf{2021}
  (2021) 2 023B06},
  \MYhref[eprintLinks]{http://arxiv.org/abs/2009.06025}{{\ttfamily
  arXiv:2009.06025 [hep-ph]}}.

\bibitem{Asaka:2020tmo}
T.~Asaka, Y.~Heo and T.~Yoshida, \emph{{Lepton flavor model with modular $A_4$
  symmetry in large volume limit}},
  \MYhref[journalLinks]{http://dx.doi.org/10.1016/j.physletb.2020.135956}{Phys.
  Lett. B
  }\MYhref[journalLinks]{http://dx.doi.org/10.1016/j.physletb.2020.135956}{\textbf{811}
  (2020) 135956},
  \MYhref[eprintLinks]{http://arxiv.org/abs/2009.12120}{{\ttfamily
  arXiv:2009.12120 [hep-ph]}}.

\bibitem{Okada:2020brs}
H.~Okada and M.~Tanimoto, \emph{{Spontaneous CP violation by modulus $\tau$ in
  $A_4$ model of lepton flavors}},
  \MYhref[journalLinks]{http://dx.doi.org/10.1007/JHEP03(2021)010}{JHEP
  }\MYhref[journalLinks]{http://dx.doi.org/10.1007/JHEP03(2021)010}{\textbf{03}
  (2021) 010}, \MYhref[eprintLinks]{http://arxiv.org/abs/2012.01688}{{\ttfamily
  arXiv:2012.01688 [hep-ph]}}.

\bibitem{Okada:2021qdf}
H.~Okada, Y.~Shimizu, M.~Tanimoto and T.~Yoshida, \emph{{Modulus
  \ensuremath{\tau} linking leptonic CP violation to baryon asymmetry in
  A$_{4}$ modular invariant flavor model}},
  \MYhref[journalLinks]{http://dx.doi.org/10.1007/JHEP07(2021)184}{JHEP
  }\MYhref[journalLinks]{http://dx.doi.org/10.1007/JHEP07(2021)184}{\textbf{07}
  (2021) 184}, \MYhref[eprintLinks]{http://arxiv.org/abs/2105.14292}{{\ttfamily
  arXiv:2105.14292 [hep-ph]}}.

\bibitem{deMedeirosVarzielas:2021pug}
I.~de~Medeiros~Varzielas and J.~a. Louren\c{c}o, \emph{{Two $A_4$ modular
  symmetries for Tri-Maximal 2 mixing}},
  \MYhref[journalLinks]{http://dx.doi.org/10.1016/j.nuclphysb.2022.115793}{Nucl.
  Phys. B
  }\MYhref[journalLinks]{http://dx.doi.org/10.1016/j.nuclphysb.2022.115793}{\textbf{979}
  (2022) 115793},
  \MYhref[eprintLinks]{http://arxiv.org/abs/2107.04042}{{\ttfamily
  arXiv:2107.04042 [hep-ph]}}.

\bibitem{Nomura:2021pld}
T.~Nomura, H.~Okada and Y.-h. Qi, \emph{{Zee model in a modular $A_4$
  symmetry}},
  \MYhref[journalLinks]{http://dx.doi.org/10.1140/epjc/s10052-025-13864-0}{Eur.
  Phys. J. C
  }\MYhref[journalLinks]{http://dx.doi.org/10.1140/epjc/s10052-025-13864-0}{\textbf{85}
  (2025) 2 134},
  \MYhref[eprintLinks]{http://arxiv.org/abs/2111.10944}{{\ttfamily
  arXiv:2111.10944 [hep-ph]}}.

\bibitem{Kobayashi:2021pav}
T.~Kobayashi, H.~Otsuka, M.~Tanimoto and K.~Yamamoto, \emph{{Modular symmetry
  in the SMEFT}},
  \MYhref[journalLinks]{http://dx.doi.org/10.1103/PhysRevD.105.055022}{Phys.
  Rev. D
  }\MYhref[journalLinks]{http://dx.doi.org/10.1103/PhysRevD.105.055022}{\textbf{105}
  (2022) 5 055022},
  \MYhref[eprintLinks]{http://arxiv.org/abs/2112.00493}{{\ttfamily
  arXiv:2112.00493 [hep-ph]}}.

\bibitem{Kobayashi:2022jvy}
T.~Kobayashi, H.~Otsuka, M.~Tanimoto and K.~Yamamoto, \emph{{Lepton flavor
  violation, lepton $(g - 2)_{\mu, e}$ and electron EDM in the modular
  symmetry}},
  \MYhref[journalLinks]{http://dx.doi.org/10.1007/JHEP08(2022)013}{JHEP
  }\MYhref[journalLinks]{http://dx.doi.org/10.1007/JHEP08(2022)013}{\textbf{08}
  (2022) 013}, \MYhref[eprintLinks]{http://arxiv.org/abs/2204.12325}{{\ttfamily
  arXiv:2204.12325 [hep-ph]}}.

\bibitem{Kang:2022psa}
D.~W. Kang, J.~Kim, T.~Nomura and H.~Okada, \emph{{Natural mass hierarchy among
  three heavy Majorana neutrinos for resonant leptogenesis under modular
  A$_{4}$ symmetry}},
  \MYhref[journalLinks]{http://dx.doi.org/10.1007/JHEP07(2022)050}{JHEP
  }\MYhref[journalLinks]{http://dx.doi.org/10.1007/JHEP07(2022)050}{\textbf{07}
  (2022) 050}, \MYhref[eprintLinks]{http://arxiv.org/abs/2205.08269}{{\ttfamily
  arXiv:2205.08269 [hep-ph]}}.

\bibitem{Gogoi:2022jwf}
J.~Gogoi, N.~Gautam and M.~K. Das, \emph{{Neutrino masses and mixing in minimal
  inverse seesaw using A4 modular symmetry}},
  \MYhref[journalLinks]{http://dx.doi.org/10.1142/S0217751X23500227}{Int. J.
  Mod. Phys. A
  }\MYhref[journalLinks]{http://dx.doi.org/10.1142/S0217751X23500227}{\textbf{38}
  (2023) 03 2350022},
  \MYhref[eprintLinks]{http://arxiv.org/abs/2207.10546}{{\ttfamily
  arXiv:2207.10546 [hep-ph]}}.

\bibitem{Abbas:2022slb}
M.~Abbas and S.~Khalil, \emph{{Modular $A_4$ symmetry with three moduli and
  flavor problem}},
  \MYhref[journalLinks]{http://dx.doi.org/10.1142/S0217751X24501379}{Int. J.
  Mod. Phys. A
  }\MYhref[journalLinks]{http://dx.doi.org/10.1142/S0217751X24501379}{\textbf{39}
  (2024) 32 2450137},
  \MYhref[eprintLinks]{http://arxiv.org/abs/2212.10666}{{\ttfamily
  arXiv:2212.10666 [hep-ph]}}.

\bibitem{Devi:2023vpe}
M.~R. Devi, \emph{{Retrieving texture zeros in 3+1 active-sterile neutrino
  framework under the action of $A_4$ modular-invariants}}  (2023),
  \MYhref[eprintLinks]{http://arxiv.org/abs/2303.04900}{{\ttfamily
  arXiv:2303.04900 [hep-ph]}}.

\bibitem{Nomura:2023usj}
T.~Nomura, H.~Okada and H.~Otsuka, \emph{{Texture zeros realization in a
  three-loop radiative neutrino mass model from modular $A_4$ symmetry}},
  \MYhref[journalLinks]{http://dx.doi.org/10.1016/j.nuclphysb.2024.116579}{Nucl.
  Phys. B
  }\MYhref[journalLinks]{http://dx.doi.org/10.1016/j.nuclphysb.2024.116579}{\textbf{1004}
  (2024) 116579},
  \MYhref[eprintLinks]{http://arxiv.org/abs/2309.13921}{{\ttfamily
  arXiv:2309.13921 [hep-ph]}}.

\bibitem{Kobayashi:2023qzt}
T.~Kobayashi, T.~Nomura, H.~Okada and H.~Otsuka, \emph{{Modular flavor models
  with positive modular weights: a new lepton model building}},
  \MYhref[journalLinks]{http://dx.doi.org/10.1007/JHEP01(2024)121}{JHEP
  }\MYhref[journalLinks]{http://dx.doi.org/10.1007/JHEP01(2024)121}{\textbf{01}
  (2024) 121}, \MYhref[eprintLinks]{http://arxiv.org/abs/2310.10091}{{\ttfamily
  arXiv:2310.10091 [hep-ph]}}.

\bibitem{Kumar:2023moh}
R.~Kumar et~al., \emph{{Predictions from scoto-seesaw with $A_4$ modular
  symmetry}},
  \MYhref[journalLinks]{http://dx.doi.org/10.1016/j.physletb.2024.138635}{Phys.
  Lett. B
  }\MYhref[journalLinks]{http://dx.doi.org/10.1016/j.physletb.2024.138635}{\textbf{853}
  (2024) 138635},
  \MYhref[eprintLinks]{http://arxiv.org/abs/2310.02363}{{\ttfamily
  arXiv:2310.02363 [hep-ph]}}.

\bibitem{Ding:2024fsf}
G.-J. Ding et~al., \emph{{Pati-Salam models with A$_{4}$ modular symmetry}},
  \MYhref[journalLinks]{http://dx.doi.org/10.1007/JHEP08(2024)134}{JHEP
  }\MYhref[journalLinks]{http://dx.doi.org/10.1007/JHEP08(2024)134}{\textbf{08}
  (2024) 134}, \MYhref[eprintLinks]{http://arxiv.org/abs/2404.06520}{{\ttfamily
  arXiv:2404.06520 [hep-ph]}}.

\bibitem{Nomura:2024ghc}
T.~Nomura and H.~Okada, \emph{{Quasi two-zero texture in Type-II seesaw at
  fixed points from modular $A_4$ symmetry}}  (2024),
  \MYhref[eprintLinks]{http://arxiv.org/abs/2407.13167}{{\ttfamily
  arXiv:2407.13167 [hep-ph]}}.

\bibitem{Singh:2024imk}
L.~Singh, M.~Kashav and S.~Verma, \emph{{Minimal type-I Dirac seesaw and
  leptogenesis under $A_4$ modular invariance}},
  \MYhref[journalLinks]{http://dx.doi.org/10.1016/j.nuclphysb.2024.116666}{Nucl.
  Phys. B
  }\MYhref[journalLinks]{http://dx.doi.org/10.1016/j.nuclphysb.2024.116666}{\textbf{1007}
  (2024) 116666},
  \MYhref[eprintLinks]{http://arxiv.org/abs/2405.07165}{{\ttfamily
  arXiv:2405.07165 [hep-ph]}}.

\bibitem{Kalita:2024vlt}
R.~Kalita and M.~Patgiri, \emph{{Neutrino Model in Left-Right Symmetric Linear
  Seesaw Augmented with $A_4$ Modular Group}}  (2024),
  \MYhref[eprintLinks]{http://arxiv.org/abs/2409.10195}{{\ttfamily
  arXiv:2409.10195 [hep-ph]}}.

\bibitem{Nomura:2024ctl}
T.~Nomura and H.~Okada, \emph{{Lepton seesaw model in a modular $A_4$
  symmetry}}  (2024),
  \MYhref[eprintLinks]{http://arxiv.org/abs/2409.10912}{{\ttfamily
  arXiv:2409.10912 [hep-ph]}}.

\bibitem{Nomura:2025bph}
T.~Nomura and H.~Okada, \emph{{A new type of lepton seesaw model in a modular
  $A_4$ symmetry}}  (2025),
  \MYhref[eprintLinks]{http://arxiv.org/abs/2503.19251}{{\ttfamily
  arXiv:2503.19251 [hep-ph]}}.

\bibitem{Petcov:2024vph}
S.~T. Petcov and M.~Tanimoto, \emph{{$A_4$ modular invariance and the strong CP
  problem}},
  \MYhref[journalLinks]{http://dx.doi.org/10.1140/epjc/s10052-024-13272-w}{Eur.
  Phys. J. C
  }\MYhref[journalLinks]{http://dx.doi.org/10.1140/epjc/s10052-024-13272-w}{\textbf{84}
  (2024) 9 914},
  \MYhref[eprintLinks]{http://arxiv.org/abs/2404.00858}{{\ttfamily
  arXiv:2404.00858 [hep-ph]}}.

\bibitem{Pathak:2024sei}
G.~Pathak, P.~Das and M.~K. Das, \emph{{Neutrino mass genesis in scoto-inverse
  seesaw with modular $A_4$}},
  \MYhref[journalLinks]{http://dx.doi.org/10.1140/epjc/s10052-025-14263-1}{Eur.
  Phys. J. C
  }\MYhref[journalLinks]{http://dx.doi.org/10.1140/epjc/s10052-025-14263-1}{\textbf{85}
  (2025) 5 569},
  \MYhref[eprintLinks]{http://arxiv.org/abs/2411.13895}{{\ttfamily
  arXiv:2411.13895 [hep-ph]}}.

\bibitem{Moreno-Sanchez:2025bzz}
A.~Moreno-S\'anchez and A.~Palavri\'c, \emph{{Leptonic Flavor from Modular
  $A_4$: UV Mediators and SMEFT Realizations}}  (2025),
  \MYhref[eprintLinks]{http://arxiv.org/abs/2505.01535}{{\ttfamily
  arXiv:2505.01535 [hep-ph]}}.

\bibitem{Pathak:2025zdp}
G.~Pathak and M.~K. Das, \emph{{Matter-antimatter asymmetry in minimal inverse
  seesaw framework with $A_4$ modular symmetry}}  (2025),
  \MYhref[eprintLinks]{http://arxiv.org/abs/2505.03000}{{\ttfamily
  arXiv:2505.03000 [hep-ph]}}.

\bibitem{Qu:2024rns}
B.-Y. Qu and G.-J. Ding, \emph{{Non-holomorphic modular flavor symmetry}},
  \MYhref[journalLinks]{http://dx.doi.org/10.1007/JHEP08(2024)136}{JHEP
  }\MYhref[journalLinks]{http://dx.doi.org/10.1007/JHEP08(2024)136}{\textbf{08}
  (2024) 136}, \MYhref[eprintLinks]{http://arxiv.org/abs/2406.02527}{{\ttfamily
  arXiv:2406.02527 [hep-ph]}}.

\bibitem{Nomura:2024vzw}
T.~Nomura, H.~Okada and O.~Popov, \emph{{Non-holomorphic modular $A_4$
  symmetric scotogenic model}},
  \MYhref[journalLinks]{http://dx.doi.org/10.1016/j.physletb.2024.139171}{Phys.
  Lett. B
  }\MYhref[journalLinks]{http://dx.doi.org/10.1016/j.physletb.2024.139171}{\textbf{860}
  (2025) 139171},
  \MYhref[eprintLinks]{http://arxiv.org/abs/2409.12547}{{\ttfamily
  arXiv:2409.12547 [hep-ph]}}.

\bibitem{Nomura:2024atp}
T.~Nomura and H.~Okada, \emph{{Type-II seesaw of a non-holomorphic modular
  $A_4$ symmetry}}  (2024),
  \MYhref[eprintLinks]{http://arxiv.org/abs/2408.01143}{{\ttfamily
  arXiv:2408.01143 [hep-ph]}}.

\bibitem{Nomura:2024nwh}
T.~Nomura and H.~Okada, \emph{{Zee model in a non-holomorphic modular $A_4$
  symmetry}},
  \MYhref[journalLinks]{http://dx.doi.org/10.1016/j.physletb.2025.139618}{Phys.
  Lett. B
  }\MYhref[journalLinks]{http://dx.doi.org/10.1016/j.physletb.2025.139618}{\textbf{867}
  (2025) 139618},
  \MYhref[eprintLinks]{http://arxiv.org/abs/2412.18095}{{\ttfamily
  arXiv:2412.18095 [hep-ph]}}.

\bibitem{Kobayashi:2025hnc}
T.~Kobayashi, H.~Okada and Y.~Orikasa, \emph{{Zee-Babu model in a
  non-holomorphic modular $A_4$ symmetry and modular stabilization}}  (2025),
  \MYhref[eprintLinks]{http://arxiv.org/abs/2502.12662}{{\ttfamily
  arXiv:2502.12662 [hep-ph]}}.

\bibitem{Nomura:2025ovm}
T.~Nomura, H.~Okada and X.-Y. Wang, \emph{{A radiative neutrino mass model with
  leptoquarks under non-holomorphic modular $A_4$ symmetry}}  (2025),
  \MYhref[eprintLinks]{http://arxiv.org/abs/2504.21404}{{\ttfamily
  arXiv:2504.21404 [hep-ph]}}.

\bibitem{Loualidi:2025tgw}
M.~A. Loualidi, M.~Miskaoui and S.~Nasri, \emph{{Non-holomorphic $A_4$ modular
  invariance for fermion masses and mixing in SU(5) GUT}}  (2025),
  \MYhref[eprintLinks]{http://arxiv.org/abs/2503.12594}{{\ttfamily
  arXiv:2503.12594 [hep-ph]}}.

\bibitem{Nomura:2025raf}
T.~Nomura and H.~Okada, \emph{{Neutrino mass model at a three-loop level from a
  non-holomorphic modular $A_4$ symmetry}}  (2025),
  \MYhref[eprintLinks]{http://arxiv.org/abs/2506.02639}{{\ttfamily
  arXiv:2506.02639 [hep-ph]}}.

\bibitem{Okada:2025jjo}
H.~Okada and Y.~Orikasa, \emph{{A radiative seesaw in a non-holomorphic modular
  $S_3$ flavor symmetry}}  (2025),
  \MYhref[eprintLinks]{http://arxiv.org/abs/2501.15748}{{\ttfamily
  arXiv:2501.15748 [hep-ph]}}.

\bibitem{Ding:2024inn}
G.-J. Ding, J.-N. Lu, S.~T. Petcov and B.-Y. Qu, \emph{{Non-holomorphic modular
  S$_{4}$ lepton flavour models}},
  \MYhref[journalLinks]{http://dx.doi.org/10.1007/JHEP01(2025)191}{JHEP
  }\MYhref[journalLinks]{http://dx.doi.org/10.1007/JHEP01(2025)191}{\textbf{01}
  (2025) 191}, \MYhref[eprintLinks]{http://arxiv.org/abs/2408.15988}{{\ttfamily
  arXiv:2408.15988 [hep-ph]}}.

\bibitem{Li:2024svh}
C.-C. Li, J.-N. Lu and G.-J. Ding, \emph{{Non-holomorphic modular A$_{5}$
  symmetry for lepton masses and mixing}},
  \MYhref[journalLinks]{http://dx.doi.org/10.1007/JHEP12(2024)189}{JHEP
  }\MYhref[journalLinks]{http://dx.doi.org/10.1007/JHEP12(2024)189}{\textbf{12}
  (2024) 189}, \MYhref[eprintLinks]{http://arxiv.org/abs/2410.24103}{{\ttfamily
  arXiv:2410.24103 [hep-ph]}}.

\bibitem{Ding:2020zxw}
G.-J. Ding, F.~Feruglio and X.-G. Liu, \emph{{Automorphic Forms and Fermion
  Masses}},
  \MYhref[journalLinks]{http://dx.doi.org/10.1007/JHEP01(2021)037}{JHEP
  }\MYhref[journalLinks]{http://dx.doi.org/10.1007/JHEP01(2021)037}{\textbf{01}
  (2021) 037}, \MYhref[eprintLinks]{http://arxiv.org/abs/2010.07952}{{\ttfamily
  arXiv:2010.07952 [hep-th]}}.

\bibitem{Kumar:2025bfe}
B.~Kumar and M.~K. Das, \emph{{Leptogenesis, $0\nu\beta\beta$ and lepton flavor
  violation in modular left-right asymmetric model with polyharmonic $Maa\beta$
  forms}}  (2025),
  \MYhref[eprintLinks]{http://arxiv.org/abs/2504.21701}{{\ttfamily
  arXiv:2504.21701 [hep-ph]}}.

\bibitem{Wang:2024qhe}
Z.~Wang, Y.~Reyimuaji and N.~Yalikun, \emph{{A $Z_4$ symmetric inverse seesaw
  model for neutrino masses and FIMP dark matter}}  (2024),
  \MYhref[eprintLinks]{http://arxiv.org/abs/2412.15672}{{\ttfamily
  arXiv:2412.15672 [hep-ph]}}.

\bibitem{Novichkov:2019sqv}
P.~P. Novichkov, J.~T. Penedo, S.~T. Petcov and A.~V. Titov, \emph{{Generalised
  CP Symmetry in Modular-Invariant Models of Flavour}},
  \MYhref[journalLinks]{http://dx.doi.org/10.1007/JHEP07(2019)165}{JHEP
  }\MYhref[journalLinks]{http://dx.doi.org/10.1007/JHEP07(2019)165}{\textbf{07}
  (2019) 165}, \MYhref[eprintLinks]{http://arxiv.org/abs/1905.11970}{{\ttfamily
  arXiv:1905.11970 [hep-ph]}}.

\bibitem{Esteban:2024eli}
I.~Esteban et~al., \emph{{NuFit-6.0: updated global analysis of three-flavor
  neutrino oscillations}},
  \MYhref[journalLinks]{http://dx.doi.org/10.1007/JHEP12(2024)216}{JHEP
  }\MYhref[journalLinks]{http://dx.doi.org/10.1007/JHEP12(2024)216}{\textbf{12}
  (2024) 216}, \MYhref[eprintLinks]{http://arxiv.org/abs/2410.05380}{{\ttfamily
  arXiv:2410.05380 [hep-ph]}}.

\bibitem{Xing:2007fb}
Z.-z. Xing, H.~Zhang and S.~Zhou, \emph{{Updated Values of Running Quark and
  Lepton Masses}},
  \MYhref[journalLinks]{http://dx.doi.org/10.1103/PhysRevD.77.113016}{Phys.
  Rev. D
  }\MYhref[journalLinks]{http://dx.doi.org/10.1103/PhysRevD.77.113016}{\textbf{77}
  (2008) 113016},
  \MYhref[eprintLinks]{http://arxiv.org/abs/0712.1419}{{\ttfamily
  arXiv:0712.1419 [hep-ph]}}.

\bibitem{FlavorPy}
A.~Baur, \emph{{FlavorPy}} (2024),
  \urlprefix\url{https://doi.org/10.5281/zenodo.11060597}.

\bibitem{Planck:2018vyg}
N.~Aghanim et~al. (Planck), \emph{{Planck 2018 results. VI. Cosmological
  parameters}},
  \MYhref[journalLinks]{http://dx.doi.org/10.1051/0004-6361/201833910}{Astron.
  Astrophys.
  }\MYhref[journalLinks]{http://dx.doi.org/10.1051/0004-6361/201833910}{\textbf{641}
  (2020) A6}, [Erratum: Astron.Astrophys. 652, C4 (2021)],
  \MYhref[eprintLinks]{http://arxiv.org/abs/1807.06209}{{\ttfamily
  arXiv:1807.06209 [astro-ph.CO]}}.

\bibitem{Katrin:2024tvg}
M.~Aker et~al. (Katrin), \emph{{Direct neutrino-mass measurement based on 259
  days of KATRIN data}}  (2024),
  \MYhref[eprintLinks]{http://arxiv.org/abs/2406.13516}{{\ttfamily
  arXiv:2406.13516 [nucl-ex]}}.

\bibitem{KamLAND-Zen:2024eml}
S.~Abe et~al. (KamLAND-Zen), \emph{{Search for Majorana Neutrinos with the
  Complete KamLAND-Zen Dataset}}  (2024),
  \MYhref[eprintLinks]{http://arxiv.org/abs/2406.11438}{{\ttfamily
  arXiv:2406.11438 [hep-ex]}}.

\bibitem{LEGEND:2021bnm}
N.~Abgrall et~al. (LEGEND), \emph{{The Large Enriched Germanium Experiment for
  Neutrinoless $\beta\beta$ Decay}: {LEGEND-1000 Preconceptual Design Report}}
  (2021), \MYhref[eprintLinks]{http://arxiv.org/abs/2107.11462}{{\ttfamily
  arXiv:2107.11462 [physics.ins-det]}}.

\bibitem{nEXO:2021ujk}
G.~Adhikari et~al. (nEXO), \emph{{nEXO: neutrinoless double beta decay search
  beyond 10$^{28}$ year half-life sensitivity}},
  \MYhref[journalLinks]{http://dx.doi.org/10.1088/1361-6471/ac3631}{J. Phys. G
  }\MYhref[journalLinks]{http://dx.doi.org/10.1088/1361-6471/ac3631}{\textbf{49}
  (2022) 1 015104},
  \MYhref[eprintLinks]{http://arxiv.org/abs/2106.16243}{{\ttfamily
  arXiv:2106.16243 [nucl-ex]}}.

\bibitem{GAMBITCosmologyWorkgroup:2020rmf}
P.~St\"ocker et~al. (GAMBIT Cosmology Workgroup), \emph{{Strengthening the
  bound on the mass of the lightest neutrino with terrestrial and cosmological
  experiments}},
  \MYhref[journalLinks]{http://dx.doi.org/10.1103/PhysRevD.103.123508}{Phys.
  Rev. D
  }\MYhref[journalLinks]{http://dx.doi.org/10.1103/PhysRevD.103.123508}{\textbf{103}
  (2021) 12 123508},
  \MYhref[eprintLinks]{http://arxiv.org/abs/2009.03287}{{\ttfamily
  arXiv:2009.03287 [astro-ph.CO]}}.

\bibitem{DUNE:2020ypp}
B.~Abi et~al. (DUNE), \emph{{Deep Underground Neutrino Experiment (DUNE), Far
  Detector Technical Design Report, Volume II: DUNE Physics}}  (2020),
  \MYhref[eprintLinks]{http://arxiv.org/abs/2002.03005}{{\ttfamily
  arXiv:2002.03005 [hep-ex]}}.

\bibitem{Hyper-Kamiokande:2018ofw}
K.~Abe et~al. (Hyper-Kamiokande), \emph{{Hyper-Kamiokande Design Report}}
  (2018), \MYhref[eprintLinks]{http://arxiv.org/abs/1805.04163}{{\ttfamily
  arXiv:1805.04163 [physics.ins-det]}}.

\bibitem{KATRIN:2022ayy}
M.~Aker et~al. (KATRIN), \emph{{KATRIN: status and prospects for the neutrino
  mass and beyond}},
  \MYhref[journalLinks]{http://dx.doi.org/10.1088/1361-6471/ac834e}{J. Phys. G
  }\MYhref[journalLinks]{http://dx.doi.org/10.1088/1361-6471/ac834e}{\textbf{49}
  (2022) 10 100501},
  \MYhref[eprintLinks]{http://arxiv.org/abs/2203.08059}{{\ttfamily
  arXiv:2203.08059 [nucl-ex]}}.

\bibitem{Project8:2022wqh}
A.~A. Esfahani et~al. (Project 8), \emph{{The Project 8 Neutrino Mass
  Experiment}}, in \emph{{Snowmass 2021}} (2022)
  \MYhref[eprintLinks]{http://arxiv.org/abs/2203.07349}{{\ttfamily
  arXiv:2203.07349 [nucl-ex]}}.

\bibitem{deAdelhartToorop:2011re}
R.~de~Adelhart~Toorop, F.~Feruglio and C.~Hagedorn, \emph{{Finite Modular
  Groups and Lepton Mixing}},
  \MYhref[journalLinks]{http://dx.doi.org/10.1016/j.nuclphysb.2012.01.017}{Nucl.
  Phys. B
  }\MYhref[journalLinks]{http://dx.doi.org/10.1016/j.nuclphysb.2012.01.017}{\textbf{858}
  (2012) 437--467},
  \MYhref[eprintLinks]{http://arxiv.org/abs/1112.1340}{{\ttfamily
  arXiv:1112.1340 [hep-ph]}}.

\bibitem{2008The}
J.~H. Bruinier, G.~V.~D. Geer, G.~Harder and D.~Zagier, \emph{The 1-2-3 of
  modular forms}, Universitext  (2008) 3.

\bibitem{Qu:2025ddz}
B.-Y. Qu, J.-N. Lu and G.-J. Ding, \emph{{Non-holomorphic modular flavor
  symmetry and odd weight polyharmonic Maa{\ss} form}}  (2025),
  \MYhref[eprintLinks]{http://arxiv.org/abs/2506.19822}{{\ttfamily
  arXiv:2506.19822 [hep-ph]}}.

\bibitem{Ishiguro:2020tmo}
K.~Ishiguro, T.~Kobayashi and H.~Otsuka, \emph{{Landscape of Modular Symmetric
  Flavor Models}},
  \MYhref[journalLinks]{http://dx.doi.org/10.1007/JHEP03(2021)161}{JHEP
  }\MYhref[journalLinks]{http://dx.doi.org/10.1007/JHEP03(2021)161}{\textbf{03}
  (2021) 161}, \MYhref[eprintLinks]{http://arxiv.org/abs/2011.09154}{{\ttfamily
  arXiv:2011.09154 [hep-ph]}}.

\bibitem{Gonzalo:2018guu}
E.~Gonzalo, L.~E. Ib{\'a}{\~n}ez and {\'A}.~M. Uranga, \emph{{Modular
  symmetries and the swampland conjectures}},
  \MYhref[journalLinks]{http://dx.doi.org/10.1007/JHEP05(2019)105}{JHEP
  }\MYhref[journalLinks]{http://dx.doi.org/10.1007/JHEP05(2019)105}{\textbf{05}
  (2019) 105}, \MYhref[eprintLinks]{http://arxiv.org/abs/1812.06520}{{\ttfamily
  arXiv:1812.06520 [hep-th]}}.

\bibitem{Novichkov:2022wvg}
P.~P. Novichkov, J.~T. Penedo and S.~T. Petcov, \emph{{Modular flavour
  symmetries and modulus stabilisation}},
  \MYhref[journalLinks]{http://dx.doi.org/10.1007/JHEP03(2022)149}{JHEP
  }\MYhref[journalLinks]{http://dx.doi.org/10.1007/JHEP03(2022)149}{\textbf{03}
  (2022) 149}, \MYhref[eprintLinks]{http://arxiv.org/abs/2201.02020}{{\ttfamily
  arXiv:2201.02020 [hep-ph]}}.

\bibitem{Kachru:2003aw}
S.~Kachru, R.~Kallosh, A.~D. Linde and S.~P. Trivedi, \emph{{De Sitter vacua in
  string theory}},
  \MYhref[journalLinks]{http://dx.doi.org/10.1103/PhysRevD.68.046005}{Phys.
  Rev. D
  }\MYhref[journalLinks]{http://dx.doi.org/10.1103/PhysRevD.68.046005}{\textbf{68}
  (2003) 046005},
  \MYhref[eprintLinks]{http://arxiv.org/abs/hep-th/0301240}{{\ttfamily
  arXiv:hep-th/0301240}}.

\bibitem{Knapp-Perez:2023nty}
V.~Knapp-Perez et~al., \emph{{Matter matters in moduli fixing and modular
  flavor symmetries}},
  \MYhref[journalLinks]{http://dx.doi.org/10.1016/j.physletb.2023.138106}{Phys.
  Lett. B
  }\MYhref[journalLinks]{http://dx.doi.org/10.1016/j.physletb.2023.138106}{\textbf{844}
  (2023) 138106},
  \MYhref[eprintLinks]{http://arxiv.org/abs/2304.14437}{{\ttfamily
  arXiv:2304.14437 [hep-th]}}.

\bibitem{Leedom:2022zdm}
J.~M. Leedom, N.~Righi and A.~Westphal, \emph{{Heterotic de Sitter beyond
  modular symmetry}},
  \MYhref[journalLinks]{http://dx.doi.org/10.1007/JHEP02(2023)209}{JHEP
  }\MYhref[journalLinks]{http://dx.doi.org/10.1007/JHEP02(2023)209}{\textbf{02}
  (2023) 209}, \MYhref[eprintLinks]{http://arxiv.org/abs/2212.03876}{{\ttfamily
  arXiv:2212.03876 [hep-th]}}.

\bibitem{King:2023snq}
S.~F. King and X.~Wang, \emph{{Modulus stabilization in the multiple-modulus
  framework}},
  \MYhref[journalLinks]{http://dx.doi.org/10.1103/PhysRevD.110.076026}{Phys.
  Rev. D
  }\MYhref[journalLinks]{http://dx.doi.org/10.1103/PhysRevD.110.076026}{\textbf{110}
  (2024) 7 076026},
  \MYhref[eprintLinks]{http://arxiv.org/abs/2310.10369}{{\ttfamily
  arXiv:2310.10369 [hep-ph]}}.

\end{thebibliography}

\end{document}